\documentclass[useAMS,usenatbib]{mn2e}

\def \isc{\textsc{i}}
\def \iscs{\textsc{i} }
\def \ii{\textsc{ii}}
\def \iis{\textsc{ii} }

\def \iiis{\textsc{iii} }

\newcommand{\lyxdot}{.}

\usepackage{ifpdf}
\usepackage{graphicx}
\usepackage{amsmath}
\usepackage{amsfonts}
\usepackage{amssymb}
\usepackage{times}
\usepackage{array}
\usepackage{supertabular}
\usepackage{url} 
\usepackage[figuresright]{rotating}
\usepackage{array}
\usepackage{caption} 
\usepackage{afterpage}
\usepackage{fixltx2e} 

\newcolumntype{:}{>{\global\let\currentrowstyle\relax}}
\newcolumntype{;}{>{\currentrowstyle}}
\newcommand{\rowstyle}[1]{\gdef\currentrowstyle{#1}%
  #1\ignorespaces
}

\title[Spatial variation in the fine-structure constant]{Spatial variation in the fine-structure constant -- new results from VLT/UVES}
\author[J. A. King et al.]{Julian A. King$^{1}$\thanks{E-mail:
jking.phys@gmail.com}, John K. Webb$^{1}$\thanks{E-mail: jkw@phys.unsw.edu.au}, Michael T. Murphy$^{2}$\thanks{E-mail: mmurphy@swin.edu.au}, Victor V. Flambaum$^{1}$, \newauthor Robert F. Carswell$^{3}$, Matthew B. Bainbridge$^{1}$, Michael R. Wilczynska$^{1}$\newauthor and F. Elliot Koch$^{1}$.\\
$^{1}$School of Physics, University of New South Wales, Sydney, NSW, 2052, Australia\\
$^{2}$Centre for Astrophysics and Supercomputing, Swinburne University of Technology, Victoria, 3122, Australia\\
$^{3}$Institute of Astronomy, University of Cambridge CB3 0HA, UK}
\begin{document}

\date{19 February 2012}

\pagerange{\pageref{firstpage}--\pageref{lastpage}} \pubyear{2012}

\maketitle

\label{firstpage}

\begin{abstract} Quasar absorption lines provide a precise test of whether the fine-structure constant, $\alpha$, is the same in different places and through cosmological time. We present a new analysis of a large sample of quasar absorption-line spectra obtained using UVES (the Ultraviolet and Visual Echelle Spectrograph) on the VLT (Very Large Telescope) in Chile. We apply the many-multiplet method to derive values of $\Delta\alpha/\alpha \equiv (\alpha_z-\alpha_0)/\alpha_0$ from 154 absorbers, and combine these values with 141 values from previous observations at the Keck Observatory in Hawaii. In the VLT sample, we find evidence that $\alpha$ increases with increasing cosmological distance from Earth. However, as previously shown, the Keck sample provided evidence for a smaller $\alpha$ in the distant absorption clouds. Upon combining the samples an apparent variation of $\alpha$ across the sky emerges which is well represented by an angular dipole model pointing in the direction $\mathrm{RA}=(17.3\pm1.0)\,\mathrm{hr}$, $\mathrm{dec.}=(-61\pm 10)\deg$, with amplitude $0.97^{+0.22}_{-0.20}\times 10^{-5}$. The dipole model is required at the 4.1$\sigma$ statistical significance level over a simple monopole model where $\alpha$ is the same across the sky (but possibly different to the current laboratory value). The data sets reveal remarkable consistencies: \emph{i)} the directions of dipoles fitted to the VLT and Keck samples separately agree; \emph{ii)} the directions of dipoles fitted to $z<1.6$ and $z>1.6$ cuts of the combined VLT+Keck samples agree; \emph{iii)} in the equatorial region of the dipole, where both the Keck and VLT samples contribute a significant number of absorbers, there is no evidence for inconsistency between Keck and VLT.  The amplitude of the dipole is clearly larger at higher redshift. Assuming a dipole-only (i.e.\ no-monopole) model whose amplitude grows proportionally with `lookback-time distance' ($r=ct$, where $t$ is the lookback time), the amplitude is $(1.1\pm0.2)\times 10^{-6}\,\mathrm{GLyr}^{-1}$ and the model is significant at the $4.2\sigma$ confidence level over the null model ($\Delta\alpha/\alpha \equiv 0$). We apply robustness checks and demonstrate that the dipole effect does not originate from a small subset of the absorbers or spectra. We present an analysis of systematic effects, and are unable to identify any single systematic effect which can emulate the observed variation in $\alpha$. To the best of our knowledge, this result is not in conflict with any other observational or experimental result.  \end{abstract}


\section{Introduction}

Quasar absorption lines, generated by the absorption of light along the line of sight to quasars by intervening gas clouds, provide a method through which variation in certain fundamental dimensionless constants can be constrained. For high enough absorption redshifts ($z$), certain UV metal transitions are redshifted into the optical and therefore are observable with ground-based telescopes. The wavelengths of these transitions depend to various degrees on the fine-structure constant, $\alpha \equiv e^2/(4 \pi \epsilon_0 \hbar c)$. By comparing observations of these transitions with precision laboratory measurements, one can determine if the fine-structure constant is the same in different places in the universe.

In this section, we describe the many-multiplet method and discuss its previous application. In section \ref{s_data}, we discuss the origin and treatment of the spectral data used. We describe our analysis methodology in section \ref{s_methodology}. We present our new VLT results in section \ref{s_VLT_results}. We combine the VLT results with previous Keck results in section \ref{s_results_combine}, and explore different models for the observed values of $\Delta\alpha/\alpha$. We discuss potential systematic errors in section \ref{s_systematic_errors}.

\subsection{The many-multiplet method (MM method)}

Fine structure splitting of atomic transitions is observed as a consequence of $\alpha$ being non-zero. For alkali-doublet (AD) type transitions, the separation in the wavelength of transitions is proportional to $\alpha^2$. More generally, one can derive the observed change in the fine-structure constant, 
\begin{equation}
\Delta\alpha/\alpha \equiv (\alpha_z - \alpha_0)/\alpha_0 
\end{equation}
where $\alpha_z$ is the value measured at absorption redshift $z$ and $\alpha_0$ is the present value of $\alpha$. For a given transition, the observed wavenumber, $\omega_z$, depends on $\alpha$ as 
\begin{equation}
\omega_z = \omega_0 + q x
\end{equation}
where
\begin{equation}
x = \left(\frac{\alpha_z}{\alpha_0}\right)^2-1
\end{equation}
provided that $\lvert \Delta\alpha/\alpha \ll 1 \rvert$. $q$ is the ``sensitivity coefficient'' which determines how sensitive a particular transition is to a change in $\alpha$ \citep{Dzuba:99:01,Murphy:03}. 

Comparing the wavelengths of AD absorption transitions to laboratory wavelength values for those transitions yields the AD method. The AD method does not, however, make full use of the available information in the spectrum, as it only compares transitions which arise from the same ground state of the same atomic species. By examining transitions from the ground state of different atomic species, one gains a statistical improvement (from the use of more transitions) and a sensitivity improvement (as transitions from different atomic species may display widely differing sensitivities to a change in $\alpha$). Comparing many transitions from different multiplets, possibly from different atoms/ions, yields the MM method \citep{Dzuba:99:01,Dzuba:99:02,Webb:99}, which typically yields an order of magnitude improvement in sensitivity over the AD method, as well as significant resistance to certain systematics.

The transitions used by the MM method have $q$ coefficients that are approximately zero (``anchor transitions''), positive (``positive shifters'') or negative (``negative shifters''). It is not the absolute values of the $q$ coefficients which are important, but instead the differences in $q$ between the transitions used; transitions with greater differences in $q$ provide increased sensitivity to detect a change in $\alpha$. The values of $q$ must be calculated using quantum many-body methods \citep{Dzuba:99:01,Dzuba:99:02}. 

\subsection{Previous results}

In the first application of the MM method, \citet{Webb:99} examined 30 absorbers using spectra from HIRES (High Resolution Echelle Spectrometer) on the Keck Telescope, in Hawaii. This yielded evidence that $\alpha$ may have been smaller in the absorption systems (compared to the current laboratory value) at the $3\sigma$ level using Fe\,\iis and Mg\,\isc/\iis transitions. The addition of more absorbers by 2001 \citep{Webb:01,Murphy:01b} increased the significance of the detection to $4\sigma$. In addition to the use of Fe\,\iis and Mg\,\isc/\iis transitions, these works made use of the Ni\,\ii/Cr\,\ii/Zn\,\iis transitions. The two different sets of transitions yielded consistent evidence for a change in $\alpha$. The sample size was increased to 128 absorbers by \citet{Murphy:03}, who found $>5\sigma$ evidence for a change in $\alpha$. By 2004, this was increased to 143 absorption systems with further data from the Keck telescope \citep{Murphy:04:LNP}. \citet{Murphy:04:LNP} give a weighted mean over all the systems of $\Delta\alpha/\alpha = (-0.57 \pm 0.11) \times 10^{-5}$ -- evidence that the $\alpha$ may have been smaller at high redshift at the $5\sigma$ level. However, as the Keck telescope is situated in the northern hemisphere (at $\sim 20^\circ$N), it is unable to observe much of the southern sky, and therefore $\Delta\alpha/\alpha$ is preferentially measured in one celestial hemisphere. Additionally, the use of only a single telescope and instrument is undesirable; one can postulate that the measured results are due to some instrumental systematic. 

Although it is possible to conjecture that the Keck results are due to some instrumental systematic, we note that there have been thorough analyses of many potential systematic effects \citep{Murphy:01c,Murphy:03}. At present, there is no evidence or analysis in the literature which identifies any specific systematic of sufficient magnitude and character to explain the Keck results. 

The only other statistical sample to search for a change in $\alpha$ using the MM method is the work of \citet{Chand:2004}. They analysed 23 absorption systems using spectra from UVES (the Ultraviolet and Visual Echelle Spectrograph) on the VLT (Very Large Telescope), in Chile. Although they reported a statistical error on $\Delta\alpha/\alpha$ of just $0.06\times 10^{-5}$, it has been shown that the errors were under-estimated by a factor of 6 and that the optimisation algorithm was demonstrably unreliable, failing to reach a solution in most of the absorbers \citep{Murphy:07-2,Murphy:08}. \citet{Srianand:07a} reply to \citet{Murphy:07-2}, and state that an updated analysis of the \citeauthor{Chand:2004} sample shows adequate performance of the optimisation algorithm, though no updated values of $\Delta\alpha/\alpha$ for individual absorbers are provided and so we are unable to comment further on the robustness or otherwise of that work. 

\subsection{Objective \& overview}

In section \ref{s_data}, we describe the data that form the basis of our analysis. In section \ref{s_methodology}, we describe our analysis methodology in detail, with a particular focus on improvements made to our application of the MM method since the results of \citet{Murphy:04:LNP}. In that section, we also discuss certain aspects of the MM method which we believe should be re-emphasised. We further introduce a modification of the Least Trimmed Squares method -- a robust statistical method which assists with the detection of outliers in a dataset. In section \ref{s_VLT_results} we present the results of our MM analysis, giving estimates of $\Delta\alpha/\alpha$ for our sample. We also give weighted mean and dipole fits to our VLT data. In section \ref{s_results_combine}, we combine our results with those of \citet{Murphy:04:LNP} to derive the main results of this work. In section \ref{s_systematic_errors}, we discuss potential systematic errors.

This paper analyses the same data set presented in \citet{Webb:10a}, however we provide substantially more detail here.

\section{Quasar spectra and laboratory wavelength transitions}\label{s_data}

We have used publically available spectra from VLT/UVES, in Chile (at $25^\circ$S), to generate a MM sample similar in size to the Keck sample. To do this, we have searched the ESO Science Archive for VLT/UVES observations which might be used to constrain $\Delta\alpha/\alpha$. Priority was given to spectra with high expected signal-to-noise, although obtaining a sufficient number of high redshift objects and maximising sky coverage was also a consideration. The ESO \textsc{midas} pipeline was used for the first stage of the data reduction, including wavelength calibration, although enhancements were made to derive a more robust and accurate wavelength solution from an improved selection of thorium-argon calibration lamp transitions \citep{Murphy:07b}. In the next stage, echelle spectral orders from several exposures for each quasar were combined using the program \textsc{uves\_popler}\footnote{Murphy, M.~T., \url{http://astronomy.swin.edu.au/~mmurphy/UVES_popler.html}}, which was specifically designed for this purpose. A separate paper will be presented which describes the data reduction and processing in considerable detail (Murphy et. al., in prep).

The signal-to-noise ratio (SNR) of the combined data is generally significantly better than the Keck data used by \citet{Murphy:04:LNP}. The SNR per 2.5 km\,s$^{-1}$ pixel in the VLT sample ranges from $\sim 8$ up to $\sim 190$ (see figure \ref{fig_SNR_VLT}). The higher SNR for the VLT sample allows better determination of the velocity structure required for a particular absorber, and reduces the statistical error on derived values of $\Delta\alpha/\alpha$. 

UVES has three CCD chips (1 in the blue arm and 2 in the red arm), which allows almost complete wavelength coverage ($\sim 3,000\mathrm{\AA}$ to $\sim 10,000\mathrm{\AA}$) to be obtained with a single setting for the echelle grating. A gap occurs at either $\sim4,500\mathrm{\AA}$ or $\sim5,500\mathrm{\AA}$ depending on which dichroic is used. The Keck spectra in the sample of \citet{Murphy:04:LNP} were taken when HIRES had only 1 chip, thereby necessitating multiple exposures to achieve good wavelength coverage. In many of the Keck spectra, wavelength coverage gaps are visible, whereas in the VLT spectra there are almost no spectral gaps. In principle, this allows the use of a greater number of absorbers per sightline. Similarly, this in principle allows the use of more transitions per absorber. 

\subsection{Extraction problems}\label{s_extractionproblems}

\begin{figure}
\ifpdf
\includegraphics[bb=132 54 556 722,angle=-90,width=82.5mm]{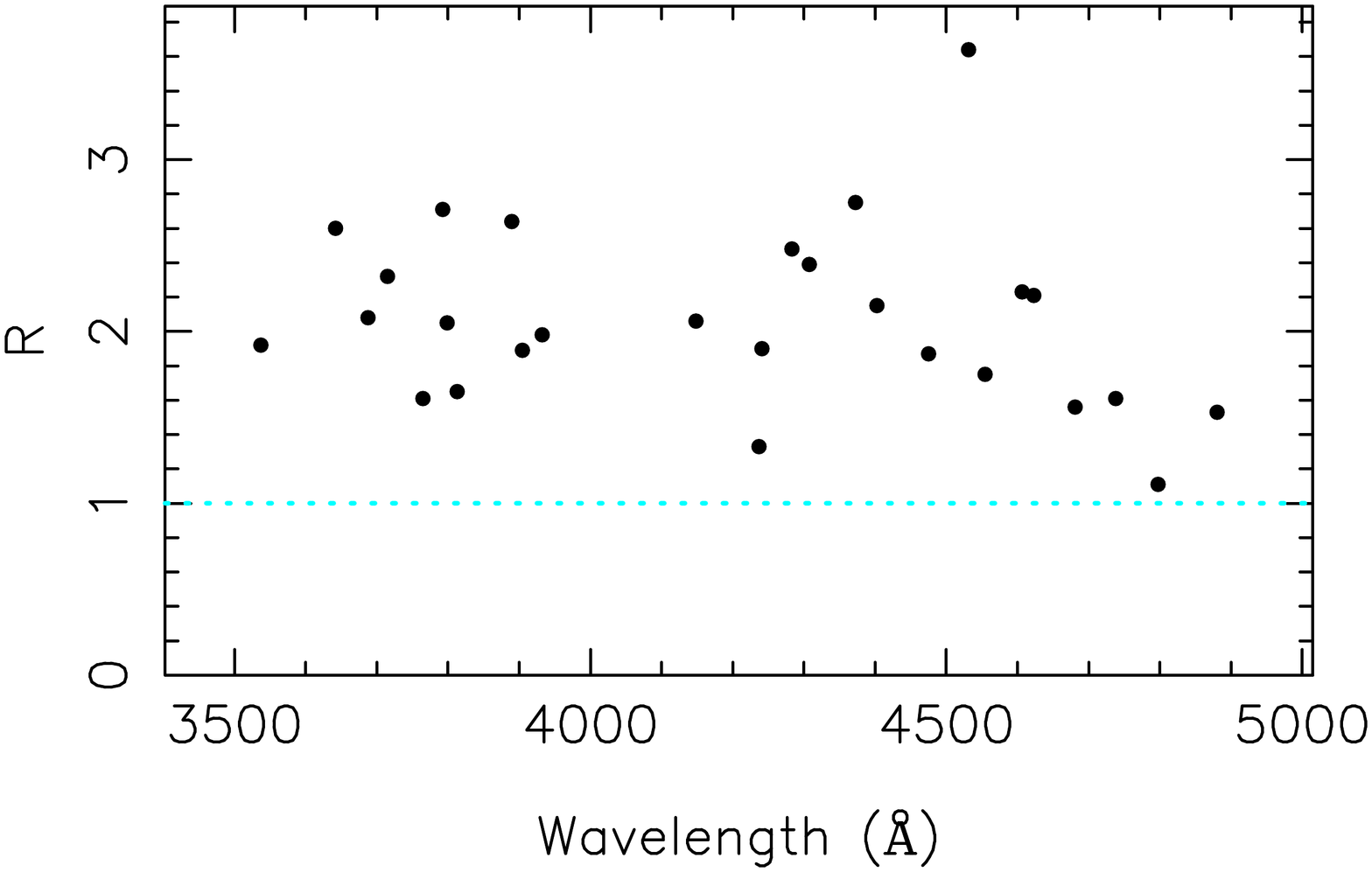}
\else
\includegraphics[bb=132 54 556 722,angle=-90,width=82.5mm]{images/Q0405_errorproblem.eps}
\fi
  \caption{Example of the under-estimation of flux uncertainties in the base of saturated lines for the spectrum of J040718$-$441013. The quantity $R$ plotted on the vertical axis is the ratio of the RMS of the normalised flux array to the average value of the RMS array in selected regions of apparently zero flux. The RMS array is a modified version of the error array produced by \textsc{uves\_popler} which attempts to account for inter-pixel correlations. The quantity $R$ should be approximately 1 if the RMS array is a good representation of the statistical uncertainty, but will be significantly greater than 1 if the errors are under-estimated. One can easily see from this plot that all the values are greater than $1$. Here, the mean value is $R=2.08$, with a standard error of $0.10$, demonstrating inconsistency with $R=1$ at the $>10\sigma$ level.\label{Q0405_errorproblem}}
\end{figure}

The \textsc{midas} pipeline appears to incorrectly estimate the errors associated with the flux data points in the base of saturated lines. In particular, the dispersion of the flux data points is too large to be accounted for by the statistical error. Fitting a straight line through the base of saturated lines typically produces $\chi^2_\nu \gtrsim 2$, where $\chi^2_\nu$ is $\chi^2$ measured with respect to the weighted mean flux and divided by the number of degrees of freedom, $\nu$. The problem is somewhat more noticeable in the blue end of the spectra. Although it is difficult to determine precisely what happens in regions of low, but non-zero flux, we believe that the errors there are also underestimated. The effect of this is to give falsely high precision on any quantity derived from these data points (including $\Delta\alpha/\alpha$). Additionally, one cannot fit plausible models to data involving regions of low or negligible flux. We demonstrate this effect in figure \ref{Q0405_errorproblem}, and give more details there.

When we combine individual exposures into a final spectrum using \textsc{uves\_popler}, \textsc{uves\_popler} provides a check on the consistency of the exposures contributing to each pixel. As the combined value of each flux pixel is given as a weighted mean of pixels from the contributing spectra, this consistency check is a value of $\chi^2_\nu$ for each pixel about the weighted mean. We attempt to correct for the problem described earlier, and any other problems which cause inconsistency between contributing spectra, by applying the following algorithm. For each pixel in the combined spectrum, we take a region of five pixels centred on that point. We then take the median of the $\chi^2_\nu$ values just described that are associated with those five points. We then multiply the error estimate for that spectral pixel by the square root of that median value (that is, $\sigma_i \rightarrow \sigma_i \times \sqrt{\mathrm{median}[\chi^2_\nu]}$). 

However, we only increase the error array if $\mathrm{median}(\chi^2_\nu) > 1$. This is because the spectral errors are generated from photon-counting. Poisson statistics requires that the error from photon-counting is a lower limit on the true error. Thus, it is difficult to justify decreasing the error array without particular evidence that the spectral uncertainty is systematically under-estimated. 

\subsection{Other data and fitting problems}

Inspection of the spectra which contribute to our exposures reveal problems which could prevent acceptable fits being achieved, and which might interfere with the determination of $\Delta\alpha/\alpha$. We discuss these here.

\emph{Cosmic rays.} When the raw spectra are combined, \textsc{uves\_popler} attempts to automatically remove cosmic rays through $\sigma$-clipping. That is, pixels from contributing spectra which are more than some number of standard deviations from the weighted mean (calculated excluding the pixel that is being tested) are discarded from the combination. The default rejection threshold is $3\sigma$, although this is user-adjustable. A second round of spectral cleaning is done manually. More details will be provided in Murphy et al. (in preparation). 

\emph{Sky absorption lines.} Sky absorption lines are seen with varying intensities depending on the orientation of the telescope. Transitions from quasar absorbers sometimes fall in regions affected by sky absorption. In principle, atmospheric absorption could be fitted simultaneously with the absorption caused by the quasar absorber, allowing the use of these transitions. However, our spectra have been converted to heliocentric wavelengths. As a result, the atmospheric transitions are shifted by velocities of up to $\pm \approx 30\,\mathrm{km\,s^{-1}}$. The co-addition of exposures taken at significantly different times of the year therefore creates a complex absorption pattern. Rather than attempt to model this, we simply do not use transitions which fall in regions of spectra which appear to be affected by sky absorption.

\emph{Sky emission lines.} The \textsc{midas} program should automatically remove sky emission as part of the pipeline. However, we have noticed that strong sky emission lines are not necessarily properly subtracted. This can be seen visually in the spectrum, with sharp dips and spikes at the location of strong sky emission lines. Transitions from the quasar absorbers are sometimes affected by this problem. Where we suspect that the spectral data are affected by sky emission lines, we either clip out the affected pixels (if they cover a small region of spectrum we want to use) or do not use the affected transition.

\emph{Gravitational lenses}. We discovered a small number of absorbers for which the velocity structure appeared to be the same for transtions arising from the same ground state, but where the line intensities differed substantially. A particular example of this is the $z=0.82$ absorber along the line of sight to J081331+254503. Further investigation demonstrated that this quasar is a known gravitational lens. The complex line-of-sight geometry causes the effect described. We discarded any system for which this problem appeared, and where we could identify the quasar as a known gravitationally lensed system.

\subsection{Atomic data}

In table \ref{tab:atomdata} we present the atomic data and $q$ coefficients used for our analysis. Our $q$-coefficients are from \citet{Dzuba:99:01,Dzuba:99:02} and \citet{Berengut:04a,Berengut:04b} and we provide all references for the other atomic data in the table caption.

\section{Analysis methodology}\label{s_methodology}

\subsection{Voigt profile fitting}

For each absorption system, we attempt to model all available MM transitions simultaneously (see appendix \ref{appendix_atomicdata} for a list of these transitions). We have used the non-linear least squares Voigt profile fitting program \textsc{vpfit}\footnote{\url{http://www.ast.cam.ac.uk/~rfc/vpfit.html}}, which was used to produce the Keck $\Delta\alpha/\alpha$ results. $\Delta\alpha/\alpha$ is included explicitly as a free parameter in the fit. Statistical uncertainties are determined from the diagonal terms of the covariance matrix at the best-fitting solution, but are multiplied by $\sqrt{\chi^2_\nu}$ to yield confidence intervals \citep{NumericalRecipes:92}. We have verified through Markov Chain Monte Carlo simulations that the statistical uncertainties produced by \textsc{vpfit} for $\Delta\alpha/\alpha$, $\sigma_{\Delta\alpha/\alpha}$, are robust \citep{King:09}. We emphasise that the statistical inference techniques used to measure $\Delta\alpha/\alpha$ from each spectrum are standard and well-established.

\subsection{Modelling the velocity structure}\label{s_modellingvelstructure}

\begin{figure}
\ifpdf
\includegraphics[bb=34 58 554 727,angle=-90,width=82.5mm]{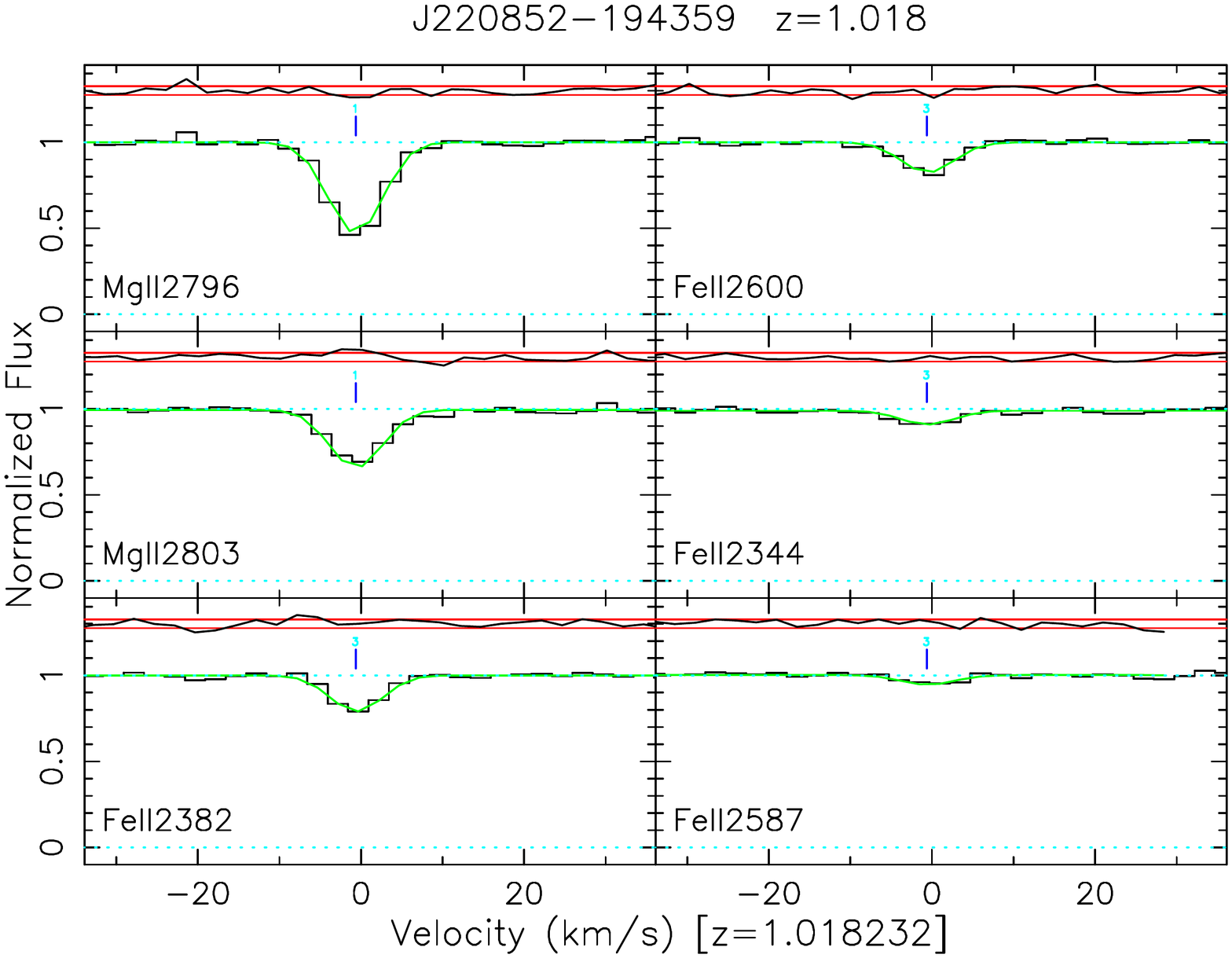}
\else
\includegraphics[bb=34 58 554 727,angle=-90,width=82.5mm]{images/J220852-194359-z1_018.eps}
\fi
  \caption{\label{J220852-194359-absorber}Our MM fit to the absorber at $z_\mathrm{abs}=1.018$ toward J220852$-$1934359, which is apparently well fitted by a single component. The horizontal scale indicates the velocity difference from the arbitrary redshift stated at the bottom for the given data points. The black line indicates the observed normalised flux, with the green line indicating our best fit solution. At the top of each box, the black line indicates the standardised residuals (that is, [data - model]/error), with the red lines indicating $\pm 1\sigma$. The position of the blue tick marks indicates the fitted position of the single component. }
\end{figure}

Some absorption complexes are well-modelled by a single Voigt profile, such as the one shown in figure \ref{J220852-194359-absorber}. However, it is clear that almost all absorption profiles are not well modelled by a single Voigt profile. However, it is possible to obtain statistically acceptable fits by using multiple Voigt profiles (``velocity components''). Unfortunately, there is no way of knowing \emph{a priori} how many components must be used to obtain a statistically acceptable fit. Thus, the process of modelling the absorption profile amounts to adding components until a physically realistic, statistically acceptable fit is reached. 

Although we tie corresponding velocity components for species with similar ionisation potentials in our modelling procedure, this does not equate to the assumption that our results rely inherently on an assumption of identical velocity structure amongst all species fitted. In fact the assumptions we make in deriving the results presented in this paper are: (a) that the velocity structures for species with similar ionisation potentials are \emph{sufficiently similar} such that it is statistically valid to model multiple species simultaneously with one velocity structure (that is, any kinematic segregation which may be present is small relative to the observed linewidths), \emph{and} (b) that any kinematic segregation between different species is random in nature \emph{between different absorbers}
i.e. the sign of any velocity shift which may be present as a result of kinematic segregation is \emph{random} between any pair of species in different absorbers). Assumption (b) is justified because we are viewing absorption systems, where the quasar light passes through the absorption cloud along the line of sight, rather than optically thick emission systems, where we might receive light preferentially from specific components of the emitting region. If assumption (b) were found to be violated this would mean that kinematic segregation in different absorbers would be spatially correlated over cosmological scales, implying that the atoms in different absorbers must ``know'' about the presence and location of the atoms in the other absorbers, and that there is coherent organisation of the different atoms over cosmological distances. We do not consider such an arrangement to be plausible.

We have three criteria for a statistically acceptable fit.

\emph{i) Normalised $\chi^2$ of order unity}. As a basic statistical test, we require that $\chi^2_\nu \sim 1$. This follows directly from the fact that the $\chi^2$ distribution with $\nu$ degrees of freedom has mean $\nu$. However, this criterion is \emph{not} the only criterion that must be used. Adding components until $\chi^2_\nu \lesssim 1$ only suggests that the dispersion of the spectral data about the model is what one would expect for a good model. 

\citet{Murphy:08} demonstrate through simulations that, at least for one synthetic spectrum considered, ``underfitting'' of spectra may lead to significant differences between the input and output $\Delta\alpha/\alpha$ values, whereas ``overfitting'' seems not to induce a significant difference. We are therefore particularly cautious about underfitting spectra. 

A model with $\chi^2_\nu \lesssim 1$ may be a reasonable model, but this test cannot determine whether it is the optimal model. Under the maximum likelihood method, if one can find a model which better explains the data, then this model should be preferred. The addition of components will always improve $\chi^2$, and therefore one considers whether the improvement is significant. A rigorous way to proceed is to to perform a statistical signifiance test on every component added, and only accept components which are statistically significant. However, this is laborious. Instead, other heuristics are available which tend to lead to reasonable choices of models. One of these is to try to find the model which minimises $\chi^2_\nu$. If one adds a component and $\chi^2_\nu$ increases, this suggests that the extra component is not supported by the data.

Other methods are available, which penalise free parameters more or less strongly. We have chosen to use the Akaike Information Criterion (AIC) \citep{Akaike:74}, defined as $\mathrm{AIC} = \chi^2 + 2p$, where $p$ here is the number of free parameters. In fact, the AIC is only correct in the limit of large $n/p$ (where $n$ is the number of spectral points included in the fit). This is not generally true for our fits. Thus, we use the AIC corrected for finite sample sizes, defined as
\begin{equation}
 \mathrm{AICC} = \chi^2 + 2p + \frac{2p(p+1)}{n-p-1}
\end{equation}
\citep{Sigiura:1978}. A significant advantage of the AICC is that it allows multiple models to be compared simultaneously. If several competing models are being considered, one chooses the model which has the lowest AICC. The actual value of the AICC is not important; only relative differences matter. The suggested interpretation scale for the AICC is the Jeffreys' scale \citep{Jeffreys:1961}, where $\Delta \mathrm{AICC} = 5$ and $\Delta \mathrm{AICC}=10$ are considered strong and very strong evidence against the weaker model respectively. $\Delta \mathrm{AICC}<5$ means that the difference between the models is ``barely worth mentioning''. These considerations lead to our second and third criteria:

\emph{ii) Choose the model which has the lowest AICC}. In this way, we attempt to find a model which best explains the data. 

\emph{iii) No substantial correlations in the residuals}. For a well-fitting model, the residuals of the fit $r_i$ (defined as $r_i = [\mathrm{data} - \mathrm{model}]/\mathrm{error}$) should be random. This means that no significant correlations should occur in the residuals. Such correlations suggest a deficient model. Although statistically improbable runs of residuals can be detected through formal tests \citep{Wald:40a}, in practice this is not necessary. The addition of components in regions where such correlations are visible generally removes the effect, and decreases the AICC as well. We accept the addition of components which decreases the AICC. 

\begin{figure*}
\ifpdf 
\includegraphics[bb=34 58 554 727,width=145mm]{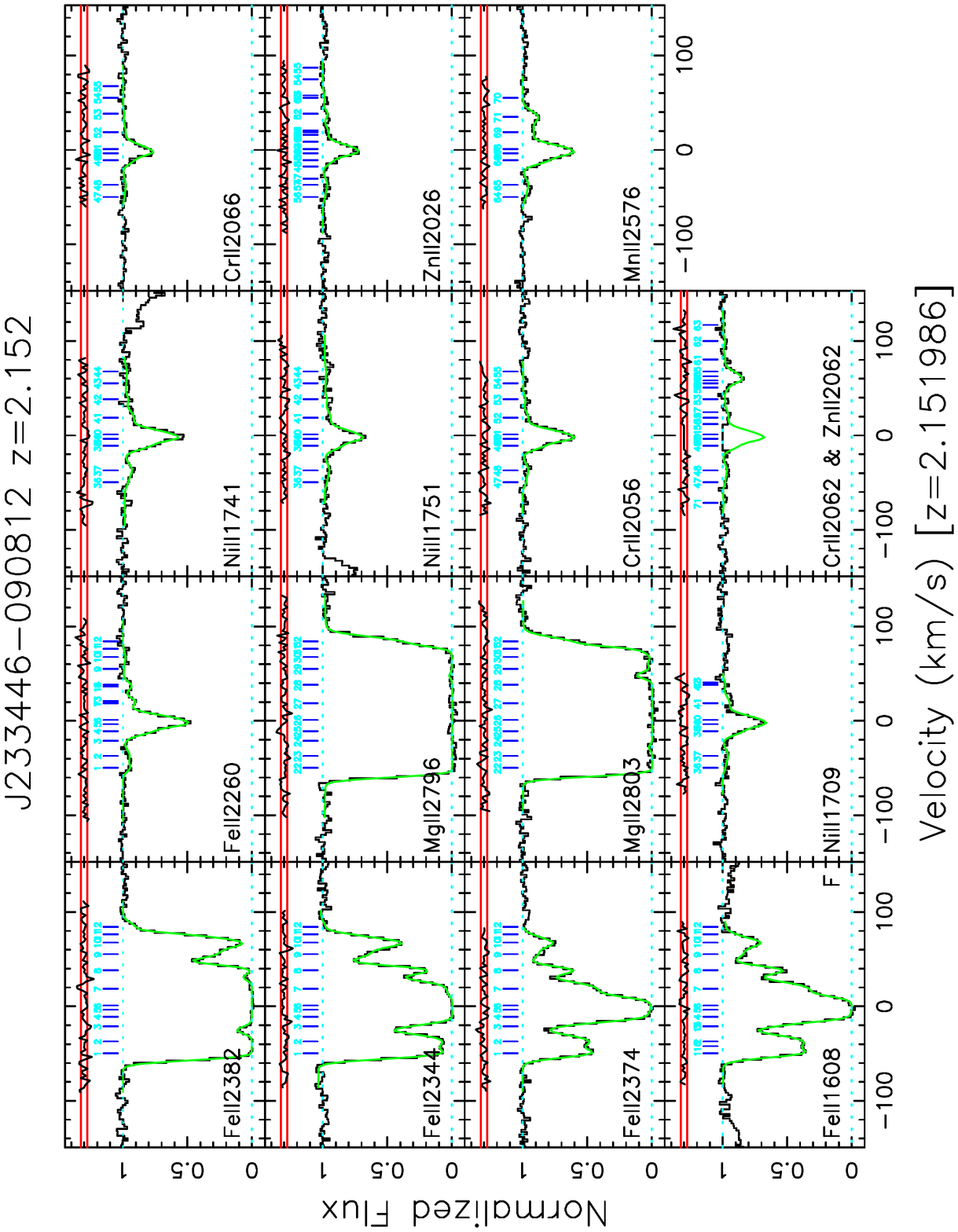}
\else
\includegraphics[bb=34 58 554 727,width=145mm]{images/J233446-090812-z2_152-noAlIII.eps}
\fi
  \caption{\label{J223446-090812-absorber}Part of our MM fit to the absorber at $z_\mathrm{abs}=2.152$ toward J233446$-$090812. This is a complex absorption system, requiring many components in order to achieve a statistically acceptable ($\chi^2_\nu \sim 1)$ fit. See figure \ref{J220852-194359-absorber} for a description of the various lines, marks and labels. We draw the reader's attention to the presence of a wide range of transitions, some with relatively small magnitude $q$ coefficients (Ni\,\textsc{ii} $\lambda 1709$ and the two Mg\,\textsc{ii} transitions), some with large magnitude, positive $q$ coefficients (Fe\,\iis $\lambda 2382,2344,2374,2260$, the Zn\,\iis transitions and the Mn\,\iis $\lambda2576$ transition) and some with large magnitude, negative $q$ coefficients (Fe\,\iis $\lambda 1608$, Ni\,\iis $\lambda 1741,1751$ and the Cr\,\iis transitions). The Al\,\iiis transitions fitted are not shown, but these have a minimal impact on the value of and precision for $\Delta\alpha/\alpha$. Note that for the stronger species, such as the Fe\,\iis $\lambda 2382$ and Mg\,\iis transitions, the centre regions of the profile are saturated, and thus a constraint on $\Delta\alpha/\alpha$ only comes from the optically thin wings. Conversely, for the weaker species (for example, Fe\,\textsc{ii} $\lambda 1608,2260$ and the Ni\,\textsc{ii} transitions) most or all of the profile is optically thin, and thus a constraint on $\Delta\alpha/\alpha$ is derived across the whole profile. Importantly, a single velocity structure model provides a good model to all the observed MM transitions. This serves to validate an assumption underlying the MM method, namely that kinematic segregation of the different species -- if present -- must be relatively small. We note the presence of two weak interlopers in Fe\,\iis $\lambda2260$ and one in Fe\.\iis $\lambda1608$, yielding an additional three tick marks. The regions of the Mg\,\iis transitions near $v\approx0$ have one less component than the corresponding regions in Fe\,\iis. This is because this region is saturated, and \textsc{vpfit} would not support so many strong components in the saturated region. This does not significantly affect $\Delta\alpha/\alpha$ because changes in parameters for lines in the middle of saturated regions have a marginal impact on $\chi^2$ and therefore on $\Delta\alpha/\alpha$. We have clipped pixels out of the spectrum for the region containing Cr\,\iis $\lambda 2062$ and Zn\,\iis $\lambda 2062$ because of a problem affecting the spectrum in that region.}
\end{figure*}

Thus, in fitting the observed absorption profiles, we attempt to obtain a fit which has $\chi^2_\nu \sim 1$ \emph{and} the minimum AICC possible \emph{and} has no substantial correlations in the residuals. We regard this fit as our final fit. We show in figure \ref{J223446-090812-absorber} an example of a complicated absorption system, which requires many components to achieve a statistically acceptable fit. We do not include $\Delta\alpha/\alpha$ as a parameter while establishing the final fit. That is, we avoid biasing $\Delta\alpha/\alpha$ by only adding it once the final model is found.

Determining the best-fitting number of velocity components, i.e. minimising the AICC, can be difficult. Considerable effort was expended to find the lowest possible AICC; that is, for each absorber we could not find a model with a lower AICC. Thus, we can say that our models should be close to the best model for the data. Any inaccuracy in the measurement of $\Delta\alpha/\alpha$ which is induced by our model being sub-optimal will have a random magnitude and sign, and therefore average out when considering a large number of $\Delta\alpha/\alpha$ values.

\subsection{Random and systematic errors}\label{s_randsyseffect}

Unfortunately, the Voigt profile decomposition of any given absorption system, with any particular choice of fitted transitions, is not unique. Errors in modelling the velocity structure may impact the fitted value of $\Delta\alpha/\alpha$. We thus distinguish between three different types of errors which may affect $\Delta\alpha/\alpha$:

\emph{Statistical errors.} These errors are simply the errors on $\Delta\alpha/\alpha$ which derive from the propagation of uncertainty from the flux error array via the Voigt profile model. These errors are simply the errors produced by \textsc{vpfit} from the covariance matrix at the best-fitting solution.

\emph{Random errors}. Random errors are any effects which might cause $\Delta\alpha/\alpha$ to be measured inaccurately when considering a single absorber. Significant errors made in determining the correct velocity structure could cause an error of this type. \citet{Murphy:08} demonstrated that an under-fitted spectrum (i.e one with a deficient model for the velocity structure) gives erroneous values of $\Delta\alpha/\alpha$. Other potential causes of random errors include: \emph{i)} kinematic segregation of different elements (which cannot be preferentially biased along the radial sightline over a large number of absorbers); \emph{ii)} random blends with other transitions arising in different absorbers (at different redshifts); \emph{iii)} random departures of the wavelength calibration solution from the true wavelength scale; \emph{iv)} cosmic rays and other uncleaned data glitches; \emph{v)} incorrect determination of the broadening mechanism (turbulent or thermal) for any component. Importantly, the effect of random errors will average to zero when considering an ensemble of velocity components in a given absorber, and also an ensemble of absorbers. This is because these errors will displace $\Delta\alpha/\alpha$ to be more positive as often as they will displace it to be more negative. When considering only a single absorber, this type of effect must be considered a systematic error. However, when considering an ensemble of absorbers, this effect is a \emph{random} effect, which adds extra scatter into the data. Any small kinematic shifts between different species in corresponding velocity components are of random sign and magnitude, and therefore any effect this has on $\Delta\alpha/\alpha$ will tend to average out even when considering many components within one absorber. When considering many absorbers, each with multiple velocity components, any bias introduced into $\Delta\alpha/\alpha$ must average to zero. This point was also discussed by \citet{Murphy:01c} and \citet{Murphy:03}. We explicitly constrain potential kinematic segregation of different species for a subsample of absorbers used in our analysis in Section \ref{s_testforkinematicsegregation}.

\emph{Systematic errors}. This is any error which systematically affects the values of $\Delta\alpha/\alpha$ measured over an ensemble of absorbers. Such effects would include: \emph{i)} inaccuracies in the laboratory wavelengths for particular MM transitions; \emph{ii)} a different heavy isotope abundance for Mg in the clouds relative to terrestrial values; \emph{iii)} systematic blends with nearby but unmodelled transitions in the same absorber; \emph{iv)} time-invariant differential light paths through the telescope for different wavelengths; \emph{v)} atmospheric dispersion for spectra taken without an image rotator; \emph{vi)} differential isotopic saturation \citep[see][]{Murphy:03}, and; \emph{vii)} wavelength miscalibration due to thorium-argon line list inaccuracies. \citet{Murphy:03} consider many possible systematic effects in detail. Note that these effects will not necessarily produce the same spurious shift in $\Delta\alpha/\alpha$ in every absorber. For example, if some transitions have inaccurate laboratory wavelengths, the effect on $\Delta\alpha/\alpha$ will depend on which other transitions are fitted in the same absorber. Nevertheless, the above effects are considered systematic errors because they will cause similar or correlated shifts in certain subsets of absorbers.

There is a crucial distinction between these effects: some effects may be considered systematics in single absorbers, but are not systematics in an ensemble of absorbers. For this reason, we are cautious against placing too much emphasis on the interpretation of the value of $\Delta\alpha/\alpha$ from any individual absorber. We demonstrate later that the impact of random effects is non-negligible. We explain our treatment of random effects in section \ref{LTS_method}.

\subsubsection{Physical constraints}\label{s_physicalconstraints}

As in the previous Keck analyses, we require that the velocity width ($b$-parameter, with $b=\sqrt{2}\sigma$) of each modelled component for a particular species in the fit is related to the corresponding components for other species. The two extreme cases are wholly thermal broadening ($b^2 = 2kT/M$ for a species with atomic mass $M$) and wholly turbulent broadening ($b^2 = b^2_{\mathrm{turb}}$). In general, there will be contributions from the two mechanisms ($b^2 = 2kT/M + b^2_{\mathrm{turb}}$), however we have found that most systems are generally well-fitted with turbulent broadening. As noted by \citet{Murphy:03}, it is possible to explicitly determine the degree of thermal and turbulent broadening, however in this circumstance the $b$-parameters are generally poorly determined, which makes the optimisation difficult. 

It turns out that the turbulent fit is preferred on the basis of the AICC in 71 percent of the absorbers, and the thermal fit in 29 percent of the absorbers. However, it should be noted that the fits were initially constructed with turbulent broadening and then converted to thermal broadening. It may be that if the fits were constructed thermally and then converted to turbulent that these figures might change significantly. We emphasise that mistakes made in choosing turbulent or thermal fitting may bias $\Delta\alpha/\alpha$ for a single absorber, but these effects must average to zero over a large number of absorbers due to the random nature of the bias from absorber to absorber. 

The previous analyses of the Keck results required that $\Delta\alpha/\alpha$ calculated using both thermal and turbulent fits differed by no more than $1\sigma$ for that absorber to be included in their final analysis, where the difference is considered only in terms of the statistical error. However, the generally higher SNR of the VLT spectral data (compared to the Keck spectral data) often leads to very precise statistical bounds on $\Delta\alpha/\alpha$. This makes the $1\sigma$-difference criterion difficult to fulfil in a significant number of cases. We describe below how we resolve any potential inconsistency between $\Delta\alpha/\alpha$ values from the thermal and turbulent fits.

In determining how to resolve any potential inconsistency, there are three cases to consider. \emph{i)} Where the difference between the fits is substantial, as measured by the AICC, one wants to take the statistically preferred fit. \emph{ii)} Where the quality of the fits is similar ($\mathrm{AICC}_{\mathrm{turbulent}} \approx  \mathrm{AICC}_\mathrm{thermal}$), and the values of $\Delta\alpha/\alpha$ are the same, then it does not matter which fit is used. \emph{iii)} If the values of the AICC for the thermal and turbulent fits are similar, but the values of $\Delta\alpha/\alpha$ produced by those fits differ significantly, then the statistical precision accorded to $\Delta\alpha/\alpha$ should be reduced to account for the conflicting evidence, and value of $\Delta\alpha/\alpha$ should be somewhere between the two cases.

To resolve this problem, and to be more accurate for our meaning in ``similar'' and ``significant'' above, we use a method-of-moments estimator which takes into account the relative differences in the AICC and the agreement, or otherwise, of the values of $\Delta\alpha/\alpha$. We estimate the underlying probability distribution of $\Delta\alpha/\alpha$ for the absorber in question as the weighted sum of two Gaussian distributions (one for the thermal result, one for the turbulent), with centroids given by the best fit value of $\Delta\alpha/\alpha$ for each fit, and $\sigma$ equal to $\sigma_{\Delta\alpha/\alpha}$ for each fit. We weight the sum by the penalised likelihood of the fits, via the AICC \citep[see][]{Liddle:07}. That is, if 
\begin{align}
k &= \exp( -\mathrm{AICC}_\mathrm{turbulent}/2 ) + \exp( -\mathrm{AICC}_\mathrm{thermal}/2 ),\\ 
j_1 &= \exp( -\mathrm{AICC}_\mathrm{turbulent}/2 )/k,\\
j_2 &= \exp( -\mathrm{AICC}_\mathrm{thermal}/2 )/k, \\
a_1 &= \Delta\alpha/\alpha_\mathrm{turbulent}, \\
a_2 &= \Delta\alpha/\alpha_\mathrm{thermal}, \\
s_1 &= \sigma(\Delta\alpha/\alpha_\mathrm{turbulent}), \quad \mathrm{and}\\
s_2 &= \sigma(\Delta\alpha/\alpha_\mathrm{thermal})
\end{align}
then matching the first two moments of our weighted sum of distributions with a Gaussian yields
\begin{align}
m = \Delta\alpha/\alpha &=  j_1 a_1 + j_2 a_2 ,\quad\mathrm{and}\\
\sigma_{\Delta\alpha/\alpha} &= \sqrt{j_1 s_1^2 + j_2 s_2^2 + j_1 a_1^2 + j_2 a_2^2 - m^2}. \label{eq_mom_s}
\end{align}
This covers all the cases described above. In particular, where the AICC is similar but $\Delta\alpha/\alpha$ differs significantly between the turbulent and thermal fits, the estimated error increases with the difference between them, providing resistance to incorrectly determining the line broadening mechanism. To see this, note that with $j_1 + j_2 = 1$, equation \ref{eq_mom_s} reduces to 
\begin{equation}
\sigma_{\Delta\alpha/\alpha} = \sqrt{j_1 s_1^2 + (1-j_1) s_2^2 + j_1 (1 - j_1)(a_1  - a_2)^2}. 
\end{equation}
Thus, errors only ever increase from our smallest error estimate, and therefore this method could be considered conservative. In the event where one broadening mechanism is significantly preferred, then our result will be effectively the same as if only that broadening mechanism was considered. For the case where the fits are statistically indistinguishable ($j_1 = j_2$), $\Delta\alpha/\alpha$ is given by the simple mean of the two values of $\Delta\alpha/\alpha$, and the variance is the simple mean of the individual variances plus $0.25(a_1 - a_2)^2$.

\subsubsection{Al\,\textsc{iii}}
In principle the Al\textsc{\, iii} transitions can be included in a MM fit, however its ionisation potential is somewhat different to the other MM transitions described. Therefore, the Al\,\textsc{iii} transitions may not arise from the same location, and therefore velocity, as the other MM transitions. If the Al\,\textsc{iii} transitions arise at  significantly different velocities to the other MM transitions then an error would be introduced into $\Delta\alpha/\alpha$ for a system with Al\,\textsc{iii} included (although this effect must average to zero over a large number of absorbers, as there is no reason for a systematic bias in the centroid of the Al\,\textsc{iii} transitions with respect to the other MM transitions along a line of sight to Earth). 

Generally, the profiles for different transitions for the other MM transitions used correlate well with each other. By this, we mean that the relative column densities  between corresponding velocity components are similar for different MM transitions. However, we have noticed that the absorption profiles for some Al\textsc{\, iii} transitions in some absorbers differ significantly in the relative line strengths between components, when compared to other MM transitions. Importantly, we found some absorbers where it was difficult to apply the same velocity structure model to Al\,\iiis transitions and the other MM transitions simultaneously. For this reason, we are therefore cautious in fitting Al\textsc{\, iii} together with the other MM transitions. 

Therefore, we include and model Al\textsc{\, iii} if both the transitions are available, and allow the spectral data to contribute to $\Delta\alpha/\alpha$ derived from the other MM transitions for that absorber, but do not constrain the modelled structure with the velocity structure from other MM transitions. Given the small difference in the $q$ coefficients between the two Al\textsc{\, iii} transitions ($\Delta q \sim 250$) the statistical contribution of Al\textsc{\, iii} to $\Delta\alpha/\alpha$ is low, however given that the exposures have already been obtained it is prudent to try to maximise our use of the existing data.

As an example: if Al\textsc{\, iii} is not utilised in the $z=1.857$ absorber towards J013105$-$213446, the turbulent fit value of $\Delta\alpha/\alpha$ changes from $(0.30 \pm 1.43) \times 10^{-5}$ to $(0.43 \pm 1.44) \times 10^{-5}$. Similarly, if Al\textsc{\, iii} is not utilised in the $z=1.71$ absorber towards J014333$-$391700, the turbulent fit value of $\Delta\alpha/\alpha$ changes from $(-2.20 \pm 2.67) \times 10^{-5}$ to $(-2.32 \pm 2.68) \times 10^{-5}$.

We note that previous works have included the Al\textsc{\, iii} transitions as part of the MM analysis, and it was demonstrated by \citet{Murphy:03} that the inclusion of these transitions did not significantly alter the Keck results. Nevertheless, the approach we have adopted is conservative.

We have also observed less substantial relative line strength differences between the Mg\,\iscs transitions and other MM transitions, but in no case did we find a system where we could not apply the same velocity structure model to the Mg\,\iscs and MM transitions, and so we include the Mg\,\iscs transitions in the full MM analysis.

\subsection{Interlopers}\label{s_interlopers}

Some transitions display excess absorption beyond what is predicted from other transitions from the same atomic species. This excess absorption is caused by another absorber located along the line of sight to the quasar, which is usually extragalactic. Even where a prediction cannot be made from another transition of the same atomic species, interlopers can still be detected when many transitions are fitted together, as the redshifts and $b$-parameters of each component are constrained. Although the contaminated sections of spectrum can be discarded, it is often possible to adequately model the contamination, thereby maximising use of the spectral data.

We distinguish two types of interlopers: identifiable and unidentifiable.

\subsubsection{Identifiable interlopers}

In some cases, the interlopers can be identified and modelled simultaneously with the MM transitions. By ``identified'' we mean that the redshift, atomic species and wavelength of the transition which causes the excess absorption can be determined. In principle, one needs accurate rest wavelengths and $q$-coefficients for the interloping transitions. This means that the interloping transition can be modelled if it is an MM transition from an absorber at a different redshift, or if it is Si\,\textsc{iv} $\lambda\lambda 1393$ or $1402$. If the interloper is from the C\,\textsc{iv} doublet, the contamination can also be modelled despite the fact that the rest wavelengths for this doublet are relatively poorly known. This is done by allowing the C\,\textsc{iv} transitions to have a separate value of $\Delta\alpha/\alpha$, which is then discarded. This extra parameter effectively absorbs any error introduced through inaccurate knowledge of the rest wavelengths.

\subsubsection{Unidentifiable interlopers}

\begin{figure}
\ifpdf
\includegraphics[bb=34 58 554 727,angle=-90,width=82.5mm]{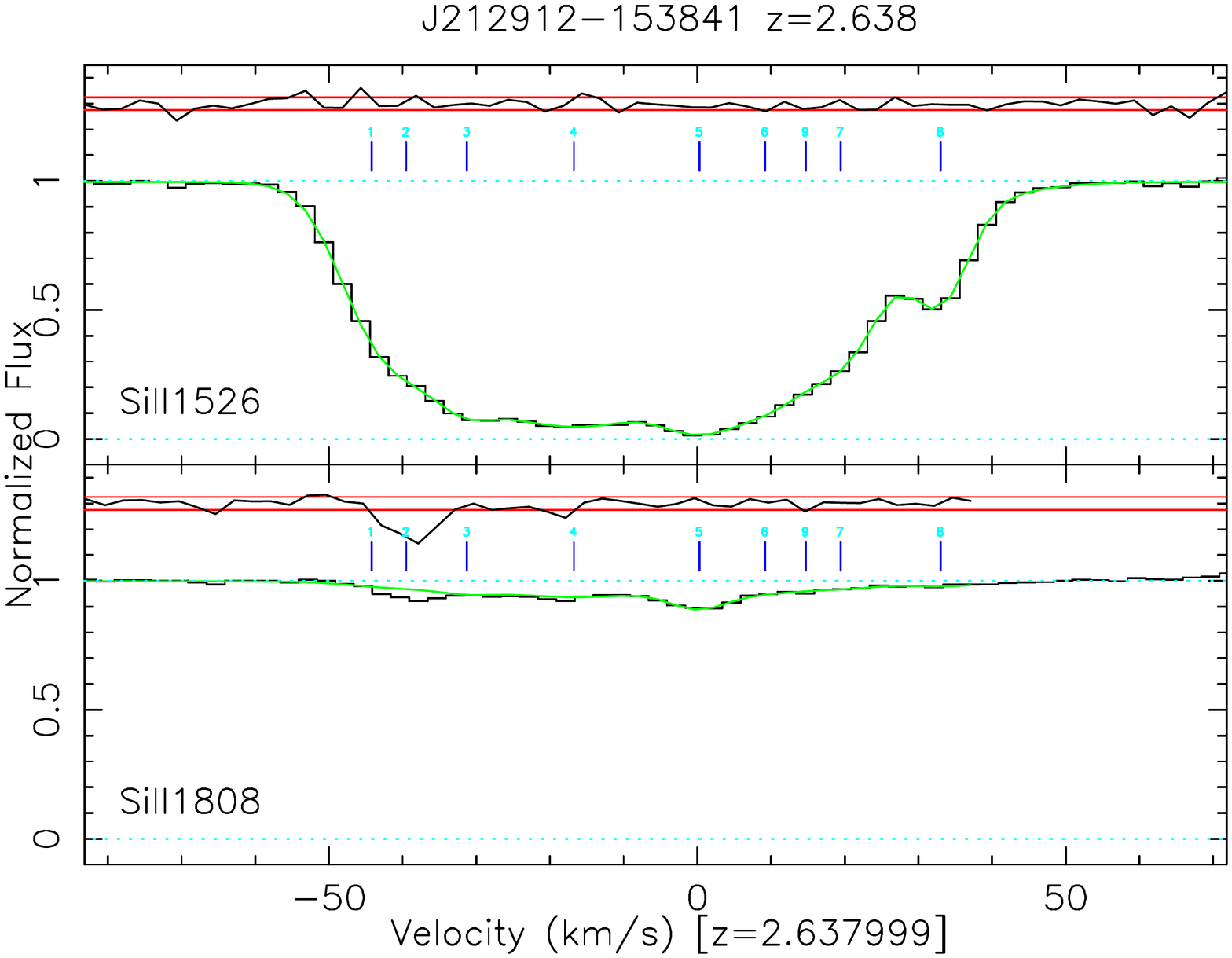}
\else
\includegraphics[bb=34 58 554 727,angle=-90,width=82.5mm]{images/J212912-153841-z2_638-nointerloper.eps}
\fi
  \caption{\label{J212912-153841-absorber}The Si\,\iis transitions from our MM fit to the absorber at $z_\mathrm{abs}=2.638$ toward J212912$-$153848, shown without an included interloper. The horizontal scale indicates the velocity difference from the arbitrary redshift stated at the bottom for the given data points. The black line indicates the observed normalised flux, with the green line indicating our best fit solution. At the top of each box, the black line indicates the standardised residuals (that is, [data - model]/error), with the red lines indicating $\pm 1\sigma$. The position of the blue tick marks indicates the fitted position of the single component. The strength of the Si\,\iis $\lambda1808$ transition can be predicted from the model for Si\,\iis$\lambda 1526$. Si\,\iis $\lambda1808$ shows excess absorption at $v \approx -40\,\mathrm{km\,s^{-1}}$ which cannot be explained by Si\,\iis$\lambda 1526$. To account for this, we include a single, unconstrained interloper. After including the interloper, the fit is statistically acceptable. }
\end{figure}

In many cases, however, the interloping transition can not be identified. In this case, our decision as to how to proceed depends on the degree of contamination. If the degree of contamination is small, and confined to a small area of the observed profile, we can include unknown interloping transitions where the residuals of the fit ([data - model]/error) are bad until a statistically acceptable fit is achieved. Doing this provides a statistically acceptable model of the contamination. Note that the contribution to $\Delta\alpha/\alpha$ of the affected MM transition will be reduced as a result of this, as the interlopers included are unconstrained by other spectral regions. We show an example of this in figure \ref{J212912-153841-absorber}.

As the degree of contamination begins to increase, the potential error introduced into $\Delta\alpha/\alpha$ may grow. Our treatment of transitions affected by significant contamination depends on whether there are other transitions from the same species available which can be used to constrain the velocity structure for that transition. In the case of Fe \textsc{ii}, a wide variety of transitions are often available. Effectively, a fit to several other transitions of the same species may allow the structure of the contamination to be determined, particularly if the SNR is high. However, in some cases, there may be no other transitions which can be used to obtain the velocity structure. This may occur as a result of all of the transitions suffering from contamination from different absorbers, or because the spectra only includes one of the transitions of that species (due to gaps in the spectral coverage), or because the species has no other transitions which could be used for that purpose. The last case is particularly problematic for Al\,\textsc{ii}, for which we only use the Al\,\textsc{ii} $\lambda1670$ transition -- there are no other Al\,\textsc{ii} transitions which can be used to directly constrain the Al\,\textsc{ii} structure. This issue also occurs for Si\,\textsc{ii} -- although in theory both the Si\,\textsc{ii} $\lambda1526$ and $1808$ transitions can be used to constrain the velocity structure of Si\,\textsc{ii}, the oscillator strength for the $\lambda 1526$ transition ($f\approx 0.13$) is much larger than for the $\lambda 1808$ transition ($f \approx 0.002$). For many systems observed to have Si\,\textsc{ii} $\lambda 1526$ absorption, the column density of Si\,\textsc{ii} is not large enough to detect the $\lambda 1808$ transition. Even if the $\lambda 1808$ transition is detected, it may be too weak to provide a meaningful constraint on the Si\,\textsc{ii} structure. 

\begin{figure}
\ifpdf
\includegraphics[bb=34 58 554 727,angle=-90,width=82.5mm]{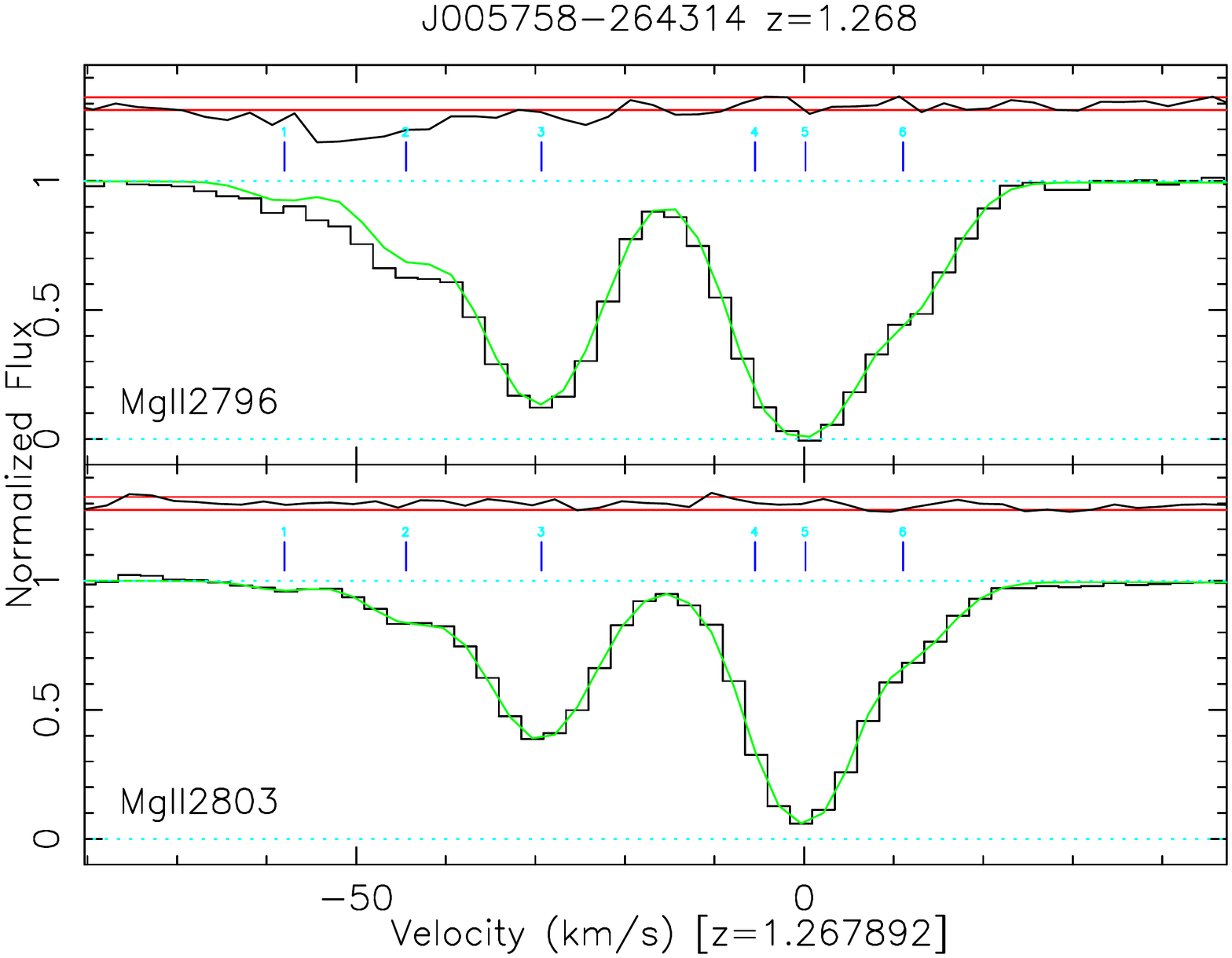}
\else
\includegraphics[bb=34 58 554 727,angle=-90,width=82.5mm]{images/J005758-264314-z1_2679-noclipping.eps}
\fi
  \caption{\label{J005758-264314-absorber}Part of our MM fit to the absorber at $z_\mathrm{abs}=1.268$ toward J005758$-$264314. The horizontal scale indicates the velocity difference from the arbitrary redshift stated at the bottom for the given data points. The black line indicates the observed normalised flux, with the green line indicating our best fit solution. At the top of each box, the black line indicates the standardised residuals (that is, [data - model]/error), with the red lines indicating $\pm 1\sigma$. The position of the blue tick marks indicates the fitted position of the single component. We show here the model from Mg\,\iis $\lambda2803$ plotted over the spectral region of Mg\,\iis $\lambda2796$. One can see that there is significant excess absorption in Mg\,\iis $\lambda2796$ for $v\lesssim -40\,\mathrm{km\,s^{-1}}$. There also appears to be excess absorption in the range $-40\,\mathrm{km\,s^{-1}} \lesssim v \lesssim -20\,\mathrm{km\,s^{-1}}$. Due to the wide range in velocity over which the absorption occurs, instead of attempting to model the absorption we clip away all pixels for $v \lesssim -20\,\mathrm{km\,s^{-1}}$ in the Mg\,\iis $\lambda2796$ region for our actual fit.}
\end{figure}

In cases where we are unable to obtain a good constraint on the velocity structure of a particular species from other unaffected transitions of the same species, and the degree of contamination is not small, we are very cautious about inserting interlopers, due to the potential bias this could introduce into $\Delta\alpha/\alpha$ for this absorber. Note that any bias introduced here is a \emph{random} effect, and therefore will average out when considering an ensemble of absorbers. Nevertheless, we wish to avoid introducing extra scatter into the $\Delta\alpha/\alpha$ values where possible. In these cases, we clip out the pixels which appear to be affected by contamination, leaving a wide buffer on either side. Note that we can only do this where another transition from the same species exists. Otherwise, components situated in the middle of the clipped pixels might have very little, if any spectral data to constrain them, and thus their column densities could take on values which would not be consistent with the general model used for the absorber. If there is no other transition for the species in this case, we simply do not use the transition. In figure \ref{J005758-264314-absorber} we show an example of where we have clipped out pixels because of contamination by an interloper.

Fitting contamination has the potential to introduce a \emph{random} bias for individual absorbers, however this may nevertheless reduce \emph{systematic} effects. To see this consider a hypothetical absorber where the Mg\,\textsc{ii} $\lambda\lambda 2796,2803$, Al\,\textsc{ii} $\lambda 1670$ and the Fe\,\textsc{ii} $\lambda\lambda\lambda 2383, 2600, 2344$ transitions are available, but that the Al\,\textsc{ii} $\lambda 1670$ transition suffers from some minor contamination in part of the observed profile, and the absorber of the contaminating transition cannot be identified. One could simply ignore the Al\,\textsc{ii} $\lambda 1670$ transition and fit the Mg\,\textsc{ii} and Fe\,\textsc{ii} transitions. Deriving $\Delta\alpha/\alpha$ from just the Mg\,\textsc{ii}/Fe\,\textsc{ii} combination is not very robust to a simple stretching or compression of the wavelength scale \citep{Murphy:03}. The Mg\,\textsc{ii}/Fe\,\textsc{ii}/Al\,\textsc{ii} combination just described is much more robust against this effect, as the Mg\,\textsc{ii} and Al\,\textsc{ii} anchor transitions are positioned on other side of the high-$q$ Fe\,\textsc{ii} transitions. Modelling the contamination of the Al\,\textsc{ii} profile here may introduce a random error, but this can be averaged out in the context of many absorbers. The potential introduction of this random error is easily justified by the increased resistance to systematic effects (such as wavelength scale distortions, which in principle could be common to many absorbers or spectra). In the situation just described, the only choices are to discard the Al\,\textsc{ii} transition, or to model the contamination. We choose the latter option where the degree of contamination is minor, and the former where the contamination is severe.

\subsubsection{Transitions in the Lyman-$\alpha$ forest}

It also happens that some transitions fall in the Lyman-$\alpha$ forest (``the forest''), a dense series of absorption lines blueward of the quasar Lyman-$\alpha$ emission line. These transitions are caused by H\,\textsc{i} absorption along the line of sight to the quasar. We are cautious about using MM transitions which fall in the forest, due to the uncertainties in determining the structure of the forest. Neverthless, the use of MM transitions in the forest may afford significantly better constraints on $\Delta\alpha/\alpha$. Although this often occurs with low-$z$ Mg\,\textsc{ii}/Fe\,\textsc{ii} absorbers, with some Fe\,\textsc{ii} transitions falling in the forest, it also occurs in high-$z$ systems, where the Si\,\textsc{ii} $\lambda 1526$/Al\,\textsc{ii} $\lambda 1670$/Fe\,\textsc{ii} $\lambda 1608$ combination is common. Where the SNR is high for the transitions which fall in the forest, we model the forest structure with H\,\textsc{i} absorption. If the SNR ratio is low, determination of the forest structure can be difficult, and therefore we do not utilise the contaminated transitions. Again, we emphasise that although this has the potential to introduce bias for a single absorber, because the contamination is random from absorber to absorber, it must average out over a large number of systems. We have used transitions which fall in the forest in 27 of the 1142 spectral fitting regions in the VLT sample (2.4 percent).

\subsubsection{Cr\,\textsc{ii} $\lambda 2062$, Zn\,\textsc{ii} and Mg\,\textsc{i} $\lambda2026$}

The Cr \iis $\lambda2062$ $(2062.24\mathrm{\AA})$ and Zn \iis $\lambda 2062$ $(2062.66\mathrm{\AA})$ lines are relatively closely spaced, being separated by $\approx 62\,\mathrm{km\,s^{-1}}$. For narrow absorption systems, one can distinguish between these transitions as they do not overlap. However, these transitions are most commonly associatd with damped Lyman-$\alpha$ absorbers, where the velocity structure is generally complicated, and the system displays absorption over tens to hundreds of km/s. In this case, the Cr \iis $\lambda2062$ and Zn \iis $\lambda 2062$ transitions often overlap. The velocity structure for these transitions can be determined by simultaneously modelling these transitions with the Cr \iis $\lambda2052,2056,2066$ and Zn \iis $\lambda 2062$ transitions. 

There is one point of caution here. For high column density systems, a potential blend exists with Mg\,\textsc{i} $\lambda 2026$. Mg\,\textsc{i} $\lambda 2026$ is weak, with oscillator strength $f=0.113$, and is rarely seen. One can in principle use the Mg\,\textsc{i} $\lambda 2852$ ($f=1.83$) information to constrain the Mg\,\textsc{i} structure. In this case, a joint fit of Zn \iis$\lambda2026$ and Mg\,\textsc{i} $\lambda\lambda2026,2852$ will ensure that the Zn \iis $\lambda2026$ results are not biased by any absorption due to Mg\,\textsc{i} $\lambda 2026$. However, the absorbers for which Mg\,\textsc{i} $\lambda 2026$ might be detected are often at high redshift, in which case Mg\,\textsc{i} $\lambda 2852$ is often unusable, either due to heavy contamination by sky emission or absorption, or because it is located out of the red end of the spectral coverage. In this particular circumstance, we are generally cautious about fitting Zn \iis $\lambda 2026$. Where we consider that the Zn \iis $\lambda 2026$ transition might be affected by Mg\,\textsc{i} $\lambda 2026$, and we are unable to utilise Mg\,\textsc{i} $\lambda 2852$, we do not include the Zn\,\textsc{ii} $\lambda 2026$ transition.

\subsection{Aggregation of $\Delta\alpha/\alpha$ values from many absorbers}

\subsubsection{Weighted mean}

If one assumes that all the $\Delta\alpha/\alpha$ values are described by a constant offset of $\Delta\alpha/\alpha$ from the laboratory values, one can combine the $\Delta\alpha/\alpha$ values together using a weighted mean. This process is valid provided that the $\Delta\alpha/\alpha$ values support a constant value of $\Delta\alpha/\alpha$. If $\Delta\alpha/\alpha \neq 0$, then this implies that there must be a transition at some point from laboratory conditions ($\Delta\alpha/\alpha \equiv 0$), and therefore the $\Delta\alpha/\alpha$ values should be inspected to see if a transition point can be identified. Additionally, one must examine the residuals about the fit for a weighted mean, plotted against various parameters of interest (e.g. redshift and sky position) to determine if unmodelled trends exist. 

\subsubsection{Dipole fit}

A dipole+monopole model constitutes the first two terms of the spherical harmonic expansion. The simplest dipole model is of the form
\begin{equation}
\Delta\alpha/\alpha = A\cos(\Theta) + m,\label{dipole_eq}
\end{equation}
where $\Theta$ is the angle between the pole of the dipole and the sky position under consideration, $A$ is an angular amplitude and $m$ (the monopole) represents a possible offset of $\Delta\alpha/\alpha$ from the laboratory value. An equivalent (and more computationally convenient) form is 
\begin{equation}
 \Delta\alpha/\alpha = \mathbf{c}.\mathbf{\hat{x}} + m, \label{dipole_eq2}
\end{equation}
where $\mathbf{\hat{x}}$ is a unit vector pointing towards the direction under consideration and $\mathbf{c}$ contains the amplitude and direction information of the dipole. The components of $\mathbf{c}$ ($c_x$, $c_y$ and $c_z$) are easily related to the right ascension (RA) and declination (dec.)\ of the direction of the dipole, and $\lvert \mathbf{c} \rvert$ gives the magnitude of the dipole. In this form, $\Delta\alpha/\alpha$ is linear in the $c_{i}$ and so the $c_{i}$ may be determined through weighted linear least squares.

Although naively we might expect that $m=0$, some theories contemplate otherwise. This could be possible if $\alpha$ depends on the local gravitational potential or density \citep{Khoury:04a,Mota:07a,Olive:08a} -- laboratory conditions differ quite significantly in this regard to the conditions in the quasar absorbers. Note that by including the $m$ term, one gets an explicit statistical test of this idea. Additionally, in the presence of temporal evolution of $\alpha$, $m$ amounts to its average value. In any particular redshift slice, $m$ therefore represents the average angle-independent value of $\Delta\alpha/\alpha$.

Equations \ref{dipole_eq} and \ref{dipole_eq2} take no account of any redshift dependence. Clearly, $A = A(z)$. Nevertheless, a model which includes angular dependence is useful because it provides a method of detecting spatial variations in $\alpha$ which does not require the specification of a functional form for $A(z)$. Use of this model is valid under several possible circumstances. One is where any variation in $\alpha$ with redshift in the sample along a particular direction is small compared to variation in $\alpha$ in the opposite direction. This might be possible in our sample, depending on how $\alpha$ might vary, as we typically probe lookback times of greater than 5 gigayears. Another is where $\alpha$ does vary significantly with redshift in our sample and the distribution of absorber redshifts does not vary greatly with sky position. With enough data, one could simply take redshift slices and apply this model to each redshift slice, thereby building up the functional form of $A(z)$ in model-independent manner. However, given that we have only $\sim 300$ absorbers between the Keck and VLT samples, we simply cannot slice the data enough to do this for more than $\sim$two redshift bins. Another consideration is the effect of choosing a particular form of $A(z)$. An incorrect choice of $A(z)$ may reduce sensitivity to detect an effect, and could lead to the wrong conclusion if the choice is sufficiently bad. As a result, we explore an angle-dependence model initially, and later consider explicitly including distance dependence. 

We derive our uncertainties on the direction of dipole vectors by tranforming the covariance matrix from rectilinear coordinates $(c_{x},c_{y},c_{z})$ to spherical coordinates $(r,\phi,\theta)$, using the standard Jacobian matrix. That is, if $\mathbfss{J}$ is the Jacobian matrix of the transformation from rectilinear to spherical coordinates, and $\mathbfss{C}$ is the covariance matrix calculated from the fit, then $\mathbfss{C'} = \mathbfss{J} \cdot \mathbfss{C} \cdot \mathbfss{J}^\mathbfss{T}$ gives the approximate covariance matrix in spherical coordinates. The radial component, $r$, corresponds to the amplitude of the dipole. Our errors given on RA and dec.\ are thus linearised approximations based on the covariance matrix at the best-fitting solution, and should be regarded only as approximate. These error estimates will be inaccurate if they subtend a large fraction of the sky. 

Note that, by virtue of the fact that $r \geq 0$ in spherical coordinates, the dipole amplitude, $A$, is not Gaussian. Thus, we perform a resampling bootstrap analysis \citep{NumericalRecipes:92} to derive an uncertainty for a dipole amplitude. Similarly, one cannot use a $t$-test to determine if $A$ is significantly different from zero. Thus, we calculate the statistical signifiance, $1-p$, of the dipole model over the monopole model using a bootstrap method where we randomise $\Delta\alpha/\alpha$ over sightlines, and from the observed distribution of $\chi^2$ determine the probability that a value of $\chi^2$ as good or better than that given by our observed dipole fit would occur by chance. One can also use analytic methods \citep{Cooke:09}, if desired. These methods should yield similar answers for large sample sizes. However, for small sample sizes, the results may differ somewhat (especially if the statistical uncertainties vary significantly in magnitude between the $\Delta\alpha/\alpha$ values). As the dipole model will always improve the fit over a monopole, the statistical test is one-tailed, and so when we state the $\sigma$-equivalence of a statistical significance, this is calculated as $\lvert \mathrm{probit}(p/2) \rvert$ in order to accord with the conventional usage. $\mathrm{probit}$ is the inverse normal cumulative distribution function. 

Unless otherwise mentioned, we multiply uncertainty estimates on monopole values and sky coordinates by $\sqrt{\chi^2_\nu}$ as a first-order correction for over- or under-dispersion about the fitted model \citep{NumericalRecipes:92}. 

\subsubsection{Robust estimate of random errors}\label{LTS_method}

It may happen that the values of $\Delta\alpha/\alpha$ demonstrate excess scatter about a particular model. Although this might be due to model misspecification, it is also due to the random effects described in section \ref{s_randsyseffect}. Previous works \citep[e.g.][]{Murphy:03,Murphy:04:LNP} have added a constant, $\sigma_\mathrm{rand}$, in quadrature with the uncertainty estimates of each $\Delta\alpha/\alpha$ value, $\sigma_\mathrm{stat}$, until $\chi^2_\nu = 1$ about the fitted model (i.e. $\sigma_\mathrm{tot}^2=\sigma_\mathrm{rand}^2+\sigma_\mathrm{stat}^2$), thereby making the data statistically consistent with each other under that model. This constant is an estimate of the aggregation of any errors which average to zero over a large number of systems (i.e. random errors), under the assumption that the magnitude of this effect is the same in each absorber. 

We are cautious, however, about not over-estimating the term $\sigma_\mathrm{rand}$ required on account of apparent outliers in the $\Delta\alpha/\alpha$ values. We define an outlier as any point which has a standardised residual $\lvert r_i \rvert = \lvert (\Delta\alpha/\alpha - \mathrm{model\ prediction})/\sigma_\mathrm{tot}\rvert \ga 3$, even after adding an appropriate $\sigma_\mathrm{rand}$. The existence of outliers even after increasing the error bars implies that these systems may be affected by random or systematic errors which occur infrequently, or with significantly lower magnitude in most of the systems. Adding a constant in quadrature with the error estimates until $\chi^2_\nu=1$ will therefore over-estimate the random error associated with most points, and under-estimate the random error associated with high scatter points, increasing the probability of a false negative (a failure to detect $\Delta\alpha/\alpha \neq 0$ when it is in fact $\neq 0$). 

To alleviate this problem, we use a modification of the Least Trimmed Squares (LTS) method \citep{Rousseeuw:84}. Instead of fitting all $n$ $\Delta\alpha/\alpha$ values using weighted least squares, the LTS method traditionally fits only $k=(n+p+1)/2$ $\Delta\alpha/\alpha$ values (where $p$ is the number of parameters fitted) using weighted least squares, and searches for the combination of $k$ data points and fitted model that yields the lowest weighted sum of squared residuals ($\chi^2$). If one is more confident that the data contain only a small number of outliers, then one can choose $k$ somewhere between $(n+p+1)/2$ and $n$. For $n\rightarrow \infty$, the use of $k=(n+p+1)/2$ will produce a very robust fit, where the fit is resistant to $\lesssim 50$ percent of the data being erroneous. However, for small $n$ (e.g. $n\lesssim 20$) the resultant fit parameters can be significantly affected by small perturbations to the fitted data points because of the strong sub-sample sensitivity of the LTS in small sample sizes. To help remedy this, we choose $k = 0.85n$ when applying the LTS method. We utilise the Fast-LTS method \citep{Rousseeuw:02}, an algorithm which determines the LTS solution fairly rapidly.  

The LTS method has traditionally been applied to fit models to data for which uncertainty estimates are not initially available, leading to the assignment of constant magnitude error estimates to all data points. However, our $\Delta\alpha/\alpha$ values already have statistical error estimates,  which serve as lower bounds on the true errors. We propose a new variant of the LTS method to increase the error bars in quadrature with some $\sigma_\mathrm{rand}$, but that addresses our concerns about false negatives. This method was inspired by \citet{Pitsoulis:09}, who use linear scaling of existing errors rather than our quadrature error addition. 

To obtain our robust estimate of $\sigma_\mathrm{rand}$, we increase the error bars until a robust scatter measure is what we would expect for a Gaussian distribution. We calculate our robust scatter measure, $\chi^2_\nu(k)$, as the sum of squared residuals per degree of freedom about a fit taken over only the $k$ $\Delta\alpha/\alpha$ values with the smallest squared residuals, rather than over all $n$ $\Delta\alpha/\alpha$ values. To estimate $\sigma_\mathrm{rand}$, we slowly increase $\sigma_\mathrm{rand}$ from zero and add $\sigma_\mathrm{rand}$ in quadrature with the existing error bars. For each value of $\sigma_\mathrm{rand}$, we calculate the LTS fit, then calculate $\chi^2_\nu(k)$. We continue to increase $\sigma_\mathrm{rand}$ until $\chi^2_\nu(k)$ is equal to the expected value for a Gaussian distribution with large numbers of degrees of freedom, $\langle\chi^2_\nu(k)\rangle$. If we define $f$ as $k/n$, then $\langle\chi^2_\nu(k)\rangle$ is calculated as
\begin{equation}
\langle\chi^2_\nu(k)\rangle = \int_{-a}^{a} \frac{x^2 e^{-x^2/2}}{\sqrt{2\pi}}\,\mathrm{d}x
\end{equation}
where $a = \mathrm{probit}[(1+f)/2]$. We take the value of $\sigma_\mathrm{rand}$ derived in this way as our estimate of the additional random error for the data given the model. For the standard least squares case (given by $k=n$), this yields $\langle\chi^2_\nu(n)\rangle = 1$, the well-known result that the expected value of $\chi^2_\nu$ is 1 for large numbers of degrees of freedom. Thus, if the data are contaminated by a few outliers, these will not impact the estimate of the random error which affects most points. 

After applying the LTS method to estimate the random error term, we then discard all points with $\lvert r_i \rvert > 3$ about the LTS fit, but only if we are applying the method to a full sample of $\Delta\alpha/\alpha$ values (i.e.\ the whole VLT or Keck sample, or a combination of the two). This is because in small-$n$ fits one does not have much data, and so it is not clear whether outliers would become inliers with more data. 

If we remove outliers, we then reapply the LTS method to check that no more outliers are unmasked, and to re-estimate $\sigma_\mathrm{rand}$. The LTS fit is statistically inefficient because it ignores some good data (15 percent for $k = 0.85n$ if all the remaining points are inliers). Therefore, after we discard high residual points, we apply a normal weighted least squares fit to the remaining data to estimate the parameters and achieve the best possible confidence limits on our modelled parameters.  

The benefits of the LTS method can be summarised as follows.

\emph{i) Robust estimate of $\sigma_\mathrm{rand}$.} If we calculate $\sigma_\mathrm{rand}$ by increasing it until $\chi^2_\nu=1$, then even a single, arbitrarily large outlier can increase $\sigma_\mathrm{rand}$ without bound. This is much less likely with the LTS method. A more appropriate estimate of $\sigma_\mathrm{rand}$ means that false negatives are less likely. 

\emph{ii) Robust detection of outliers.} In a standard $\chi^2$ minimisation fit, residuals with larger magnitude $|r_i|$ are weighted as $r_i^2$, which distorts the fit towards them. This tends to mask outliers. By distorting the fit, one might incorrectly decide that some good points are in fact outliers. Similarly, the existence of one outlier tends to conceal the existence of additional outliers (a masking effect). 

\emph{iii) Objectivity.} Manual outlier rejection is often characterised as subjective. The rule provided here provides an objective method of classifying data points as outliers, thereby removing this objection.

\emph{iv) More robust parameter estimates.} Even a few outliers can substantially distort the fit. This biases parameter estimates away from their underlying values. We are interested in the underlying values, not the values given by a blind least squares fit. The rate of false positives (detections of $\Delta\alpha/\alpha \neq 0$ when $\Delta\alpha/\alpha = 0$) should also be decreased, as false positives can be caused by outliers.

\section{VLT results}\label{s_VLT_results}

We present here the results of our analysis of the VLT MM absorbers. The values of $\Delta\alpha/\alpha$ for each absorber are given in table \ref{VLT_daoa_results}. The estimated additional error term, $\sigma_\mathrm{rand}$, depends on the model used, and we give values of $\sigma_\mathrm{rand}$ used below according to the model in question. The key for the transitions given is defined in table \ref{tab_freqtransitions}. 

\begin{figure}
\begin{center}
\ifpdf
\includegraphics[bb=79 88 540 766,width=82.5mm]{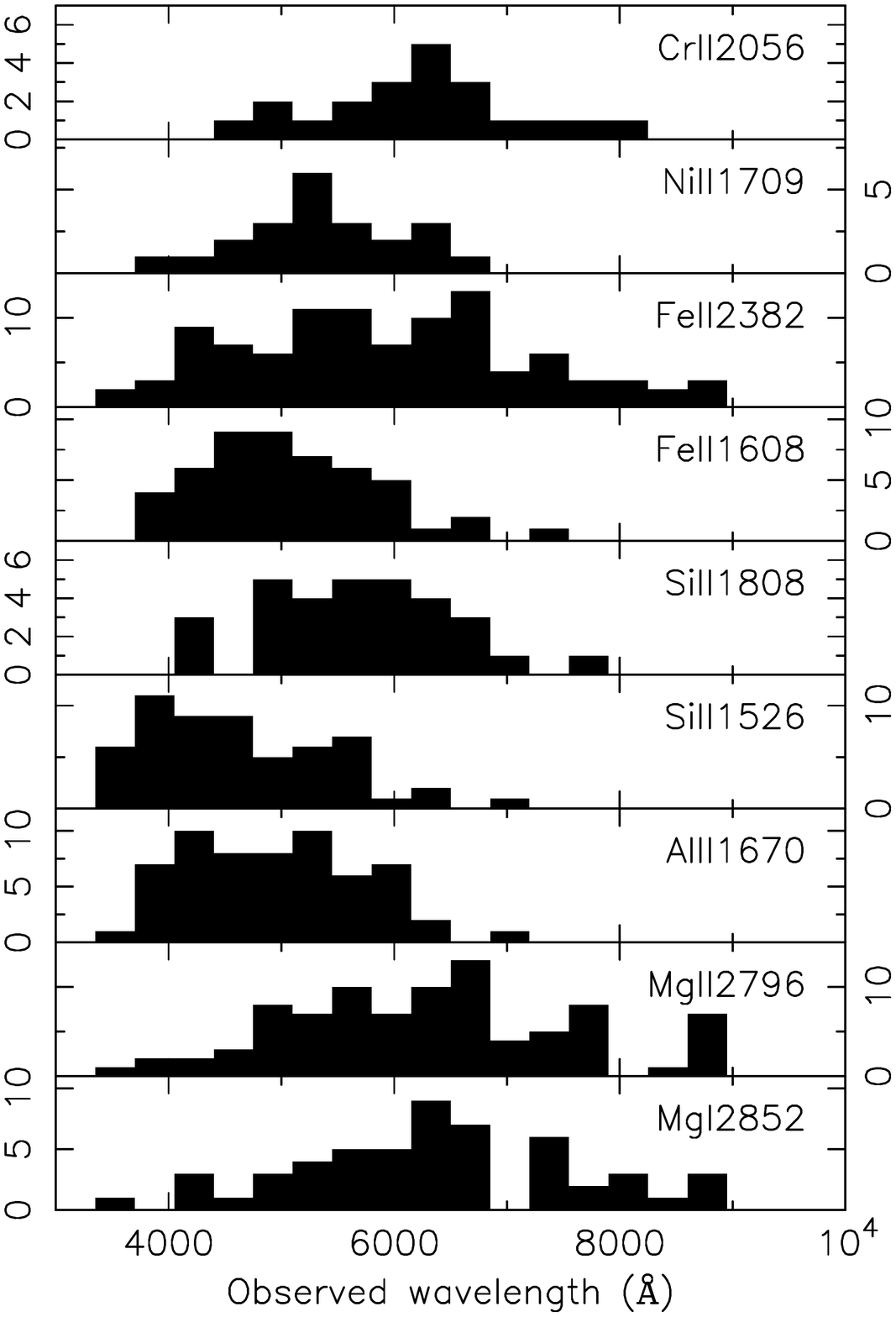}
\else
\includegraphics[bb=79 88 540 766,width=82.5mm]{images/distribution_sometransitions.eps}
\fi
\end{center}
\caption{Distribution of the observed wavelength of certain representative transitions utilised in the MM fits for the VLT sample. The vertical scales alternate. \label{fig_dist_sometrans}}
\end{figure}

\begin{figure}
\begin{center}
\ifpdf
\includegraphics[bb=50 92 556 777,angle=-90,width=82.5mm]{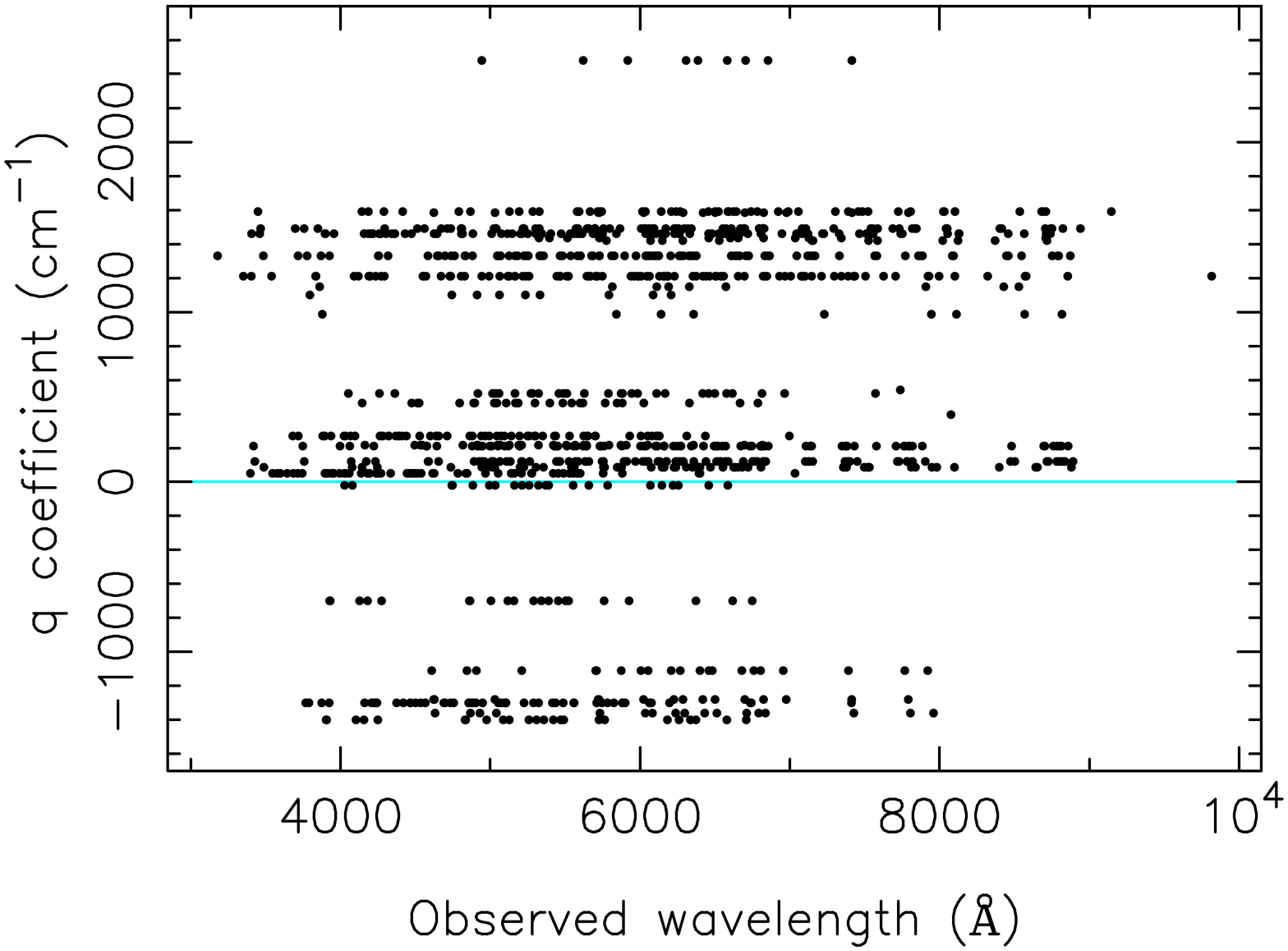}
\else
\includegraphics[bb=50 92 556 777,angle=-90,width=82.5mm]{images/wavelength_vs_q.eps}
\fi
\end{center}
\caption{Relationship between $q$-coefficients and observed wavelength for all utilised transitions in all absorbers in the VLT sample. Although the low-$z$ Fe\ \ii/Mg\ \iis combination is sensitive to low-order wavelength distortions because the $q$-coefficients for this combination are correlated with wavelength \citep[see fig.\ 1 of][]{Murphy:03}, one can see that for the full sample there is little correlation between observed wavelength and $q$, making the MM method resistant to systematics when many absorbers at different redshifts are used. \label{fig_wavelength_vs_q}}
\end{figure}

The frequency with which certain transitions are fitted is also given in table \ref{tab_freqtransitions}. We show the distribution of observed wavelengths for certain representative MM transitions in figure \ref{fig_dist_sometrans}. In figure \ref{fig_wavelength_vs_q}, we show the relationship between the $q$-coefficients and the observed wavelength of the transitions used. This figure demonstrates that, although $q$ is correlated with wavelength for the low-$z$ Fe\,\ii/Mg\,\iis combination (and therefore this combination is susceptible to low-order wavelength scale distortions), when the full MM sample is considered there is a wide variety of combinations of $q$ and observed wavelength used. This means that the MM method is generally resistant to simple wavelength scale distortions. We give the distribution of the signal-to-noise ratio (SNR) for the spectral regions used in the VLT sample in figure \ref{fig_SNR_VLT}.

\begin{figure*}
\begin{center}
\ifpdf
\includegraphics[bb=78 79 540 727,angle=-90,width=160mm]{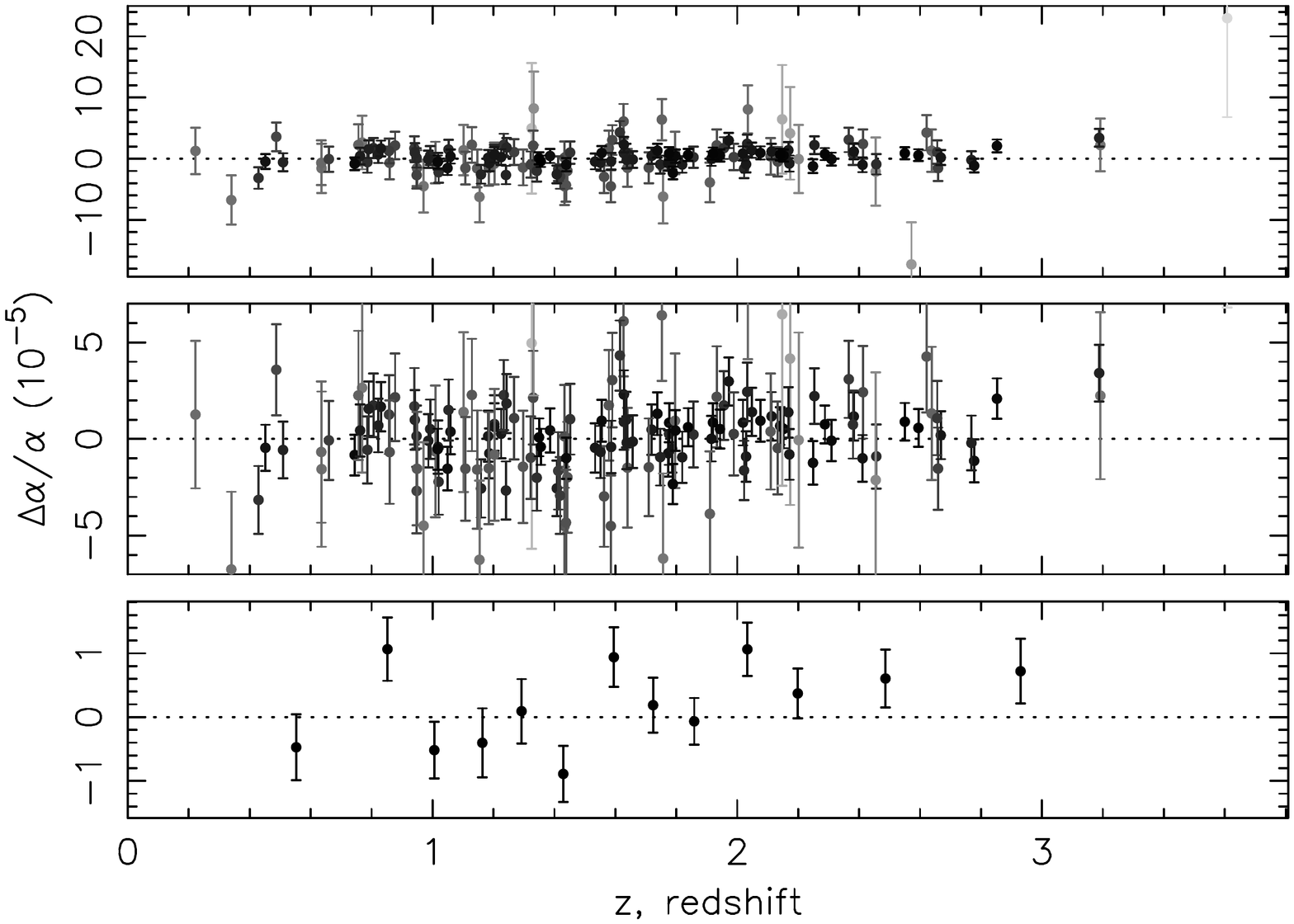}
\else
\includegraphics[bb=78 79 540 727,angle=-90,width=160mm]{images/zstack_VLT.eps}
\fi
\end{center}
\caption{Values of $\Delta\alpha/\alpha$ for the VLT sample. The top panel shows all values of $\Delta\alpha/\alpha$ with error bars increased in quadrature with $\sigma_\mathrm{rand} = 0.905 \times 10^{-5}$. The midel panel shows the same data as the top panel, with with the vertical range restricted for better viewing of the higher statistical weight points. Both of the panels have been shaded according to a greyscale as the logarithm of the uncertainty estimate, with lower uncertainty points being darker.
The bottom panel shows binned values of $\Delta\alpha/\alpha$ where approximately 12 points contribute to each bin. At $z>1.5$, $\Delta\alpha/\alpha>0$. This contrasts with \citet{Murphy:04:LNP} (see fig.\ 6 of that paper), where $\Delta\alpha/\alpha<0$ for $z>1.5$. This suggests that a weighted mean model is not a good description of the $\Delta\alpha/\alpha$ values. \label{fig_zstack_VLT}}
\end{figure*}

\begin{figure}
\begin{center}
\ifpdf
\includegraphics[bb=27 61 550 770,angle=-90,width=82.5mm]{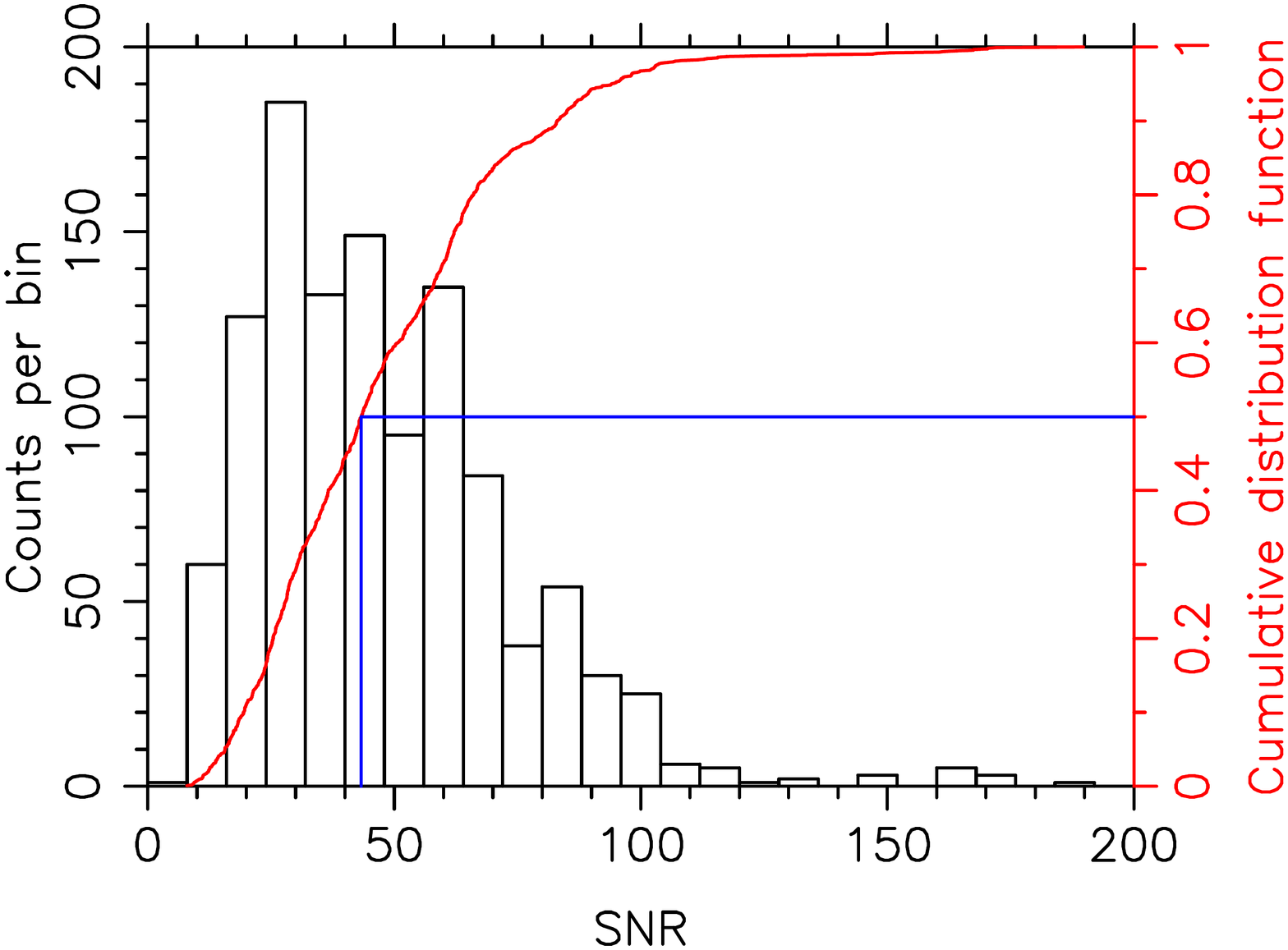}
\else
\includegraphics[bb=27 61 550 770,angle=-90,width=82.5mm]{images/SNRhist.eps}
\fi
\end{center}
\caption{Distribution of the signal-to-noise ratio (SNR) for different spectral regions in the VLT sample. The histogram gives the distribution of SNR values, and the red line gives the cumulative distribution function. The blue line indicates the median SNR of $\approx 43$. The SNR ratio was calculated as the inverse of the median of the ratio of the flux array to the error array for the (up to) 11 pixels on either side of the spectral region. By ``up to'', we note that for some spectral regions less than 11 pixels are available on either side due to gaps in the spectral coverage, or the fact that pixels had been clipped out. The error array has been modified according to the algorithm described in section \ref{s_extractionproblems}, and so the SNR estimates given here are more appropriate than what would result from considering the raw error array. Note that the median SNR is somewhat higher than for the sample in \citet{Murphy:03}, which was $\approx 31$. \label{fig_SNR_VLT}}
\end{figure}

Our VLT results are summarised as samples 3, 5 and 6 in table \ref{combineresults_dipole}. We show the values of $\Delta\alpha/\alpha$ for the VLT sample against redshift in figure \ref{fig_zstack_VLT}.

Appendix \ref{appendix_VPfits} (in figures \ref{fig_VPexample1} and \ref{fig_VPexample2}) gives two examples of the Voigt profile fits for the VLT absorbers. The fits to all of the absorbers and an ASCII version of table \ref{VLT_daoa_results} can be found at \url{http://astronomy.swin.edu.au/~mmurphy/pub.html}.

\begin{table}
\begin{center}
\caption{The frequency of occurrence for each MM transition in our fits. Note that the Mg\,\textsc{i} $\lambda 2026$ transition is fairly weak compared Mg\,\iscs $\lambda2852$, and so it is included in few fits. Where we have included Mg\,\textsc{i} $\lambda2052$ in our fit, and Zn\,\textsc{ii} $\lambda 2026$, is included in our fit, Mg\,\textsc{i} $\lambda2026$ will also be modelled, although the contribution may be extremely minor. Nevertheless, we count this as an occurrence of Mg\,\textsc{i} $\lambda2026$, as that transition is included in our model. The transition key provides a convenient, short-hand way of referring to a particular transition. This key is used in table \ref{VLT_daoa_results}. \label{tab_freqtransitions}}
\begin{tabular}{lccc}
\hline
\multicolumn{1}{c}{Transition}& $q$ (cm$^{-1}$) & Key & Frequency of occurrence \\\hline
Mg{\sc \,i} $\lambda$2026     & 87                 &  $a_1$  &  3 \\
Mg{\sc \,i} $\lambda$2852     & 86                 &  $a_2$ &   53 \vspace{0.2cm}\\
Mg{\sc \,ii} $\lambda$2796    & 211                  &  $b_1$ &  88  \\
Mg{\sc \,ii} $\lambda$2803    & 120                   &  $b_2$ &   86  \vspace{0.2cm}\\
Al{\sc \,ii} $\lambda$1670    & 270                  &  $c_1$ &  60  \vspace{0.2cm}\\
Al{\sc \,iii} $\lambda$1854   & 464                  &  $d_1$ &  25  \\
Al{\sc \,iii} $\lambda$1862   & 216                 &  $d_2$ &   25  \vspace{0.2cm}\\
Si{\sc \,ii} $\lambda$1526    & 50                  &  $e_1$ &   57 \\
Si{\sc \,ii} $\lambda$1808    & 520                  &  $e_2$ &   31 \vspace{0.2cm}\\
Cr{\sc \,ii} $\lambda$2056    & -1110                 &  $h_1$ &  21  \\
Cr{\sc \,ii} $\lambda$2062    & -1280                 &  $h_2$ &   15 \\
Cr{\sc \,ii} $\lambda$2066    & -1360                  &  $h_3$ &  17   \vspace{0.2cm}\\
Fe{\sc \,ii} $\lambda$1608    & -1300                  &  $j_1$ &   50\\
Fe{\sc \,ii} $\lambda$1611    & 1100                  &  $j_2$ &   9 \\
Fe{\sc \,ii} $\lambda$2260    & 1435                  &  $j_3$ &   12 \\
Fe{\sc \,ii} $\lambda$2344    & 1210                  &  $j_4$ &   97 \\
Fe{\sc \,ii} $\lambda$2374    & 1590                  &  $j_5$ &   51\\
Fe{\sc \,ii} $\lambda$2382    & 1460                   &  $j_6$ &   100 \\
Fe{\sc \,ii} $\lambda$2587    & 1490                  &  $j_7$ &    74\\
Fe{\sc \,ii} $\lambda$2600    & 1330                   &  $j_8$ &   97 \vspace{0.2cm}\\
Mn{\sc \,ii} $\lambda$2576    & 1420               & $i_1$ &    13 \\
Mn{\sc \,ii} $\lambda$2594    & 1148                 &  $i_2$ &  9  \\
Mn{\sc \,ii} $\lambda$2606    & 986               &  $i_3$ &    9 \vspace{0.2cm}\\
Ni{\sc \,ii} $\lambda$1709    & -20                 &  $k_1$ &  22 \\
Ni{\sc \,ii} $\lambda$1741    & -1400                  &  $k_2$   &24  \\
Ni{\sc \,ii} $\lambda$1751    & -700                  &  $k_3$ &  21 \vspace{0.2cm}\\
Ti{\sc \,ii} $\lambda$3067    & 791                 &  $g_1$ &  0\\
Ti{\sc \,ii} $\lambda$3073    & 677                 &  $g_2$ &  0\\
Ti{\sc \,ii} $\lambda$3230    & 673                 &  $g_3$ &  0\\
Ti{\sc \,ii} $\lambda$3342    & 541                 &  $g_4$ &  1\\
Ti{\sc \,ii} $\lambda$3384    & 396                 &  $g_5$ &  1\vspace{0.2cm}\\
Zn{\sc \,ii} $\lambda$2026    & 2479                  &  $l_1$ &  9\\
Zn{\sc \,ii} $\lambda$2062    & 1584                  &  $l_2$ &  13 \\\hline
\end{tabular}
\end{center}
\end{table}

\begin{table*}
\centering
\begin{minipage}{173mm}
  \centering
  \caption{\emph{First table:} Constraints from different models for $\Delta\alpha/\alpha$ from Keck and VLT spectra. $I$ gives a number to identify the sample + model. The samples are as described in the text. $N_\mathrm{abs}$ gives the number of absorption systems used for the fit. $m$ gives the monopole component of the dipole fit for dipole models, and the weighted mean of the $\Delta\alpha/\alpha$ values for a weighted mean model. The RA and dec.\ columns specify the right ascension and declination of the fitted pole. The $A$ column gives the amplitude of the dipole, and the $\delta(A)$ column gives the $1\sigma$ confidence limits on the dipole amplitude. Raw statistical uncertainties have been multiplied by $\sqrt{\chi^2_\nu}$ to obtain confidence limits \citep{NumericalRecipes:92}, with the exception of the uncertainty on the dipole amplitude, which is derived from bootstrapping. $A$ is unitless, except for the $r$-dipole where $A$ has units of $\mathrm{GLyr}^{-1}$. The column labelled ``significance'' gives the significance of the dipole model over the monopole model, as both a probability and its $\sigma$ equivalent, assessed using a bootstrap method. However, for models with no monopole, the significance is given with respect to the null model ($\Delta\alpha/\alpha = 0$). \emph{Second table:} The values of $\sigma_\mathrm{rand}$ used for each model. For the Keck $\Delta\alpha/\alpha$ values, LC and HC refer to low contrast and high contrast respectively. \label{combineresults_dipole}}
  \begin{tabular}{cllrccccc}
  \hline
   $I$  & Sample + model           & $N_\mathrm{abs}$ & $m\ (10^{-5})$ & RA (hr)           & dec.\ ($^\circ$) & $A\ (10^{-5})$         & $\delta A\ (10^{-5})$       & significance \\
  \hline
   1           & Keck04-dipole           &    $140$   & $-0.465\pm0.145$     & $16.0\pm 2.7$ &  $-47\pm29$   &  $0.41$   &    $[0.29, 0.78]$    &  $36$ percent ($0.5\sigma$)\\
   2           & \# 1 with no monopole   &    $140$   &  N/A                 & $16.4\pm 1.2$ &  $-56\pm12$   &  $1.06$   &    $[0.82, 1.34]$    &  $72$ percent ($1.1\sigma$)\\ 
   3           & VLT-weighted mean     &    $153$     &   $0.208\pm0.124$  &   N/A            & N/A                  &  N/A        &  N/A               &  N/A \\
   4           & Combined weighted mean &   $293$     &   $-0.216 \pm 0.086$ &  N/A           & N/A                  &  N/A        &  N/A               & N/A\\
   5           & VLT-dipole            &    $153$     &   $-0.109\pm0.180$  &   $18.3\pm1.2$  &  $-62\pm13$  & 1.18  & $[0.80,1.66]$  &   97.1 percent ($2.19\sigma$)\\
   6           & \#5 with no monopole  &    $153$     &   N/A                &  $18.4\pm1.3$  &  $-58\pm15$  & 0.99  & $[0.70,1.37]$  &   98.4 percent ($2.39\sigma$)\\
   7           & Combined dipole       &    $293$     &   $-0.178\pm0.084$  & $17.3\pm1.0$    &   $-61\pm10$   &   $0.97$     &  $[0.77,1.19]$ &  $99.995$ percent ($4.06\sigma$)\\
   8           & \#7 with no monopole &    $293$     &   N/A             &  $17.4 \pm 0.9$   &   $-58 \pm 9$  &   $1.02$     &  $[0.83, 1.24]$ & $99.996$ percent ($4.14\sigma$)\\
   9           & Combined $r$-dipole   &    $293$     &   $-0.187\pm0.084$  & $17.5\pm1.0$     &   $-62\pm10$  &  $0.11$    & $[0.09,0.13]$   &  $99.997$ percent ($4.15\sigma$)\\
   10           & \#9 with no monopole   &    $293$     &   N/A  & $17.5\pm0.9$     &   $-58\pm9$  &  $0.11$    & $[0.09,0.14]$   &  $99.998$ percent ($4.22\sigma$)\\
   11           & $z^\beta$ dipole, $\beta = 0.46\pm0.49$   &    $293$     &   $-0.184 \pm 0.085$  & $17.5\pm1.1$     &   $-62\pm10$  &  $0.81$    & $[0.55,1.09]$   &  $99.99$ percent ($3.9\sigma$)\\
  \hline
 \\
  \end{tabular}
  \begin{tabular}{cllll}
  \hline
  $I$ & Sample + model            & $\sigma_\mathrm{rand}$(VLT) $(10^{-5})$ & $\sigma_\mathrm{rand}$(Keck LC) $(10^{-5})$ & $\sigma_\mathrm{rand}$(Keck HC) $(10^{-5})$\\
  \hline
  1   & Keck04-dipole             & N/A                         & 0            & 1.630  \\
  2   & \#1 with no monopole      & N/A                         & 0            & 1.668  \\
  3   & VLT-weighted mean         & 0.905                       & N/A          & N/A \\
  4   & Combined weighted mean    & 0.905                       & 0            & 1.743\\
  5   & VLT-dipole                & 0.905                       & N/A          & N/A \\
  6   & \#5 with no monopole      & 0.882                       & N/A          & N/A \\
  7   & Combined dipole           & 0.905                       & 0            & 1.630 \\
  8   & \#7 with no monopole      & 0.882                       & 0            & 1.668 \\
  9   & Combined $r$-dipole       & 0.858                       & 0            & 1.630\\
  10   & \#9 with no monopole      & 0.858                       & 0            & 1.630\\
  11  & $z^\beta$ dipole          & 0.812                       & 0            & 1.592\\
  \hline
  \end{tabular}
\end{minipage}
  
\end{table*}

\subsection{Weighted mean for the VLT $\Delta\alpha/\alpha$ values}\label{s_VLTwmean}

The LTS method applied to a weighted mean model indicates that the $z=1.542$ absorber toward J000448$-$415728 is an outlier, with a residual of $4.2\sigma$ about the LTS fit, and so we remove this $\Delta\alpha/\alpha$ value. If we do not remove this value, the weighted mean after increasing errors is $\Delta\alpha/\alpha = (0.154 \pm 0.132) \times 10^{-5}$, with $\chi^2_\nu=1.06$ ($\sigma_{\mathrm{rand}}=0.951 \times 10^{-5}$). 

After removing this value when calculating our weighted mean result for the VLT sample, a weighted mean fit with our raw statistical errors yields $\Delta\alpha/\alpha = (0.229 \pm 0.095) \times 10^{-5}$, with $\chi^2_\nu = 1.78$. Applying the LTS method to this data set yields a random error estimate of $\sigma_\mathrm{rand} = 0.905 \times 10^{-5}$. 

After accounting for this extra random error, the weighted mean becomes $\Delta\alpha/\alpha = (0.208 \pm 0.124) \times 10^{-5}$, with $\chi^2_\nu = 0.99$. This result differs from that of \citet{Murphy:04:LNP} at the $\sim 4.7\sigma$ level. 

\subsubsection{Distribution of $\Delta\alpha/\alpha$ values with redshift and validity of a weighted mean model}

In the bottom panel of figure \ref{fig_zstack_VLT} we show binned values of $\Delta\alpha/\alpha$ plotted against redshift for the VLT sample.  For $z<1.5$, 3 of the 5 binned points fall in the region $\Delta\alpha/\alpha < 0$. For $z>1.5$, 6 of 7 points in the binned plot fall in the region $\Delta\alpha/\alpha > 0$. This trend with redshift is different to that seen in fig.\ 6 of \citet{Murphy:04:LNP}; for $z<1.6$, all 7 points fall in the region $\Delta\alpha/\alpha < 0$, whereas for $z>1.5$ all 6 points also fall in the region $\Delta\alpha/\alpha < 0$. The apparent change in sign of $\Delta\alpha/\alpha$ with $z$ in the VLT sample suggests that a weighted mean model is not a good description of the VLT data.

\subsection{Angular dipole fit applied to the VLT $\Delta\alpha/\alpha$ values}\label{s_dipolefit}

In this section, we fit the angular dipole model of equation \ref{dipole_eq} to the VLT sample.  

Inspection of the residuals about the LTS angular dipole fit, plotted as a function of redshift, reveals no obvious trend for higher scatter at higher redshifts and therefore we treat all absorbers the same in attempting to estimate $\sigma_\mathrm{rand}$. We again identify the $z = 1.542$ system toward J000448$-$415728 as an outlier, with a residual of $4.6\sigma$ about the LTS fit, even after increasing the error bars. Thus, we remove this system from our sample when fitting the dipole model, and re-estimate $\sigma_\mathrm{rand} = 0.905 \times 10^{-5}$. We call this sample ``VLT-dipole''.

Our dipole fit parameters after adding $\sigma_\mathrm{rand} = 0.905 \times 10^{-5}$ in quadrature to all error bars are: $m = (-0.109 \pm 0.180) \times 10^{-5}$, $A = 1.18 \times 10^{-5}$ ($1\sigma$ confidence limits $[0.80, 1.66] \times 10^{-5}$), $\mathrm{RA}= (18.3 \pm 1.2)\,  \mathrm{hr}$ and $\mathrm{dec.}= (-62 \pm 13) ^\circ$. For this fit, $\chi^2 = 141.8$ and $\chi^2_\nu = 0.95$.

\begin{figure}
\begin{center}
\ifpdf
\includegraphics[bb=77 92 556 742,angle=-90,width=82.5mm]{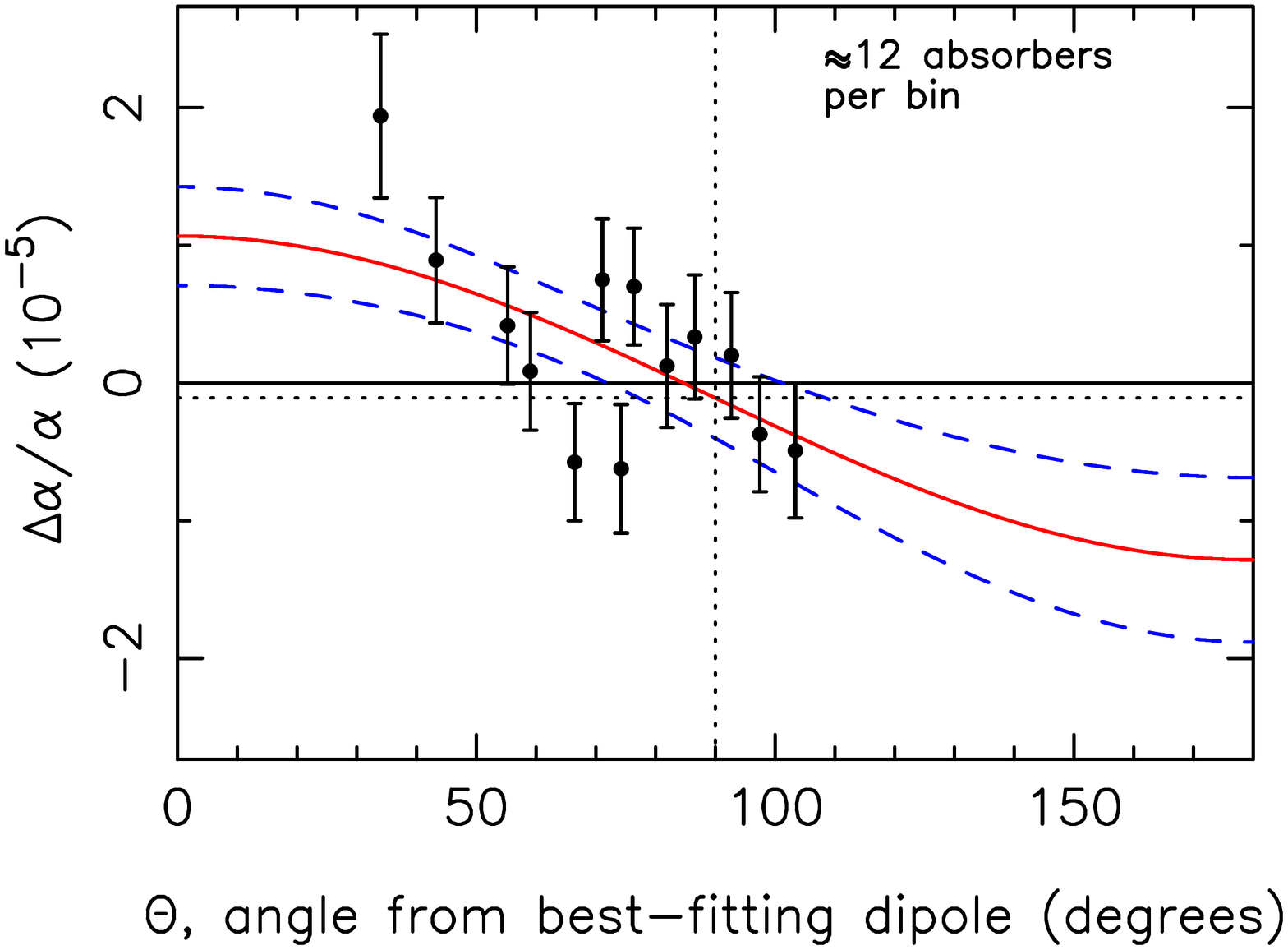}
\else
\includegraphics[bb=77 92 556 742,angle=-90,width=82.5mm]{images/VLT_dipole.eps}
\fi
\end{center}
\caption{Binned values of $\Delta\alpha/\alpha$ plotted against angle to the best-fitting dipole for the VLT sample. The red, solid line is the model $\Delta\alpha/\alpha = A\cos(\Theta) + m$, and the dashed, blue lines indicate the $1\sigma$ uncertainty on the dipole fit. Statistical errors have been increased prior to binning as described in the text. The dipole+monopole model is preferred over monopole-only model at the $2.2\sigma$ level (see section \ref{s_dipolefit}). The parameters for this fit are: $m = (-0.109 \pm 0.180) \times 10^{-5}$, $A = 1.18 \times 10^{-5}$ ($1\sigma$ confidence limits $[0.80, 1.66] \times 10^{-5}$), $\mathrm{RA}= (18.3 \pm 1.2)\,  \mathrm{hr}$ and $\mathrm{dec.}= (-62 \pm 13) ^\circ$. \label{fig_VLT_dipole}}
\end{figure}

To assess the dipole fit compared to a monopole-only (weighted mean) fit, we compare a weighted mean fit with errors adjusted according to \emph{the same} $\sigma_\mathrm{rand}$ as used for the dipole fit, in order to ensure consistency of the data points used when calculating the statistical significance. As the weighted mean fit has $\chi^2 = 149.8$, the dipole fit yields a reduction in $\chi^2$ of $7.9$ for an extra 3 degrees of freedom, when a reduction of $\sim3$ would be expected by chance. Our bootstrap method yields a significance for the dipole+monopole model over the monopole-only model at the 97.1 percent confidence level ($2.19\sigma$), indicating marginal evidence for the existence of a dipole when considering only the VLT data. We demonstrate this fit in figure \ref{fig_VLT_dipole}.

We also give the parameters for a dipole-only (no monopole) fit in table \ref{combineresults_dipole}. 

\subsubsection{Effect of the choice of the method-of-moments estimator}

In section \ref{s_physicalconstraints} we suggested that a method-of-moments estimator was preferable in attempting to reconcile $\Delta\alpha/\alpha$ values from turbulent and thermal fits. It is legitimate to ask whether our results differ if we simply choose that fit (turbulent or thermal) which has the lowest $\chi^2_\nu$, instead of applying our method-of-moments estimator. The results for a VLT dipole model if we do this are: $\sigma_\mathrm{rand}=0.928\times 10^{-5}$, $m = (-0.112\pm0.184)\times 10^{-5}$, $A=1.15 \times 10^{-5}$ ($1 \sigma$ confidence limits $[0.76,1.63]\times 10^{-5}$), $\mathrm{RA} = (18.2\pm 1.2)\,\mathrm{hr}$, $\mathrm{dec.} = (-62 \pm 14)^\circ$. The dipole model is preferred over the monopole model at the 96 percent level ($2.1\sigma$). Thus our choice of the method-of-moments estimator does not change the results significantly, although $\sigma_\mathrm{rand}$ is mildly larger if we simply choose those fits which have the lowest $\chi^2_\nu$.  

\subsection{VLT results summary}

In this section, we have described the analysis of 154 new MM absorbers. The $\Delta\alpha/\alpha$ values from the VLT sample at $z\gtrsim 1.5$ tend to show $\Delta\alpha/\alpha > 0$, which contrasts with $\Delta\alpha/\alpha < 0$ from the Keck sample \citep{Murphy:04:LNP}. We showed that, in the VLT sample, an angular dipole model is preferred over a weighted mean model at the $2.2\sigma$ level, which seems to suggest angular (and therefore spatial) variations in $\alpha$. The direction of maximal increase in $\alpha$ is found to be $\mathrm{RA}= (18.3 \pm 1.2)\,  \mathrm{hr}$ and $\mathrm{dec.}= (-62 \pm 13) ^\circ$ under a dipole+monopole model. We therefore explore the consistency of our $\Delta\alpha/\alpha$ values and model parameters with those derived from the same models applied to the Keck sample, and a Keck + VLT sample, in the next section.

We have shown that the VLT $\Delta\alpha/\alpha$ values display excess scatter ($\chi^2_\nu > 1$) about the simple models described. This is likely due to both model mis-specification (from the use of simple weighted mean and angular dipole models) as well as unmodelled uncertainties. We described in section \ref{s_randsyseffect} a number of potential random effects which could give rise to excess scatter in the data, even if our model for $\Delta\alpha/\alpha$ were correct. It is difficult to determine the contribution of each of these effects to the error budget, and so we have assumed that all absorbers are affected by the same processes, and therefore increased our error bars conservatively in quadrature with a $\sigma_\mathrm{rand}$ term. If the extra scatter in the $\Delta\alpha/\alpha$ values is due to inaccuracies in modelling the velocity structure of the absorbers, it may be that observations at higher signal-to-noise ratios and higher resolving powers might help reduce the scatter. On the other hand, if the inter-component spacing is comparable to the intrinsic line widths then this may not be the case. 

We consider the specific effect of wavelength scale distortions on the VLT sample in sections \ref{s_dvtest} and \ref{s_intraorder_distortions}, and show there how such distortions can give rise to extra scatter in the $\Delta\alpha/\alpha$ values.

\section{Combination and comparison with previous Keck results}\label{s_results_combine}

We give a summary of all of the results presented here in table \ref{combineresults_dipole}.

\subsection{Previous Keck results}


We note briefly that we noticed significant wavelength calibration problems in the spectrum of Q2206$-$1958 (J220852$-$194359) from sample 3 of \citet{Murphy:03} for $\lambda \gtrsim 5000\mathrm{\AA}$ of the order of $\sim 10\,\mathrm{kms}^{-1}$ at the time of that analysis. The two absorbers contributed by this spectrum were erroneously included in that paper, and so we remove them from the sample. 

\citet{Murphy:04:LNP} divide their high-$z$ sample into two portions, a high-contrast sample and a low-contrast sample. The high-contrast sample was defined by 27 absorbers where there were significant differences between the optical depth in the transitions used. \citet{Murphy:03} give arguments as to why this might be expected to generate extra scatter in the $\Delta\alpha/\alpha$ values. Due to the fact that many of the high redshift ($z>1.8$) absorbers considered in \citet{Murphy:04:LNP} are associated with damped Lyman-$\alpha$ systems, this effect manifests itself as extra scatter in the $\Delta\alpha/\alpha$ values about a weighted mean at high redshifts. For the VLT sample, we note that there is no evidence for excess scatter at higher redshifts compared to lower redshifts. 

We can examine the differences between the Keck and VLT samples in terms of the prevalence of weak species as follows. Firstly, define the following transitions as weak: Mg\,\textsc{i} $\lambda 2026$, Si\,\iis $\lambda 1808$, the Cr\,\iis transitions, Fe\,\iis $\lambda \lambda \lambda 1608, 1611, 2260$, the Mn\,\iis transitions, the Ni\,\iis transitions, the Ti\,\iis transitions and the Zn\,\iis transitions. From the table of the frequency of occurrence of these transitions in \citet{Murphy:03}, at $z<1.8$ these transitions constitute about 3 percent of the total number of transitions used. On the other hand, in the VLT sample these transitions constitute about 13 percent of the sample used. The significantly greater prevalence of these weak transitions at low redshifts in the VLT sample may explain the lack of evidence for differential scatter between high and low redshifts. Effectively, the greater prevalence of weak species in the low-$z$ VLT sample may increase the scatter at low redshifts in that sample, making any low-$z$/high-$z$ difference appear smaller. We retain the high/low contrast distinction when analysing the Keck sample.

\subsubsection{LTS method applied to the \citet{Murphy:04:LNP} results}\label{results_Keck04_LTS}

If we apply the LTS method to the high contrast sample to estimate the extra error needed about a dipole model, we find that an extra error term of $\sigma_\mathrm{rand} = 1.630 \times 10^{-5}$ is needed. With this extra term, $\chi^2_\nu = 1.13$, indicating that the distribution is mildly leptokurtic (fat-tailed). The low-contrast sample data are already consistent under a dipole model with the LTS method ($\sigma_\mathrm{rand} = 0$). 

We then combine the high-contrast $\Delta\alpha/\alpha$ values (with error bars increased) with the low-contrast $\Delta\alpha/\alpha$ values to form a new sample under a dipole model (equation \ref{dipole_eq}). The LTS method applied to this set reveals that the $\Delta\alpha/\alpha$ values are consistent about dipole model. Additionally, $\chi^2_\nu = 1.04$. Nevertheless, we identify one possible outlier from this set: the absorber with $z\approx 2.84$ towards Q1946$+$7658, with $\Delta\alpha/\alpha = (-4.959 \pm 1.334) \times 10^{-5}$, and remove this absorber from the sample. This point has a residual of $-3.6\sigma$ about the LTS fit. We refer to this sample as ``Keck04-dipole''. 

A dipole fitted to this sample\footnote{We note that in \citet{Murphy:03} the reported position of a dipole model fitted to that data set was $(5\,\mathrm{hr},-48^\circ)$. The RA value quoted there is erroneous due to a typographical error. The value should have been stated as $(17\,\mathrm{hr},-48^\circ)$. The value reported there in Galactic coordinates (the coordinate system in which the computations were done) was correct.} yields $\mathrm{RA} = (16.0 \pm 2.7)\,\mathrm{hr}$, $\mathrm{dec.} = (-47 \pm 29)^\circ$, and $A = 0.41 \times 10^{-5}$. $1\sigma$ confidence limits on $A$ are $[0.29, 0.78] \times 10^{-5}$. The monopole is $m = (-0.465 \pm 0.145) \times 10^{-5}$. This fit has $\chi^2_\nu = 0.96$. The dipole model is preferred over the weighted mean model at the 36 percent confidence ($0.47\sigma$). 

The monopole offset appears to be significant at the $3.2\sigma$ confidence level, but this is related to the fact that the Keck results alone do not clearly support a dipole interpretation.

For dipole model with no monopole ($\Delta\alpha/\alpha = A\cos\Theta$), the fitted parameters are $A=1.06\times 10^{-5}$ ($1\sigma$ confidence limits $[0.82,1.34]\times 10^{-5}$), $\mathrm{RA} = (-16.4\pm 1.2)\,\mathrm{hr}$, $\mathrm{dec.} = (-56\pm 12)^\circ$. This model is significant at the 72 percent confidence level ($1.1\sigma$).

\subsection{Combined weighted mean}\label{s_combinewmean}

We create a combined weighted mean fit by combining the VLT-dipole sample with the Keck04-dipole sample. The VLT sample has had errors increased in quadrature with $\sigma_\mathrm{rand} = 0.905 \times 10^{-5}$, whereas the Keck high-contrast sample has had errors increased in quadrature with $\sigma_\mathrm{rand} = 1.743 \times 10^{-5}$. The same points identified as outliers have been removed.

This leads to a weighted mean of $(\Delta\alpha/\alpha)_w = (-0.216 \pm 0.086) \times 10^{-5}$, with $\chi^2_\nu = 1.03$. However, a weighted mean model does not appear to adequately capture all the information in the data (see figure \ref{fig_zstack}). Comparing the weighted mean of the $z>1.6$ points for both samples yields a simple demonstration of north/south difference. For the VLT sample, $\Delta\alpha/\alpha_w(z>1.6) = (0.533 \pm  0.172) \times 10^{-5}$, whereas for the Keck sample $\Delta\alpha/\alpha_w(z>1.6) = (-0.603 \pm 0.224) \times 10^{-5}$. The difference between these weighted means is significant at the $4\sigma$ level.

\subsection{Combined dipole fit}\label{s_combined_dipole_fit}

To create our combined dipole fit, we combine the VLT-dipole sample with the Keck04-dipole sample to create the ``combined dipole'' sample, which is our main sample. This sample consists of 293 MM absorbers. Importantly, both of these sets exhibit no $\lvert r_i \rvert \geq 3$ residuals, and thus a combined fit is unlikely to exhibit any large residuals provided that both data sets are well described by the same model. 

For an angular dipole fit to these $\Delta\alpha/\alpha$ values ($\Delta\alpha/\alpha = A\cos\Theta + m$), we find that $m = (-0.178 \pm 0.084) \times 10^{-5}$, $A = 0.97 \times 10^{-5}$ ($1\sigma$ confidence limits $[0.77, 1.19] \times 10^{-5}$), $\mathrm{RA} = (17.3 \pm 1.0)\,\mathrm{hr}$, $\mathrm{dec.} = (-61 \pm 10)^\circ$, with $\chi^2 = 280.6$ and $\chi^2_\nu = 0.97$.  A weighted mean (i.e. monopole only) fit to the same $\Delta\alpha/\alpha$ values and uncertainties yields $\chi^2 = 303.8$, and so a dipole model yields a reduction in $\chi^2$ of $23.2$ for an extra 3 free parameters. With our bootstrap method, we find that the dipole model is preferred over the weighted mean fit at the 99.995 percent confidence level ($4.06\sigma$), thus yielding significant evidence for the existence of angular variations in $\alpha$. Using the method of \citet{Cooke:09}, the significance of the dipole is  found to be $4.07\sigma$. 

Importantly, the combination of the Keck04-dipole $\Delta\alpha/\alpha$ values with the VLT-dipole $\Delta\alpha/\alpha$ values yields $\chi^2_\nu \sim 1$ about a dipole model. If inter-telescope systematics were present, we would expect the combination of the Keck and VLT data to yield a $\chi^2_\nu$ that is significantly greater than unity under the dipole model, despite $\chi^2_\nu$ being $\sim 1$ when that model is fitted to the samples individually. Thus, there is no significant evidence based on $\chi^2$ that inter-telescope systematics are present. 

\begin{figure*}
\begin{center}
\ifpdf
\includegraphics[bb=44 88 517 727,width=140mm]{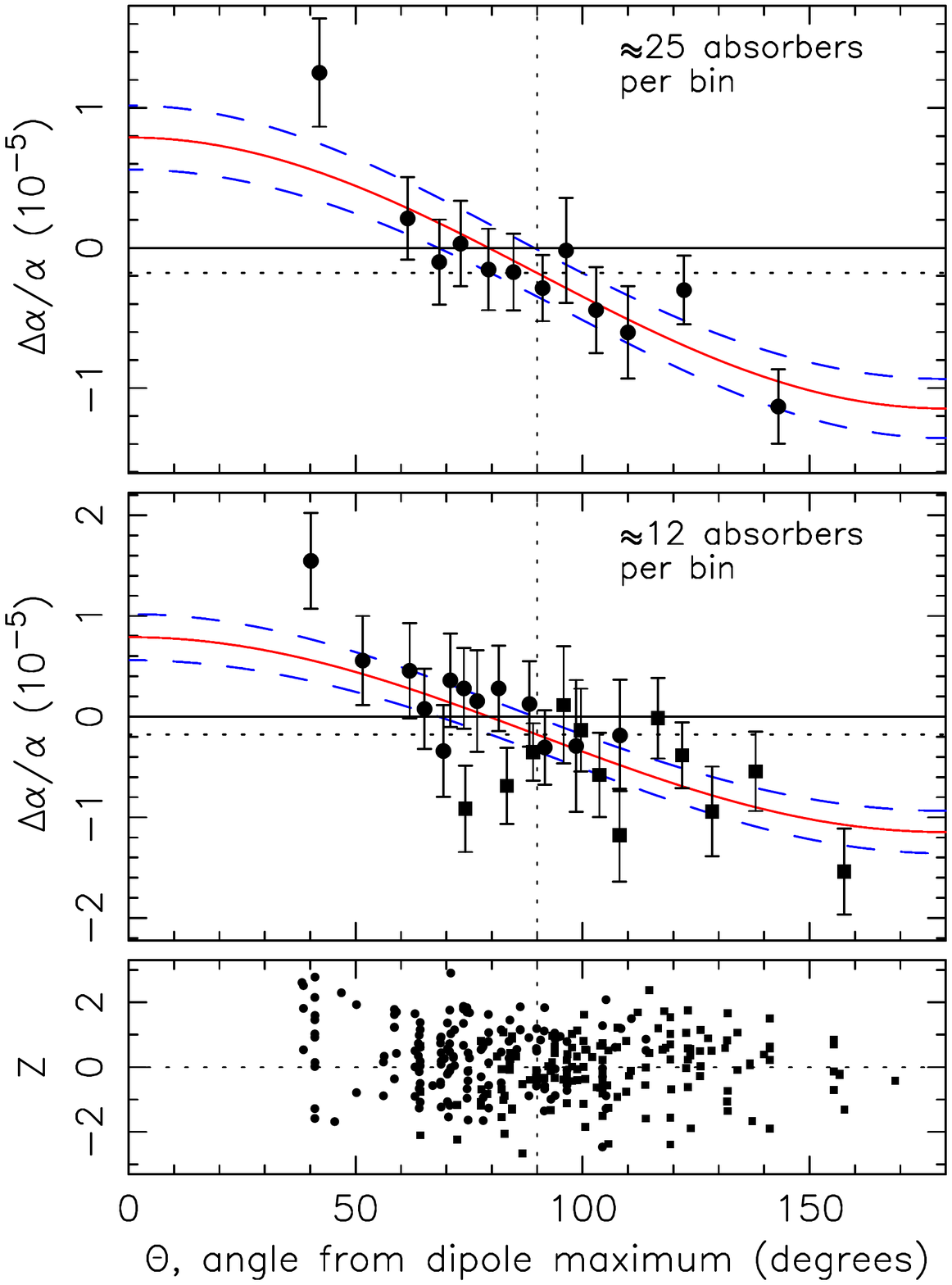}
\else
\includegraphics[bb=44 88 517 727,width=140mm]{images/VLT+Keck_dipolestack.eps}
\fi
\end{center}

\caption{The top panel shows the combined results for the Keck and VLT samples, plotting $\Delta\alpha/\alpha$ against angle from the fitted dipole location for the combination of the Keck and VLT $\Delta\alpha/\alpha$ values, binned together. The middle panel shows the data for the top panel, but separating out the different telescopes. Points in the top panel contain approximately 25 absorbers per bin, whereas the points in the middle panel contain approximately 12. In the middle and bottom panels, Keck points are indicated by squares and VLT points are indicated by circles. The model shown (red, solid line) is $\Delta\alpha/\alpha = A\cos(\Theta) + m$. The parameters for this model are: $m = (-0.178 \pm 0.084) \times 10^{-5}$, $A = 0.97 \times 10^{-5}$ ($1\sigma$ confidence limits $[0.77, 1.19] \times 10^{-5}$), $\mathrm{RA} = (17.3 \pm 1.0)\,\mathrm{hr}$, $\mathrm{dec.} = (-61 \pm 10)^\circ$.  The dashed, blue lines indicate the $1\sigma$ uncertainty on the fit, including the uncertainty in determining the position of the dipole, the amplitude of the dipole and the monopole value. In both the top and middle panels, the dotted horizontal line indicates the monopole value. The bottom panel indicates the standardised residuals ($r_i = [\mathrm{data} - \mathrm{model}]/\mathrm{error}$) about fitted model. The vertical dotted line in all panels show $90^\circ$ from the dipole pole. The presence of no points with $\lvert r_i \rvert > 3$ indicates that the fit is not being dominated by a small number of large residual points. Note that because of the direction of the dipole axis in relation to the positions of the telescopes, many of the $\Delta\alpha/\alpha$ values fall near the equatorial region of the dipole ($\Theta = 90^\circ$).\label{fig_combinedresults}}
\end{figure*}

We show in figure \ref{fig_combinedresults} the values of $\Delta\alpha/\alpha$ for both Keck and VLT against the best-fitting dipole model. We give binned values there, which yields a visual demonstration of the dipole effect. We also give there a plot of the standardised residuals about the fit, which demonstrates that the fit is statistically reasonable. We also show binned values of $\Delta\alpha/\alpha$ for the Keck, VLT and combined samples in figure \ref{fig_zstack}. We show an unbinned version of these data for $|\Delta\alpha/\alpha| < 5 \times 10^{-5}$ in figure \ref{fig_anglefromdipole_size}.

For a model with no monopole ($\Delta\alpha/\alpha = A\cos\Theta$), the fitted parameters are $A=1.02\times 10^{-5}$ ($1\sigma$ confidence limits $[0.83, 1.24]\times 10^{-5}$), $\mathrm{RA} = (17.4\pm 0.9)\times 10^{-5}$, $\mathrm{dec.} = (-58\pm 9)^\circ$. This model is significant at the 99.996 percent level ($4.14\sigma$).

\begin{figure*}
\begin{center}
\ifpdf
\includegraphics[bb=41 65 468 728,angle=-90,width=160mm]{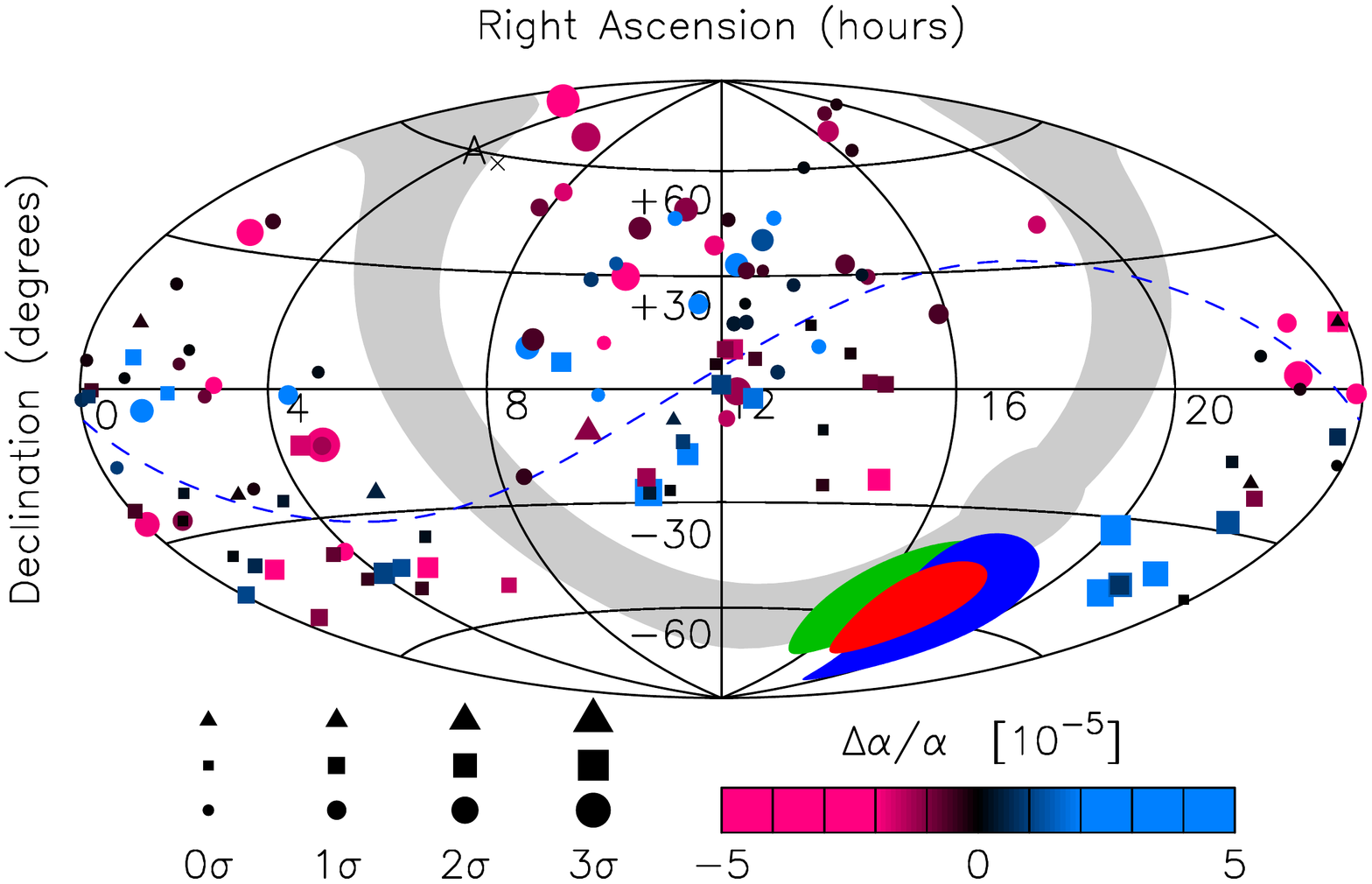}
\else
\includegraphics[bb=41 65 468 728,angle=-90,width=160mm]{images/compressed_sightlines_mult.eps}
\fi
\end{center}
\caption{Sky map in equatorial (J2000) coordinates showing $\Delta\alpha/\alpha$ values aggregated for each sightline to a quasar, where the value of $\Delta\alpha/\alpha$ along a sightline is taken as the weighted mean of the $\Delta\alpha/\alpha$ values for that sightline. This necessarily obscures any $z$ dependence that may be present. The symbols show the residual of the value of $\Delta\alpha/\alpha$ with respect to the null model, that is $r = (\Delta\alpha/\alpha)/\sigma_{\Delta\alpha/\alpha}$. Larger symbols indicate greater deviation. Circles represent Keck quasars, squares represent VLT quasars and triangles represent quasars common to both samples. The colour coding indicates the difference between the sightline value of $\Delta\alpha/\alpha$ and zero. The green shaded region indicates the $1\sigma$ confidence limit for a dipole of the form $\Delta\alpha/\alpha = A\cos(\Theta)$ fitted to values of $\Delta\alpha/\alpha$ from all Keck absorbers. The blue and red regions show the same for dipole models fitted to the VLT and VLT+Keck sample respectively. The anti-pole is marked with an ``A''. The blue, dashed line indicates the equatorial region of the dipole. The grey shaded region indicates the galactic plane, with the bulge indicating the galactic centre. The dipole is visible as more large, blue points near the pole and more large, pink points near the anti-pole. \label{fig_sightlines}}
\end{figure*}

\begin{figure*}
\begin{center}
\ifpdf
\includegraphics[bb=77 79 527 727,angle=-90,width=160mm]{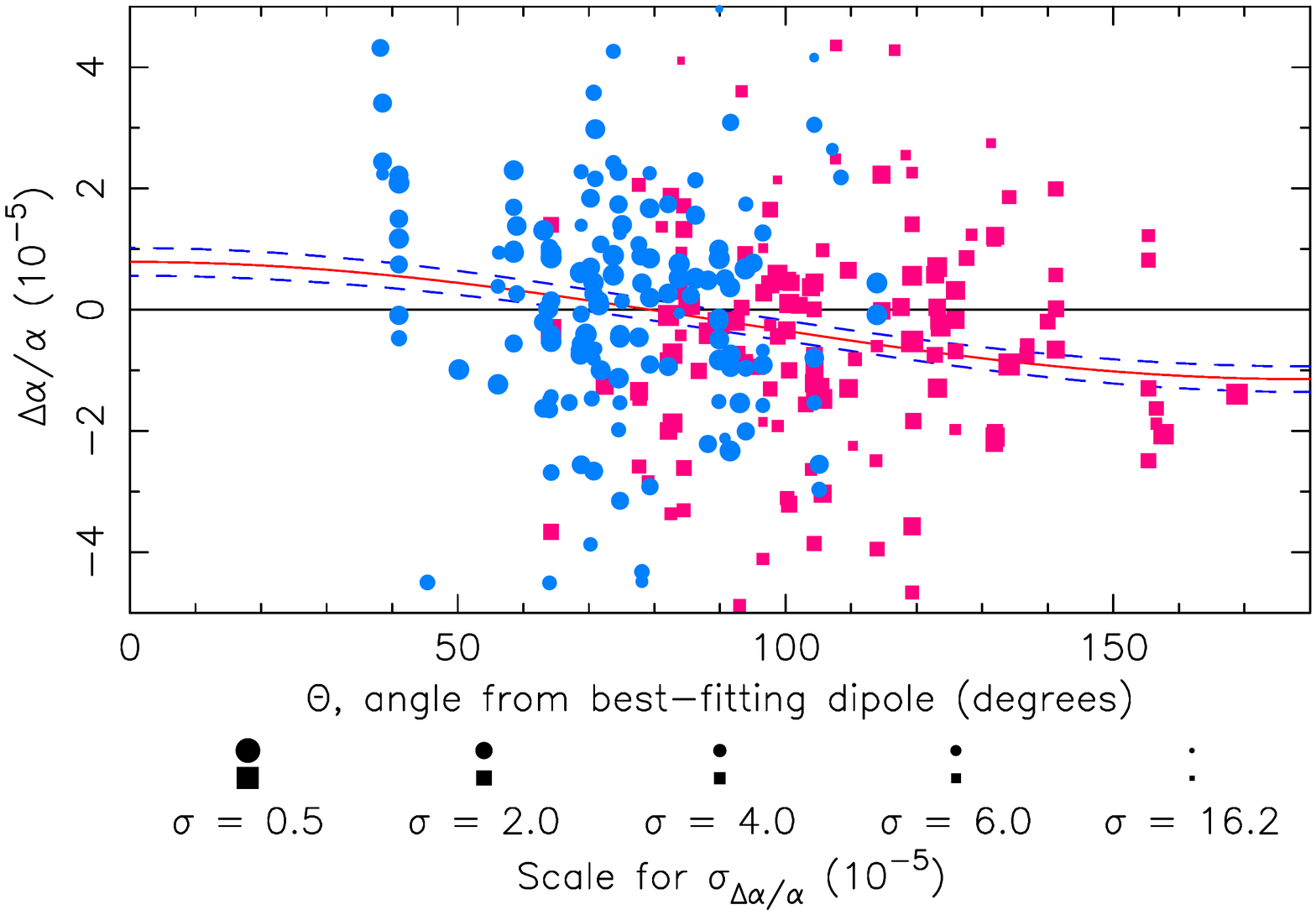}
\else
\includegraphics[bb=77 79 527 727,angle=-90,width=160mm]{images/anglefromdipole_coloursize.eps}
\fi
\end{center}
\caption{$\Delta\alpha/\alpha$ against the angle from the fitted dipole location under the model $\Delta\alpha/\alpha = A\cos(\Theta) + m$ for the VLT and Keck $\Delta\alpha/\alpha$ values. In contrast to figure \ref{fig_combinedresults}, these data are not binned. Blue circles are VLT absorbers and pink squares are Keck absorbers. Error bars have been omitted; instead, larger points indicate points with greater statistical weight, according to the key provided. The precision includes the effect of $\sigma_\mathrm{rand}$. The dipole trend is visible as the presence of more and larger points in the upper left and lower right quadrants. The visual cluster of points at $\Theta < 47^\circ$ is due to 4 quasars which contribute 14 values of $\Delta\alpha/\alpha$ (2 points not shown because they lie outside the vertical range of the plot). One can investigate the consistency of the VLT and Keck $\Delta\alpha/\alpha$ values in the region near the dipole equator (defined here as $80^\circ < \Theta < 100^\circ$) by comparing the weighted mean of the $\Delta\alpha/\alpha$ values. In this case, $\Delta\alpha/\alpha_w(\mathrm{VLT}) - \Delta\alpha/\alpha_w(\mathrm{Keck}) = (0.32 \pm 0.19) \times 10^{-5}$, giving no significant evidence for a difference between the two samples. In this region, the VLT sample contributes 39 points and the Keck sample contributes 43 points. The difference here is calculated so as to include the effect of $\sigma_\mathrm{rand}$. \label{fig_anglefromdipole_size} }
\end{figure*}

\begin{figure*}
\begin{center}
\ifpdf
\includegraphics[bb=78 40 526 751,angle=-90,width=160mm]{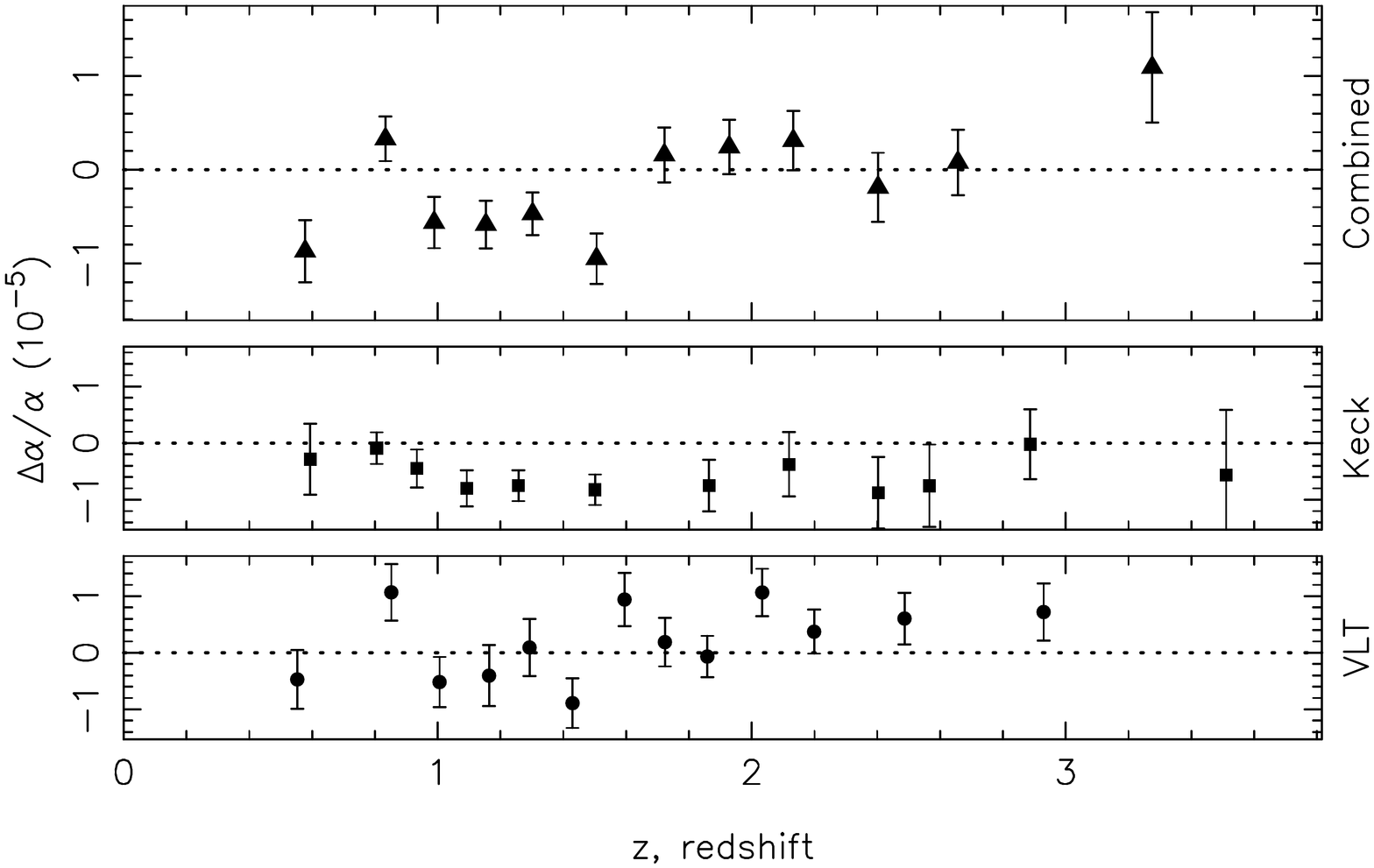}
\else
\includegraphics[bb=78 40 526 751,angle=-90,width=160mm]{images/VLT+Keck_zstack.eps}
\fi
\end{center}
\caption{Binned values of $\Delta\alpha/\alpha$ by redshift in the VLT-dipole sample (bottom panel, circles), the Keck04-dipole sample (middle panel, squares) and the combination of the two (top panel, triangles). The value of $\Delta\alpha/\alpha$ for each bin is calculated as the weighted mean of the values of $\Delta\alpha/\alpha$ from the contributing absorbers. The statistical errors for certain points have been increased prior to binning, as described in the text. Note that for $z \ga 1.5$, the Keck values generally indicate $\Delta\alpha/\alpha < 0$, whereas the VLT values indicate $\Delta\alpha/\alpha > 0$. As Keck is located in the northern hemisphere, and VLT is in the south, this is a rough visual demonstration of the dipole effect. However, given the overlap between the samples, the proper procedure is to directly fit a dipole (see figure \ref{fig_combinedresults}). Interestingly, both the VLT and Keck values seem to to support $\Delta\alpha/\alpha < 0$ for $z \la 1.5$. This effect is considered in more detail in section \ref{s_monopole}. \label{fig_zstack}}
\end{figure*}

In figure \ref{fig_sightlines}, we show the confidence limits on the dipole location for the model $\Delta\alpha/\alpha = A\cos\Theta$ for the Keck, VLT and combined samples. The individual symbols illustrate the weighted mean of $\Delta\alpha/\alpha$ along each sightline. 

There are several significant points to consider from these results:

\emph{i) The dipole is statistically significant.} Even after accounting for random errors in a conservative fashion, the statistical significance of the dipole is greater than $4\sigma$. This is strong statistical evidence for angular and therefore spatial variation in $\alpha$.

\emph{ii) Dipole models fitted to the Keck and VLT $\Delta\alpha/\alpha$ values yield consistent estimates for the pole direction.} This is important, and would be very surprising if one assumes that a dipole effect is not present. If two different systematic effects were operating in each telescope so as to produce a trend in $\Delta\alpha/\alpha$, then: \emph{a)} it is unlikely that these effects would be correlated with sky position (the most likely systematics relate to problems with wavelength calibration), and \emph{b)} even if systematic effects existed in both telescopes which were correlated with sky position, it is very unlikely that such effects would occur in such a way as to yield very consistent estimates of the dipole position between the two telescopes, with a similar amplitude, particularly when the two telescopes are independently constructed and separated by $\sim 45^\circ$ in latitude. Any attempt to ascribe the observed variation in $\alpha$ to systematics must account for the good alignment of the dipole vectors from dipole models fitted independently to the Keck and VLT samples. Note that telescope or instrumental systematics which depend only on wavelength \emph{cannot} produce observed angular variation in $\alpha$ for a sufficiently large sample of absorbers. 

\emph{iii) The VLT and Keck $\Delta\alpha/\alpha$ values appear consistent near the equatorial region of the dipole.} From the middle panel of figure \ref{fig_combinedresults}, both the VLT and Keck results show large variation from $\Delta\alpha/\alpha=0$ near the pole ($\Theta=0^\circ$) and anti-pole ($\Theta=180^\circ$) of the dipole, but show much less variation in the equatorial region ($\Theta=90^\circ$). So, at least visually, the Keck and VLT points are not inconsistent in the region where they overlap. This issue is addressed quantitatively in the caption to figure \ref{fig_anglefromdipole_size}.

\emph{iv) The dipole effect is not being caused by large residual points.} The bottom panel of figure \ref{fig_combinedresults} clearly shows that there are no $|r_i|>3\sigma$ points present.

\subsection{Potential effect of differences in atomic data and $q$ coefficients}

If the atomic data or $q$ coefficients we used were significantly different to those used by \citet{Murphy:04:LNP}, this could spuriously create differences in $\Delta\alpha/\alpha$ between VLT and Keck. This has the potential to mimic spatial variation in $\alpha$. To check the influence of this, we re-fit the VLT spectra using the same atomic data used by \citeauthor{Murphy:04:LNP}, and then combine the $\Delta\alpha/\alpha$ values with the Keck values. Where we use transitions that were not available to \citeauthor{Murphy:04:LNP} (e.g. Mn~\iis and Ti~\ii) we make no modification to the atomic data or $q$ coefficients. The frequency of occurrence of these transitions in the sample is small and therefore this is of little consequence. When we proceed in this way, the parameters for the model $\Delta\alpha/\alpha=A\cos\Theta+m$ are: $A=0.97\times 10^{-5}$, $\mathrm{RA}=(17.5\pm1.0)\,\mathrm{hr}$, $\mathrm{dec.}=(-60\pm 10)^\circ$ and $m=(-0.168\pm0.084)\times 10^{-5}$. The significance of the dipole+monopole model with respect to the monopole-only model is $4.15\sigma$. We conclude that the impact of any variations between atomic data or the $q$ coefficients used for our fits and those used by \citet{Murphy:04:LNP} is negligble.

\subsection{Alignment by chance between Keck and VLT}\label{s_alignmentbychance}

One can pose the question: ``Given the distribution of sightlines and values of $\Delta\alpha/\alpha$ in each sample, what is the probability of observing alignment as good or better than that observed between the Keck and VLT samples by chance?'' To assess this, we undertake a bootstrap analysis, where at each bootstrap iteration we randomly reassign the values of $\Delta\alpha/\alpha$ in both the Keck and VLT samples to different sightlines within those samples, keeping the redshifts of the absorbers fixed. That is, we do not mix the two samples. We then calculate the best-fitting dipole vectors for each sample, and calculate the angle between them. We then assess over many iterations in what percentage of cases is the fitted angle smaller than the angle for our actual data. 

\begin{figure}
\begin{center}
\ifpdf
\includegraphics[bb=101 54 556 696,angle=-90,width=82.5mm]{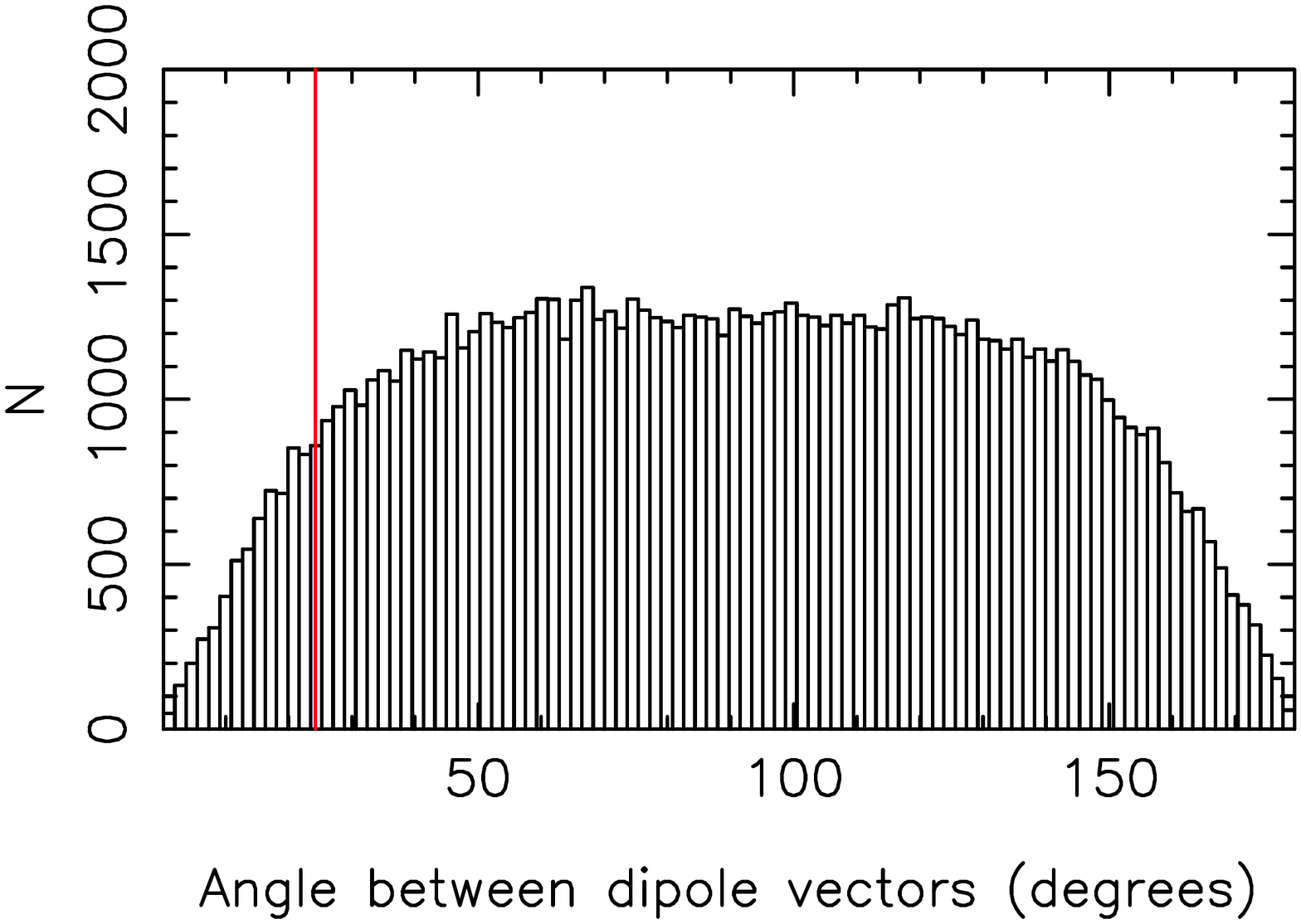}
\else
\includegraphics[bb=101 54 556 696,angle=-90,width=82.5mm]{images/doublebootstrap.eps}
\fi
\end{center}
\caption{Results of the bootstrap analysis described in section \ref{s_alignmentbychance} to assess the probability of obtaining alignment of the dipole vectors between the VLT and Keck samples as good as we have observed by chance. The vertical red line shows the angle between the dipole vectors of our sample (24 degrees). The area to the left of the red line indicates the probability of interest, namely 6 percent. \label{fig_alignmentbychance}}
\end{figure}

For our actual Keck and VLT samples, the angle between the fitted dipole vectors is 24 degrees, and the chance probability is $\approx 6$ percent. We show the results of this bootstrap analysis in figure \ref{fig_alignmentbychance}. Thus, it seems unlikely that inter-telescope systematics are responsible for the observed effect. The good consistency between the results also qualitatively supports the notion that the measured effect is real.

\subsection{Low-$z$ vs high-$z$}\label{s_lowz_vs_highz}

We divide our sample into low-$z$ and high-$z$ absorbers to examine the contribution of the different redshifts to the dipole detection. Although there is no clear delineation between which transitions are fitted for a given redshift, we can generally say that the low-$z$ sample is dominated by the Mg/Fe combination, that intermediate redshifts display a wide range of transitions, and that high redshift systems are dominated by the Si \ii/Al \ii/Fe\,\textsc{ii} $\lambda 1608$ combination. In particular, Mg\,\ii, Mg\.\iscs $\lambda2852$ and the Fe\,\iis transitions with $\lambda_0\gtrsim 2200$ are not generally used when fitting absorbers at high $z$ because they are either beyond the red cut-off in the observed spectral range, or the transitions are affected by sky absorption or emission.

If the observed dipole effect was due to chance or caused by a systematic effect which affects some combination of transitions, then we would not expect dipole fits to absorbers from high and low redshift to yield the same location on the sky. Conversely, if dipole models fitted to high and low redshift samples point in a similar direction, this lends support to the dipole interpretation of the data. 

\begin{figure}
\begin{center}
\ifpdf
\includegraphics[viewport=77 78 455 727,angle=-90,width=82.5mm]{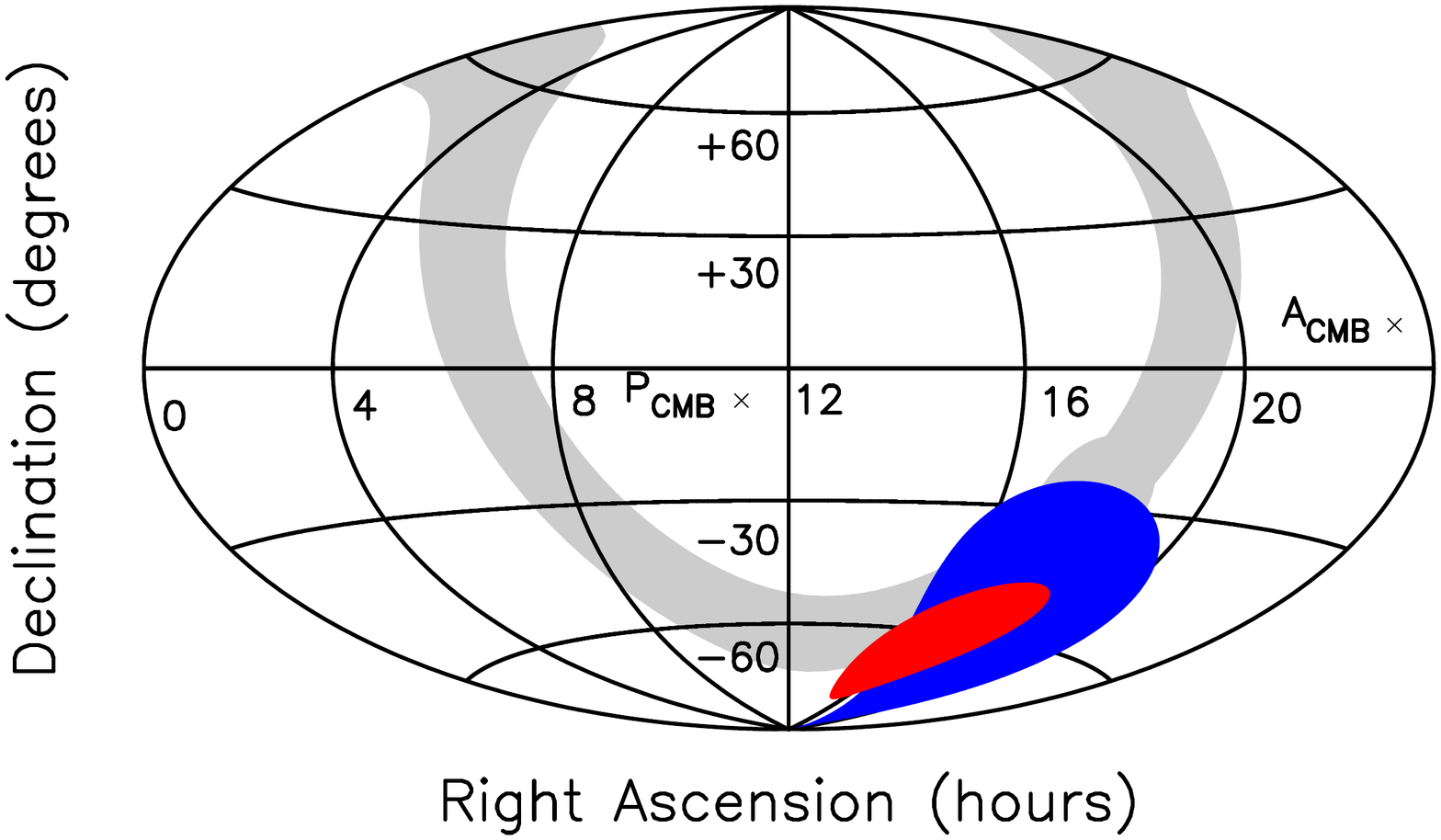}
\else
\includegraphics[bb=77 78 455 727,angle=-90,width=82.5mm]{images/skymap_lowvshigh.eps}
\fi
\end{center}
\caption{Sky map in equatorial (J2000) coordinates showing the 68.3 percent ($1\sigma$ equivalent) confidence limits of the location of the pole of the dipole for the $z<1.6$ combined sample (blue region) and $z > 1.6$ combined sample (red region) under the model $\Delta\alpha/\alpha = A\cos(\Theta) + m$. The location of the CMB dipole and anti-pole are marked as P$_\mathrm{CMB}$ and A$_\mathrm{CMB}$ respectively for comparison \citep{Lineweaver:97}. This figure demonstrates that the low-$z$ and high-$z$ absorbers produce consistent estimates of the dipole location, despite generally using significantly different combinations of transitions. The dipole vectors for the $z<1.6$ and $z>1.6$ sample are separated by 13 degrees. The probability of getting alignment this good or better by chance is 2 percent. \label{fig_skymap_lowvshigh}}
\end{figure}

To check the consistency of the estimates for low and high redshift samples, we cut the data into a $z<1.6$ sample (low-$z$) and a $z>1.6$ sample (high-$z$). This divides the data approximately in half, with $148$ points in the low-$z$ sample and $145$ in the high-$z$ sample. We show in figure \ref{fig_skymap_lowvshigh} the results of separate fits to the low-$z$ and high-$z$ sample, and demonstrate that they yield consistent estimates of the dipole location. We give the parameters to the model $\Delta\alpha/\alpha = A\cos(\Theta) + m$ in table \ref{tab_zsplit}. In particular, the dipole vectors are separated by 13 degrees on the sky. 

To assess the probability of obtaining alignment as good or better than is seen by chance, we use a bootstrap method where we randomly re-assign values of $\Delta\alpha/\alpha$ to different absorbers within the $z<1.6$ and $z>1.6$ samples. That is, we do not allow values of $\Delta\alpha/\alpha$ to mix between the $z<1.6$ and $z>1.6$ samples. Given the distribution of $\Delta\alpha/\alpha$ values, absorbers and sightlines in each sample, the chance probability is $\approx 2$ percent. Given that the transitions used at low and high redshift are significantly different, this consistency further supports the dipole interpretation of the data. It is also clear that the dipole signal is significantly larger at high redshift, although the low redshift sample contributes. 

There is no significant evidence for a high-$z$ monopole, but the low-$z$ monopole is significant at the $3.6\sigma$ level. We discuss the significance of the low-$z$ monopole in section \ref{s_monopole}.

\begin{table*}
\caption{Parameters for the model $\Delta\alpha/\alpha = A\cos(\Theta) + m$ for $z<1.6$ and $z>1.6$ samples. The column ``$\delta A$'' gives $1 \sigma$ confidence limits on $A$. The column labelled ``significance'' gives the significance of the dipole model with respect to the monopole model. Although it is clear that most of the significance comes from the $z>1.6$ sample, the $z<1.6$ sample also contributes. Additionally, a dipole model for the $z<1.6$ sample points in a similar direction to that of the $z>1.6$ sample.\label{tab_zsplit}}
 \begin{center}
\begin{tabular}{cccccccc}
\hline 
Sample & $A$ ($10^{-5}$) & $\delta A$ ($10^{-5}$) & RA (hr) & dec.\ ($^\circ$) & $m$ ($10^{-5}$) & significance\\
\hline
$z<1.6$     & 0.56  & $[0.38, 0.85]$ & $(18.1 \pm 1.8)$ & $(-57\pm22)$ & $(-0.390 \pm 0.108)$ & $1.4\sigma$ \\
$z>1.6$     & 1.38  & $[1.12, 1.74]$ & $(16.5 \pm 1.4)$ & $(-63\pm11)$ & $(0.097 \pm 0.138)$ & $3.5\sigma$  \\
\hline
 \end{tabular}
  \end{center}
\end{table*} 

\subsection{Joint probability}

The probability of obtaining alignment between the dipole vectors for dipole models fitted to the Keck and VLT samples separately as good or better than is seen by chance is about 6 percent. The chance probability of obtaining alignment between the dipole vectors for dipole models fitted to low- and high-redshift samples is about 2 percent. Through a bootstrap method we have calculated the joint probability of obtaining alignment that is at least as good as seen for both of these conditions by chance, and it is $\approx 0.1$ percent.

\subsection{Significance of the low-$z$ monopole}\label{s_monopole}

In section \ref{s_lowz_vs_highz} we noted that the low-$z$ sample shows evidence for a statistically significant monopole at the $3.6\sigma$ level. In figure \ref{fig_zstack}, the existence of the monopole in both the Keck and VLT samples can be seen at low $z$: the mean value of $\Delta\alpha/\alpha$ at $z\la1.6$ is significantly negative in both samples. Note in particular the top panel in figure \ref{fig_zstack}, showing the combined results from both telescopes, where the mean value of $\Delta\alpha/\alpha$ shifts from approximately zero at high $z$ to negative values at low $z$.

An obvious question is whether the monopole arises from one of the Keck or VLT samples. For a dipole+monopole model ($\Delta\alpha/\alpha = A\cos(\Theta) + m$), the Keck sample yields a $z<1.6$ monopole of $m= (-0.404 \pm 0.171) \times 10^{-5}$, which differs from zero at the $2.4\sigma$ level. However, the same model fitted to the VLT $\Delta\alpha/\alpha$ values at $z<1.6$ yields $m = (-0.373 \pm 0.295) \times 10^{-5}$, differing from zero at the $1.3\sigma$ level. There are three important considerations from these values: \emph{i)} Both data sets yield consistent monopole values for $\Delta\alpha/\alpha$ at low redshift, differing at only the $0.09\sigma$ level. Therefore, whatever is generating the monopole appears to affect both the Keck and VLT samples. \emph{ii)} Because there is no significant difference between the monopole values in the Keck and VLT samples, the monopole cannot be responsible for mimicing angular variation in $\alpha$ across the sky. \emph{iii)} Additionally, the dipole signal is dominated by $\Delta\alpha/\alpha$ values at $z>1.6$; there, the dipole+monopole model is significant at $3.5\sigma$ over the monopole-only model. As such, the significance of the dipole signal is not strongly affected by the presence of the low-$z$ monopole. That is, to first order, the dipole and low-$z$ monopole signals appear decoupled.

There are several possible explanations for this, and we discuss each of them in turn:

\emph{i) Telescope systematics}. Perhaps the most obvious and important concern is that there is some simple, wavelength-dependent systematic effect between the Keck or VLT telescopes or spectrographs. However, such wavelength-dependent telescope systematics seem difficult to support given the inter-telescope consistency of the best-fitting dipole directions on the sky. Nevertheless, such effects are difficult to rule out entirely. Therefore, we address this potential concern in detail in section \ref{s_systematic_errors} by directly comparing spectra of quasars taken with \emph{both} the VLT and Keck telescopes.

\emph{ii) Errors in the laboratory wavelengths}. Errors in the laboratory wavelengths of transitions which feature predominantly at low redshifts could cause a statistically significant monopole at low redshifts. However, this seems particularly unlikely. The Mg \isc/\iis wavelengths have been accurately measured on an absolute scale generated using a frequency-comb calibration system. The Fe \iis wavelengths used at $z<1.6$ have also been precisely measured (the $\lambda 1608,1611$ transitions are more difficult to measure accurately, but these transitions are used infrequently at low redshifts due to their short rest wavelengths). For instance, the absolute velocity uncertainty in the Fe \iis $\lambda 2382$ transition is $\approx 14\,\mathrm{m\,s^{-1}}$, which is significantly smaller than the $\approx 82\,\mathrm{m\,s^{-1}}$ which would be needed to generate a monopole value of $-0.39 \times 10^{-5}$. This implies a systematic error some six times larger than the existing error budget, which seems unlikely. Additionally, the relative wavelength scales of the different experiments which measured the transitions used at lower redshifts are likely to be significantly better than this.

\emph{iii) Time evolution of $\alpha$}. The functional form for variation of $\alpha$ (if $\alpha$ varies) is unknown. Recent observations confirm the apparent acceleration of the universe at late times \citep{Astier:06}, for which dark energy is posited as an explanation. For $z \la 0.5$, dark energy dominates over matter and radiation \citep{Riess:04}. If $\alpha$ couples to dark energy, then late-time evolution of $\alpha$ might be possible. Monotonic evolution of $\alpha$ cannot by itself be an explanation for a low-$z$ monopole, because this would imply that $\Delta\alpha/\alpha$ should approach zero for $z\rightarrow 0$, with the greatest divergence of $\Delta\alpha/\alpha$ from zero at high redshift. If $\alpha$ oscillates with time then a pattern such as that observed in figure \ref{fig_zstack} could arise. However, this would require the period of oscillations to be $\sim$twice the age of the Universe, with the present day at a node of the oscillation, in order to obtain $\langle \Delta\alpha/\alpha_{z>1.6}\rangle \sim 0$, $\langle \Delta\alpha/\alpha_{0.2<z<1.6} \rangle \sim -0.4\times 10^{-5}$ and $\Delta\alpha/\alpha_{z=0}=0$. This may be possible, but this case seems rather contrived.

\emph{iv) Dependence of $\alpha$ on the local environment}. If the value of $\alpha$ depends on the local environment (e.g. matter density, gravitational potential, or gradient of the gravitational potential) then this could produce an offset between the value of $\alpha$ measured in the quasar absorbers and the value measured in the laboratory, even as $z \rightarrow 0$. If this was the case, we would expect a similar magnitude monopole to also be present at high redshift, which is not seen. It is possible that a combination of the time evolution of $\alpha$ \emph{and} dependence of $\alpha$ on the local environment could explain the low-$z$ monopole, but this requires two different mechanisms. Additionally, in this circumstance the magnitude of the environmental dependence must be very similar to the magnitude of the time evolution from $z\sim 4$ to $z=0$ in order to obtain the observed distribution of $\Delta\alpha/\alpha$ with $z$, requiring significant fine-tuning.

\emph{v) Non-terrestrial isotopic abundances in the absorbers}. The isotopic splitting scales as $\Delta \omega_i \propto \omega_0 / m_i^2$, where $m_i$ is the mass of the species under consideration. Mg is the lightest atom used in the MM method, and therefore the isotopic splitting for the Mg transitions is relatively large. If the abundance of the three Mg isotopes differs significantly in the quasar absorbers to terrestrial values, this would mimic a change in $\alpha$. The low-$z$ sample is dominated by the Mg \textsc{ii}/Fe \textsc{ii} combination, which is particularly sensitive to the effect of differences in the abundance of the Mg isotopes \citep{Murphy:01c}. At higher redshifts, the effect on $\Delta\alpha/\alpha$ of isotopic abundance variations in other light species (e.g. Si) has been shown to be negligible \citep{Murphy:01c}. The fact that we do not see a significant monopole signal at high redshift lends additional credence to this interpretation.

On balance, this last possibility -- evolution in the abundance of the Mg isotopes -- seems like the most likely of the explanations above. We therefore explore this isotopic abundance effect in more detail in section \ref{s_isotopic_abundances}.

\subsection{Iterative clipping of potentially outlying $\Delta\alpha/\alpha$ values}\label{s_sigmaclipping}

We have attempted to be conservative in presenting our results when accounting for extra scatter in the $\Delta\alpha/\alpha$ values about a model by adding a term, $\sigma_\mathrm{rand}$, in quadrature with the error bars. This effectively functions as an interpolation between a $\chi^2$ fit where the error bars are believed to be correct and an unweighted fit, where the error bars are unknown. 

However, another option is to assume that the statistical error bars on most $\Delta\alpha/\alpha$ values are a good representation of the total uncertainty for those absorbers, and then remove points one-by-one (``clipping'') until $\chi^2_\nu = 1$. In our sample it is difficult to determine to what extent different random processes affect different absorbers, and therefore to determine to what extent clipping is justified. Adding some $\sigma_\mathrm{rand}$ in quadrature with all $\Delta\alpha/\alpha$ values, as we have done, is a conservative option. Nevertheless, we explore the effect of clipping here to investigate the robustness of our results to the removal of $\Delta\alpha/\alpha$ values.

Traditionally, data clipping involves iteratively removing the point with the largest residual and then re-fitting. However, for the reasons given in section \ref{LTS_method}, this has the potential to incorrectly remove points. Therefore, we use a modified method. At each iteration, we calculate the LTS fit using the model $\Delta\alpha/\alpha=A\cos\Theta+m$ to the $\Delta\alpha/\alpha$ values with their raw statistical errors, and then remove the point with the largest residual about the fit. However, we choose $k = n-1$ in this case. Effectively, at each stage, we want to identify only one point to remove, and therefore it makes sense to calculate a fit to $n-1$ points. We then calculate a weighted fit, using only the statistical error bars, and calculate the significance of the dipole model. For efficiency of calculation, we avoid bootstrapping, and so we use the method of \citet{Cooke:09} to calculate the significance of the dipole fit with respect to the monopole fit. We then repeat the process. At any iteration, if $\chi^2_\nu > 1$, we multiply all entries of the covariance matrix by $\chi^2_\nu$ in order to account for excess scatter about the model. If $\chi^2_\nu < 1$, we do not adjust the covariance matrix.

Initially, one expects the significance of the fit to improve, as one discards a few points which are not consistent with the general trend of the fit. Eventually, one will remove enough $\Delta\alpha/\alpha$ values that the significance must decline. If the significance declines rapidly, this implies that the dipole effect is dominated by a few points. Conversely, if the significance of the fit is sustained or improved for the removal of small fractions of data (e.g.\ $\sim 10$ percent), this qualitatively implies robustness of the dipole effect.

\begin{figure}
\begin{center}
\ifpdf
\includegraphics[viewport=44 23 556 805,angle=-90,width=82.5mm]{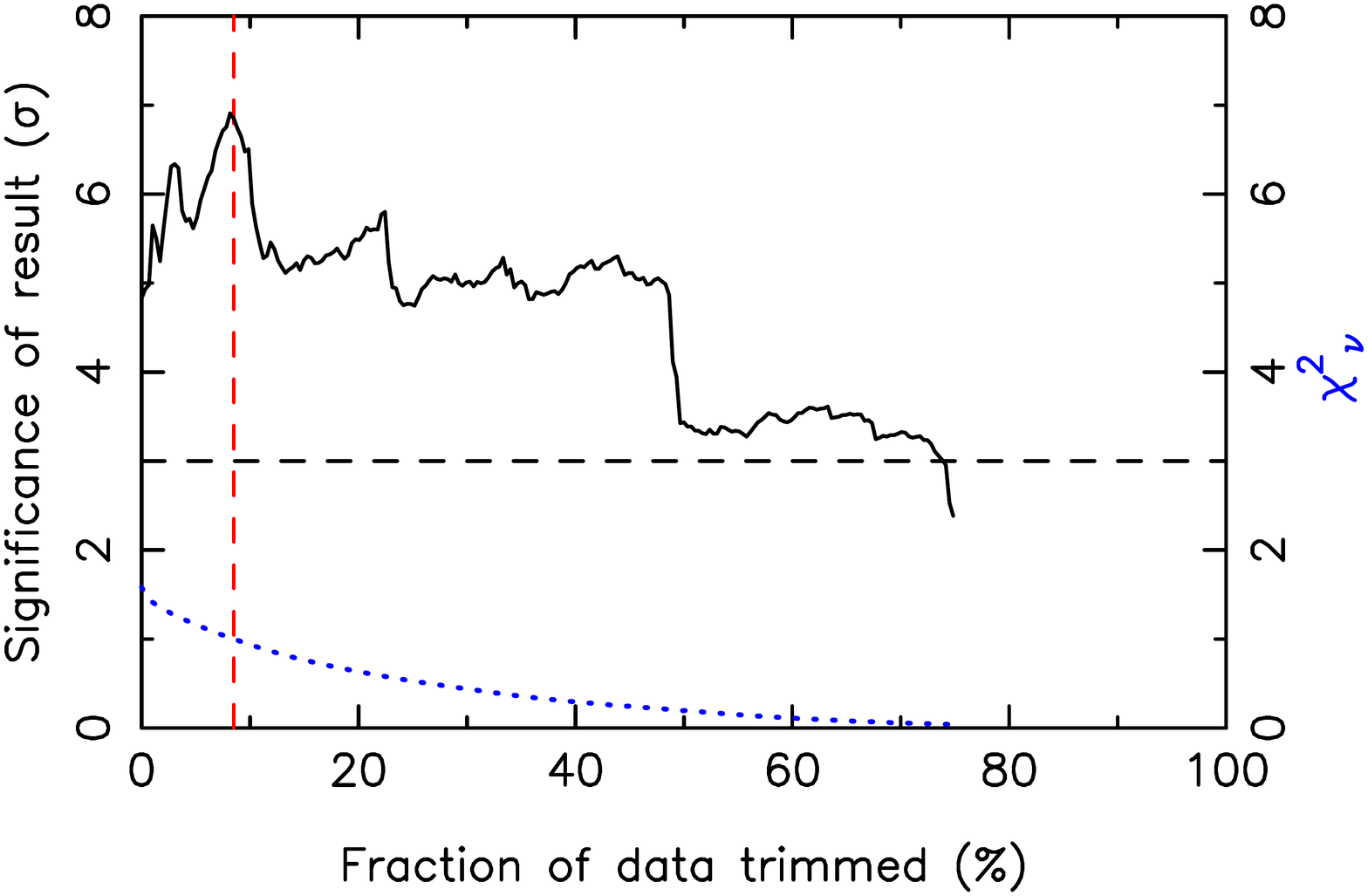}
\else
\includegraphics[bb=44 23 556 805,angle=-90,width=82.5mm]{images/trimfrac_vs_significance.eps}
\fi
\end{center}
\caption{Effect of iteratively clipping $\Delta\alpha/\alpha$ values on the statistical significance of the dipole for the combined sample, as described in section \ref{s_sigmaclipping}, for the model $\Delta\alpha/\alpha = A\cos(\Theta) + m$. The vertical axis shows the statistical significance of the dipole as determined by the method of \citet{Cooke:09} given in terms of $\sigma$ (solid line) and $\chi^2_\nu$ at that point (blue, dotted line). A dashed horizontal line is drawn at $3\sigma$ for reference. The vertical red (dashed) line indicates the point at which our clipping method reduces $\chi^2_\nu$ to below unity. We note that we have to remove more than 40 percent of absorbers before the significance of the detection drops to about $3\sigma$. As there is no good reason to remove so much data, this implies that the dipole result is not being drive by a few outlying $\Delta\alpha/\alpha$ values. The actual significance given here is probably overstated compared to the ``true'' significance, given that no attempt has been made to account for systematic or random errors. \label{fig_sigmaclipping}}
\end{figure}

We show in figure \ref{fig_sigmaclipping} the results of this process. We find that we must remove large numbers of absorbers to destroy the significance of the dipole. In particular, the significance does not decrease rapidly with the number of $\Delta\alpha/\alpha$ values clipped initially, suggesting that the observed dipole effect is not being caused by a few outlying points. If we clip until $\chi^2_\nu = 1$, the significance of the dipole is almost $7 \sigma$. 

\begin{figure}
\begin{center}
\ifpdf
\includegraphics[viewport=50 23 556 750,angle=-90,width=82.5mm]{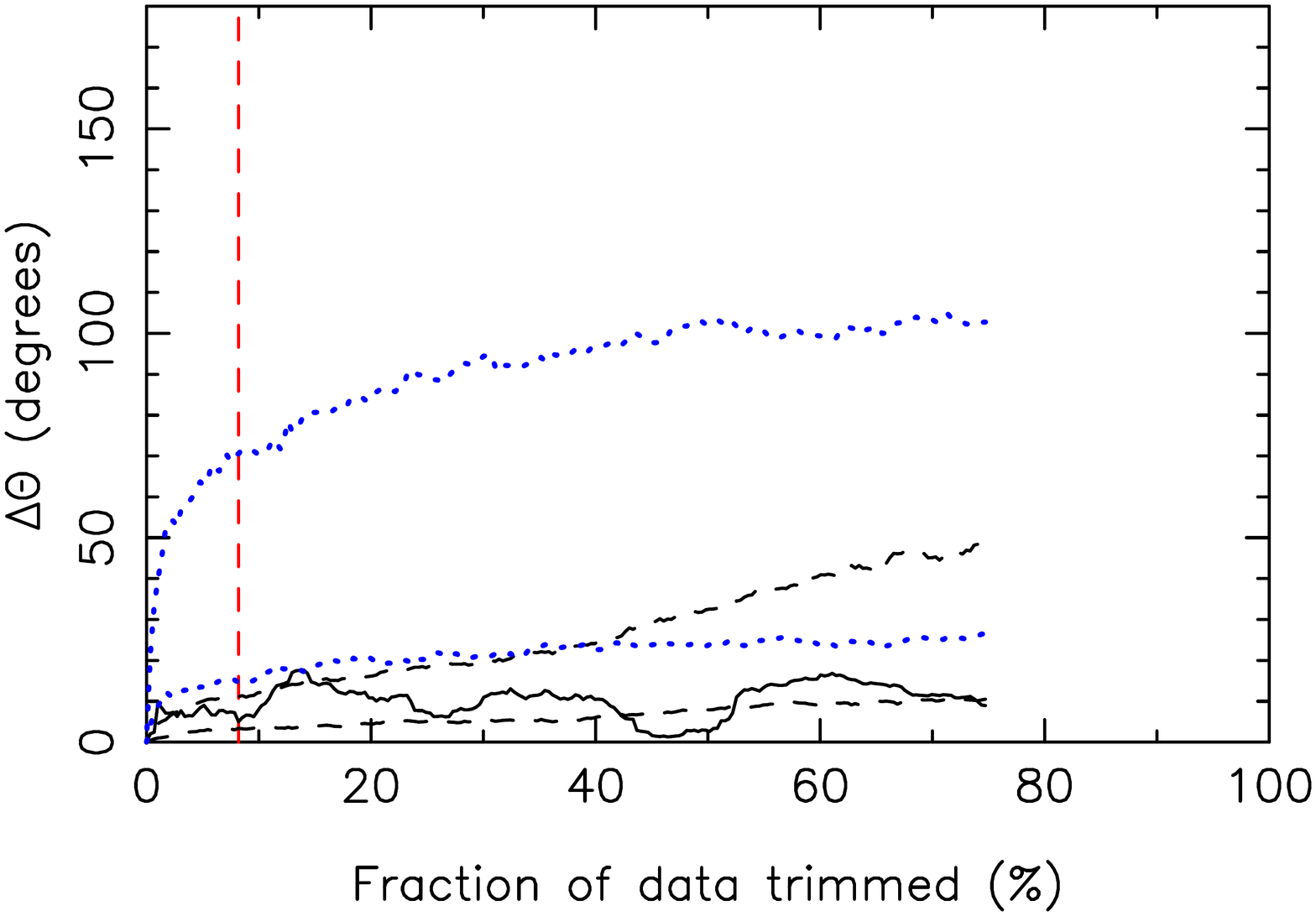}
\else
\includegraphics[bb=50 23 556 750,angle=-90,width=82.5mm]{images/angular_trimfrac_vs_dtheta.eps}
\fi
\end{center}
\caption{Effect of iteratively clipping $\Delta\alpha/\alpha$ values on the location of the dipole, as described in section \ref{s_sigmaclipping}. The vertical axis shows the deviation of the fitted angle from the untrimmed model ($\Delta \Theta$) as a function of the percentage of absorbers removed. This figure compares the results of trimming on our actual $\Delta\alpha/\alpha$ values, bootstrapped samples designed to emulate our data, and random samples. \emph{Actual data:} The solid black line shows the results of trimming for our $\Delta\alpha/\alpha$ values. If the fit is stable and not due to the presence of a small number of highly significant points, we expect to see that $\Delta\Theta$ should not grow rapidly with the amount of data removed. This is what is seen. \emph{Bootstrapped samples:} The dashed lines show the $1\sigma$ range for 300 bootstrap samples (generated as described in the text). This shows the typical range of variation at fraction of absorbers removed given distribution of sightlines, values of $\Delta\alpha/\alpha$, statistical errors and distribution of residuals in the sample. The region given reflects the $1\sigma$ range for the bootstrapped samples at each point; each individual sample may wander substantially more than is indicated by this range, and so the deviation of the path for our actual sample outside the region is not indicative of any problem with our $\Delta\alpha/\alpha$ values. \emph{Random samples:} The blue, dotted lines show the $1\sigma$ range for 300 samples where we have randomised $\Delta\alpha/\alpha$ over the sightlines. We see that $\Delta \Theta$ in this case grows rapidly with increased trimming for these samples. Our real sample does not do this, which suggests that our real sample is significantly dissimilar from a random sample. \label{fig_sigmaclipping_dtheta_bootstrap}}
\end{figure}

In figure \ref{fig_sigmaclipping_dtheta_bootstrap} we show the effect of clipping $\Delta\alpha/\alpha$ values on the location of the dipole. One expects that if the dipole effect is real, then the position of the dipole should not change dramatically with the removal of small amounts of data (that is, $\Delta \Theta = \Theta_i - \Theta_0$ should be small).  To assess how likely it is that this seemingly restricted path is typical for our distribution of data, we apply a bootstrap method to generate and iteratively trim 300 new samples, and examine the distribution of $\Delta\Theta$ at each point. We cannot use a traditional bootstrap, which resamples the data with replacement, because how the data is trimmed depends crucially on the distribution of residuals. Therefore, we resample the residuals of the fit to generate new samples. To do this, we use the following process to generate one sample: \emph{i)} calculate the model prediction for each absorber given the model, $p_i = A\cos(\Theta)+m$; \emph{ii)} calculate the residuals about the fit for each absorber, $r_i = (\Delta\alpha/\alpha_i - p_i)/\sigma_i$; \emph{iii)} randomly reassign the calculated $r_i$ to different absorbers, generating $r'_j$; \emph{iv)} generate a new set of $\Delta\alpha/\alpha$ values as $\Delta\alpha/\alpha'_j = p_j + r'_j \sigma_j$. In this way, we generate new values of $\Delta\alpha/\alpha$ which represent different possible realisations of our sample where the actual distribution of residuals is preserved. This is demonstrated in figure \ref{fig_sigmaclipping_dtheta_bootstrap}. We see that the bootstrapped samples do not wander very far even when much of the data is removed ($\Delta\Theta \lesssim 20^\circ$). 

To contrast this with the effect on a random sample, we also show in figure \ref{fig_sigmaclipping_dtheta_bootstrap} the effect of trimming random samples of data. To do this, we generate 300 new samples by randomly reassigning values of $\Delta\alpha/\alpha$ to different sightlines, and iteratively trimming under the model $\Delta\alpha/\alpha = A\cos(\Theta) + m$. We see here that our actual sample is not typical of the random samples, therefore suggesting that the actual sample is significantly dissimilar to random samples. 

\subsection{Removal of spectra}\label{s_removal_of_spectra}

A further question one might ask is how sensitive our results are to the inclusion of particular spectra. We would like to know whether the dipole result could be dominated by a small number of spectra which, if removed, would destroy the result. 

\begin{figure}
\begin{center}
\ifpdf
\includegraphics[bb=61 23 556 795,angle=-90,width=82.5mm]{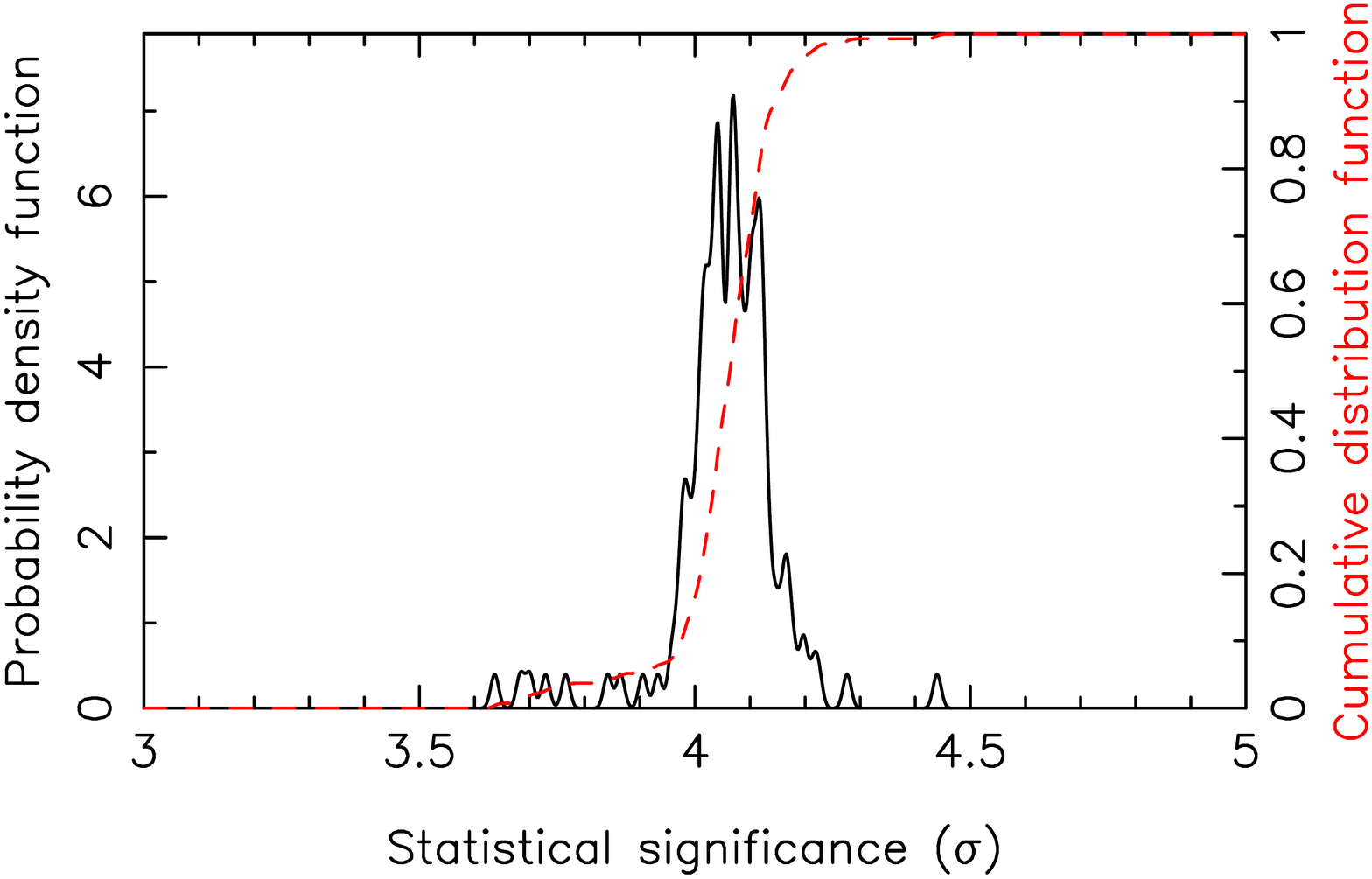}
\else
\includegraphics[bb=61 23 556 795,angle=-90,width=82.5mm]{images/quasar_jackknife_significance.eps}
\fi
\end{center}
\caption{Effect of removing quasar spectra on the statistical significance of the dipole, as assessed through a jack-knife method. Each spectrum is removed one at a time, and the value of the statistical significance of the model $\Delta\alpha/\alpha = A\cos(\Theta) + m$ is calculated with respect to the monopole model, using the bootstrap method. We use a Gaussian kernel density estimator to construct the approximate probability density function of the effect of quasar spectrum removal, where the width of the Gaussian basis functions has been chosen to be the inverse of the number of spectra. The cumulative distribution function is plotted as a dashed, red line. This demonstrates that the angular dipole effect is not due to a single spectrum. \label{quasar_jackknife} }
\end{figure}

\begin{figure}
\begin{center}
\ifpdf
\includegraphics[bb=61 23 556 795,angle=-90,width=82.5mm]{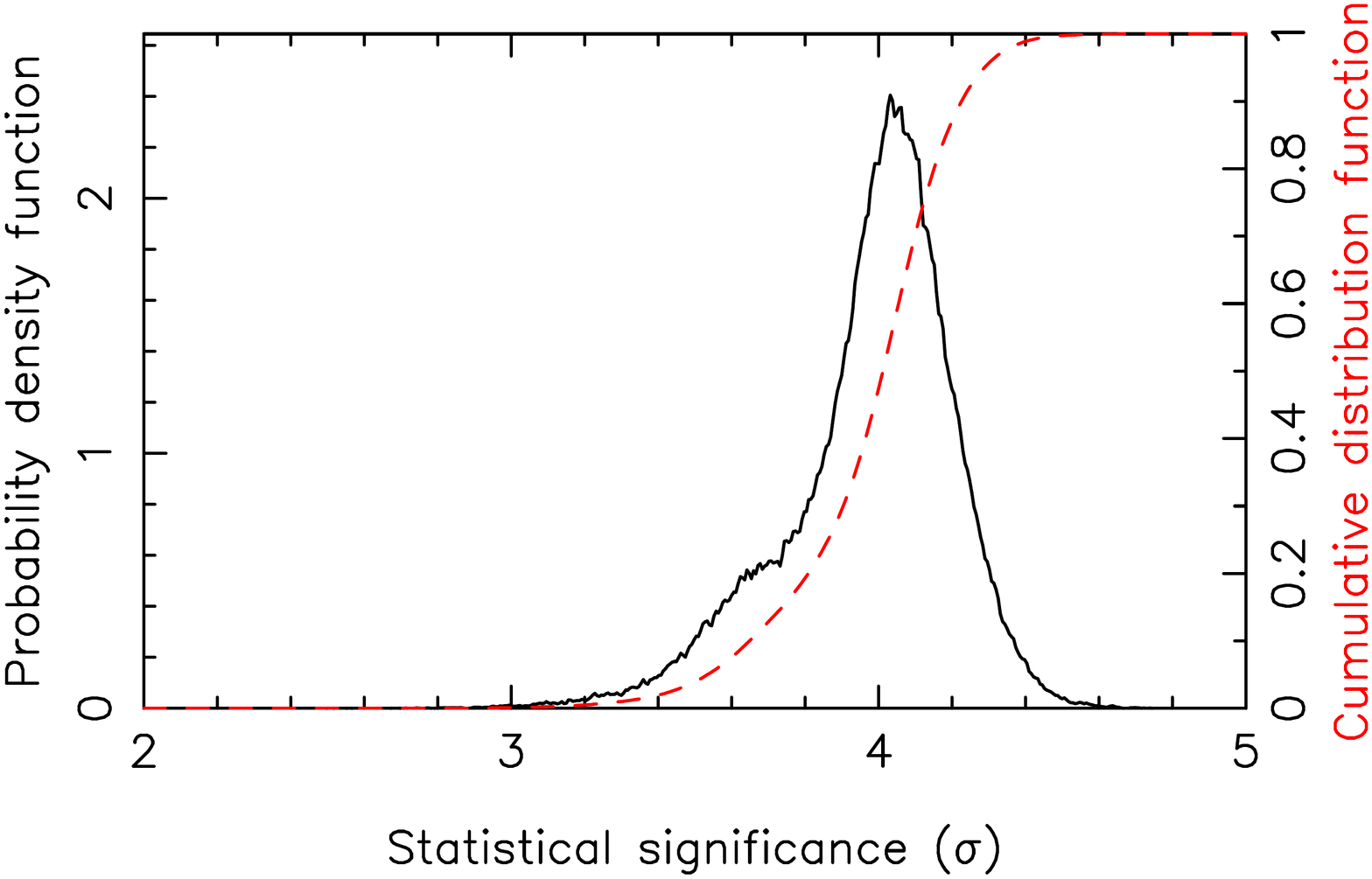}
\else
\includegraphics[bb=61 23 556 795,angle=-90,width=82.5mm]{images/quasar_jackknife_n5.eps}
\fi
\end{center}
\caption{Effect of removing quasar spectra on the statistical significance of the dipole, as assessed through a sampling method. For 100,000 samples, 5 quasar spectra are randomly removed from the combined Keck + VLT sample, and the statistical significance of the dipole model $\Delta\alpha/\alpha = A\cos(\Theta) + m$ is calculated through the method of \citet{Cooke:09}. Gaussian basis functions with width $\approx 1/316$ (=$10^{5/2}$) are used. This graph demonstrates that, in the absence of particular knowledge about problematic spectra, the chance of obtaining a dipole model where the statistical significance of the dipole is less than $3\sigma$ is small as a result of randomly removing 5 spectra. \label{quasar_jackknife_n5} }
\end{figure}

We therefore explore this question through a jack-knife method, where we remove one quasar at a time and recalculate the statistical significance of the fit. We show the results of this exploration in figure \ref{quasar_jackknife}. The figure clearly demonstrates that, unsurprisingly, our result is not due to a single quasar spectrum. We extend this in figure \ref{quasar_jackknife_n5} to show the effect of removing 5 spectra at random. We chose 5 as a number somewhat larger than 1, and also in order to potentially include the cluster of 5 quasars at $\mathrm{RA}\sim 22\mathrm{hr}$, $\mathrm{dec.}\sim -45^\circ$, where all of these sightlines demonstrate $\Delta\alpha/\alpha > 0$. Under this circumstance,the probability of obtaining a dipole result which is insignificant ($<3\sigma$) is small. This suggests that the dipole effect is not being created by a small number of spectra.

\subsection{Translation from angular variation to a physical model including a distance measure}\label{s_zdipolemodel}

We now explore simple phenomenological parameterisations of the dipole effect which attempt to account for distance dependence. In all of these models, the same $\Delta\alpha/\alpha$ values identified as outliers have been removed from consideration.

\subsubsection{$z$-dipole fit}\label{s_zdipole}

To model potential distance dependence directly with the observable quantity, $z$, we fit a power-law relationship of the form
\begin{equation}
\Delta\alpha/\alpha = C z^\beta \cos(\Theta) + m\label{dipole_z_beta_eq},
\end{equation}
for some $\beta$ and amplitude $C$ to the combined Keck + VLT data, where the $\Delta\alpha/\alpha$ values have been increased in quadrature by the amounts set out for sample 11 in table \ref{combineresults_dipole}. This sample is the ``$z^\beta$ dipole'' sample.

\begin{figure}
\begin{center}
\ifpdf
\includegraphics[bb=77 92 556 727,angle=-90,width=82.5mm]{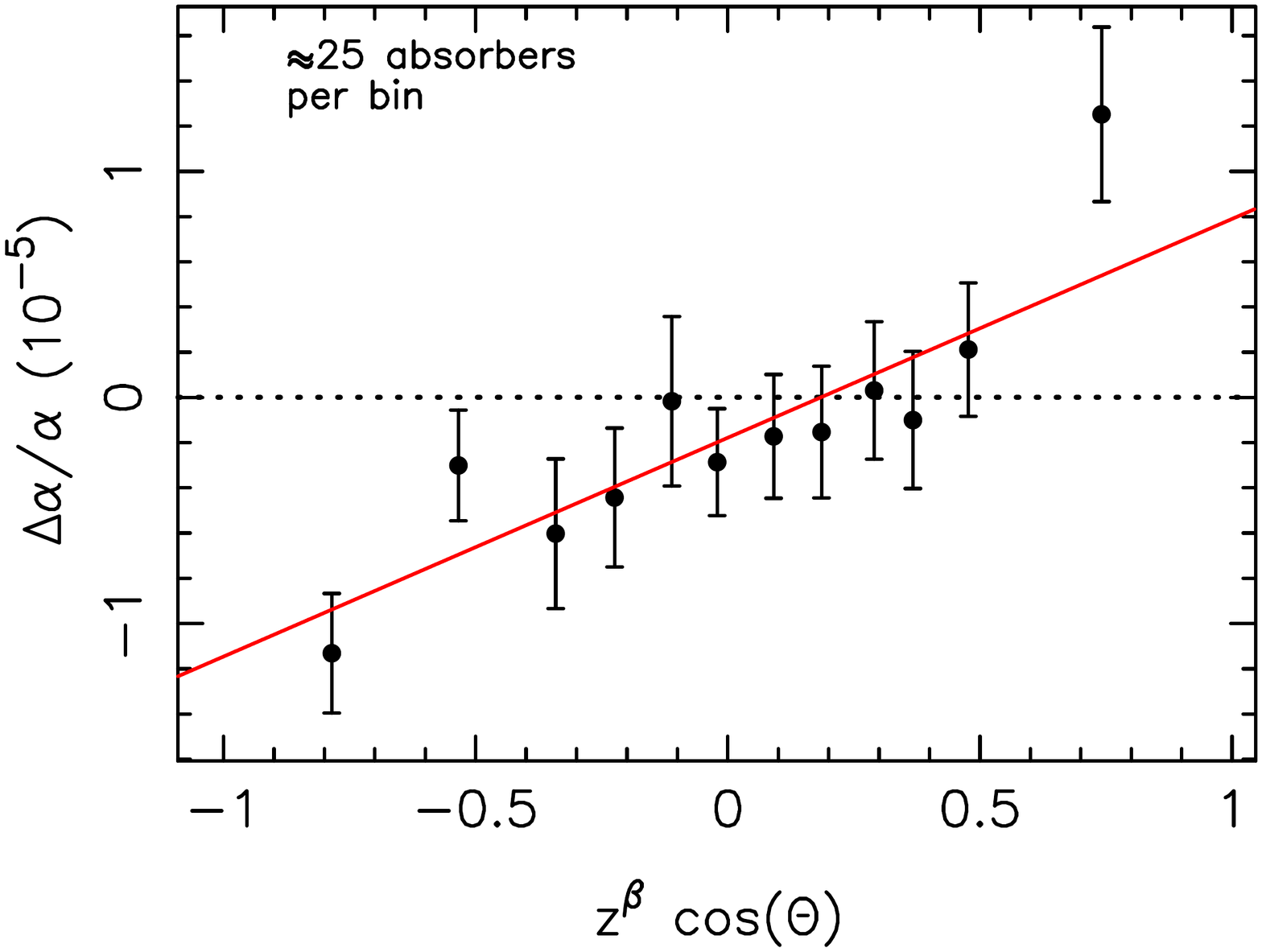}
\else
\includegraphics[bb=77 92 556 727,angle=-90,width=82.5mm]{images/zbetacostheta.eps}
\fi
\end{center}
\caption{Binned values of $\Delta\alpha/\alpha$ plotted against $z^\beta \cos \Theta$, for $\beta = 0.41$. The $z^\beta$ dipole+monopole model is preferred over the monopole-only model at the $3.9\sigma$ level. Importantly, this graph only covers $\lvert z^\beta \cos(\Theta) \rvert \lesssim 1$. Given that it is possible to probe up to redshift $z \lesssim 4$ with the MM method, judicious choice of observational targets close to the dipole axis might be able to extend this horizontal range of this graph up to $\sim \pm 2$, thereby potentially increasing sensitivity to the effect substantially if it is real. \label{da_vs_zbetacostheta} }
\end{figure}

We use the Levenberg-Marquardt algorithm \citep{NumericalRecipes:92} to fit equation \ref{dipole_z_beta_eq} to the $z^\beta$ dipole sample. This fit yields $\mathrm{RA} = (17.5 \pm 1.0)\,\mathrm{hr}$, $\mathrm{dec.} = (-62 \pm 10)^\circ$, $C = 0.81$ ($1\sigma$ confidence limits $[0.55,1.09] \times 10^{-5}$), $m = (-0.184 \pm 0.085) \times 10^{-5}$ and $\beta = 0.46 \pm 0.49$. The fact that the amplitude grows as a low power of $z$, and the fact that it is statistically consistent with zero, is the reason that the approximation $A \sim Cz^0$ yields reasonable results earlier. We show the results of this fit in figure \ref{da_vs_zbetacostheta}. This dipole + monopole model is statistically preferred over the monopole-only model at the 99.99 percent confidence level ($3.9\sigma$). The reduction in significance relative to the angular dipole model occurs as a result of the uncertainty in determining $\beta$, but is relatively small. 

\subsubsection{$r$-dipole fit}\label{s_tdipole}

Another plausible alternative is to try to relate the amplitude of the dipole to some explicit distance metric. For simplicity, we use the lookback-time distance. This is defined by $r = ct$, where $c$ is the speed of light and $t$ is the lookback time to the absorber. Thus, we try a fit of the form
\begin{equation}
\Delta\alpha/\alpha = Br\cos(\Theta) + m\label{dipole_t_eq}.
\end{equation}
To calculate lookback times, we use the standard $\Lambda$CDM model ($\Lambda$ Cold Dark Matter), with parameters given by the 5-year WMAP (Wilkinson Microwave Anisotropy Probe) results \citep{WMAP5yr}. We note that this calculation is derived from the FLRW metric, which assumes isotropy of the universe. Our model implies anisotropy of the universe, and therefore use of the FLRW (Friedmann-Lema\^{i}tre-Robertson-Walker) metric is strictly incorrect. Nevertheless, as $\Delta\alpha/\alpha \ll 1$ we assume that the FLRW metric is a good approximation to the actual metric, and therefore that our lookback times are approximately correct. The $\Lambda$CDM parameters used are: ($H_0$, $\Omega_M$, $\Omega_\Lambda$) = (70.5 km\,s$^{-1}$\,Mpc$^{-1}$, 0.2736, 0.726).

\begin{figure*}
\begin{center}
\ifpdf
\includegraphics[bb=78 79 488 727,angle=-90,width=160mm]{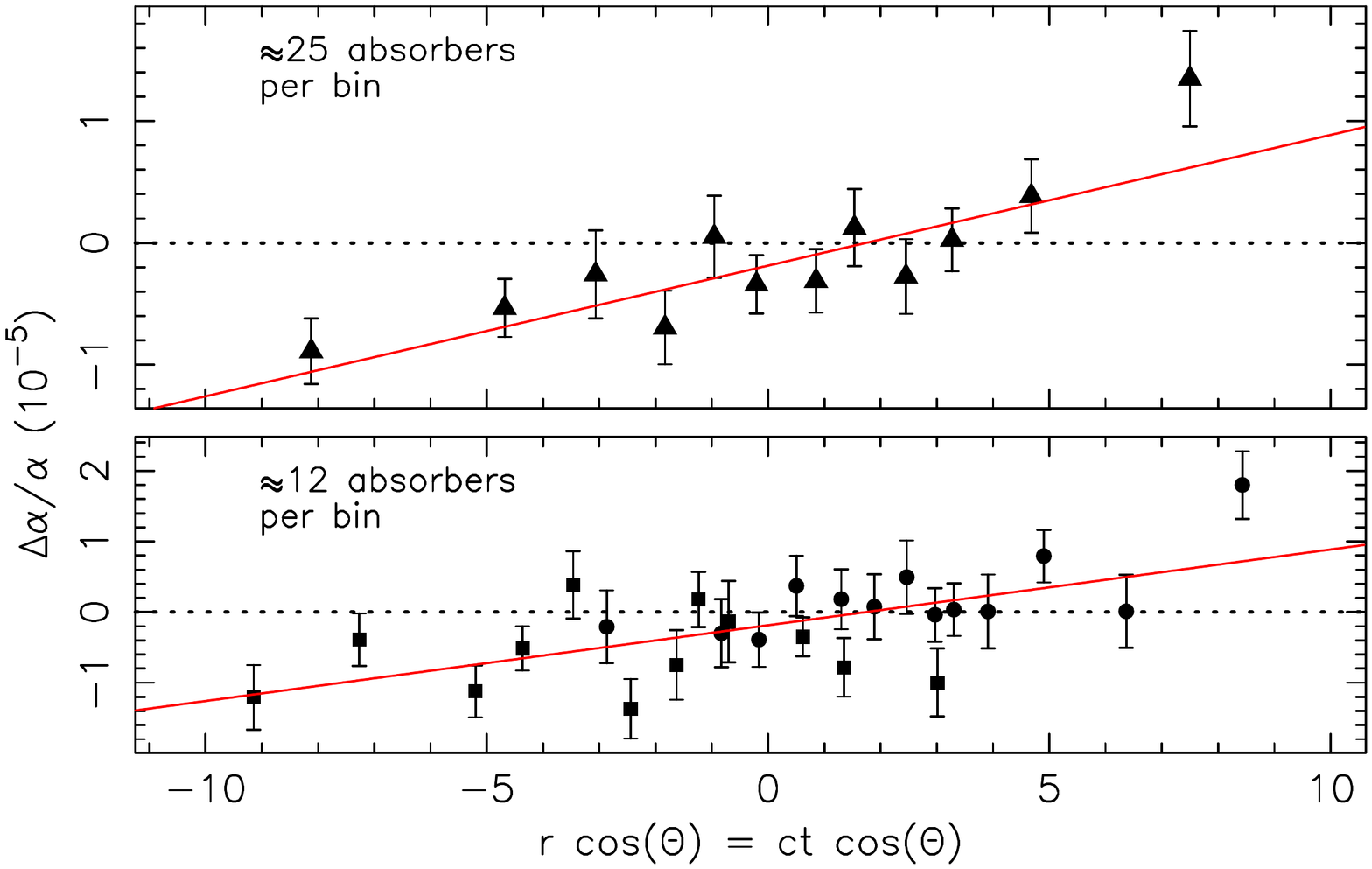}
\else
\includegraphics[bb=78 79 488 727,angle=-90,width=160mm]{images/ctdipole.eps}
\fi
\end{center}
\caption{Binned values of $\Delta\alpha/\alpha$ plotted against $r\cos(\Theta) \equiv ct\cos \Theta$, where $t$ is the look-back time to a redshift $z$. $\Lambda$CDM parameters are from \citet{WMAP5yr}. The top panel (triangles) shows the combined sample, binned with approximately 25 absorbers per bin. The bottom panel shows VLT (circles) and Keck (squares) points, binned with approximately 12 absorbers per bin. The red (solid) line in both cases shows the model, $\Delta\alpha/\alpha = Br\cos(\Theta) + m$. The parameters for the fit are:  $B = 1.1\times 10^{-6}\,\mathrm{GLyr}^{-1}$ ($1\sigma$ confidence limits $[0.9, 1.3] \times 10^{-6} \,\mathrm{GLyr}^{-1}$), $\mathrm{RA} = (17.5 \pm 1.0)\,\mathrm{hr}$, $\mathrm{dec.} = (-62 \pm 10)^\circ$ and $m = (-0.187 \pm 0.084) \times 10^{-5}$. It is interesting that this simple model is a reasonable representation of the data.\label{da_vs_ctcostheta} }
\end{figure*}

\begin{figure}
\begin{center}
\ifpdf
\includegraphics[bb=77 78 455 727,angle=-90,width=82.5mm]{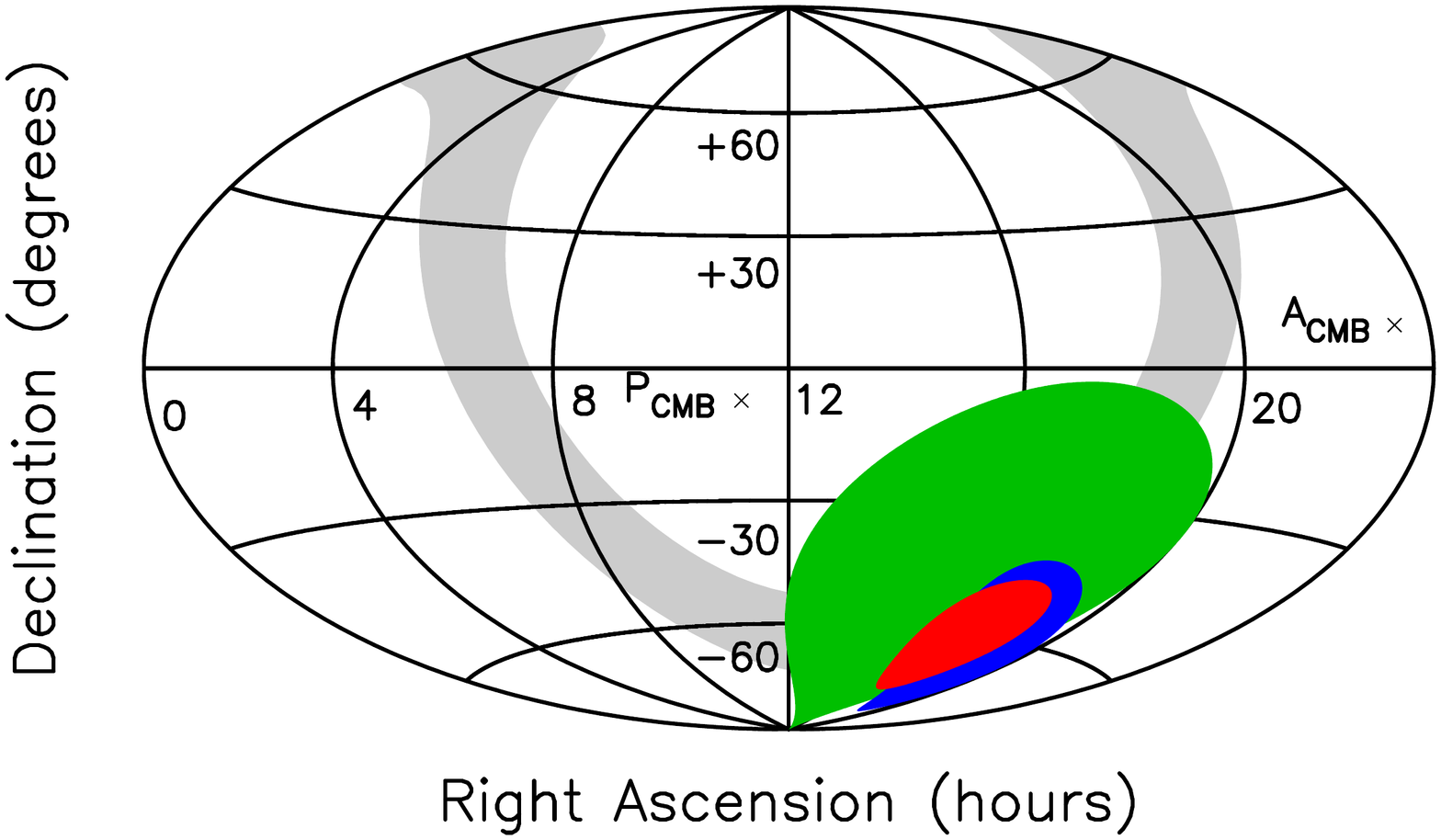}
\else
\includegraphics[bb=77 78 455 727,angle=-90,width=82.5mm]{images/skymap_mult_ct.eps}
\fi
\ifpdf
\includegraphics[bb=77 78 455 727,angle=-90,width=82.5mm]{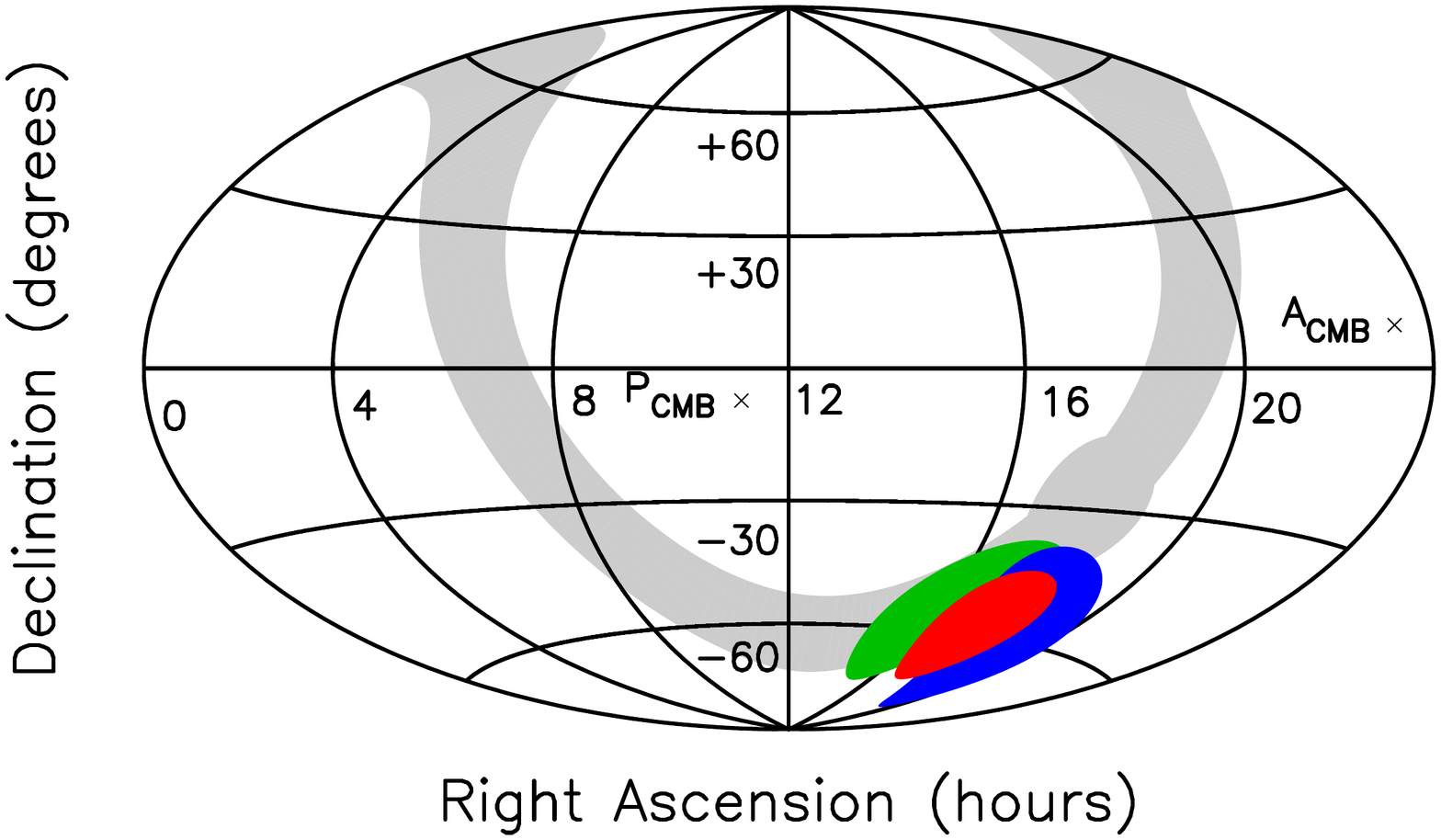}
\else
\includegraphics[bb=77 78 455 727,angle=-90,width=82.5mm]{images/skymap_mult_ct_nomonopole.eps}
\fi
\end{center}
\caption{Sky map in equatorial coordinates showing the 68.3 percent ($1\sigma$ equivalent) confidence limits of the location of the pole of the dipole for a fit to the Keck $\Delta\alpha/\alpha$ values (green region), VLT $\Delta\alpha/\alpha$ values (blue region) and combined $\Delta\alpha/\alpha$ values (red region), for a fit of $\Delta\alpha/\alpha = Br\cos(\Theta) + m$ (\emph{top figure}) and $\Delta\alpha/\alpha = Br\cos(\Theta)$ (\emph{bottom figure}), where $r=ct$ and $t$ is the lookback time to the absorber. The pole and anti-pole of the CMB dipole are marked as $P_\mathrm{CMB}$ and $A_\mathrm{CMB}$ respectively. In the model which includes a monopole (top figure), the Keck confidence region is large due to a relative degeneracy with the monopole; the region is much smaller in the bottom figure on account of no monopole term being included.\label{fig_skymap_ct}}
\end{figure}

We show in figure \ref{da_vs_ctcostheta} the fit of $\Delta\alpha/\alpha$ for a combined VLT + Keck sample (``combined $r$-dipole sample'') against $r\cos(\Theta)=ct\cos(\Theta)$. The parameters for this fit are: $B = 1.1\times 10^{-6}\,\mathrm{GLyr}^{-1}$ ($1\sigma$ confidence limits $[0.9, 1.3] \times 10^{-6} \,\mathrm{GLyr}^{-1}$), $\mathrm{RA} = (17.5 \pm 1.0)\,\mathrm{hr}$, $\mathrm{dec.} = (-62 \pm 10)^\circ$ and $m = (-0.187 \pm 0.084) \times 10^{-5}$. Using the bootstrap method we assess the statistical significance of this fit with respect to the monopole-only fit as $4.15\sigma$. In figure \ref{fig_skymap_ct}, we show the confidence regions on the dipole location for the VLT, Keck and combined samples on the sky.  

In galactic coordinates, the pole of this fit is at approximately $(l,b) = (330, -15)^\circ$. The fact that the pole and anti-pole are close to the Galactic Plane explains the relative lack of absorbers near to the pole and anti-pole in both the Keck and VLT samples, a fact made obvious in figure \ref{fig_sightlines} earlier. 

If we adopt a dipole-only model,
\begin{equation}
\Delta\alpha/\alpha = Br\cos(\Theta)\label{dipole_t_eq_nomonopole},
\end{equation}
we derive parameter estimates $B=1.1 \times 10^{-6}\,\mathrm{GLyr}^{-1}$, $\mathrm{RA} = (17.5 \pm 0.9)\,\mathrm{hr}$, $\mathrm{dec.} = (-58 \pm 9)^\circ$. $1\sigma$ confidence limits on $B$ are $[0.9, 1.3] \times 10^{-5}$. The statistical significance of the dipole model is 99.998 percent ($4.22\sigma$). The confidence limits on the dipole location for this fit for the VLT, Keck and combined samples are shown in figure \ref{fig_skymap_ct}.  

\section{Systematic errors}\label{s_systematic_errors}

\subsection{General effect of systematic errors on monopole and dipole terms}

\citet{Murphy:01c,Murphy:03} considered a number of potential systematic effects, including: quasar absorber isotopic abudances for Mg that are different from terrestrial values; differential isotopic saturation; spectrograph temperature variations; the possibility that a single transition could be spuriously generating a non-zero $\Delta\alpha/\alpha$ (which could occur if the laboratory wavelength for this transition was significantly inaccurate); atmospheric dispersion; stretching or compression of the quasar wavelength scale relative to the ThAr calibration scale; absorption cloud kinematics; systematic blends; random blends; and others. 

Generating a significant angular (dipolar) variation in $\alpha$ from systematic effects is, however, rather harder than producing an offset from $\Delta\alpha/\alpha = 0$. Whatever this effect is -- if it exists -- it must be well correlated with sky position \emph{or} must be a combination of systematics that by coincidence mimics angular variation in $\alpha$. 

The most significant potential concern relates to inter-telescope wavelength calibration systematics, and we treat this issue in detail in section \ref{s_dvtest}. However, in section \ref{s_monopole} we raised the presence of a consistent offset from $\Delta\alpha/\alpha=0$ in both the low-$z$ Keck and VLT samples (the low-$z$ monopole), and suggested that this may be due to evolution in the abundance of the Mg isotopes. We deal with this concern in the next section.

\subsection{Effect of isotopic abundances}\label{s_isotopic_abundances}

Most of the atomic species we use have a number of stable isotopes, and each of these isotopes exhibits a slightly different rest wavelength for a given transition. The isotopic spacing depends on the transition and species under consideration, but scales according to the inverse square of the mass. That is, $\Delta\omega_i \propto \omega_0 / m_i^2$. The isotopic shifts of Mg, as the lightest of the elements species under consideration, are relatively significant. The terrestrial abundance of the Mg isotopes is $^{24}$Mg:$^{25}$Mg:$^{26}$Mg = 79:10:11 \citep{Rosman:98}. We define the heavy isotope fraction as $\Gamma = ({}^{25}\mathrm{Mg} + {}^{26}\mathrm{Mg})/\mathrm{Mg}$, which has a terrestrial value of $\Gamma_t=0.21$.

We have assumed for our final fits that the quasar absorber isotopic abundances are the same as the terrestrial abundances. However, if the abundances in the absorbers differ from the terrestrial abundances, this will introduce a small but potentially significant shift in the quasar absorption lines compared to laboratory measurements. Mg will be most affected by this, due to its low atomic mass compared to the other species. Previous work \citep{Murphy:03} noted that the effect could be particularly significant for low-$z$ absorbers, as these predominantly consist of the Fe/Mg combination. High-$z$ absorbers are less likely to be affected due to the use of more massive anchors (Si and Al), for which this effect is less relevant. Additionally, the use of many transitions with differing $q$ coefficients at high redshift will tend to reduce the importance of this effect \citep{Murphy:04:LNP}. 

Both observations \citep{Gay:2000} and theoretical estimates \citep{Timmes:1995} of stellar abundances for Mg suggest that the heavy isotope abundance of Mg (i.e. the $^{25}$Mg and $^{26}$Mg isotopes) decreases with decreasing metallicity. \citet{Murphy:03} noted that the low-$z$ Mg/Fe systems considered in the Keck sample have relative metal abundances, [Fe/H], in the range $-2.5$ to $0.0$, whereas the high-$z$ DLA systems have relative metal abundances of about $-1.0$. Therefore, the quasar absorbers we consider may also have sub-solar metallicities. However, observations of some low metallicity red giants show significant enrichment of the heavy Mg isotopes. \cite{Ashenfelter:04a,Ashenfelter:04b} considered a ``modest'' enhancement of the stellar initial mass function (IMF) for intermediate mass stars ($M\approx 5M_\odot$), and showed that this could produce $\Gamma \sim 0.4$ for [Fe/H] $\sim -1.5$. \citet{Fenner:05a} argued that such an IMF would substantially overproduce nitrogen relative to observations, and therefore that this mechanism of creating $\Gamma > \Gamma_t$ does not seem possible. 

However, it appears that the link between stellar evolution and the likely nitrogen abundance in quasar absorbers is not fully understood. \citet{Centurion:03a} described observations of extremely low relative abundances of nitrogen in DLAs, and thus argued that nitrogen production cannot be dominated by massive stars. In a detailed study, \citet{Dessauges-Zavadsky:07a} argued that ``no single star formation history explains the diverse sets of abundance patterns in DLAs''. \citet{Melendez:07a} claimed (in contrast to previous analyses) that heavy Mg isotope enrichment due to AGB stars in the Galaxy halo does not occur until [Fe/H] $\gtrsim -1.5$. \citet{Levshakov:09a} examined 11 metal-rich, high-redshift ($1.5 < z < 2.9$) quasar absorbers and argued that the nitrogen abundance is uncorrelated with the metallicity, which implies that nitrogen enrichment has several sources. 

From the arguments above, it appears that the observational situation concerning $\Gamma$ at high redshift is uncertain. There are no stringent, independent observations which constrain $\Gamma$ in our sample. We therefore treat $\Gamma$ as unknown and explore what happens if we vary it.

\begin{figure}
\begin{center}
\ifpdf
\includegraphics[bb=77 79 455 727,angle=-90,width=82.5mm]{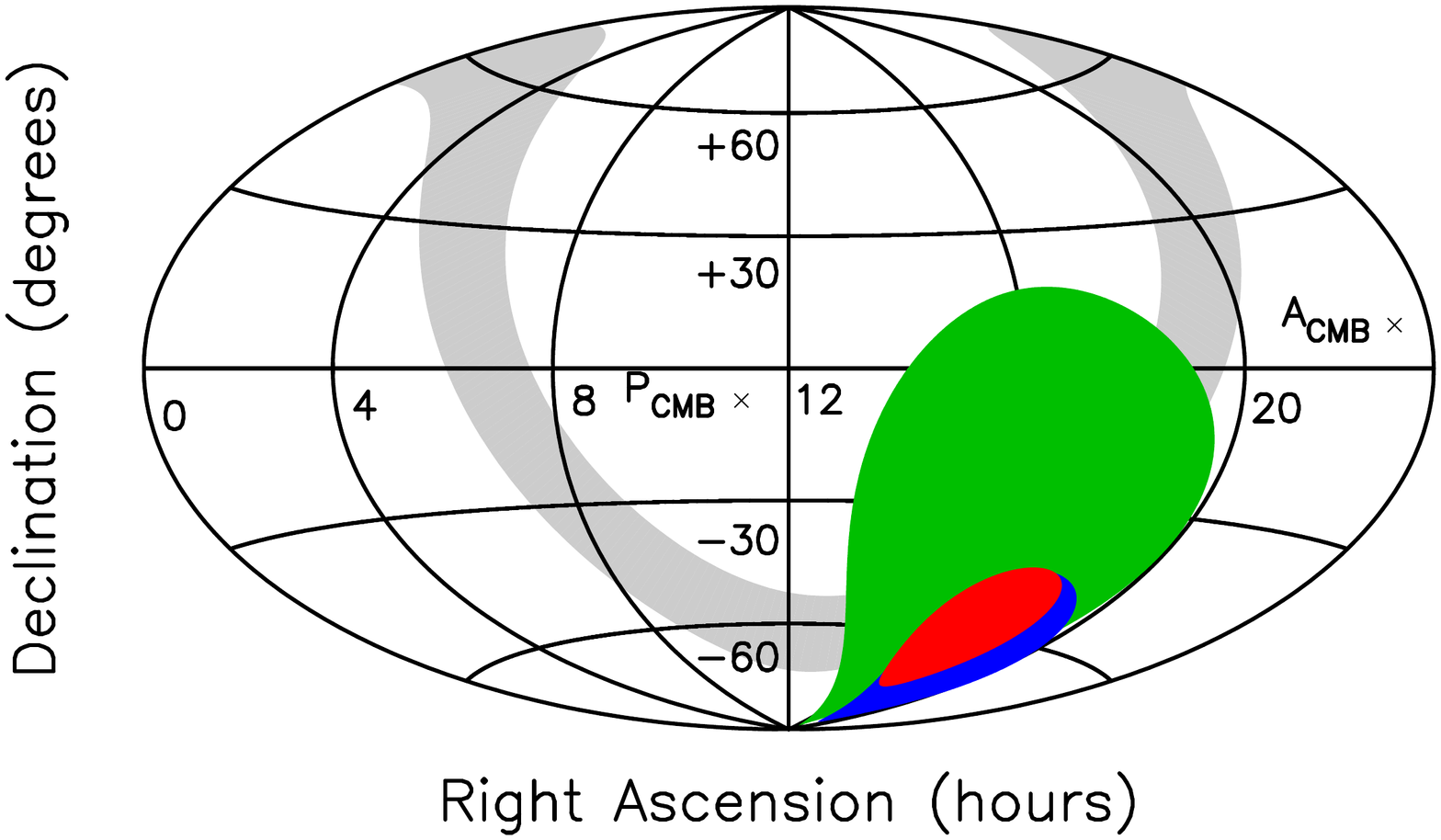}
\else
\includegraphics[bb=77 79 455 727,angle=-90,width=82.5mm]{images/skymap_nomg.eps}
\fi
\end{center}
\caption{$1\sigma$ confidence regions for the Keck (green), VLT (blue) and combined (red) dipoles in the circumstance where absorbers containing Mg are fitted with no $^{25}$Mg or $^{26}$Mg, to mimic the maximum possible effect of a lower heavy isotope fraction in the quasar absorbers compared to terrestrial values. Although the confidence regions are enlarged as a result of extra scatter introduced into the data, the reasonable alignment between the samples is still maintained. The separation between the Keck and VLT dipole vectors increases to $32^\circ$, which has a chance probability of 11 percent. \label{fig_skymap_noheavymg}}
\end{figure}

We first consider $\Gamma < \Gamma_t$. To place an upper limit on the effect of $\Gamma<\Gamma_t$, we re-fit all the VLT absorbers with no $^{25}$Mg or $^{26}$Mg, and similarly re-fit the absorbers in \citet{Murphy:04:LNP} using no $^{25}$Mg or $^{26}$Mg. We give the parameters for the fits to the Keck, VLT and combined samples in this situation in table \ref{tab_noheavymg}. The confidence regions on the dipole location are shown in figure \ref{fig_skymap_noheavymg}. Importantly, the dipole model remains statistically significant at the $3.5\sigma$ level. The reduction in significance from $4.1\sigma$ is primarily due to extra scatter introduced into the $\Delta\alpha/\alpha$ values about the model. The extra scatter implies that the $\Gamma=0$ model is not a good model for the absorbers. Additionally, the monopole becomes statistically significant at the $5.7\sigma$ level. Thus, a lower heavy isotope abundance in the quasar absorbers is unable to explain the dipole effect, and additionally increases the significance of the monopole term. The increase in significance of the monopole term mirrors the result in \citet{Murphy:03}. 

\begin{table*}
\caption{Effect of removal of $^{25}$Mg and $^{26}$Mg isotopes on the model $\Delta\alpha/\alpha = A\cos(\Theta) + m$. Generally speaking the effect is to push $\Delta\alpha/\alpha$ to more negative values. The column ``$\delta A$'' gives $1 \sigma$ confidence limits on $A$. The column labelled ``significance'' gives the significance of the dipole+monopole model with respect to the monopole model. This also introduces extra scatter into the $\Delta\alpha/\alpha$ values about the dipole model, which implies (unsurprisingly) that fits with no heavy Mg isotopes are not a good representation of the absorbers. Despite the extra scatter, the dipole model is still significant at the $3.5\sigma$ level. Additionally, the monopole term here becomes significant at the $5.7\sigma$ level. $\sigma_\mathrm{rand}$ is given for the different samples; HC refers to the Keck high-contrast sample. \label{tab_noheavymg}}
 \begin{center}
\begin{tabular}{clccccccc}
\hline 
Sample & $\sigma_\mathrm{rand}$ $(10^{-5})$ &  $A$ ($10^{-5}$) & $\delta A$ ($10^{-5}$) & RA (hr) & dec.\ ($^\circ$) & $m$ ($10^{-5}$) & significance\\
\hline
VLT      & 1.04           & 1.20  & $[0.78, 1.75]$ & $18.1 \pm 1.4$ & $-65\pm14$ & $-0.439 \pm 0.197$ & $1.9\sigma$ \\
Keck     & 1.63 for HC    & 0.42  & $[0.32, 0.88]$ & $16.6 \pm 2.2$ & $-35\pm35$ & $-0.835 \pm 0.156$ & $0.4\sigma$  \\
VLT + Keck\ & As above       & 0.98  & $[0.77, 1.23]$ & $17.3 \pm 1.0$ & $-59\pm10$ & $-0.528 \pm 0.092$ & $3.5\sigma$ \\ 
\hline
 \end{tabular}
 \end{center}
\end{table*} 

We now consider the impact of increasing the Mg heavy isotope fraction ($\Gamma > \Gamma_t$). In section \ref{s_monopole}, we discussed the presence of a low-$z$ monopole in both samples, where the difference between the two samples is remarkably small. Explaining this result via alterations to the Mg isotope abundance would require enrichment of the heavy isotope fraction relative to terrestrial values. If we assume that all of the $z<1.6$ monopole is due to relative enrichment of the heavy Mg isotopes, we can extrapolate from the $\Gamma=\Gamma_t$ and $\Gamma=0$ cases to estimate $\langle \Gamma_{z<1.6} \rangle$ using a simple linear model. A linear model may be used as the response of $\Delta\alpha/\alpha$ to changes in $\Gamma$ is linear \citep{Murphy:04:LNP}. This model assumes that the ratio of $^{25}$Mg/$^{26}$Mg is fixed. For $z<1.6$, $m = (-0.390 \pm 0.108) \times 10^{-5}$ for $\Gamma=\Gamma_t$, and $m = (-0.884 \pm 0.115) \times 10^{-5}$ for the case $\Gamma=0$. Under our linear model, $\langle \Gamma_{z<1.6} \rangle \approx 0.32$ in order to make $m=0$. If we take $\sigma_m = 0.108 \times 10^{-5}$ as a representative error, this yields $ \langle \Gamma_{z<1.6} \rangle = 0.32 \pm 0.03$. 

In summary, variations in the magnesium heavy isotope fraction have the potential to significantly impact the monopole component of the angular dipole + monopole model, but cannot explain angular variations in $\alpha$. 

\subsection{A test for systematic kinematic segregation of species}\label{s_testforkinematicsegregation}

In the absorption system modelling described earlier in this paper we have tied 
corresponding velocity components to have the same redshift.  We are thus assuming that 
the velocity structure for each complex is the same in those species which are tied.  
The justification, apart from empirical, is that we restrict the assumption to 
species with similar ionisation potentials. If there were gross departures from
this assumption we would be unable to derive statistically acceptable fits in all
cases. However, this is not found.  Thus, if there are kinematic segregations between
species, these must therefore be small compared to the parameter error estimates for individual 
redshifts.

Kinematic segregation between different ions might in principle result from several 
causes, including ionisation potential differences, mass segregation due to 
gravitational effects, and a possible dependence on the formation site.
Kinematic segregation clearly applies only to different species or potentially
to different ionisation stages of the same element. Pairs of transitions 
in the same species are immune to the effect.  Thus we formed two groups of pairs of 
transitions, intra- and inter-ion using a previously studied subset of our entire sample,
 that present in \citet{Chand:2004}.  We chose that sample for this experiment because 
it was initially selected on the basis of simple velocity structure, which usually means that
all the transitions studied are not saturated.
However, two of the absorption systems from the \citeauthor{Chand:2004} sample towards the same 
quasar (J000344$-$232355, $z_{\mathrm{abs}}=2.1854, 2.1872$) were not used because this complex 
failed the initial selection tests described in Section \ref{s_modellingvelstructure}.

There is an additional advantage of using this particular sample for
a kinematic segregation test.  In this test we measure velocity shifts, $\Delta v$, between 
different absorption transitions to identify kinematic segregation.  A non-zero $\Delta \alpha/\alpha$ may 
manifest itself as such a velocity shift. The redshift distribution and location
on the sky of these absorption systems happens to be such that the 
values of $\Delta \alpha/\alpha$ predicted by our dipole model are small. The velocity splitting for each pair is then
\begin{equation}
\Delta v = v_1 - v_2 \approx -2c \Delta q \frac{\Delta \alpha}{\alpha},
\end{equation}
where $\Delta q = q_2/\omega_2 - q_1/\omega_1$.  Here the subscripts 1 and 2 refer to the
different members of the pair. Column 7 in table \ref{tab_velseptest} shows that the
dipole-predicted values are small for these absorption systems.  
Hence, for this sample, whether or not there is a cosmological dipole in $\alpha$, these particular
absorption systems provide a clean test for kinematic segregation.

In order to measure the velocity shifts between pairs of transitions,
we fitted all species simultaneously.  Redshifts and $b$-parameters of corresponding
velocity components were tied in the usual way.  $\Delta \alpha/\alpha$ was
fixed at zero. All column densities of all velocity components were
free to vary.  However, in order to allow complete freedom for one transition to
shift relative to another, i.e.~to allow us to identify any kinematic segregation,
if present, for a system having $m$ different absorbing transitions,
$m-1$ new free parameters were introduced into each fit, where each of the $m-1$ 
parameters allowed for a potential relative velocity shift for that transition relative to the 
mean system redshift.  (There were $m-1$ velocity shifts to avoid
degeneracy with redshift). 

In doing it this way, we can allow for, and measure most sensitively,
bulk shifts between one ion and another,
which, if present, could emulate a non-zero $\Delta \alpha/\alpha$ for
any particular absorption system.  However, we note we would require an extremely unusual 
universe if kinematic segregation were to emulate a
spatial dipole in $\Delta \alpha/\alpha$ since each gas
cloud would need to be aligned with some particular direction in
the universe.

The $m-1$ measured velocity shifts were paired 
in order to obtain a roughly equal number of inter- and intra-ion
pairs.  The way in which this was done depended on transitions available.
However, the systematic pairing procedure we followed was to match pairs
of transitions of different strengths, such that at least one member of the
pair provided a good estimate of the global velocity structure.  
In order to derive a consistent velocity sign convention over all pairs, 
the velocity difference for each pair is in all cases measured relative to the bluest 
line in the pair.  

\begin{figure}
\includegraphics[bb=36 31 687 512,width=80mm]{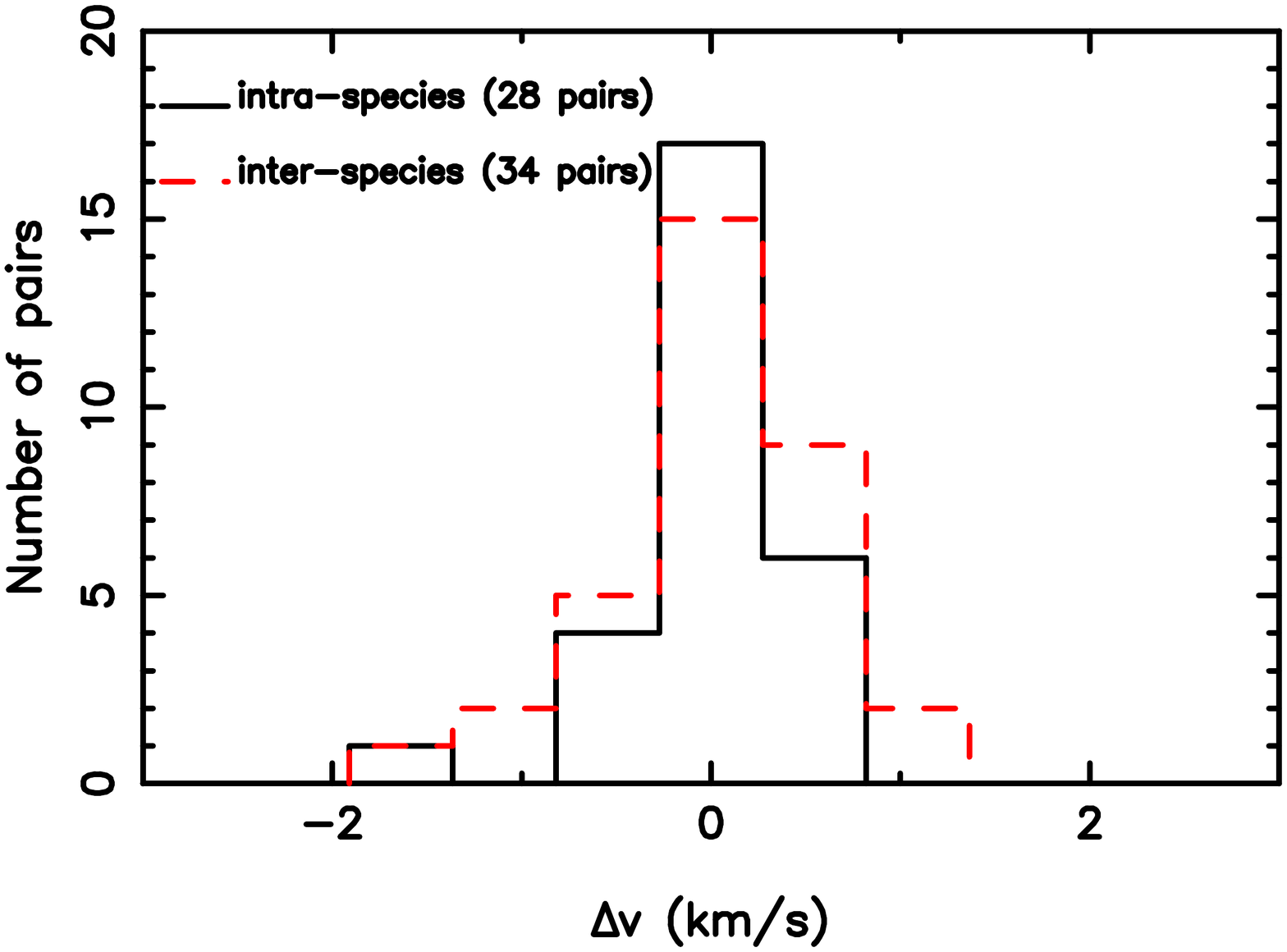}
\caption{\label{fig:dvhistv3} Histograms of velocity differences between intra-ion (solid line) and inter-ion pairs (dashed line). A Mann-Whitney $U$ test and the $F$ test find no significant differences between these 2 distributions (see text).  Thus, we have not found any evidence for kinematic segregation between the species used.}
\end{figure}

The individual velocity measurements are presented in Table \ref{tab_velseptest},
giving 34 inter-ion and 28 intra-ion pair measurements.
In Figure \ref{fig:dvhistv3}, we show the distributions for intra- and inter-ion
velocities.  A random kinematic segregation between inter-ion pairs
would be revealed as a broadening of the inter-ion distribution
relative to the intra-ion distribution.  The $F$ test is sensitive to such a 
broadening and we use it to test the equality of the variances of the intra- 
and inter-ion distributions. We calculate an $F$ statistic of 1.91, which implies 
that the probability that the variances of the two distributions differ is only $\approx$9\%.  
In the unlikely case that kinematic segregation is non-random the effect would 
be revealed as a shifting of the  mean of the inter-ion distribution with respect 
to the intra-ion distribution. The Mann-Whitney $U$ test is sensitive to such a shift. 
Comparing these 2 distributions the Mann-Whitney $U$ statistic is 403, which for a 
two-sided test implies that the probability that the two distributions differ is $\approx$30\%. 
We thus detect no significant difference between the inter- and intra-ion distributions.

In summary, we have devised a method which checks for
kinematic segregation and which is similar to the way in which we have
measured $\Delta \alpha/\alpha$.  Although we have tied species together, we 
allow for bulk velocity shifts between species by introducing the
additional velocity shift parameter.  It is exactly this
kind of bulk kinematic segregation which, if present, could emulate
a variation in $\alpha$.  An $\alpha$ spatial dipole may only be produced by 
kinematic segregation if and only if the kinematic segregation is non-random.
Our Mann-Whitney $U$ test results show no evidence for non-random segregation.
If random kinematic segregation were present in the sample used for
this test (our $F$ test), that would be indicated by an increase in $\sigma_{rand}$
(see Section \ref{LTS_method} earlier in this paper).
At present there is no evidence for any kinematic 
segregation amongst the atomic species used in the current sample and thus we
find no evidence that the dipole signal we have found 
is a result of kinematic segregation.

\begin{table*}
\caption{Results for $\Delta v$, the velocity separations between inter-ion and intra-ion pairs. The emission redshift of the quasar and the absorption redshift are given by $z_{\mathrm{em}}$ and $z_{\mathrm{abs}}$ respectively. Columns five and six show $\Delta v$ and the corresponding 1$\sigma$ error for each pair. We can predict the expected $\Delta v$ for each pair given a model of $\alpha$ variation. Column seven shows the predicted value of $\Delta v$ corresponding to $\Delta \alpha/\alpha=A\cos(\Theta)+m$.\label{tab_velseptest}}
\begin{tabular}{llllrrr}
\hline
Quasar Name & $z_{\mathrm{em}}$ & $z_{\mathrm{abs}}$ & Species                       & $\Delta v$ (km\,s$^{-1}$) & $\sigma_{\Delta v}$ (km\,s$^{-1}$)& $\Delta v_{\mathrm{predicted}}$ (km\,s$^{-1}$) \\
\hline
  
J134427$-$103541 & 2.35     & 0.873     & Fe\,\textsc{ii}\,2344/Mg\,\textsc{ii}\,2796   & $-$0.446    & 0.557 & 0.031  \\
               &          &           & Mg\,\textsc{ii}\,2803/Mg\,\textsc{i}\,2852    & $-$1.808    & 0.744 & 0.001  \\
               &          &           & Fe\,\textsc{ii}\,2374/Fe\,\textsc{ii}\,2383   & 0.553     & 0.867 & 0.003  \\
               &          & 1.278     & Fe\,\textsc{ii}\,2344/Mg\,\textsc{ii}\,2796   & $-$0.446    & 0.557 & 0.031  \\
               &          &           & Mg\,\textsc{ii}\,2803/Mg\,\textsc{i}\,2852    & 1.171     & 0.992 & 0.001  \\
               &          &           & Fe\,\textsc{ii}\,2374/Fe\,\textsc{ii}\,2600   & 0.034     & 0.950 & 0.003  \\
               &          &           & Mn\,\textsc{ii}\,2576/Mn\,\textsc{ii}\,2606   & $-$1.476    & 3.759 & 0.012  \\
               &          & 1.915     & Si\,\textsc{ii}\,1526/Fe\,\textsc{ii}\,2600   & 0.104     & 0.268 & $-$0.041 \\
               &          &           & Fe\,\textsc{ii}\,2344/Fe\,\textsc{ii}\,2383   & 0.135     & 0.390 & $-$0.004 \\
J012417$-$374423 & 1.91     & 0.822     & Fe\,\textsc{ii}\,2600/Mg\,\textsc{ii}\,2796   & 0.237     & 0.325 & 0.017  \\
               &          &           & Mg\,\textsc{ii}\,2803/Mg\,\textsc{i}\,2852    & $-$0.792    & 0.401 & 0.000  \\
               &          &           & Fe\,\textsc{ii}\,2344/Fe\,\textsc{ii}\,2383   & $-$0.189    & 0.883 & $-$0.002 \\
               &          & 0.859     & Fe\,\textsc{ii}\,2600/Mg\,\textsc{ii}\,2796   & 0.252     & 0.473 & 0.017  \\
               &          &           & Mg\,\textsc{ii}\,2803/Mg\,\textsc{i}\,2852    & 0.561     & 0.958 & 0.000  \\
               &          &           & Fe\,\textsc{ii}\,2344/Fe\,\textsc{ii}\,2383   & 0.191     & 0.708 & $-$0.002 \\
               &          & 1.243     & Fe\,\textsc{ii}\,2374/Mg\,\textsc{ii}\,2796   & $-$0.413    & 0.409 & 0.019  \\
               &          &           & Fe\,\textsc{ii}\,2260/Mg\,\textsc{i}\,2852    & 0.545     & 0.790 & 0.017  \\
               &          &           & Fe\,\textsc{ii}\,2344/Fe\,\textsc{ii}\,2383   & $-$0.124    & 0.542 & $-$0.002 \\
               &          &           & Al\,\textsc{iii}\,1854/Al\,\textsc{iii}\,1862 & $-$0.065    & 0.532 & 0.002  \\
J024008$-$230915 & 2.23     & 1.635     & Mg\,\textsc{ii}\,2803/Mg\,\textsc{i}\,2852    & 0.584     & 0.478 & $-$0.004 \\
               &          &           & Si\,\textsc{ii}\,1526/Mg\,\textsc{ii}\,2803   & 0.328     & 0.252 & 0.003  \\
               &          &           & Fe\,\textsc{ii}\,2344/Fe\,\textsc{ii}\,2587   & $-$0.164    & 0.471 & 0.009  \\
               &          &           & Al\,\textsc{iii}\,1854/Al\,\textsc{iii}\,1862 & 0.292     & 0.480 & $-$0.005 \\
               &          & 1.637     & Fe\,\textsc{ii}\,2344/Mg\,\textsc{ii}\,2796   & $-$0.126    & 0.292 & $-$0.033 \\
               &          &           & Mg\,\textsc{ii}\,2803/Mg\,\textsc{i}\,2852    & $-$1.347    & 0.712 & $-$0.001 \\
               &          &           & Fe\,\textsc{ii}\,2587/Fe\,\textsc{ii}\,2600   & 0.662     & 0.453 & $-$0.004 \\
               &          & 1.657     & Al\,\textsc{ii}\,1670/Mg\,\textsc{ii}\,2803   & 0.131     & 0.244 & $-$0.006 \\
               &          &           & Si\,\textsc{ii}\,1526/Mg\,\textsc{i}\,2852    & 0.332     & 0.780 & 0.002  \\
               &          &           & Fe\,\textsc{ii}\,2344/Fe\,\textsc{ii}\,2383   & $-$0.211    & 0.451 & 0.005  \\
J000344$-$232355 & 2.28     & 0.452     & Fe\,\textsc{ii}\,2600/Mg\,\textsc{ii}\,2796   & 0.001     & 0.142 & 0.005  \\
               &          &           & Mg\,\textsc{ii}\,2803/Mg\,\textsc{i}\,2852    & 0.163     & 0.659 & 0.000  \\
               &          &           & Fe\,\textsc{ii}\,2344/Fe\,\textsc{ii}\,2383   & 0.436     & 0.280 & $-$0.001 \\
J011143$-$350300 & 2.40     & 1.182     & Fe\,\textsc{ii}\,2600/Mg\,\textsc{ii}\,2796   & $-$0.038    & 0.297 & 0.013  \\
               &          &           & Mg\,\textsc{ii}\,2803/Mg\,\textsc{i}\,2852    & $-$0.229    & 0.788 & 0.000  \\
               &          &           & Fe\,\textsc{ii}\,2344/Fe\,\textsc{ii}\,2383   & $-$0.424    & 0.653 & $-$0.002 \\
               &          & 1.350     & Fe\,\textsc{ii}\,2600/Mg\,\textsc{ii}\,2796   & 0.006     & 0.090 & 0.013  \\
               &          &           & Mg\,\textsc{ii}\,2803/Mg\,\textsc{i}\,2852    & 0.252     & 0.164 & 0.000  \\
               &          &           & Fe\,\textsc{ii}\,2344/Fe\,\textsc{ii}\,2383   & $-$0.134    & 0.134 & $-$0.002 \\
J222006$-$280323 & 2.41     & 0.942     & Fe\,\textsc{ii}\,2600/Mg\,\textsc{ii}\,2796   & $-$0.015    & 0.225 & 0.047  \\
               &          &           & Fe\,\textsc{ii}\,2587/Mg\,\textsc{ii}\,2803   & 0.357     & 0.278 & 0.056  \\
               &          &           & Fe\,\textsc{ii}\,2344/Fe\,\textsc{ii}\,2383   & 0.312     & 0.529 & $-$0.006 \\
               &          & 1.556     & Al\,\textsc{ii}\,1670/Mg\,\textsc{ii}\,2796  & $-$0.285    & 0.224 & 0.004  \\
               &          &           & Fe\,\textsc{ii}\,2383/Fe\,\textsc{ii}\,2600   & 0.135     & 0.170 & 0.000  \\
               &          &           & Fe\,\textsc{ii}\,2344/Fe\,\textsc{ii}\,2587   & 0.344     & 0.458 & $-$0.011 \\
J135038$-$251216 & 2.61     & 1.439     & Fe\,\textsc{ii}\,1608/Mg\,\textsc{ii}\,2796   & 0.622     & 0.802 & $-$0.055 \\
               &          &           & Si\,\textsc{ii}\,1808/Mg\,\textsc{ii}\,2803   & 0.919     & 0.824 & 0.014  \\
               &          &           & Zn\,\textsc{ii}\,2026/Mn\,\textsc{ii}\,2594   & 0.335     & 1.272 & 0.052  \\
               &          &           & Cr\,\textsc{ii}\,2056/Cr\,\textsc{ii}\,2066   & $-$0.802    & 1.319 & 0.012  \\
               &          &           & Fe\,\textsc{ii}\,2260/Fe\,\textsc{ii}\,2344   & 0.163     & 1.153 & 0.000  \\
J045523$$-$$421617 & 2.67     & 0.908     & Mg\,\textsc{ii}\,2803/Mg\,\textsc{i}\,2852    & 0.234     & 0.293 & 0.000  \\
               &          &           & Fe\,\textsc{ii}\,2344/Fe\,\textsc{ii}\,2600   & 0.150     & 0.175 & 0.000  \\
               &          & 1.858     & Al\,\textsc{ii}\,1670/Mg\,\textsc{ii}\,2796   & -0.507    & 0.818 & 0.000  \\
               &          &           & Fe\,\textsc{ii}\,2383/Fe\,\textsc{ii}\,2600   & $-$0.030    & 1.220 & 0.000  \\
J000448$-$415728 & 2.76     & 1.542     & Fe\,\textsc{ii}\,2383/Fe\,\textsc{ii}\,2600   & $-$0.102    & 0.896 & 0.000  \\
               &          &           & Fe\,\textsc{ii}\,2344/Fe\,\textsc{ii}\,2374   & $-$0.199    & 1.125 & $-$0.010 \\
               &          & 2.168     & Si\,\textsc{ii}\,1526/Al\,\textsc{ii}\,1670   & 0.171     & 0.199 & $-$0.012 \\
               &          &           & Mg\,\textsc{ii}\,2796/Mg\,\textsc{ii}\,2803   & 0.002     & 0.091 & 0.004  \\
               &          &           & Fe\,\textsc{ii}\,1608/Fe\,\textsc{ii}\,2344   & $-$0.753    & 0.624 & $-$0.078 \\
               &          & 2.302     & Si\,\textsc{ii}\,1526/Fe\,\textsc{ii}\,1608   & $-$0.205    & 0.117 & 0.030  \\
               &          &           & Ni\,\textsc{ii}\,1709/Ni\,\textsc{ii}\,1741   & 0.142     & 0.540 & 0.037  \\
J212912$-$153841 & 3.28     & 2.022     & Si\,\textsc{ii}\,1808/Mg\,\textsc{ii}\,2803   & 0.533     & 0.401 & $-$0.001 \\
               &          &           & Fe\,\textsc{ii}\,2344/Fe\,\textsc{ii}\,2587   & $-$0.395    & 0.578 & $-$0.017 \\

\hline
\end{tabular}
\end{table*}

\subsection{UVES, a dual-arm spectrograph}

UVES is a dual-arm spectrograph, where the incoming light is split into a red arm and a blue arm using a dichroic mirror. In principle, misalignment of the slits in the blue arm relative to the red arm would produce a distortion of the wavelength scale between the two arms, which could mimic a change in $\alpha$ if transitions were fitted simultaneously from the blue and red arms. \citet{Molaro:08} investigated the possibility that such misalignment might cause velocity shifts between the blue and red arms, using measurements of asteroids, and argue that the two arms do not show separation by more than $\sim 30\,\mathrm{ms}^{-1}$ in the situation where the science exposures are bracketed by the ThAr exposures. 

However, we note that \citeauthor{Molaro:08} used a slit with of $0.5\arcsec$, which is rather different to the $\sim0.7\arcsec$ to $\sim 1.0\arcsec$ typical of the quasar exposures. The UVES archive indicates that, for  the observations of \citeauthor{Molaro:08}, the seeing was always poorer than the slit width. If the slits for the blue and red arms are misaligned, one would expect the induced effect on wavelength calibration to depend on slit size. In the seeing-limited regime, the slit is relatively uniformly illuminated, and therefore the observed science wavelengths should be well calibrated through the ThAr exposure. On the other hand, when the seeing is much better than the slit, one might expect to see larger differences, if such differences exist. 

\subsection{Inter-telescope systematics \& the $\Delta v$ test}\label{s_dvtest}

We have previously argued that the observed angular variation in $\alpha$ is unlikely to be caused by systematic effects. Any such systematic must be well correlated with sky position in order to explain the good alignment between dipole directions in dipole models fitted to the Keck and VLT samples. Nevertheless, accurate wavelength calibration for quasar spectra remains an issue of importance given the limitations of the existing ThAr calibration process. Therefore, in this section we directly explore wavelength-dependent systematics using common observations of quasars from both Keck and VLT.

One can ask whether there is some systematic effect which is constant in time between the two telescopes which might generate the observed dipole effect. All of the Keck spectra used in the analysis in this paper were acquired whilst HIRES had only one CCD chip. In this configuration, multiple exposures are needed to yield full wavelength coverage. If the quasar image is not precisely centred in the spectrograph slit for every exposure, velocity offsets between spectral segments obtained at different times are possible. This issue should be substantially mitigated at VLT, as UVES can acquire almost the entire spectral range in a single observation. The effect could be exaccerbated in conditions of good seeing and could include an additional small effect due to the seeing profile decreasing slightly towards the red end of the spectrum.

It turns out that the VLT and Keck samples have 7 quasars in common. We give a list of the quasars common to the VLT and Keck samples in table \ref{tab_dvquasars}. The use of common sources allows one to search for problems with wavelength calibration; absorption features should be found at the same barycentric vacuum wavelength between different exposures. Note that the number of absorption lines which can be used for this purpose is much larger than is used for analysing $\Delta\alpha/\alpha$. Whilst for $\Delta\alpha/\alpha$ many absorption lines are needed to yield a single measurement of $\Delta\alpha/\alpha$, in principle each absorption line in the spectrum yields one constraint on potential wavelength distortion. 

\begin{table}
 \caption{List of quasars common to the VLT and Keck samples. Keck names are given by their B1950 designation as they appear in \citet{Murphy:03}, whereas VLT names are given by their J2000 designation. \label{tab_dvquasars}}
\begin{center}
\begin{tabular}{cc}
\hline
Keck sample name & VLT sample name\\
\hline
0216+0803   & J021857+081727\\
0237$-$2321  & J024008$-$230915\\
0940$-$1050 & J094253$-$110426\\
1202$-$0725 & J120523$-$074232\\
0528$-$250  & J053007$-$250329\\
1337+1121   & J134002+110630\\
2206$-$1958 & J220852$-$194359 \\
\hline
\end{tabular}
\end{center}
\end{table} 

To explore potential wavelength scale distortions, we propose a method which we refer to as the $\Delta v$ test. The method proceeds as follows: \emph{i)} for each common quasar, visually identify regions of non-terrestrial absorption, typically having width of a few \AA; \emph{ii)} for each of these regions, perform a Voigt profile fit to the VLT spectral data (identification of the transition responsible is unimportant); \emph{iii)} fit corresponding spectral regions of the Keck and VLT simultaneously, but with an extra free parameter, $\Delta v$, which allows for a velocity shift between the two spectral regions. The VLT spectral data for these regions were fitted using an automated Voigt profile fitting routine designed to fit regions of the forest automatically. $\Delta v$ is defined hereafter as the velocity difference $\Delta v = v(\mathrm{VLT}) - v(\mathrm{Keck})$ which must be applied to minimise $\chi^2$ between two comparable spectral regions. Each value of $\Delta v$ provides an estimate of the velocity offset between the two telescopes at that observed wavelength, giving $\Delta v(\lambda)$. One can therefore examine the functional form of $\Delta v(\lambda)_i$, where $i$ refers to the $i$th quasar pair under consideration. For each set of $\Delta v$ values from a particular spectral pair, we use the LTS method to calculate the weighted mean of that set of $\Delta v$ values, which we then subtract from the $\Delta v$ values for that spectral pair. This is to remove any constant offset resulting from mis-centering of the quasar within the slit. We use $k=0.95n$ for the LTS fit (see section \ref{LTS_method}). 

Any relative wavelength scale distortion can in principle be removed by applying an inverse function based on the observed $\Delta v$ data. To see this, consider the form of the distortion. For an absorption line with rest wavelength $\lambda_0$, observed wavelength $\lambda_i$, and velocity distortion $\Delta v$ then
\begin{equation}
\lambda_i = \lambda_0(1+z)\left(1+\Delta v/c\right),
\end{equation}
where we have assumed that $\Delta v$ is constant over the absorption profile under consideration. The effect of $\Delta\alpha/\alpha$ can be ignored -- whatever transition is being examined is the same in both spectra, and so any effect due to a change in constants will be absorbed into the determination of $z$. There are two options to attempt to remove the wavelength scale distortion given some function $\Delta v(\lambda_\mathrm{obs})$. One could modify the spectral data, changing the observed wavelengths as $\lambda_\mathrm{obs} \rightarrow \lambda_\mathrm{obs}/[1+\Delta v(\lambda_\mathrm{obs})/c)]$. When one fits a particular transition, the other possibility is to perturb the rest wavelength of the transition fitted, as
\begin{equation}
\lambda_0 \rightarrow \lambda_0[1+\Delta v(\lambda_\mathrm{obs})/c]\label{eq_inv_dv_func}. 
\end{equation}
We use the second option for ease of implementation within \textsc{vpfit}. Doing this means that the value of $\Delta\alpha/\alpha$ derived from the fit will be the same as if the  wavelength scale from the other telescope in the spectral pair had been used, thereby removing any inter-telescope differences (provided that $\Delta v$ is correctly specified).

\begin{figure*}
\begin{center}
\ifpdf
\includegraphics[bb=52 44 556 750,angle=-90,width=160mm]{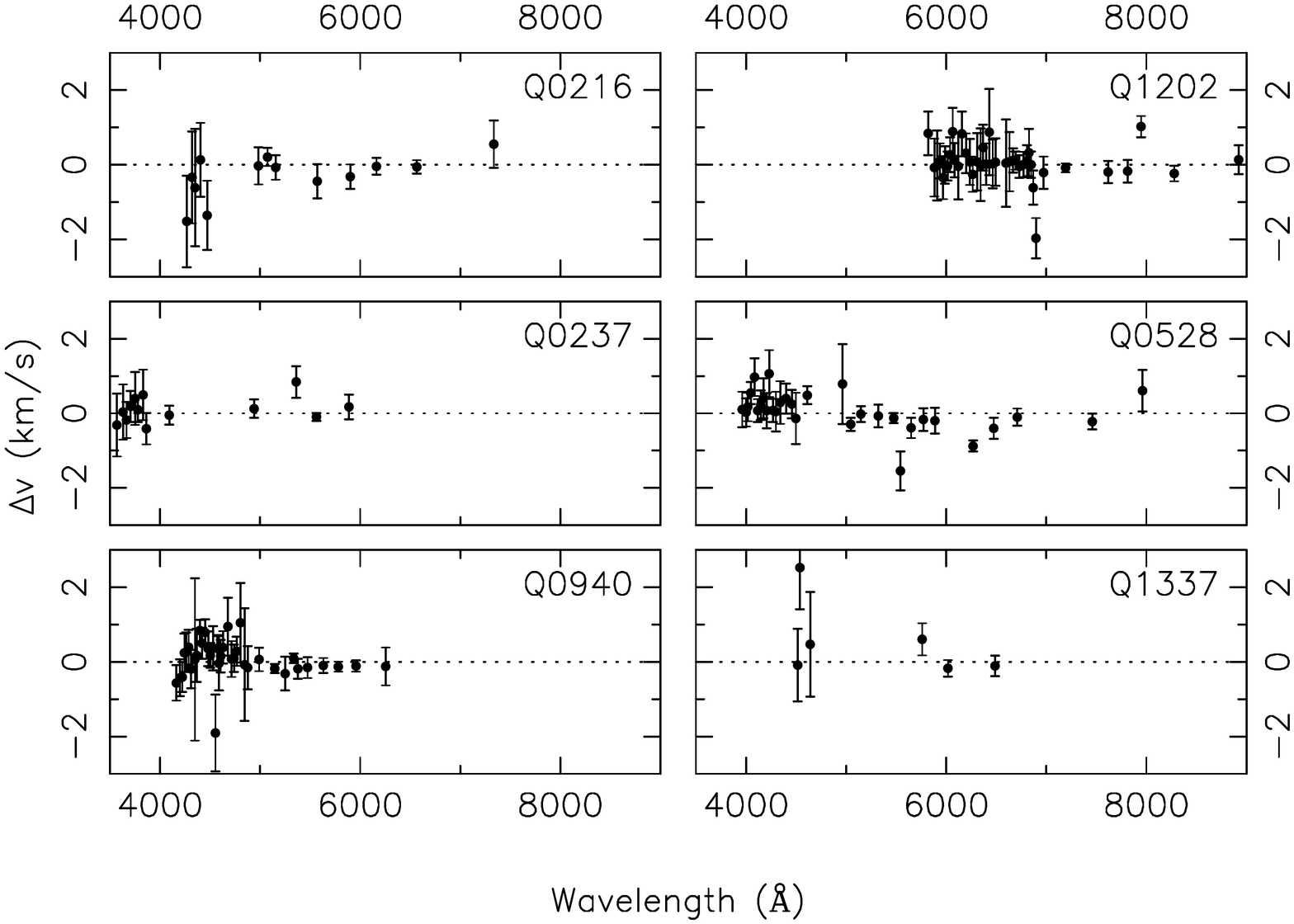}
\else
\includegraphics[bb=52 44 556 750,angle=-90,width=160mm]{images/dv_6quasars_data.eps}
\fi
\end{center}
\caption{Binned values of $\Delta v$ points for six quasar spectral pairs (``core pairs''), with vertical range restricted for clarity. Some data points lie outside the visible range due to the contribution of gross outliers; these points are automatically ignored in our robust fit. 5 values of $\Delta v$ contribute to each bin. $\Delta v$ is defined as the velocity difference which must be applied to the Keck spectrum to minimise $\chi^2$ for a joint fit. Visually, it is not clear that a common trend exists for the six quasars shown. For 0216+0803/J021857+081727, $\Delta v$ appears to increase with increasing wavelength. For 0237$-$2321/J024008$-$230915 and 1202$-$0725/J120523$-$074232, it is not clear whether a trend exists. 0940$-$1050/J094253$-$110426 shows a non-linear effect. 0528$-$250/J053007$-$250329 shows a clear trend for a decrease in $\Delta v$ with wavelength. The effect in 1337+1121/J134002$-$110630 is unclear given the limited wavelength coverage and relative lack of data. \label{fig_dv_6quasars} }
\end{figure*}

We show the $\Delta v$ data for 6 of the quasar spectral pairs (``core pairs''), which appear similar to each other, in figure \ref{fig_dv_6quasars}. We analyse the $\Delta v$ data from these quasars in the following section. We noticed a problem with the 7th pair, 2206$-$1958/J220852$-$194359, which displays variations of $\Delta v$ with wavelength which are grossly different from the other six pairs. A systematic trend in $\Delta v$ is seen, with a maximum difference in $\Delta v$ of $\sim 2.5\,\mathrm{km\,s^{-1}}$ over the range $4000 \lesssim \lambda \lesssim 6000\mathrm{\AA}$. In section \ref{Q2206P}, we apply an inverse function derived from the $\Delta v$ data seen in this spectral pair to the all the VLT spectra, and show that a distortion of this type cannot affect all the data. We consider the joint impact of the $\Delta v$ functions from the 6 core quasars and from 2206$-$1958/J220852$-$194359 in section \ref{s_dv_overalleffect}.

In all our analysis in this section we have removed those absorbers which were previously flagged as outliers from consideration in the statistical analysis (the $z = 1.542$ absorber towards J000448$-$415728 in the VLT sample and the $z\approx 2.84$ absorber towards Q1946$+$7658 in the Keck sample).

\subsubsection{$\Delta v$ data \& 6 core quasar pairs}\label{s_dv_6corepairs}

We note the presence of significant outliers within the set of $\Delta v$ values from these 6 spectral pairs. Therefore we rely wholly on robust statistical methods to estimate parameters for phenomenological models of $\Delta v(\lambda)$.

We discuss the particular trends for each spectral pair in the caption for figure \ref{fig_dv_6quasars}, but note that no common trend is seen between these spectra, as would be expected if no systematic wavelength distortions exist. Importantly, the functional form of $\Delta v$ appears to differ in both magnitude and sign between quasars. This suggests that any relative wavelength distortion is likely to average out over a large number of absorbers. Additionally, the wavelength coverage of the $\Delta v$ data for most spectra is significantly smaller than the wavelength range within which MM absorbers are fitted. This means that from each spectral pair it is impossible to tell what the wavelength distortion might be over large amounts of the spectral range.

\begin{figure}
\begin{center}
\ifpdf
\includegraphics[bb=50 59 556 766,angle=-90,width=82.5mm]{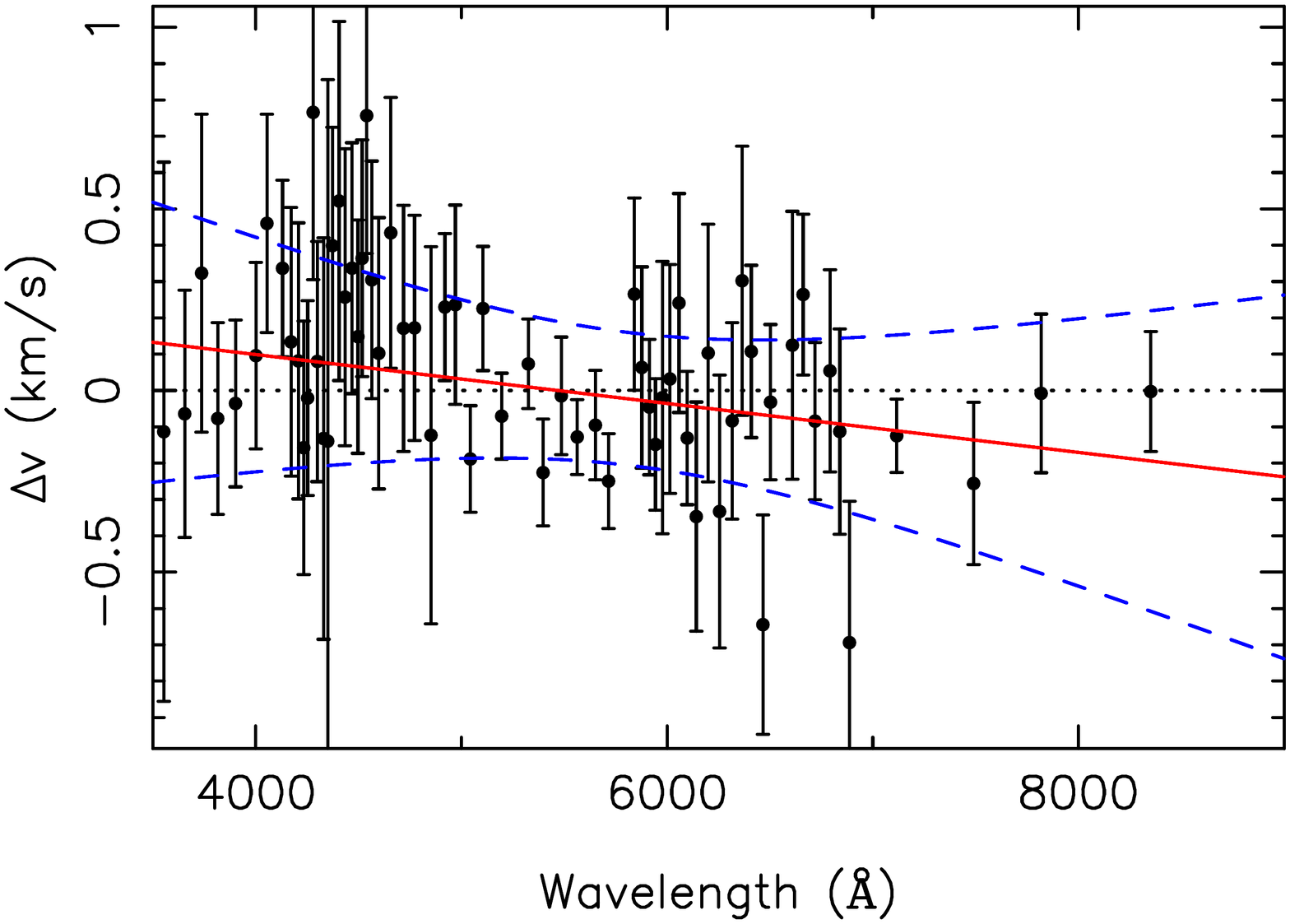}
\else
\includegraphics[bb=50 59 556 766,angle=-90,width=82.5mm]{images/dv_6q_linfit.eps}
\fi
\end{center}
\caption{LTS linear fit to the $\Delta v$ data for the 6 core quasar pairs. The dashed blue lines show the $1\sigma$ confidence limit on the linear fit (red line). We have utilised $k=0.95n$ for this fit to obtain robustness against some outliers. The data shown are binned with 5 points per bin, but only those points contributing to the fit have been included. For a fit of $\Delta v = a\lambda + b$, $a = (-7 \pm 14) \times 10^{-5}\,\mathrm{km\,s^{-1}\,\AA^{-1}}$ and $b = 0.38 \pm 0.71\,\mathrm{km\,s^{-1}}$. Most transitions used in the MM analysis fall in the range $4000 \lesssim \lambda \lesssim 7000\mathrm{\AA}$; from this graph there is no significant evidence for a significant wavelength distortion in this region. \label{fig_dv_6quasars_linfit}}
\end{figure}

Due to the fact that the $\Delta v$ values from each spectral pair do not densely span the whole spectroscopic wavelength range, we combine the $\Delta v$ values from each of the six core pairs together in order to estimate a common function which spans the full wavelength range. The functional form of this is unknown, however a high-order polynomial cannot be statistically supported. We use a linear function as a first approximation. We fit the linear function with the LTS method, using $k=0.95n$. We show this linear fit in figure \ref{fig_dv_6quasars_linfit}. For the form 
\begin{equation}
\Delta v = a\lambda + b, \label{eq_dv_linfunc}
\end{equation}
$a = (-7 \pm 14) \times 10^{-5}\,\mathrm{km\,s^{-1}\,\AA^{-1}}$ and $b = 0.38 \pm 0.71\,\mathrm{km\,s^{-1}}$. Note firstly that $a$ is statistically consistent with zero. Therefore, it is difficult to conclude that a common linear systematic exists in the $\Delta v$ data. Nevertheless, in the next section we apply an inverse function of this type to the VLT spectral data to determine the effect that a distortion of this type and magnitude would have on the detected dipole effect.

\begin{figure}
\begin{center}
\ifpdf
\includegraphics[bb=132 45 556 696,angle=-90,width=82.5mm]{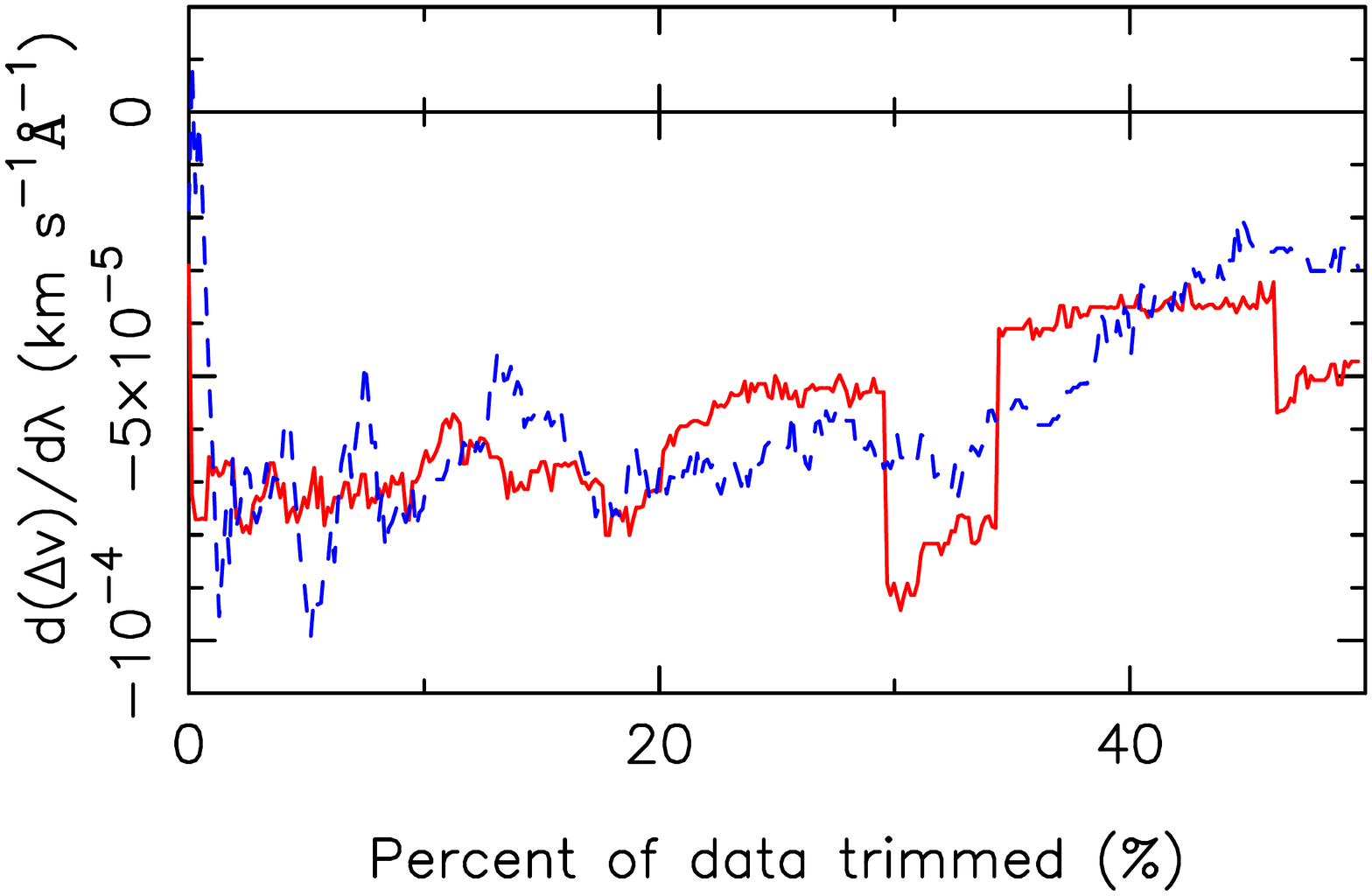}
\else
\includegraphics[bb=132 45 556 696,angle=-90,width=82.5mm]{images/dv_6q_trimgraph.eps}
\fi
\end{center}
\caption{Slope of $\Delta v$ vs $\lambda$ under the LTS method with different amounts of data excluded from the fit. The red (solid) line shows a weighted LTS fit, and the blue (dashed) line shows the unweighted fit. Importantly, the magnitude of the slope decreases as $k\rightarrow 0.5n$, implying that the slope we utilise may be an over-estimate of any underlying effect. \label{fig_dv_6quasars_trimgraph}}
\end{figure}

A legitimate question to ask is whether the choice of $k=0.95n$ is reasonable. We show the effect of difference choices of $k$ in figure \ref{fig_dv_6quasars_trimgraph}. Our estimate of the slope is not overly sensitive to a choice of $k$. With the exception of a small region around $k=0.7n$, the general trend is for the slope to decrease with decreasing $k$. The fact that the slope decreases with increased trimming implies that the underlying trend may be less than what we have estimated.

\subsubsection{Application to VLT sample}\label{s_dv_applytoVLT}

To investigate the effect of the potential wavelength distortions from the 6 core pairs, we apply an inverse $\Delta v$ function (equation \ref{eq_dv_linfunc}) to all the VLT absorbers by perturbing the rest wavelengths of the transitions fitted in each absorber, as described in section \ref{s_dvtest}. We apply the same linear function in every VLT spectrum fitted. This therefore puts the VLT and Keck spectral data on a common wavelength scale. Any observed angular variation in $\alpha$ which survives the inverse function can not be due to stable inter-telescope wavelength calibration differences.

Because we apply the inverse function by perturbing the rest wavelengths of transitions in absorbers fitted, we can only do this where each fitted transition occurs in only one spectral region in a particular fit. There are two pairs of absorbers where we have fitted both absorbers in the pair simultaneously, because a transition from one absorber in the pair overlaps with a transition from the other absorber (at a different redshift) in the pair. In this case, a particular transition can be fitted twice (in two absorbers, in two widely separated spectral regions). However, we cannot apply two different perturbations to a single rest wavelength. Therefore, we remove these two pairs of absorbers to form a ``VLT reference set''. Thus, we compare a VLT set of $\Delta\alpha/\alpha$ values where the $\Delta v$ inverse function has been applied with the $\Delta\alpha/\alpha$ values from a VLT reference set. The two pairs of absorbers which are removed are the $z \sim 2.253$ and $z \sim 2.380$ absorbers associated with J214225$-$442018 and the $z\sim 1.154$ and $z\sim 0.987$ absorbers in the same spectrum (i.e. 4 absorbers are removed to form the VLT reference set). This means that the VLT reference set contains 149 absorbers.

In table \ref{tab_dv_6quasars_results}, we give the results of applying the $\Delta v$ inverse function above to those absorbers in the VLT reference set. The effect is generally to push $\Delta\alpha/\alpha$ to more negative values. We show an updated plot of the confidence regions of the Keck, VLT and combined dipole locations in figure \ref{fig_dv_6quasars_linfit_skymap}. Although the statistical significance of the dipole+monopole model ($\Delta\alpha/\alpha = A\cos(\Theta) + m$) over the monopole-only model decreases from $3.9\sigma$ (reference set) to $3.1\sigma$, the position of the VLT (and therefore combined) dipole is effectively unchanged. This accords well with our earlier argument that because the detection of a dipole is a differential effect, it is difficult to emulate through any simple systematic. The Keck and VLT dipoles in this case are separated by $25^\circ$, which has a chance probability of 7 percent (see section \ref{s_alignmentbychance}). Also note that introducing this modification to the wavelength scale of the VLT spectra does not significantly change the good alignment between the $z<1.6$ and $z>1.6$ samples. The dipole directions in this case are separated by $13^\circ$, which has a chance probability of 2 percent.

We would expect that any good model for a wavelength-dependent systematic should quantitatively improve the fit of the dipole model to the $\Delta\alpha/\alpha$ values. To see if the $\Delta v$ model significantly improves the fit, we compare the AICC of the model $\Delta\alpha/\alpha=A\cos\Theta+m$ fitted to different sets of $\Delta\alpha/\alpha$ values: \emph{i)} $\mathrm{AICC}_{\mathrm{VLT},\Delta v}$, the AICC of the angular dipole model fitted to the VLT absorbers in the VLT reference set with the linear $\Delta v$ function described above applied; \emph{ii)} $\mathrm{AICC}_{\mathrm{VLT,ref}}$, the AICC of the angular dipole model fitted to the VLT absorbers in the VLT reference set; \emph{iii)} $\mathrm{AICC}_{\mathrm{VLT},\Delta v+\mathrm{Keck}}$, the AICC for the angular dipole model fitted to the absorbers in set (i) combined with the Keck absorbers in the Keck04-dipole set, and; \emph{iv)} $\mathrm{AICC}_{\mathrm{VLT,ref}+\mathrm{Keck}}$, the AICC for the angular dipole model fitted to the absorbers in set (ii) combined with the Keck absorbers in the Keck04-dipole set. In all cases we apply the same values of $\sigma_\mathrm{rand}$ to the VLT $\Delta\alpha/\alpha$ values in order to compare points on a like-with-like basis. We find that $\mathrm{AICC}_{\mathrm{VLT},\Delta v} - \mathrm{AICC}_{\mathrm{VLT},\mathrm{ref}} \approx -0.8$, indicating that the set of VLT absorbers with the $\Delta v$ inverse function applied is preferred, but not significantly. Comparing the VLT+Keck set to the equivalent reference set, we find that $\mathrm{AICC}_{\mathrm{VLT},\Delta v+\mathrm{Keck}} - \mathrm{AICC}_{\mathrm{VLT,ref}+\mathrm{Keck}} \approx -3.2$, indicating that the VLT+Keck set where the $\Delta v$ inverse function has been applied to the VLT absorbers is weakly preferred. However, when comparing the reference sets and the $\Delta v$ sets, the AICC does not account for the extra two parameters for the linear model of $\Delta v$ vs $\lambda$. Thus, with a two-parameter model for the $\Delta v$ function there is no significant preference for the $\Delta v$ results, and thus there is no strong evidence in the $\Delta \alpha/\alpha$ values themselves for a wavelength distortion of this type.

\begin{figure}
\begin{center}
\ifpdf
\includegraphics[bb=50 59 556 766,angle=-90,width=82.5mm]{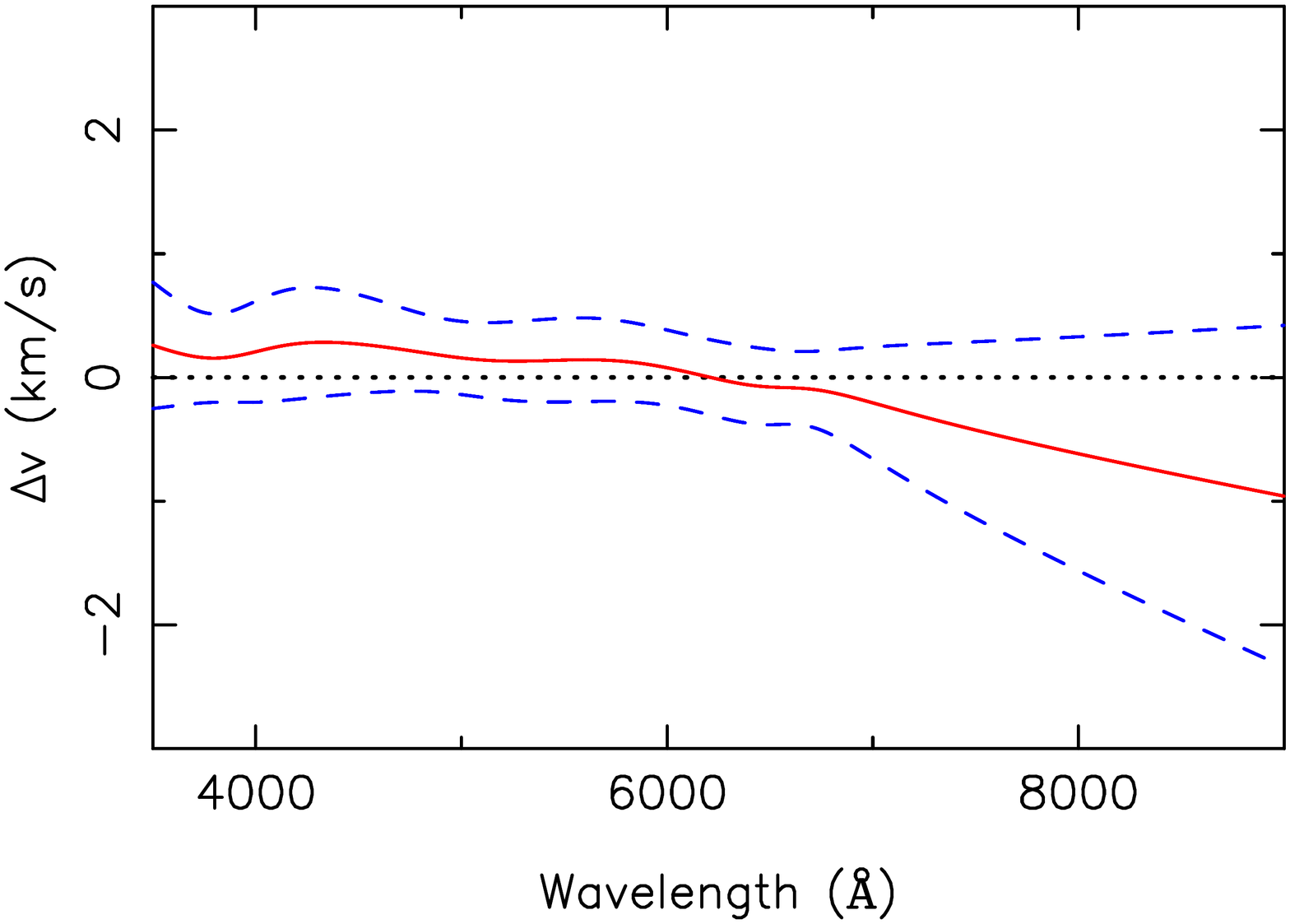}
\else
\includegraphics[bb=50 59 556 766,angle=-90,width=82.5mm]{images/dv_6q_jointpred.eps}
\fi
\end{center}
\caption{$\Delta v_w(\lambda)$, a joint estimate of $\Delta v(\lambda)$ from the 6 core quasar pairs made without combining the data into a single linear fit. The red (solid) line shows the estimate for $\Delta v$, and the dashed (blue) lines show the 95 percent confidence interval on the estimate. One can see that over the range where most of our $\Delta v$ data are obtained ($4000\mathrm{\AA} \lesssim \lambda \lesssim 7000\mathrm{\AA}$) that $\Delta v$ is relatively flat. $\Delta v$ diverges from zero for $\lambda \lesssim 3500\mathrm{\AA}$ (not shown) and for $\lambda \gtrsim 7000\mathrm{\AA}$, however there are few $\Delta v$ measurements to constrain $\Delta v$ in these regions. At no wavelength is $\Delta v$ significantly different from zero. Because the linear model used for each quasar fit is unlikely to be a true description of any underlying wavelength distortion, there is also uncertainty due to model specification, which is naturally not included in the confidence region shown above. As such, the confidence region shown is under-estimated. These considerations show that there is no statistically significant evidence for a common systematic from consideration of the 6 core quasar pairs. \label{fig_dv_6quasars_jointprediction}}
\end{figure}

\begin{figure}
\begin{center}
\ifpdf
\includegraphics[bb=77 79 455 727,angle=-90,width=82.5mm]{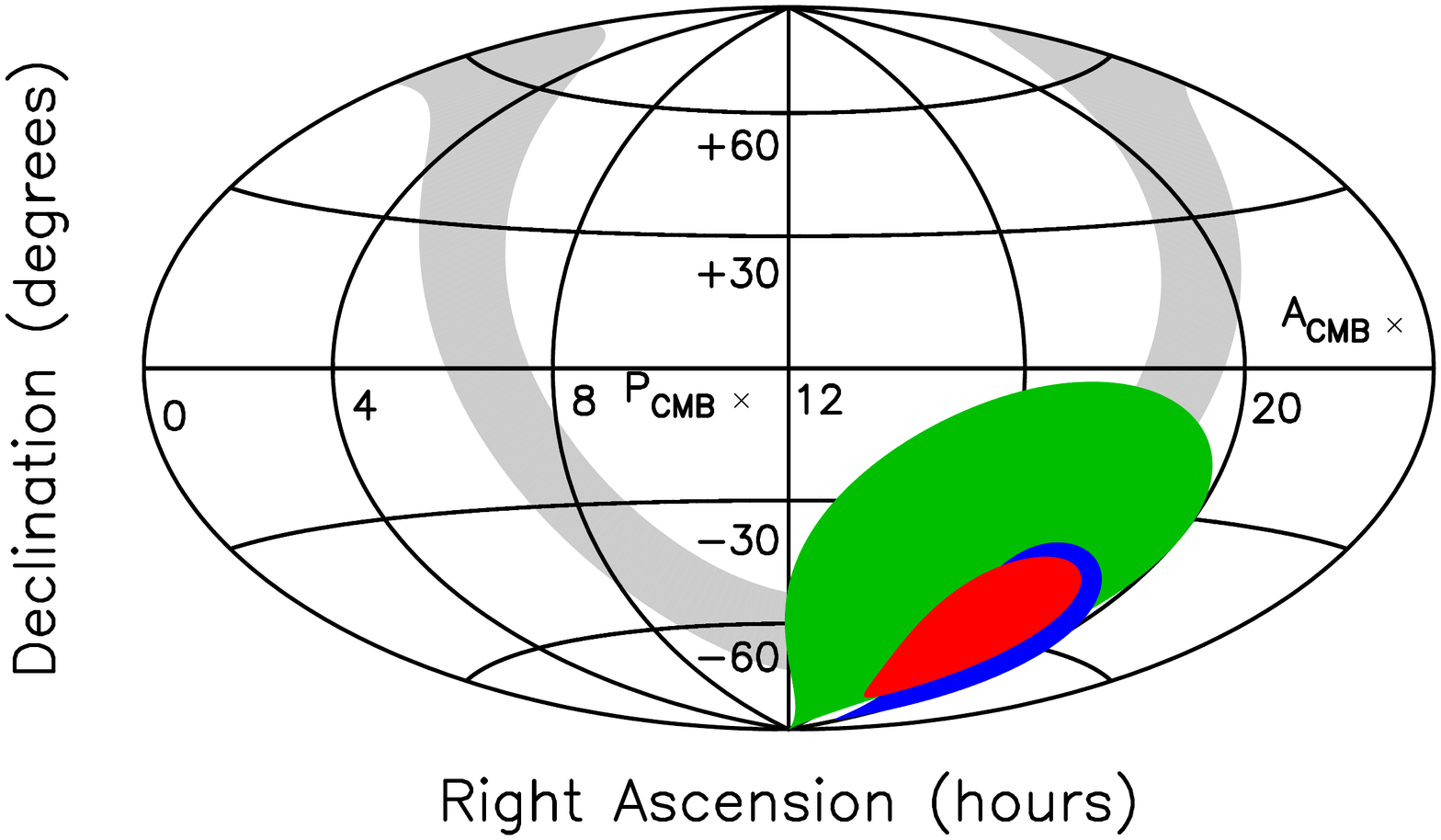}
\else
\includegraphics[bb=77 79 455 727,angle=-90,width=82.5mm]{images/dv_6q_skymap.eps}
\fi
\end{center}
\caption{$1\sigma$ confidence regions for the Keck (green), VLT (blue) and combined (red) dipoles, where the linear inverse function derived from the $\Delta v$ data from the 6 core spectral pairs has been applied to the VLT set. Although the statistical significance of the dipole decreases from $3.9\sigma$ to $3.1\sigma$, the positions of the VLT and combined dipole are effectively unchanged.  \label{fig_dv_6quasars_linfit_skymap}}
\end{figure}

In deriving the results above, we have assumed that $\Delta v$ values from different spectral pairs may be legitimately combined in order to estimate a common systematic. This may not be a good assumption, given the differences in the signal-to-noise of the spectral data, the spectral range which the $\Delta v$ values cover, the potential functional form of $\Delta v(\lambda)_i$ and number of exposures. We then proceed as follows: \emph{i)} fit a linear model to the $\Delta v$ values from each spectral pair using the LTS method; \emph{ii)} from each model, estimate $\Delta v(\lambda)$ along with an uncertainty on the estimate; \emph{iii)} for the six estimates of $\Delta v$ at each $\lambda$, form a weighted mean of the estimates, $\Delta v_w(\lambda)$, and calculate the associated uncertainty, and; \emph{iv)} plot $\Delta v_w(\lambda)$ as a function of wavelength. We show the result of this in figure \ref{fig_dv_6quasars_jointprediction}. Importantly, under this model we can find no wavelength where $\Delta v$ is statistically different from zero. 

\begin{table*}
\caption{Results of applying the inverse $\Delta v$ function from figure \ref{fig_dv_6quasars_linfit} to the VLT absorbers under the model $\Delta\alpha/\alpha = A \cos \Theta + m$. The column ``$\delta A$'' gives $1 \sigma$ confidence limits on $A$. The column labelled ``sig'' gives the significance of the dipole+monopole model with respect to the monopole model. The origin of the reference VLT set is described in section \ref{s_dv_6corepairs}. The impact of the (non-significant) linear distortion modelled from the 6 core quasar pairs overall is a reduction in the statistical significance of the dipole from $3.9\sigma$ to $3.1\sigma$.  \label{tab_dv_6quasars_results}}
 \begin{center}
\begin{tabular}{clccccccc}
\hline 
Sample & $\sigma_\mathrm{rand}$ $(10^{-5})$ & $A$ ($10^{-5}$) & $\delta A$ ($10^{-5}$) & RA (hr) & dec.\ ($^\circ$) & $m$ ($10^{-5}$) & sig\\
\hline
VLT reference                             & 0.88                & 1.21  & $[0.80, 1.72]$ & $18.3 \pm 1.1$ & $-61\pm13$ & $-0.110 \pm 0.179$ & $2.2\sigma$  \\
VLT with $\Delta v$ function applied      & 0.95                & 1.02  & $[0.65, 1.52]$ & $18.6 \pm 1.3$ & $-61\pm16$ & $-0.262 \pm 0.183$ & $1.8\sigma$ \\
VLT reference + Keck                      & Keck HC=1.63        & 0.97  & $[0.57, 1.39]$ & $17.4 \pm 1.0$ & $-61\pm10$ & $-0.177 \pm 0.085$ & $3.9\sigma$ \\
VLT with $\Delta v$ function + Keck       & Keck HC=1.63        & 0.79  & $[0.59, 1.07]$ & $17.4 \pm 1.2$ & $-60\pm12$ & $-0.273 \pm 0.085$ & $3.1\sigma$ \\ 

\hline
 \end{tabular}
 \end{center}
\end{table*}

In summary: we are unable to detect a statistically significant linear wavelength distortion common to the 6 core spectral pairs. Applying to the entire VLT spectral sample a simple linear model for $\Delta v(\lambda)$ from the six core pairs reduces the statistical significance of the dipole, but the statistical significance still remains high enough to be of interest. The systematic applied here does not destroy the good alignment between the fitted Keck and VLT dipole vectors. We are therefore unable to remove the dipole effect from the combined Keck and VLT sample.

\subsubsection{2206$-$1958/J220852$-$194359}\label{Q2206P}

\begin{figure}
\begin{center}
\ifpdf
\includegraphics[bb=50 59 556 766,angle=-90,width=82.5mm]{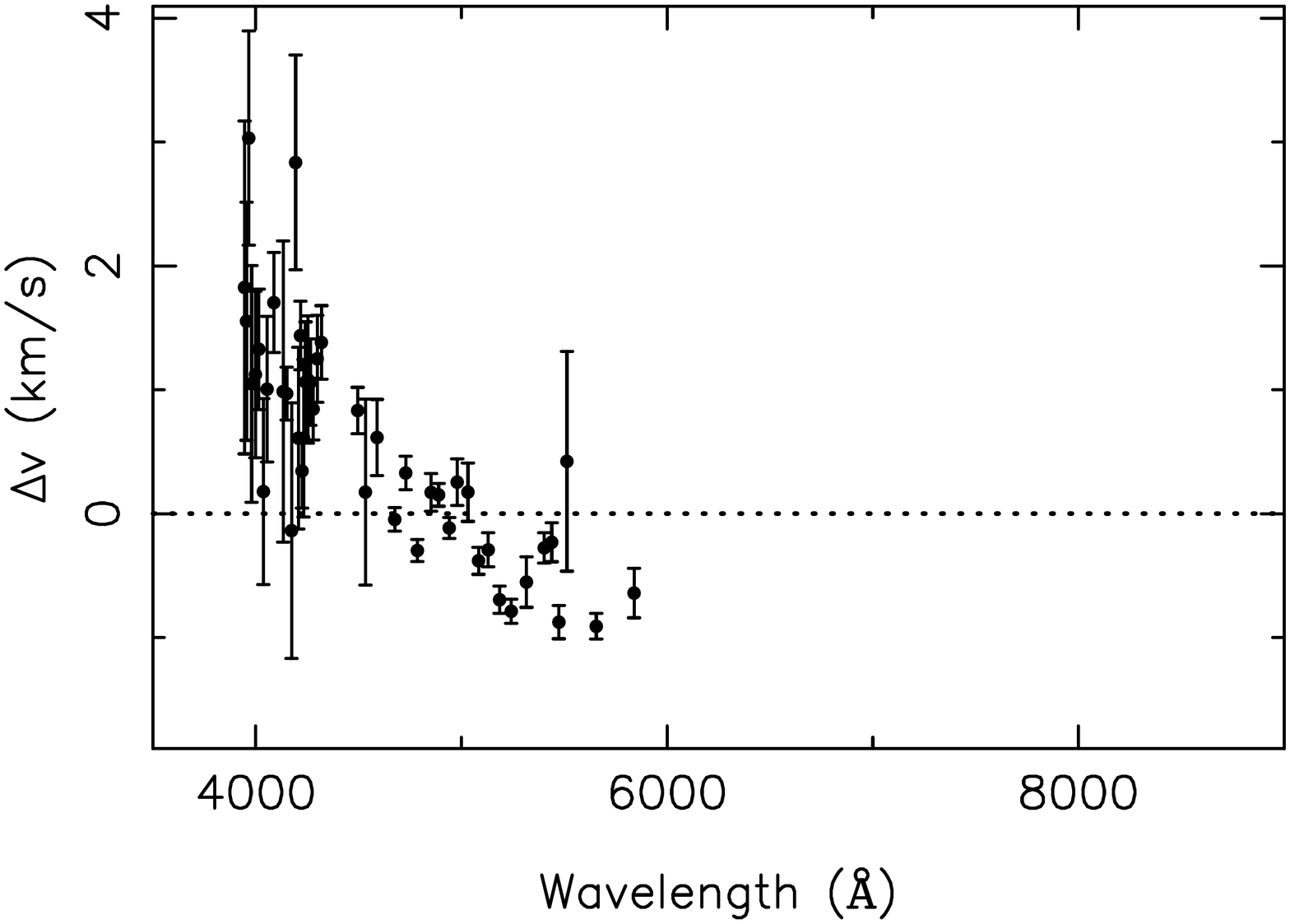}
\else
\includegraphics[bb=50 59 556 766,angle=-90,width=82.5mm]{images/dq2206_binned_bin3.eps}
\fi
\end{center}
\caption{$\Delta v$ values for the 2206$-$1958/J220852$-$194359 spectral pair. The $\Delta v$ values demonstrate a substantial distortion, with $|\delta(\Delta v)|\approx 2.5\,\mathrm{km\,s^{-1}}$ over $\sim 2000\mathrm{\AA}$. The trend is substantially larger, and different to the trends seen in the six core pairs (see figure \ref{fig_dv_6quasars}). Note that although different trends are seen individually in each of the six core pairs, no common trend is seen.  \label{fig_dv_q2206}}
\end{figure}

We show the $\Delta v$ values from the pair in figure \ref{fig_dv_q2206}. The $\Delta v$ test only examines calibration differences between Keck and VLT, and so we cannot tell whether Keck or VLT is responsible for significant trend in $\Delta v$ for this spectral pair. 

The limited spectral range ($\sim 4000 \lesssim \lambda \lesssim 6000\mathrm{\AA}$) of the $\Delta v$ data means that we simply do not know what the functional form of $\Delta v$ is for this spectral pair at $\lambda \gtrsim 6000\mathrm{\AA}$. In order to estimate the potential impact of the distortion present on $\Delta \alpha/\alpha$ values in the whole sample, we need knowledge of $\Delta v(\lambda)$ at all observed wavelengths. One possibility is to assume that the relationship is linear, but then the extrapolation over the whole spectral range results in a total change in $\Delta v$ of $\sim 5\,\mathrm{km\,s^{-1}}$, which is comparable to the velocity width of the spectrograph slit; this seems too extreme.  Additionally, the $\Delta v$ values do not appear by-eye to be linearly related to $\lambda$; the gradient of $\Delta v$ with $\lambda$ visually appears steepest in the middle of the $\Delta v$ values, and flatter near the edges of the $\Delta v$ data. We therefore try an arctangent model, 
\begin{equation}
 \Delta v = A \tan^{-1}\left[k(\lambda-\lambda_c)\right] + b.\label{eq_dv_arctan}
\end{equation}
Applying the LTS method to this fit, with $k=0.95n$, yields: $A = (-0.98 \pm 1.03)\,\mathrm{km\,s^{-1}\,\AA^{-1}}$, $k = (1.9 \pm 2.6)\times 10^{-3}\,\mathrm{\AA^{-1}}$, $\lambda_c = 4547 \pm 526 \, \mathrm{\AA}$ and $b = 0.48 \pm 0.74 \, \mathrm{km\,s^{-1}}$, where errors are derived from the diagonal terms of the covariance matrix at the best fit. The uncertainty on each $\Delta v$ values has been increased in quadrature with $0.26\,\mathrm{km\,s^{-1}}$ to account for overdispersion about the LTS fit.

The impact of this wavelength perturbation on the values of $\Delta \alpha/\alpha$ when this function is applied to the VLT absorbers is severe. In particular, large numbers of points are scattered away from $\Delta\alpha/\alpha = 0$, inducing a highly significant detection of $\Delta\alpha/\alpha$ at $z\sim 0.8$ and $z\sim 1.5$. For instance, a formal weighted mean of all the $1.3 < z < 1.8$ points yields $\Delta\alpha/\alpha = (-3.96 \pm 0.12) \times 10^{-5}$ -- a $33\sigma$ ``detection''. If one multiplies the error by $\sqrt{\chi^2_\nu}$ to account for $\chi^2_\nu = 14.7$ about the weighted mean, then one obtains $\Delta\alpha/\alpha = (-3.96 \pm 0.46) \times 10^{-5}$, still an $8.5\sigma$ ``detection''.  Such a signal is seen in neither the Keck or VLT samples, which immediately implies that this particular relative wavelength distortion can not possibly apply to the all of either the VLT or Keck spectra. That is, the distortions seen in this particular Keck/VLT spectral pair appear not to be representative of a significant fraction of the entire sample. However, the fact that we have identified this distortion demonstrates the power and utility of the quasar pair analysis in identifying systematic errors, even when their actual origin remains unknown.

\subsubsection{Overall effect of wavelength systematics using the $\Delta v$ test}\label{s_dv_overalleffect}

We now investigate whether a diluted form of the above effect (i.e. the effect from the 2206$-$1958/J220852$-$194359 pair) could exist in the spectral data in combination with the (non-significant and much smaller) effect observed from the 6 core quasars. To do this, we use a Monte Carlo approach where we apply the inverse function derived in section \ref{s_dv_6corepairs} (equation \ref{eq_dv_linfunc}) to $6/7$ of the quasar spectra selected at random in the VLT sample, and the arctan function of equation \ref{eq_dv_arctan} to the remaining $1/7$ of the spectra. We then apply the LTS method to estimate a new $\sigma_\mathrm{rand}$. We then add the Keck sample to this new VLT sample. At each iteration, we calculate the statistical significance of the dipole using the bootstrap method. The mode of the distribution obtained is $\sim2.2\sigma$, with quite substantial variation between iterations. 

\begin{figure}
\begin{center}
\ifpdf
\includegraphics[bb=132 53 550 696,angle=-90,width=82.5mm]{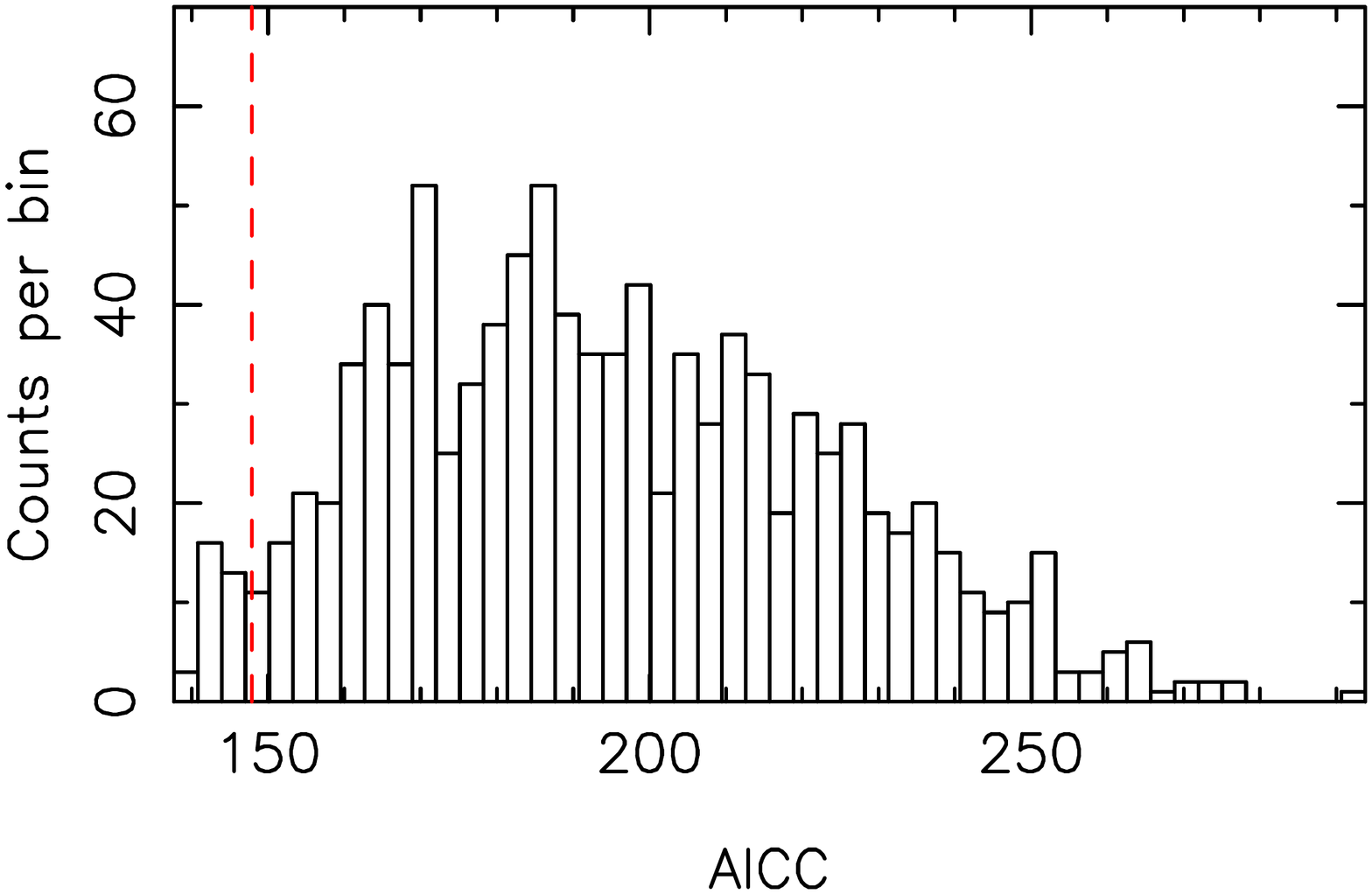}
\else
\includegraphics[bb=132 53 550 696,angle=-90,width=82.5mm]{images/dv_6q+q2206_sim_AICC.eps}
\fi
\end{center}
\caption{AICC for 1000 iterations of a Monte Carlo simulation where the $\Delta v$ function from the 6 core pairs (equation \ref{eq_dv_linfunc}) is applied to 6/7 of the VLT quasars at random, and the $\Delta v$ function from the 2206$-$1958/J220852$-$194359 pair (equation \ref{eq_dv_arctan}) is applied to the remaining 1/7 of the quasars. The AICC here is calculated for the $\Delta\alpha/\alpha$ data with respect to the model $\Delta\alpha/\alpha = A\cos(\Theta)+m$ for the VLT points only. The dashed red line shows the actual AICC for the VLT reference set (see section \ref{s_dv_applytoVLT}). Only in 3.5 percent of the iterations is the AICC lower than for the reference set, which suggests that it is unlikely that a distortion of this type is present in the VLT data.  The median value of the AICC here is 191.4, corresponding to a median $\Delta \mathrm{AICC} = 43.5$. For the median case, the odds against a wavelength distortion of this type being present in the data are $\approx 3\times 10^{9}:1$. \label{fig_dv_q6+2206_sim_AICC}}
\end{figure}

To determine whether the $\Delta v$ function significantly improves the goodness-of-fit in the VLT sample, we compare the AICC at each iteration in the Monte Carlo simulation for a dipole model ($\Delta \alpha/\alpha = A\cos\Theta + m$) fitted to the VLT $\Delta\alpha/\alpha$ values in that iteration with the AICC from a dipole model fitted to the $\Delta\alpha/\alpha$ values in the VLT reference set, where in each iteration we use $\sigma_\mathrm{rand} = 0.88\times 10^{-5}$ in order to compare $\Delta\alpha/\alpha$ values on a like-with-like basis. We show this distribution in figure \ref{fig_dv_q6+2206_sim_AICC}. In only 3.5 percent of iterations is the AICC lower than in the reference set. This implies that it is unlikely that a wavelength distortion of this type is present in our data set. However, in almost all of the iterations the AICC is much larger than the AICC from the VLT reference set; the median $\Delta \mathrm{AICC} = 43.5$.  Importantly, in \emph{no case} is the $\Delta \mathrm{AICC} > 10$. Thus in no case can we say that there is very strong evidence in favour of the model with the $\Delta v$ function applied. Additionally, the AICC does not account for the 6 parameters used in deriving the $\Delta v$ model -- we should expect a significant reduction in the AICC if the $\Delta v$ function is a good model. From this argument, we thus conclude that a wavelength distortion of this type is unlikely to be present in the VLT spectral data.

\subsection{Comment on \citet{Griest:09} I$_2$ and ThAr measurements on Keck/HIRES}

The path that the quasar light takes through the telescope is similar but not identical to that of the thorium-argon (ThAr) calibration lamp. In particular, the ThAr lamp nearly uniformly illuminates the slit, whilst the quasar image will be centrally concentrated, particularly if the seeing is smaller than the slit width used. On account of the path differences, it is possible for distortions in the wavelength scale to emerge. Such distortions may be long range or short range. 

\citeauthor{Griest:09} compared the wavelength calibration scale of exposures derived from I$_2$ absorption spectra and the standard thorium-argon (ThAr) calibration exposures. The distortions are identified by comparing the calibration of the spectrum derived from a ThAr lamp to the calibration obtained when an I$_2$ absorption cell is used. The iodine cell is placed in the quasar light path, and therefore the wavelength scale obtained for the quasar light using the iodine cell suffers from no path differences. They reported drifts between the I$_2$ and ThAr calibration scales of up to $2000\,\mathrm{m\,s^{-1}}$ over several nights, and claimed that that ``this level of systematic uncertainty may make it difficult to use Keck HIRES data to constrain the change in the fine-structure constant''. 

The $\Delta v$ test above explicitly includes the effect of any drifts in the wavelength calibration both within a single night and between observation nights.   From figure \ref{fig_dv_6quasars_linfit}, the RMS of the 66 binned points about the fit is $\approx 250\,\mathrm{m\,s^{-1}}$. The mean wavelength separation between these points is comparable to the echelle order width. This RMS can therefore be compared directly to the spread in $v_\mathrm{shift}$ seen in figure 5 of \citeauthor{Griest:09}. In contrast to their spread of $\approx 2000\,\mathrm{m\,s^{-1}}$, we see see typical wavelength distortions between VLT and Keck which are some 8 times smaller. We have directly quantified the impact of this on our measurements of $\Delta\alpha/\alpha$ in section \ref{s_dv_applytoVLT}. Our results demonstrate that it is possible to reliably use Keck/HIRES data to constrain the fine-structure constant from quasar observations. In the next section, we deal with the effect of `intra-order distortions' reported by \citeauthor{Griest:09}.

\subsection{Intra-order wavelength distortions}\label{s_intraorder_distortions}

A pattern of wavelength distortions within each echelle order has been identified both within Keck/HIRES spectra \citep{Griest:09} and VLT/UVES spectra \citep{Whitmore:10}. In this section we attempt to estimate the impact of the extra scatter that has already been introduced into the VLT $\Delta\alpha/\alpha$ values as a result of the intra-order distortions reported by \citeauthor{Whitmore:10}.

\citet{Griest:09} identified an apparently repeating pattern of distortions that occur within each echelle order (`intra-order distortions'), where the wavelength scale of pixels near the centre of the order is displaced with respect to the calibration scale for pixels at the echelle order edges. The peak-to-peak amplitude of the distortion is $\sim 500\,\mathrm{m\,s^{-1}}$ at $\sim5600\mathrm{\AA}$. As the distribution of MM transitions is random with respect to the location of the echelle orders, the effect of these distortions will be random from absorber to absorber. A distortion of this type, with no monotonic long-range component, constitutes a random effect (section \ref{s_randsyseffect}). \citet{Murphy:09} applied a model of the distortion found by \citet{Griest:09} to the 2004 Keck results, and found that the impact on the weighted mean of the $\Delta\alpha/\alpha$ values was effectively negligible. It is also worth noting that systems which utilise a large number of MM transitions will be less sensitive to an effect of this type. This is because with many transitions, the distortion is sampled in many locations; if the distortion does not have a long-range component, the average distortion must tend to zero. 

\citet{Whitmore:10} have identified similar intra-order distortions in VLT/UVES spectra. Whilst the distortions found by \citeauthor{Griest:09} display a repeating pattern, the distortions found by \citeauthor{Whitmore:10} are much more irregular, displaying little consistent pattern. It is important to note that because the VLT spectra used here are the result of the co-addition of many exposures, taken with different echelle grating settings and over many nights, it is expected that any distortions of the wavelength scale due to light path differences should reduce in magnitude. Thus, we consider the possible estimate of the impact of the effect we present to be an upper limit.

\begin{figure}
\begin{center}
\ifpdf
\includegraphics[bb=62 53 447 752,angle=-90,width=82.5mm]{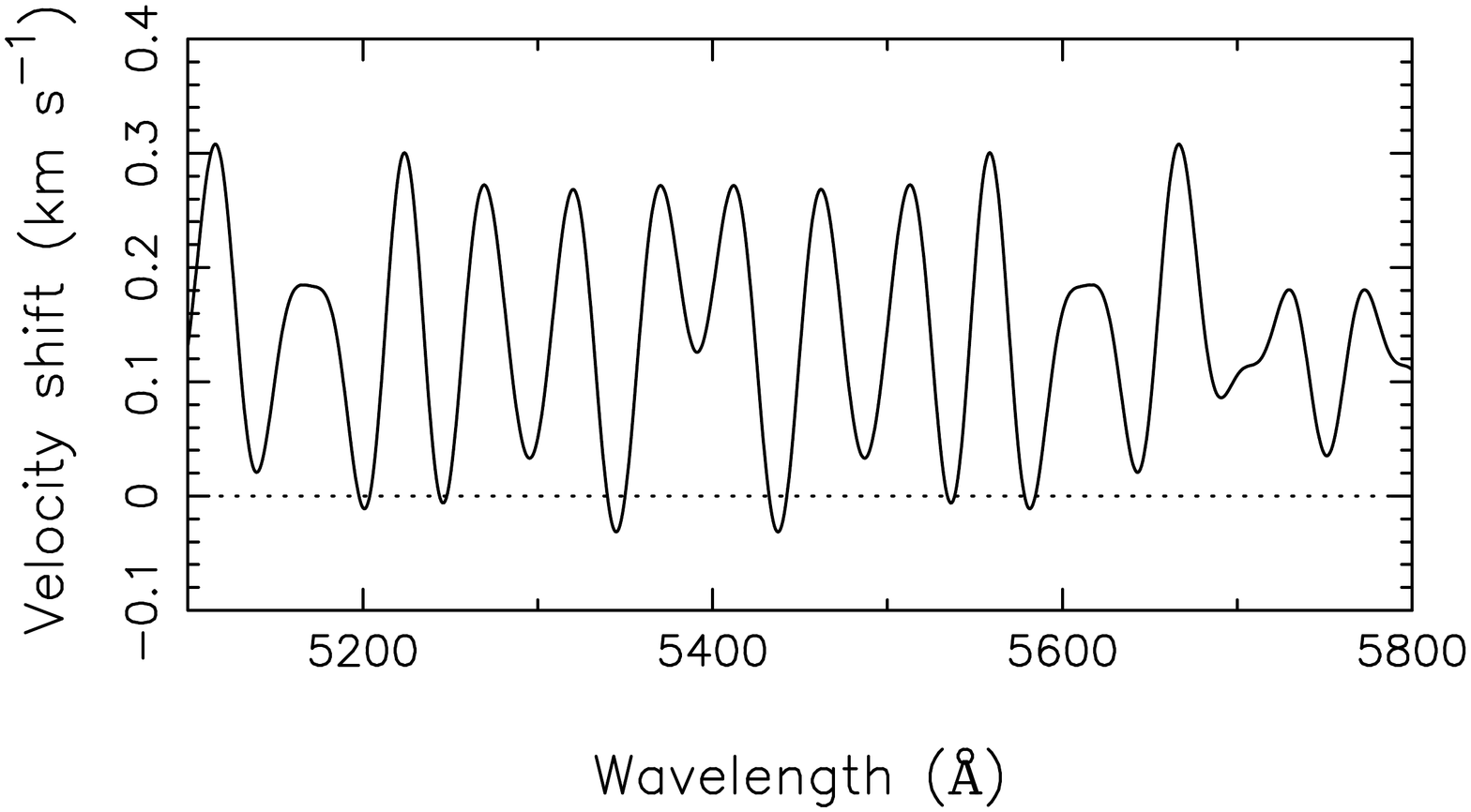}
\else
\includegraphics[bb=62 53 447 752,angle=-90,width=82.5mm]{images/murphy_wavmod.eps}
\fi
\end{center}
\caption{$\Delta v$ function used for the investigating the wavelength distortions found by \citeauthor{Whitmore:10} based on a Fourier analysis of their data. This function is repeated to longer and shorter wavelengths. \label{fig_dv_Whitmore}}
\end{figure}

To construct a model for the \citeauthor{Whitmore:10} distortions, we carried out a Fourier analysis of the velocity shift data presented in that paper to produce a $\Delta v$ model which we can apply to the VLT absorbers. The iodine cell absorption lines used to establish the intra-order distortion results only occur over the wavelength range $\sim$ 5000--6200$\mathrm{\AA}$. We are therefore forced to assume that our model of these distortions applies to much bluer and redder wavelengths as well. Clearly, the important part of the model is the amplitude rather than the period of the distortions; our model has a maximum peak-to-peak distortion of $\sim 300\,\mathrm{m\,s^{-1}}$. We show this model in figure \ref{fig_dv_Whitmore}.

In table \ref{tab_Whitmore_results}, we show the result of applying the function shown in figure \ref{fig_dv_Whitmore} to the VLT absorbers using equation \ref{eq_inv_dv_func}. The impact on the location of the dipole and the value of the monopole is minimal, as expected. However, we note that the $\sigma_\mathrm{rand}$ required is somewhat larger, which means that this model of the wavelength distortion has introduced extra scatter into the $\Delta\alpha/\alpha$ values. Any good model of the systematic should reduce, not increase, the scatter. The extra scatter reduces the significance of the dipole, but does not destroy the good alignment between Keck and VLT, nor between low and high redshift samples. In particular, the chance probability of alignment for the Keck and VLT samples (where the VLT sample has been altered with this $\Delta v$ model is 6 percent, the chance probability of alignment between low and high redshift samples is 4 percent, and the joint chance probability for these two factors is 0.3 percent. 

\begin{table*}
\caption{Results of applying the inverse $\Delta v$ function from figure \ref{fig_dv_Whitmore} to the VLT absorbers. The column ``$\delta A$'' gives $1 \sigma$ confidence limits on $A$. The column labelled ``sig'' gives the significance of the dipole model with respect to the monopole model. The origin of the reference VLT set is described in section \ref{s_dv_6corepairs}. For $I=2$, we remove the absorber at $z=1.6574$ toward J024008$-$230915, as it is identified as an outlier with the LTS method. Note that the estimates of the dipole location and monopole value do not differ greatly between samples 2 and 3. For $\sigma_\mathrm{rand}$, HC refers to the Keck high-contrast sample. Application of the Whitmore et.\ al. $\Delta v$ function causes a reduction in significance of the dipole model of $\approx 0.8\sigma$. This reflects the maximal amount by which the significance has already been reduced; the $3.3\sigma$ figure is not the significance after applying a correction for a systematic. \label{tab_Whitmore_results}}
 \begin{center}
\begin{tabular}{cccccccccc}
\hline 
$I$ & Sample & $\sigma_\mathrm{rand}$ ($10^{-5}$) & $A$ ($10^{-5}$) & $\delta A$ ($10^{-5}$) & RA (hr) & dec.\ ($^\circ$) & $m$ ($10^{-5}$) & sig\\
\hline
1 & VLT with Whitmore $\Delta v$ function z         & 1.090   & 0.87  & $[0.51, 1.43]$ & $19.0 \pm 1.6$ & $-54\pm23$ & $-0.025 \pm 0.215$ & $1.2\sigma$ \\
2 & \#1 with $z=1.6574$ absorber removed           & 1.067   & 1.03  & $[0.63, 1.56]$ & $18.4 \pm 1.1$ & $-51\pm19$ & $-0.090 \pm 0.204$ & $1.6\sigma$ \\
3 & VLT reference                                  & 0.882   & 1.21  & $[0.80, 1.72]$ & $18.3 \pm 1.1$ & $-61\pm13$ & $-0.110 \pm 0.179$ & $2.2\sigma$  \\
4 & \#2 + Keck sample                              & Keck HC=1.630    & 0.88  & $[0.68, 1.12]$ & $17.4 \pm 1.0$ & $-56\pm11$ & $-0.217 \pm 0.090$ & $3.3\sigma$ \\ 
5 & VLT reference + Keck                           & Keck HC=1.630    & 0.97           & [$0.57, 1.39]$ & $17.4 \pm 1.0$ & $-61\pm10$ & $-0.177 \pm 0.085$ & $3.9\sigma$\\
\hline
 \end{tabular}
 \end{center}
\end{table*}

The presence of intra-order wavelength distortions would serve to increase the scatter of the $\Delta\alpha/\alpha$ values about the true values. These distortions can only randomise but not bias $\Delta\alpha/\alpha$ values. They can not manufacture a dipole or monopole. Were we able to make the same quasar observations without the presence of any wavelength scale distortions, the scatter in the $\Delta\alpha/\alpha$ values about the model should be smaller (and so $\sigma_\mathrm{rand}$ would be smaller). We would therefore expect that this would increase the significance of the dipole model. Our analysis in this section suggests that the maximal reduction in statistical significance of the dipole which may have occurred as a result of intra-order wavelength distortions present is $\sim 0.6\sigma$.

\section{Discussion}

\subsection{Summary of results}

We have presented the results of a new analysis of 154 quasar absorbers over the range $0.2 \lesssim z \lesssim 3.7$ using the many-multiplet method, derived from VLT/UVES spectra of 60 quasars. These absorbers allow us to constrain changes in $\alpha$ at often better than the $10^{-5}$ level over much of the observable universe. 

Assuming that the VLT $\Delta\alpha/\alpha$ values are described by a simple weighted mean, we found that $(\Delta\alpha/\alpha)_w = (0.21\pm 0.12)\times 10^{-5}$. This appears at first to be inconsistent with the results of \citet{Murphy:04:LNP}, who reported $(\Delta\alpha/\alpha)_w = (-0.57\pm 0.11)\times 10^{-5}$ from 143 absorbers observed with Keck/HIRES.

We fitted a simple angular dipole+monopole model, $\Delta\alpha/\alpha = A\cos\Theta +m$, to the VLT $\Delta\alpha/\alpha$ values and found some ($2.2\sigma$) evidence for angular (and therefore spatial) variations in $\alpha$, with the dipole direction (i.e.~the direction of maximal deviation in $\alpha$ from the current laboratory value) towards $\mathrm{RA}=(18.3\pm 1.2)\,\mathrm{hr}$, $\mathrm{dec.} = (-62\pm 13)^\circ$, with an amplitude $A = 1.18^{+0.28}_{-0.24} \times 10^{-5}$, and a monopole term $m=(-0.109\pm0.180) \times 10^{-5}$. A dipole-only model ($\Delta\alpha/\alpha = A\cos\Theta$) points in the direction $\mathrm{RA}=(18.4\pm 1.3)\,\mathrm{hr}$, $\mathrm{dec.} = (-58\pm 15)^\circ$, with an amplitude $A = 0.99^{+0.38}_{-0.29} \times 10^{-5}$. 

A similar angular dipole+monopole model applied to the Keck $\Delta\alpha/\alpha$ values gives model parameters $\mathrm{RA}=(16.0\pm 2.7)\,\mathrm{hr}$, $\mathrm{dec.} = (-47\pm 29)^\circ$, with an amplitude $A = 1.18^{+0.37}_{-0.12} \times 10^{-5}$, and a monopole term $m = (-0.465\pm0.145) \times 10^{-5}$. The significance of the dipole+monopole model over the monopole model is low in this case, just $0.5\sigma$, and is due to the relatively low sky coverage of the Keck/HIRES quasar lines of sight, leading to a stronger degeneracy of the dipole and monopole terms. For a dipole-only model we found model parameters $\mathrm{RA}=(16.4\pm 1.2)\,\mathrm{hr}$, $\mathrm{dec.} = (-56\pm 12)^\circ$, with an amplitude $A = 1.06^{+0.52}_{-0.24} \times 10^{-5}$. This dipole-only model has a slightly higher significance ($1.1\sigma$, or 72 percent), and is consistent with the direction and amplitude found from a dipole-only model fitted to the VLT $\Delta\alpha/\alpha$ values.

For the combined Keck+VLT sample, the dipole+monopole model ($\Delta\alpha/\alpha = A\cos\Theta +m$) is preferred over the monopole-only model at the $4.06\sigma$ level, with the dipole pointing in the direction $\mathrm{RA}=(17.3\pm 1.0)\,\mathrm{hr}$, $\mathrm{dec.} = (-61\pm 10)^\circ$, with an angular amplitude $A = 0.97^{+0.22}_{-0.20} \times 10^{-5}$ and monopole term of $m = (-0.178 \pm 0.084) \times 10^{-5}$. Fitting a dipole-only model increases the significance to $4.15\sigma$ with the dipole pointing in a very similar one to the best-fit dipole+monopole model. Thus the combined VLT+Keck sample shows significant evidence for angular (and therefore spatial) variations in $\alpha$.

It is important to realise that, while the weighted mean $\Delta\alpha/\alpha$ values from the VLT and Keck samples disagree, the two samples are consistent when potential spatial variations in $\alpha$ across the sky are considered. The numerical results given above show that dipole models fitted to the VLT and Keck $\Delta\alpha/\alpha$ values point in very similar directions. This good alignment would be surprising if the dipole effect were not real. Under the model $\Delta\alpha/\alpha = A\cos\Theta +m$, the chance probability of obtaining as close an alignment as observed (or closer) between the dipole vectors in both samples is 6 percent. Additionally, dipole vectors fitted to low-redshift ($z<1.6$) and high-redshift ($z>1.6$) subsamples point in a very similar direction, with a chance probability of 2 percent under the model $\Delta\alpha/\alpha = A\cos\Theta +m$. A joint analysis for both of these conditions finds a chance probability of 0.1 percent.

In section \ref{s_monopole} we examined the presence of a statistically significant monopole ($3.6\sigma$) in the $z<1.6$ subsample of absorbers in the combined sample. There is consistent evidence that this offset of $\Delta\alpha/\alpha$ from zero exists in both the VLT and Keck samples independently. This implies not only that the low-$z$ monopole may be a real effect (i.e.~not generated by intstrumental systematic errors), but that it appears not to be responsible for the observed angular variations in $\alpha$. We suggested that evolution in the relative abundances of Mg isotopes may be responsible for the low-$z$ monopole. After examining this possibility in more detail (section \ref{s_isotopic_abundances}), we suggested that an average heavy Mg isotope fraction of $\langle \Gamma_{z<1.6} \rangle = 0.32 \pm 0.03$ could account for the observed low-$z$ monopole.

To further check the robustness of the apparent spatial variation in $\alpha$, in section \ref{s_sigmaclipping} we investigated the effect of iteratively clipping potentially outlying $\Delta\alpha/\alpha$ values from a combined Keck+VLT sample in which the raw statistical errors on $\Delta\alpha/\alpha$ were retained. When outliers are clipped iteratively from the best-fit dipole+monopole model at each iteration until $\chi^2_\nu=1$, the significance of the model (over a monopole-only model) increases to almost $7\sigma$. Importantly, its significance does not decline rapidly, even after removal of more than $\sim$20 percent of the most outlying $\Delta\alpha/\alpha$ values, suggesting that the dipole effect is robust. Similarly, the dipole direction is stable under removal of potentially outlying $\Delta\alpha/\alpha$ values.

In section \ref{s_zdipolemodel} we applied simple phenomenological models to attempt to account for the fact that the amplitude of the dipole appears larger at higher redshifts. Under a model where the amplitude is proportional to $z^\beta$ we find that $\beta = 0.46\pm 0.49$, and that the dipole+monopole model is preferred over the monopole-only model at the $3.9\sigma$ confidence level. We also investigated a model where the amplitude grows in proportion to the lookback-time distance, $r=ct$, and find that the dipole+monopole model for the combined Keck+VLT sample is significant over the monopole-only model at the $4.15\sigma$ level. Removing the monopole from consideration increases the significance to $4.22\sigma$. The fitted dipole directions are very similar to that found in the strictly angular models (i.e.~with no redshift dependence). Under the $r$-dipole+monopole model, the amplitude of the dipole is $(1.1\pm0.2) \times 10^{-6}\,\mathrm{GLyr}^{-1}$ with a monopole term of $m=(-0.187\pm 0.084)\times 10^{-5}$.

In section \ref{s_systematic_errors} we investigated various potential systematic errors. We showed that temporal evolution in the abundance of Mg isotopes cannot be responsible for the observed angular variation in $\alpha$. We demonstrated a powerful quasar pair test, where by comparing spectra of common quasars taken with both the VLT and Keck telescopes we can constrain relative wavelength scale distortions between the two telescopes. Considering 6 quasar pairs, we could not detect a statistically significant common linear wavelength distortion. Applying an estimate of the (non-significant) distortion observed, the significance of the dipole+monopole model, $\Delta\alpha/\alpha = A\cos(\Theta) + m$,  over the monopole-only model was reduced to $3.1\sigma$. The sign and functional form of the wavelength scale distortion differs in each of the quasar pairs, so any residual systematic in the full sample may be smaller than what was found. We also considered a seventh quasar pair which shows a very large relative wavelength scale distortion (which is quite unlike that seen in the other 6 pairs). We argued that it is unlikely that a distortion of this magnitude is found in a significant fraction of either the Keck or the VLT absorbers because of the large increase in the scatter of the $\Delta\alpha/\alpha$ values it would imply, a scatter grossly inconsistent with that actually observed. In section \ref{s_intraorder_distortions} we showed that the intra-order wavelength distortions reported by \citet{Griest:09} for Keck/HIRES and \citet{Whitmore:10} for VLT/UVES do not plausibly explain the observed dipole effect.

As we discussed in section \ref{s_combined_dipole_fit}, any systematic which mimics angular variation in $\alpha$ must not only be well correlated with sky position, but must do so in a way which is consistent between the two telescopes. A systematic which is correlated with zenith angle is not sufficient to produce the observed dipole effect; such an effect should produce a variation in $\alpha$ that is approximately symmetric about the latitudes of the telescopes projected onto the sky (i.e.\ dec.\ $\sim 20^\circ$ for Keck and dec. $\sim -25^\circ$ for VLT), which is not what is seen. Importantly, such an effect is unable to produce the consistency observed between the dipole locations. The fact that dipoles fitted to low-$z$ and high-$z$ subsamples of the $\Delta\alpha/\alpha$ values (combined from the two telescopes) also point in the same direction on the sky makes a simple systematic which explains the results even more difficult to identify. The low-$z$ and high-$z$ absorbers utilise different transitions, which have a different pattern of line shifts as a function of $\Delta\alpha/\alpha$. Thus, on average, the low-$z$ and high-$z$ subsamples will respond differently to simple systematic errors, making it difficult to explain consistent low-$z$ and high-$z$ dipole directions. We know of no systematic at present which is able to explain the observed dipole effect.

In summary, the results presented here provide significant and seemingly robust evidence for angular variations in $\alpha$ across the sky. We find evidence that the dipole amplitude grows with increasing distance. Taken together, these findings suggest a simple picture where there is a gradient in the value of $\alpha$ through the Universe, although this is strictly an observationally motivated interpretation and may be just the (seemingly) simplest many possible others.

\subsection{Implications}

The dipole-like variation in $\alpha$ presented here, if confirmed, would be a detection of new physics at the most fundamental level. It would directly demonstrate the existence of a preferred frame in the universe; it may be that this anisotropy could be detected in other cosmological measurements. Additionally, it would demonstrate that the Einstein Equivalence Principle is only an approximation. 

Some consider that the universe is fine-tuned for life \citep[see for instance][]{Davies:03a}, in that the values of the fundamental constants appear to be set in such a way that small variations (of order a few percent) in some fundamental constants would seemingly make some ingredients for life as we understand it (e.g.\ requiring water) unstable. If the fundamental constants vary throughout space, this has the potential to resolve this possible fine-tuning problem: the universe need not be globally fine-tuned for life. Instead, Earth may simply be located in a region of space where the constants are amenable to life. This may imply the existence of regions where the constants take on values where life as we understand it is not possible.

\subsection{Relationship with other experiments}

\citet{Berengut:10c} explicitly demonstrate that the results presented here are consistent with atomic clock measurements, the natural nuclear reactor at Oklo, and measurements of $\beta$-decay in meteorites. \citet{Berengut:10d} considered other potential cosmological dipoles (for example, in the primordial deuterium abundance and other dimensionless constants). To the best of our knowledge, the results presented here are not in conflict with the results of any experiment.

\section{Conclusions}

Our work significantly extends the work of \citet{Murphy:03,Murphy:04:LNP} from Keck/HIRES by adding a data set of comparable size using VLT/UVES. 

We find that the combination of both data sets yields a detection of spatial variation of $\alpha$. That is, we observe variation in $\alpha$ across the sky and with an increasing amplitude at greater distances from Earth. There are several lines of argument which suggest that the dipolar effect observed may be real: \emph{i)} The mild evidence ($\sim 2\sigma$) for angular variations in $\alpha$ within the VLT sample (which cannot be the result of inter-telescope systematics). \emph{ii)} The close alignment between dipole vectors in both dipole+monopole and dipole-only models fitted to the VLT and Keck samples separately. In the dipole+monopole case ($\Delta\alpha/\alpha = A\cos(\Theta) + m$), this has a chance probability of 6 percent. \emph{iii)} The close alignment between dipole+monopole models fitted to the $z<1.6$ and $z>1.6$ cuts separately, which has a chance probability of 2 percent.  \emph{iv)} The robustness of the result, both in terms of statistical significance and stability of dipole direction, to the removal of significant numbers of absorbers. \emph{v}) The general difficulty of generating angular variations in $\alpha$ through simple systematic effects.

If the observed result is due to systematics, it must be a combination of finely-tuned systematics, the nature of which is unknown. We have used spectra of a small number of quasars, all observed with both the VLT and Keck, to investigate inter-telescope wavelength-dependent systematics. We conclude that the (generally insignificant) relative wavelength distortions observed are unable to explain the dipole effect observed. The origin of the wavelength calibration issues detected at both Keck and VLT is not well understood, and as such the analysis presented here reflects our current state of knowledge. Clearly, the power of comparing spectra of the same quasars from different telescopes for ruling out -- or detecting -- relevant wavelength distortions, from any source, even unknown ones, should be exploited in a much larger sample.

The apparent dipole-like variation in $\alpha$ presented here is manifestly verifiable and falsifiable. Due to the orientation of the dipole relative to the observations, there is a relative lack of absorbers near its pole and anti-pole. Future observations toward the pole and anti-pole will have increased sensitivity to detect a significantly positive or negative $\Delta\alpha/\alpha$, respectively, if the effect is real. Furthermore, we have demonstrated the power of detecting systematics, even of unknown origin, using the quasar pair analysis; a campaign to observe quasars near the dipole pole and/or anti-pole would best be made with two telescopes targeting the same quasars so that any confirmation of any dipole-like variation in $\alpha$ is maximally compelling and reliable. Because of the orientation of the dipole, these two telescopes would have to be in the same hemisphere if one wanted to apply a similar quasar pair analysis to that demonstrated here.

We noted the presence of a statistically significant monopole in the $z<1.6$ sample. Evidence for this effect is present in both the VLT and Keck samples. Because the effect is common to both samples, it cannot be the cause of the angular variations in $\alpha$ observed. Evolution in the abundance of the heavy Mg isotope fraction may be the most likely cause for this, but further work is needed to resolve the issue.

\section*{Acknowledgments}

We would like to thank J.~C. Berengut and S.~J. Curran for advice and support in various stages of this work. JAK has been supported in part by an Australian Postgraduate Award. JKW, VVF and MTM thank the Australian Research Council for support.

\bibliographystyle{mn2e.bst}
\small
\itemindent -0.48cm
\bibliography{mnras_v01.bib}

\appendix

\section{VLT $\Delta\alpha/\alpha$ values}

We give the $\Delta\alpha/\alpha$ values from our MM analysis of the 154 VLT absorbers in table \ref{VLT_daoa_results}. An ASCII version of this table is available at \url{http://astronomy.swin.edu.au/~mmurphy/pub.html}.

\begin{sidewaystable*}
 \centering
\begin{minipage}{228mm}  
\vspace*{18cm}
\caption{Results for $\Delta\alpha/\alpha$ derived from MM absorbers. Errors given are purely statistical (see definition in section \ref{s_randsyseffect}). The emission redshift of the quasar and absorption redshift are given by $z_\mathrm{em}$ and $z_\mathrm{abs}$ respectively. $z_\mathrm{abs}$ is given as the redshift of the highest column density component in the fit. Three values of $\Delta\alpha/\alpha$ are given: the turbulent fit, the thermal fit, and the value derived from the method-of-moments estimator from section \ref{s_physicalconstraints}. The value of $\Delta\alpha/\alpha$ and its $1\sigma$ statistical uncertainty used for our statistical analyses is given in the last column. Note that the uncertainty is generally increased by adding a value of $\sigma_\mathrm{rand}$ in quadrature to the value quoted here, depending on the model under consideration (e.g\ weighted mean, dipole etc.; see text for values of $\sigma_\mathrm{rand}$).The key for the transition labels, in column 4, is given in table \ref{tab_freqtransitions}. $\chi^2_\nu$ and $\nu$, the number of degrees of freedom, are given for both the turbulent and thermal fits. The absorber marked with an asterisk (*) has been identified as an outlier and removed from the statistical analysis of particular models (e.g.e the dipole model).\label{VLT_daoa_results}}
   \begin{tabular}{@{}llll@{}c@{}rc@{}c@{}rc@{}c@{}}
  \hline
   Quasar name & $z_\mathrm{em}$ & $z_\mathrm{abs}$ & Transitions & $\Delta\alpha/\alpha_\mathrm{turb}\,(10^{-5})$ \ \ & $\nu_\mathrm{turb}$ 
  & $\chi^2_{\nu,\mathrm{turb}}$ \ \ & $\Delta\alpha/\alpha_\mathrm{therm}\,(10^{-5})$ \ \ & $\nu_\mathrm{therm}$ & $\chi^2_{\nu,\mathrm{therm}}$ \ &  $\Delta\alpha/\alpha\,(10^{-5}) $  \\
  \hline
J000344$-$232355  & 2.28  &   0.4521  &  $  a_2 b_1 b_2 j_4 j_5 j_6 j_7 j_8  $  & $-0.498 \pm 0.758$  & 533  & 0.7653 & $-0.459 \pm 0.787$ & 533 & 0.7466 &  $-0.459 \pm 0.787$   \\ 
J000344$-$232355  & 2.28  &   0.9491  &  $  a_2 b_1 b_2 j_4 j_6 j_8  $  & $-1.390 \pm 2.700$  & 530  & 0.7766 & $-4.960 \pm 2.650$ & 531 & 0.7916 &  $-1.534 \pm 2.788$   \\ 
J000344$-$232355  & 2.28  &   1.5864  &  $  a_2 j_1 j_4 j_5 j_6 j_7 j_8 c_1 e_1  $  & $-1.630 \pm 1.040$  & 380  & 0.7061 & $-0.389 \pm 0.989$ & 380 & 0.6847 &  $-0.410 \pm 1.003$   \\ 
J000448$-$415728*  & 2.76  &   1.5419  &  $  b_1 b_2 j_4 j_5 j_6 j_7 j_8  $  & $-5.270 \pm 0.906$  & 372  & 0.7682 & $-1.920 \pm 0.741$ & 372 & 1.0146 &  $-5.270 \pm 0.906$   \\ 
J000448$-$415728  & 2.76  &   1.9886  &  $  j_4 j_6 j_7 j_8 c_1 d_1 d_2  $  & $-0.952 \pm 1.510$  & 254  & 0.8991 & $0.934 \pm 1.830$ & 254 & 0.8943 &  $0.266 \pm 1.945$   \\ 
J000448$-$415728  & 2.76  &   2.1679  &  $  b_1 j_1 j_4 j_5 j_6 c_1 e_1  $  & $2.320 \pm 0.942$  & 309  & 0.7517 & $1.370 \pm 0.938$ & 309 & 0.7229 &  $1.381 \pm 0.944$   \\ 
J001210$-$012207  & 2.00  &   1.2030  &  $  b_1 b_2 j_4 j_6 j_7 j_8 c_1  $  & $0.723 \pm 1.290$  & 505  & 0.9464 & $0.772 \pm 1.190$ & 505 & 0.8959 &  $0.772 \pm 1.190$   \\ 
J001602$-$001225  & 2.09  &   0.6351  &  $  b_1 b_2 j_4 j_6 j_8  $  & $-0.719 \pm 3.580$  & 307  & 1.1152 & $-0.089 \pm 3.000$ & 307 & 1.1318 &  $-0.673 \pm 3.545$   \\ 
J001602$-$001225  & 2.09  &   0.6363  &  $  b_1 b_2 j_4 j_6 j_8  $  & $-2.130 \pm 4.310$  & 184  & 0.8184 & $-1.460 \pm 3.830$ & 184 & 0.7997 &  $-1.561 \pm 3.914$   \\ 
J001602$-$001225  & 2.09  &   0.8575  &  $  b_1 b_2 j_7 j_8  $  & $1.140 \pm 1.850$  & 191  & 0.8713 & $1.490 \pm 1.760$ & 191 & 0.8773 &  $1.266 \pm 1.826$   \\ 
J001602$-$001225  & 2.09  &   1.1468  &  $  b_1 b_2 j_6 j_8  $  & $-2.490 \pm 3.720$  & 159  & 1.1667 & $-1.580 \pm 2.920$ & 159 & 1.0859 &  $-1.581 \pm 2.922$   \\ 
J001602$-$001225  & 2.09  &   2.0292  &  $  a_2 j_1 j_4 j_6 j_7 j_8 d_1 d_2 e_1  $  & $-0.909 \pm 0.934$  & 425  & 0.9288 & $-2.780 \pm 0.877$ & 425 & 1.0167 &  $-0.909 \pm 0.934$   \\ 
J004131$-$493611  & 3.24  &   2.1095  &  $  b_1 b_2 j_7 j_8 c_1 d_1 d_2  $  & $-1.090 \pm 2.950$  & 430  & 0.6242 & $0.980 \pm 2.590$ & 430 & 0.6200 &  $0.386 \pm 2.856$   \\ 
J004131$-$493611  & 3.24  &   2.2485  &  $  j_1 j_2 j_3 j_6 j_7 j_8 c_1 d_1 d_2 e_2 h_1 h_2 h_3 l_1 l_2 k_1 i_1 i_2  $  & $-1.230 \pm 0.672$  & 756  & 0.7072 & $-0.972 \pm 0.669$ & 757 & 0.7281 &  $-1.230 \pm 0.672$   \\ 
J005758$-$264314  & 3.65  &   1.2679  &  $  a_2 b_1 b_2 j_6 j_7 j_8  $  & $1.660 \pm 2.120$  & 209  & 0.8127 & $0.416 \pm 1.430$ & 209 & 0.8139 &  $1.076 \pm 1.931$   \\ 
J005758$-$264314  & 3.65  &   1.5336  &  $  b_2 j_3 j_4 j_5 j_6 j_7 h_1 i_1 i_2  $  & $-0.456 \pm 0.903$  & 306  & 0.9050 & $-0.151 \pm 0.769$ & 306 & 0.9980 &  $-0.456 \pm 0.903$   \\ 
J010311+131617  & 2.68  &   1.7975  &  $  a_2 j_1 j_4 j_5 j_6 e_2  $  & $0.422 \pm 0.537$  & 371  & 1.1633 & $0.964 \pm 0.560$ & 372 & 1.1841 &  $0.443 \pm 0.548$   \\ 
J010311+131617  & 2.68  &   2.3092  &  $  j_1 j_2 c_1 e_1 e_2 h_1 h_2 h_3 l_1 l_2 k_1 k_2  $  & $-0.082 \pm 0.563$  & 672  & 1.0917 & $-1.690 \pm 0.521$ & 673 & 1.2312 &  $-0.082 \pm 0.563$   \\ 
J011143$-$350300  & 2.41  &   1.1827  &  $  a_2 b_1 b_2 j_4 j_6 j_7 j_8  $  & $0.150 \pm 0.953$  & 438  & 0.7192 & $-0.038 \pm 0.870$ & 438 & 0.7336 &  $0.142 \pm 0.950$   \\ 
J011143$-$350300  & 2.41  &   1.3499  &  $  b_1 b_2 j_4 j_5 j_6 j_7 j_8  $  & $0.084 \pm 0.378$  & 1037  & 0.8889 & $1.160 \pm 0.391$ & 1043 & 1.2275 &  $0.084 \pm 0.378$   \\ 
J012417$-$374423  & 2.20  &   0.8221  &  $  a_2 b_1 b_2 j_4 j_6 j_7 j_8  $  & $0.702 \pm 1.050$  & 214  & 0.7919 & $1.080 \pm 0.826$ & 214 & 1.2128 &  $0.702 \pm 1.050$   \\ 
J012417$-$374423  & 2.20  &   0.8593  &  $  a_2 b_1 b_2 j_5 j_6 j_7 j_8  $  & $0.768 \pm 1.870$  & 261  & 0.9144 & $-2.660 \pm 1.850$ & 261 & 0.9169 &  $-0.677 \pm 2.516$   \\ 
J012417$-$374423  & 2.20  &   1.2433  &  $  a_2 b_1 b_2 j_4 j_5 j_6 h_2 k_2 k_3  $  & $1.140 \pm 1.050$  & 364  & 0.8248 & $2.260 \pm 1.120$ & 365 & 0.8274 &  $1.838 \pm 1.221$   \\ 
J012417$-$374423  & 2.20  &   1.9102  &  $  j_4 j_8 c_1 d_1 d_2 e_1  $  & $-4.730 \pm 3.280$  & 248  & 0.6603 & $-3.260 \pm 2.830$ & 248 & 0.6576 &  $-3.872 \pm 3.111$   \\ 
J013105$-$213446  & 1.90  &   1.8566  &  $  j_3 j_4 j_5 j_6 d_1 d_2 e_2 h_1 k_1 k_2 k_3  $  & $0.304 \pm 1.430$  & 989  & 0.8841 & $0.212 \pm 1.450$ & 990 & 0.8834 &  $0.236 \pm 1.445$   \\ 
J014333$-$391700  & 1.81  &   0.3400  &  $  b_1 b_2 j_7 j_8  $  & $-7.010 \pm 6.020$  & 208  & 0.8698 & $-6.690 \pm 3.270$ & 208 & 0.8552 &  $-6.748 \pm 3.914$   \\ 
J014333$-$391700  & 1.81  &   1.7101  &  $  b_1 j_4 j_6 j_8 c_1 d_1 d_2 e_1  $  & $-2.200 \pm 2.670$  & 869  & 0.6519 & $-1.390 \pm 2.310$ & 869 & 0.6467 &  $-1.465 \pm 2.357$   \\ 
J015733$-$004824  & 1.55  &   0.7693  &  $  b_1 b_2 j_6  $  & $2.350 \pm 4.350$  & 89  & 1.0343 & $3.130 \pm 4.140$ & 89 & 1.0453 &  $2.647 \pm 4.288$   \\ 
J010821+062327  & 1.96  &   1.9328  &  $  j_7 e_1  $  & $2.200 \pm 2.460$  & 262  & 1.2053 & $0.782 \pm 1.230$ & 262 & 1.2392 &  $2.184 \pm 2.454$   \\ 
J024008$-$230915  & 2.22  &   1.1846  &  $  a_2 b_1 b_2 j_6 j_8  $  & $-1.680 \pm 3.620$  & 300  & 0.7229 & $-1.470 \pm 2.480$ & 300 & 0.7139 &  $-1.513 \pm 2.754$   \\ 
J024008$-$230915  & 2.22  &   1.6359  &  $  a_2 b_1 b_2 j_1 j_4 j_7 j_8 d_1 d_2 e_1  $  & $1.000 \pm 1.110$  & 388  & 0.7883 & $0.603 \pm 0.907$ & 388 & 0.8282 &  $1.000 \pm 1.110$   \\ 
J024008$-$230915  & 2.22  &   1.6373  &  $  a_2 b_1 b_2 j_1 j_4 j_7 j_8  $  & $-0.187 \pm 1.020$  & 475  & 1.1982 & $4.580 \pm 1.170$ & 475 & 1.2605 &  $-0.187 \pm 1.020$   \\ 
J024008$-$230915  & 2.22  &   1.6574  &  $  b_1 b_2 j_4 j_6 c_1 e_1  $  & $-0.137 \pm 1.010$  & 504  & 0.8087 & $2.340 \pm 1.070$ & 505 & 0.9150 &  $-0.137 \pm 1.010$   \\ 

  \hline
  \end{tabular}
\end{minipage}
\end{sidewaystable*}

\begin{sidewaystable*}
 \centering
\begin{minipage}{228mm}  
\vspace*{18cm}
  \contcaption{$\Delta\alpha/\alpha$ results.}
   \begin{tabular}{@{}llll@{}c@{}rc@{}c@{}rc@{}c@{}}
  \hline
   Quasar name & $z_\mathrm{em}$ & $z_\mathrm{abs}$ & Transitions & $\Delta\alpha/\alpha_\mathrm{turb}\,(10^{-5})$ \ \ & $\nu_\mathrm{turb}$ 
  & $\chi^2_{\nu,\mathrm{turb}}$ \ \ & $\Delta\alpha/\alpha_\mathrm{therm}\,(10^{-5})$ \ \ & $\nu_\mathrm{therm}$ & $\chi^2_{\nu,\mathrm{therm}}$ \ &  $\Delta\alpha/\alpha\,(10^{-5}) $  \\
  \hline
J033106$-$382404  & 2.42  &   0.7627  &  $  a_2 b_1 b_2 j_5 j_6 j_7 j_8  $  & $1.250 \pm 0.835$  & 395  & 0.8804 & $0.224 \pm 0.910$ & 396 & 0.8786 &  $0.440 \pm 0.988$   \\ 
J033106$-$382404  & 2.42  &   0.9709  &  $  b_1 b_2 j_8  $  & $-4.970 \pm 4.460$  & 53  & 0.5507 & $-4.030 \pm 3.920$ & 53 & 0.5483 &  $-4.485 \pm 4.216$   \\ 
J033106$-$382404  & 2.42  &   1.4380  &  $  b_1 b_2 j_4 j_8  $  & $-4.120 \pm 2.530$  & 384  & 0.7541 & $-4.380 \pm 2.580$ & 385 & 0.7524 &  $-4.323 \pm 2.571$   \\ 
J033108$-$252443  & 2.69  &   0.9925  &  $  b_1 b_2 j_4 j_8  $  & $-0.210 \pm 1.480$  & 270  & 0.7501 & $0.516 \pm 1.230$ & 270 & 0.7091 &  $0.513 \pm 1.232$   \\ 
J033108$-$252443  & 2.69  &   2.4547  &  $  j_1 e_1  $  & $-1.100 \pm 6.490$  & 203  & 0.9689 & $-2.390 \pm 5.170$ & 203 & 0.9557 &  $-2.122 \pm 5.496$   \\ 
J033244$-$445557  & 2.60  &   2.4112  &  $  j_1 j_4 j_6 j_7 c_1 e_1 e_2  $  & $-1.000 \pm 0.793$  & 340  & 0.9885 & $-0.293 \pm 0.802$ & 341 & 1.0705 &  $-1.000 \pm 0.793$   \\ 
J033244$-$445557  & 2.60  &   2.6563  &  $  j_1 j_4 j_5 j_6 c_1 e_1  $  & $1.080 \pm 1.690$  & 308  & 0.9847 & $0.846 \pm 1.550$ & 308 & 1.0179 &  $1.079 \pm 1.689$   \\ 
J040718$-$441013  & 3.00  &   2.4126  &  $  j_5 j_6 j_8 c_1 d_1 d_2  $  & $2.420 \pm 2.220$  & 250  & 0.8800 & $-5.270 \pm 3.760$ & 250 & 1.0402 &  $2.420 \pm 2.220$   \\ 
J040718$-$441013  & 3.00  &   2.5499  &  $  j_1 j_4 j_6 c_1 e_1 e_2 k_1 k_2  $  & $0.895 \pm 0.353$  & 931  & 1.0024 & $2.290 \pm 0.331$ & 934 & 1.9429 &  $0.895 \pm 0.353$   \\ 
J040718$-$441013  & 3.00  &   2.5948  &  $  j_1 j_2 j_5 j_6 c_1 d_1 d_2 e_1 e_2 h_1 h_2 h_3 l_2 k_1 k_2  $  & $0.574 \pm 0.345$  & 865  & 0.8559 & $0.566 \pm 0.361$ & 869 & 0.9963 &  $0.574 \pm 0.345$   \\ 
J040718$-$441013  & 3.00  &   2.6214  &  $  j_1 c_1 e_1  $  & $4.350 \pm 2.920$  & 325  & 0.8567 & $4.180 \pm 2.560$ & 325 & 0.8565 &  $4.264 \pm 2.744$   \\ 
J042707$-$130253  & 2.16  &   1.4080  &  $  b_1 b_2 j_1 j_4 j_5 j_6 j_7 j_8 c_1  $  & $-2.550 \pm 1.110$  & 335  & 1.0038 & $-2.600 \pm 1.120$ & 335 & 1.0261 &  $-2.551 \pm 1.110$   \\ 
J042707$-$130253  & 2.16  &   1.5632  &  $  j_4 j_6 j_8 c_1 e_1  $  & $-2.640 \pm 2.520$  & 114  & 0.7238 & $-3.030 \pm 2.430$ & 114 & 0.6951 &  $-2.967 \pm 2.449$   \\ 
J042707$-$130253  & 2.16  &   2.0351  &  $  j_1 c_1 e_2  $  & $8.060 \pm 3.830$  & 324  & 0.8410 & $5.990 \pm 3.620$ & 324 & 0.8822 &  $8.057 \pm 3.830$   \\ 
J043037$-$485523  & 1.94  &   1.3556  &  $  a_2 j_1 j_2 j_3 j_4 j_5 j_6 j_7 j_8 c_1 e_1 e_2 h_1 h_3 k_1 k_2 k_3 i_1 i_2 i_3  $  & $-0.405 \pm 0.232$  & 1039  & 0.9764 & $-0.428 \pm 0.237$ & 1041 & 1.0921 &  $-0.405 \pm 0.232$   \\ 
J044017$-$433308  & 2.86  &   1.4335  &  $  b_1 b_2 j_4 j_7 j_8  $  & $0.139 \pm 2.500$  & 472  & 1.1308 & $-1.910 \pm 2.130$ & 476 & 1.1940 &  $0.139 \pm 2.500$   \\ 
J044017$-$433308  & 2.86  &   2.0482  &  $  j_1 j_2 j_4 c_1 e_2 h_1 h_2 h_3 l_2 k_1 k_2 k_3 i_2 i_3  $  & $1.400 \pm 0.864$  & 1595  & 1.3448 & $2.510 \pm 0.773$ & 1595 & 1.3806 &  $1.400 \pm 0.864$   \\ 
J051707$-$441055  & 1.71  &   0.2223  &  $  a_2 b_1 b_2 j_8  $  & $1.380 \pm 3.850$  & 359  & 0.6694 & $1.140 \pm 3.540$ & 359 & 0.6696 &  $1.262 \pm 3.703$   \\ 
J051707$-$441055  & 1.71  &   0.4291  &  $  a_2 b_1 j_4 j_6 j_7 j_8  $  & $-2.740 \pm 1.440$  & 273  & 0.5915 & $-3.480 \pm 1.470$ & 273 & 0.5898 &  $-3.153 \pm 1.502$   \\ 
J053007$-$250329  & 2.81  &   2.1412  &  $  b_1 b_2 j_1 j_2 j_5 c_1 h_1 k_1 k_2 k_3  $  & $0.676 \pm 0.359$  & 949  & 0.8670 & $0.865 \pm 0.349$ & 949 & 0.9194 &  $0.676 \pm 0.359$   \\ 
J055246$-$363727  & 2.32  &   1.2252  &  $  a_2 b_1 b_2 j_4 j_5 j_6 j_7 c_1 e_1  $  & $1.850 \pm 1.010$  & 381  & 0.7120 & $0.269 \pm 0.895$ & 381 & 0.6147 &  $0.269 \pm 0.895$   \\ 
J055246$-$363727  & 2.32  &   1.7475  &  $  a_2 j_1 j_4 j_5 j_6 j_7 j_8 e_1  $  & $-0.795 \pm 1.080$  & 305  & 0.7899 & $-2.180 \pm 1.050$ & 305 & 0.8042 &  $-0.936 \pm 1.155$   \\ 
J055246$-$363727  & 2.32  &   1.9565  &  $  j_1 j_4 j_6 c_1 d_1 d_2 e_1  $  & $-0.104 \pm 1.500$  & 283  & 1.1209 & $1.740 \pm 1.530$ & 283 & 1.0508 &  $1.740 \pm 1.530$   \\ 
J064326$-$504112  & 3.09  &   2.6592  &  $  j_4 j_5 j_6 c_1 d_1 d_2 e_2 l_1 k_1 k_2  $  & $-1.530 \pm 1.920$  & 1101  & 1.1219 & $0.601 \pm 1.780$ & 1101 & 1.1611 &  $-1.530 \pm 1.920$   \\ 
J091613+070224  & 2.77  &   1.3324  &  $  a_2 b_1 b_2 j_4 j_6 j_7 j_8  $  & $4.230 \pm 3.380$  & 352  & 0.7171 & $12.900 \pm 4.690$ & 353 & 0.7231 &  $8.233 \pm 5.915$   \\ 
J094253$-$110426  & 3.05  &   1.0595  &  $  a_2 b_1 j_4 j_7 j_8  $  & $0.372 \pm 0.737$  & 270  & 0.7370 & $-0.478 \pm 1.060$ & 269 & 1.0524 &  $0.372 \pm 0.737$   \\ 
J094253$-$110426  & 3.05  &   1.7891  &  $  b_1 b_2 j_4 j_5 j_6  $  & $-2.330 \pm 0.495$  & 399  & 0.7420 & $-5.240 \pm 0.487$ & 404 & 2.1029 &  $-2.330 \pm 0.495$   \\ 
J103909$-$231326  & 3.13  &   1.4429  &  $  b_1 j_7 j_8  $  & $-1.980 \pm 2.720$  & 158  & 0.6943 & $-2.690 \pm 2.200$ & 158 & 0.9238 &  $-1.980 \pm 2.720$   \\ 
J103909$-$231326  & 3.13  &   2.7778  &  $  j_1 j_2 j_4 c_1 h_1 h_2 h_3 l_2 k_1 k_2 k_3  $  & $-1.130 \pm 0.660$  & 889  & 0.8394 & $-0.755 \pm 0.657$ & 891 & 0.8676 &  $-1.130 \pm 0.660$   \\ 
J103921$-$271916  & 2.23  &   0.8771  &  $  a_2 b_2 j_6 j_8  $  & $1.750 \pm 2.020$  & 139  & 1.0096 & $3.920 \pm 1.170$ & 139 & 1.0305 &  $2.159 \pm 2.071$   \\ 
J103921$-$271916  & 2.23  &   1.0093  &  $  b_1 b_2 j_8  $  & $-0.174 \pm 4.190$  & 152  & 0.8892 & $-0.652 \pm 3.260$ & 152 & 0.8370 &  $-0.643 \pm 3.280$   \\ 
J103921$-$271916  & 2.23  &   1.9721  &  $  j_4 j_5 j_6 j_7 c_1 e_1  $  & $2.650 \pm 1.030$  & 339  & 1.3837 & $2.980 \pm 0.847$ & 339 & 1.1613 &  $2.980 \pm 0.847$   \\ 
J104032$-$272749  & 2.32  &   1.3861  &  $  a_2 b_1 b_2 j_3 j_5 j_6 j_7 j_8 e_1 h_1 h_3 k_1 k_2 k_3 i_1 i_2 g_4 g_5  $  & $0.446 \pm 0.693$  & 914  & 1.2514 & $-0.565 \pm 0.734$ & 918 & 1.2919 &  $0.446 \pm 0.693$   \\ 
J104032$-$272749  & 2.32  &   1.7761  &  $  b_1 b_2 j_1 j_4 j_5 j_6 c_1 e_1  $  & $0.262 \pm 1.320$  & 430  & 1.0124 & $0.716 \pm 1.410$ & 430 & 1.0499 &  $0.262 \pm 1.320$   \\ 
J110325$-$264515  & 2.15  &   1.1868  &  $  a_2 b_1 b_2 j_4 j_5 j_6 j_7 j_8  $  & $-1.110 \pm 0.695$  & 814  & 0.8799 & $-0.155 \pm 0.945$ & 814 & 0.8811 &  $-0.745 \pm 0.925$   \\ 
J110325$-$264515  & 2.15  &   1.2029  &  $  b_1 b_2 j_4 j_6 j_7 j_8  $  & $0.622 \pm 0.831$  & 368  & 0.7472 & $0.769 \pm 0.659$ & 367 & 0.7668 &  $0.623 \pm 0.830$   \\ 
J110325$-$264515  & 2.15  &   1.5515  &  $  b_1 b_2 j_4 j_6 j_8 c_1 e_1  $  & $-0.691 \pm 1.010$  & 343  & 0.7530 & $-0.619 \pm 0.967$ & 343 & 0.7579 &  $-0.669 \pm 0.998$   \\ 
J110325$-$264515  & 2.15  &   1.8389  &  $  a_2 j_1 j_4 j_5 e_1  $  & $0.612 \pm 0.395$  & 319  & 0.9354 & $0.406 \pm 0.409$ & 318 & 1.0113 &  $0.612 \pm 0.395$   \\ 
J111113$-$080401  & 3.92  &   3.6077  &  $  j_1 e_1  $  & $7.040 \pm 6.690$  & 73  & 0.6176 & $31.400 \pm 13.000$ & 74 & 0.6323 &  $22.962 \pm 16.134$   \\ 
J112010$-$134625  & 3.96  &   1.6283  &  $  b_2 j_4 j_5 j_6  $  & $0.886 \pm 1.130$  & 177  & 1.0594 & $-1.270 \pm 1.170$ & 177 & 1.2155 &  $0.886 \pm 1.130$   \\ 
J112442$-$170517  & 2.40  &   0.8062  &  $  b_1 b_2 j_4 j_5 j_6 j_7 j_8  $  & $-1.260 \pm 0.801$  & 677  & 0.9304 & $1.900 \pm 1.200$ & 678 & 0.9240 &  $1.738 \pm 1.373$   \\ 
J112442$-$170517  & 2.40  &   1.2342  &  $  a_2 b_1 b_2 j_4 j_6 j_8 d_1 d_2  $  & $2.410 \pm 1.540$  & 467  & 0.9261 & $1.880 \pm 1.590$ & 467 & 0.9306 &  $2.271 \pm 1.571$   \\ 
 \hline
\end{tabular}
\end{minipage}  
\end{sidewaystable*}

\begin{sidewaystable*}
 \centering
\begin{minipage}{228mm}  
\vspace*{18cm}
  \contcaption{$\Delta\alpha/\alpha$ results.}
   \begin{tabular}{@{}llll@{}c@{}rc@{}c@{}rc@{}c@{}}
  \hline
   Quasar name & $z_\mathrm{em}$ & $z_\mathrm{abs}$ & Transitions & $\Delta\alpha/\alpha_\mathrm{turb}\,(10^{-5})$ \ \ & $\nu_\mathrm{turb}$ 
  & $\chi^2_{\nu,\mathrm{turb}}$ \ \ & $\Delta\alpha/\alpha_\mathrm{therm}\,(10^{-5})$ \ \ & $\nu_\mathrm{therm}$ & $\chi^2_{\nu,\mathrm{therm}}$ \ &  $\Delta\alpha/\alpha\,(10^{-5}) $  \\
  \hline
J115411+063426  & 2.76  &   1.7739  &  $  j_4 j_5 j_6 e_2 h_1 h_2 h_3 l_1 l_2 k_1 k_2 k_3 i_1 i_3  $  & $-0.740 \pm 0.784$  & 625  & 0.9578 & $-0.154 \pm 0.711$ & 627 & 0.9825 &  $-0.739 \pm 0.784$   \\ 
J115411+063426  & 2.76  &   1.8197  &  $  b_1 b_2 j_1 j_6 j_7 j_8 c_1  $  & $-0.948 \pm 0.974$  & 682  & 1.0724 & $-1.080 \pm 0.911$ & 682 & 1.1191 &  $-0.948 \pm 0.974$   \\ 
J115411+063426  & 2.76  &   2.3660  &  $  j_1 j_7 j_8 c_1 e_1  $  & $3.090 \pm 1.780$  & 136  & 1.1989 & $3.130 \pm 1.470$ & 136 & 1.3170 &  $3.090 \pm 1.780$   \\ 
J115944+011206  & 2.00  &   0.7908  &  $  b_1 b_2 j_4 j_6 j_8  $  & $1.560 \pm 1.080$  & 170  & 0.8878 & $1.720 \pm 1.000$ & 170 & 0.9531 &  $1.561 \pm 1.080$   \\ 
J115944+011206  & 2.00  &   1.3305  &  $  b_1 b_2 j_4 j_6 j_7 j_8 c_1  $  & $1.970 \pm 2.670$  & 196  & 1.0164 & $2.140 \pm 2.240$ & 196 & 0.9765 &  $2.137 \pm 2.249$   \\ 
J115944+011206  & 2.00  &   1.9438  &  $  a_2 j_1 j_2 j_3 j_5 e_1 e_2 h_1 h_3 k_1 k_2 k_3 i_1  $  & $0.518 \pm 0.442$  & 1031  & 0.9472 & $0.688 \pm 0.433$ & 1035 & 1.0251 &  $0.518 \pm 0.442$   \\ 
J120342+102831  & 1.89  &   1.3224  &  $  a_2 b_1 b_2 j_4 j_6 j_7 j_8 c_1 e_1  $  & $-0.965 \pm 1.930$  & 465  & 0.9337 & $-6.940 \pm 2.160$ & 465 & 1.0082 &  $-0.965 \pm 1.930$   \\ 
J120342+102831  & 1.89  &   1.3422  &  $  a_2 b_1 b_2 j_1 j_6 j_7 j_8 e_1  $  & $-3.210 \pm 1.530$  & 459  & 0.9903 & $-2.000 \pm 1.440$ & 459 & 0.9669 &  $-2.006 \pm 1.443$   \\ 
J120342+102831  & 1.89  &   1.5789  &  $  a_2 j_4 j_6 j_7 j_8 e_1  $  & $1.870 \pm 2.560$  & 356  & 0.9474 & $0.027 \pm 3.910$ & 356 & 0.9621 &  $1.743 \pm 2.716$   \\ 
J121140+103002  & 2.19  &   1.0496  &  $  a_2 b_1 b_2 j_4 j_5 j_6 j_7 j_8  $  & $-1.460 \pm 0.649$  & 152  & 1.0163 & $-1.540 \pm 0.673$ & 152 & 0.9691 &  $-1.538 \pm 0.672$   \\ 
J123200$-$022404  & 1.04  &   0.7569  &  $  a_2 b_1 b_2 j_4 j_6  $  & $2.250 \pm 3.220$  & 226  & 1.0794 & $3.320 \pm 2.500$ & 226 & 1.1306 &  $2.253 \pm 3.219$   \\ 
J123200$-$022404  & 1.04  &   0.8308  &  $  a_2 b_1 b_2 j_4 j_6 j_7 j_8  $  & $1.730 \pm 1.010$  & 248  & 0.8670 & $1.670 \pm 0.907$ & 248 & 0.8403 &  $1.672 \pm 0.911$   \\ 
J123437+075843  & 2.57  &   1.0201  &  $  a_2 j_4 j_6 j_7 j_8  $  & $-2.340 \pm 1.320$  & 450  & 0.9642 & $-0.046 \pm 1.700$ & 450 & 0.9768 &  $-2.213 \pm 1.442$   \\ 
J123437+075843  & 2.57  &   1.7194  &  $  j_1 j_4 j_5 j_6 d_1 d_2 e_2  $  & $0.505 \pm 0.942$  & 554  & 0.8714 & $0.467 \pm 0.944$ & 555 & 0.8740 &  $0.485 \pm 0.943$   \\ 
J133335+164903  & 2.08  &   0.7446  &  $  a_2 b_1 b_2 j_4 j_5 j_6 j_7  $  & $-0.828 \pm 0.542$  & 985  & 0.9932 & $-0.221 \pm 0.493$ & 986 & 1.0636 &  $-0.828 \pm 0.542$   \\ 
J133335+164903  & 2.08  &   1.3253  &  $  a_2 b_1 j_6 j_8 c_1  $  & $-2.670 \pm 6.750$  & 137  & 0.6816 & $11.300 \pm 8.920$ & 137 & 0.6789 &  $4.962 \pm 10.607$   \\ 
J133335+164903  & 2.08  &   1.7765  &  $  b_1 b_2 j_1 j_5 j_6 c_1 e_1 e_2 h_1 h_2 l_2 k_1 k_2 k_3 i_1  $  & $0.843 \pm 0.448$  & 1197  & 0.8669 & $0.842 \pm 0.440$ & 1199 & 0.9424 &  $0.843 \pm 0.448$   \\ 
J133335+164903  & 2.08  &   1.7863  &  $  a_2 b_1 j_6 c_1 d_1 d_2 e_1 e_2  $  & $-0.489 \pm 0.860$  & 482  & 0.8757 & $0.814 \pm 0.901$ & 496 & 1.0788 &  $-0.489 \pm 0.860$   \\ 
J134427$-$103541  & 2.13  &   1.9155  &  $  j_1 j_4 j_5 j_6 j_7 j_8 e_1 e_2  $  & $0.015 \pm 0.744$  & 470  & 0.8217 & $-2.380 \pm 0.678$ & 474 & 1.1493 &  $0.015 \pm 0.744$   \\ 
J134427$-$103541  & 2.13  &   2.1474  &  $  b_1 b_2 c_1  $  & $6.530 \pm 8.840$  & 210  & 0.8959 & $6.290 \pm 8.810$ & 210 & 0.9022 &  $6.448 \pm 8.831$   \\ 
J135038$-$251216  & 2.53  &   1.4393  &  $  a_1 b_1 b_2 j_1 j_3 j_8 d_1 d_2 h_2 h_3 l_1 l_2 k_2 k_3 i_1 i_2 i_3  $  & $-0.590 \pm 0.576$  & 935  & 0.9140 & $-0.987 \pm 0.568$ & 937 & 0.8953 &  $-0.987 \pm 0.568$   \\ 
J135038$-$251216  & 2.53  &   1.7529  &  $  b_2 j_6 j_8 c_1 d_1 d_2  $  & $6.990 \pm 3.260$  & 185  & 1.0350 & $5.850 \pm 3.160$ & 185 & 1.0341 &  $6.396 \pm 3.258$   \\ 
J141217+091624  & 2.86  &   1.4187  &  $  b_2 j_4 j_5 j_7 j_8  $  & $-2.920 \pm 1.770$  & 416  & 1.0090 & $-0.519 \pm 1.800$ & 416 & 1.0452 &  $-2.919 \pm 1.771$   \\ 
J141217+091624  & 2.86  &   2.0188  &  $  j_1 j_3 j_4 c_1 d_1 d_2 e_2 h_1 h_2 h_3 k_1 k_2 k_3  $  & $0.849 \pm 0.755$  & 610  & 1.1969 & $1.050 \pm 0.755$ & 610 & 1.2363 &  $0.849 \pm 0.755$   \\ 
J141217+091624  & 2.86  &   2.4564  &  $  j_1 j_4 j_7 e_1  $  & $-0.903 \pm 1.390$  & 340  & 1.0055 & $-1.760 \pm 1.270$ & 340 & 1.0904 &  $-0.903 \pm 1.390$   \\ 
J141217+091624  & 2.86  &   2.6682  &  $  j_1 j_5 j_6 c_1 e_1  $  & $0.199 \pm 0.849$  & 407  & 1.1119 & $0.087 \pm 0.809$ & 406 & 1.1348 &  $0.199 \pm 0.849$   \\ 
J143040+014939  & 2.11  &   0.4878  &  $  a_2 b_2 j_7 j_8 i_2 i_3  $  & $3.580 \pm 2.170$  & 320  & 1.0215 & $6.140 \pm 2.160$ & 324 & 1.1343 &  $3.580 \pm 2.170$   \\ 
J143040+014939  & 2.11  &   1.2030  &  $  b_1 b_2 j_7 j_8  $  & $-0.812 \pm 3.290$  & 328  & 0.9649 & $0.751 \pm 2.200$ & 332 & 1.1122 &  $-0.812 \pm 3.290$   \\ 
J143040+014939  & 2.11  &   1.2411  &  $  a_2 b_1 b_2 j_7 j_8 e_2 h_1 h_2 h_3 l_2 k_2 k_3 i_1 i_2 i_3  $  & $-2.660 \pm 1.200$  & 2135  & 1.1022 & $-3.580 \pm 1.080$ & 2108 & 1.1191 &  $-2.660 \pm 1.200$   \\ 
J144653+011356  & 2.21  &   0.5097  &  $  b_1 b_2 j_4 j_8  $  & $-0.574 \pm 1.140$  & 189  & 1.0588 & $0.245 \pm 1.090$ & 189 & 1.1095 &  $-0.567 \pm 1.142$   \\ 
J144653+011356  & 2.21  &   0.6602  &  $  a_2 b_1 b_2 j_6 j_7 j_8  $  & $-0.073 \pm 1.830$  & 251  & 0.8136 & $3.660 \pm 2.050$ & 251 & 0.8833 &  $-0.073 \pm 1.831$   \\ 
J144653+011356  & 2.21  &   1.1020  &  $  b_1 b_2 j_6 j_8  $  & $1.440 \pm 4.110$  & 116  & 0.8037 & $1.350 \pm 3.950$ & 116 & 0.8034 &  $1.395 \pm 4.030$   \\ 
J144653+011356  & 2.21  &   1.1292  &  $  a_2 b_2 j_4 j_6 j_8  $  & $3.700 \pm 2.970$  & 373  & 0.9208 & $2.040 \pm 2.650$ & 374 & 0.9149 &  $2.278 \pm 2.760$   \\ 
J144653+011356  & 2.21  &   1.1595  &  $  a_2 b_1 b_2 j_4 j_5 j_6 j_7  $  & $-2.560 \pm 1.200$  & 468  & 0.8479 & $-1.910 \pm 1.830$ & 468 & 0.8703 &  $-2.557 \pm 1.205$   \\ 
J145102$-$232930  & 2.21  &   1.5855  &  $  j_6 j_8 c_1 d_1 d_2 e_1  $  & $-4.590 \pm 2.500$  & 202  & 0.7543 & $-4.470 \pm 2.440$ & 202 & 0.7436 &  $-4.500 \pm 2.456$   \\ 
J200324$-$325144  & 3.77  &   2.0329  &  $  a_2 b_1 b_2 j_6 j_7 j_8  $  & $2.440 \pm 1.200$  & 339  & 1.2036 & $-1.290 \pm 1.860$ & 344 & 1.3983 &  $2.440 \pm 1.200$   \\ 
J200324$-$325144  & 3.77  &   3.1878  &  $  j_1 j_4 c_1 e_1 e_2  $  & $2.730 \pm 1.190$  & 199  & 0.8184 & $3.420 \pm 1.150$ & 199 & 0.7747 &  $3.411 \pm 1.153$   \\ 
J200324$-$325144  & 3.77  &   3.1917  &  $  j_1 e_1  $  & $2.910 \pm 4.090$  & 293  & 0.7510 & $-1.200 \pm 3.010$ & 293 & 0.7622 &  $2.238 \pm 4.217$   \\ 
J212912$-$153841  & 3.27  &   1.7380  &  $  a_2 j_4 j_5 j_6 j_7 j_8 e_1  $  & $1.310 \pm 0.636$  & 644  & 0.9611 & $2.520 \pm 0.741$ & 644 & 0.9918 &  $1.310 \pm 0.636$   \\ 
J212912$-$153841  & 3.27  &   2.0225  &  $  b_2 j_4 j_7 j_8 e_1  $  & $-1.630 \pm 1.240$  & 142  & 0.9175 & $2.850 \pm 1.190$ & 142 & 1.0258 &  $-1.628 \pm 1.244$   \\ 
J212912$-$153841  & 3.27  &   2.6378  &  $  c_1 e_1 e_2 k_1 k_2 k_3  $  & $1.320 \pm 3.330$  & 318  & 0.9206 & $0.049 \pm 2.860$ & 318 & 0.9736 &  $1.320 \pm 3.330$   \\ 
J212912$-$153841  & 3.27  &   2.7686  &  $  j_1 e_1 e_2  $  & $-0.206 \pm 1.090$  & 208  & 0.7878 & $-0.180 \pm 1.130$ & 208 & 1.0920 &  $-0.206 \pm 1.090$   \\ 

 \hline
\end{tabular}
\end{minipage}  
\end{sidewaystable*}

\begin{sidewaystable*}
 \centering
\begin{minipage}{228mm}  
\vspace*{18cm}
  \contcaption{$\Delta\alpha/\alpha$ results.}
   \begin{tabular}{@{}llll@{}c@{}rc@{}c@{}rc@{}c@{}}
  \hline
   Quasar name & $z_\mathrm{em}$ & $z_\mathrm{abs}$ & Transitions & $\Delta\alpha/\alpha_\mathrm{turb}\,(10^{-5})$ \ \ & $\nu_\mathrm{turb}$ 
  & $\chi^2_{\nu,\mathrm{turb}}$ \ \ & $\Delta\alpha/\alpha_\mathrm{therm}\,(10^{-5})$ \ \ & $\nu_\mathrm{therm}$ & $\chi^2_{\nu,\mathrm{therm}}$ \ &  $\Delta\alpha/\alpha\,(10^{-5}) $  \\
  \hline
J213314$-$464030  & 2.20  &   1.6148  &  $  j_1 j_4 j_5 j_6 c_1 e_1  $  & $4.080 \pm 1.550$  & 268  & 0.9146 & $4.550 \pm 1.550$ & 269 & 0.9206 &  $4.320 \pm 1.568$   \\ 
J214159$-$441325  & 3.17  &   2.1329  &  $  b_1 b_2 j_4 j_6 e_1  $  & $-0.454 \pm 2.210$  & 147  & 0.8312 & $-3.170 \pm 2.580$ & 147 & 0.9008 &  $-0.470 \pm 2.222$   \\ 
J214159$-$441325  & 3.17  &   2.3828  &  $  j_4 j_5 j_7 j_8 c_1 h_1 h_2 l_1 l_2 k_1 k_3 i_1 i_3  $  & $1.170 \pm 0.858$  & 1118  & 1.0153 & $1.270 \pm 0.883$ & 1124 & 1.0600 &  $1.170 \pm 0.858$   \\ 
J214159$-$441325  & 3.17  &   2.8523  &  $  j_1 j_2 j_3 j_5 c_1 e_1 e_2 h_1 h_3 k_1 k_2 k_3  $  & $2.090 \pm 0.524$  & 1020  & 0.9253 & $1.990 \pm 0.515$ & 1020 & 0.9344 &  $2.089 \pm 0.524$   \\ 
J214225$-$442018  & 3.23  &   0.9865  &  $  b_1 b_2 j_8  $  & $-0.093 \pm 1.050$  & 571  & 0.9699 & $-0.026 \pm 0.955$ & 565 & 0.9966 &  $-0.093 \pm 1.050$   \\ 
J214225$-$442018  & 3.23  &   1.0529  &  $  a_2 b_1 b_2 j_8  $  & $1.560 \pm 1.370$  & 221  & 0.8934 & $1.500 \pm 1.290$ & 223 & 0.8128 &  $1.500 \pm 1.290$   \\ 
J214225$-$442018  & 3.23  &   1.1543  &  $  a_2 b_1 b_2 j_7 j_8  $  & $-6.250 \pm 4.000$  & 571  & 0.9699 & $-5.590 \pm 3.860$ & 565 & 0.9966 &  $-6.250 \pm 4.000$   \\ 
J214225$-$442018  & 3.23  &   1.7569  &  $  b_1 b_2 j_4 j_6 j_8  $  & $-6.580 \pm 4.340$  & 162  & 0.8179 & $-4.610 \pm 3.790$ & 162 & 0.8349 &  $-6.183 \pm 4.308$   \\ 
J214225$-$442018  & 3.23  &   2.1126  &  $  a_1 a_2 b_1 b_2 j_3 j_4 j_5 j_7 j_8 e_2 h_1 h_2 h_3 l_1 l_2 k_1 k_2 k_3 i_1 i_3  $  & $0.821 \pm 1.050$  & 1225  & 1.2353 & $1.230 \pm 0.813$ & 1228 & 1.2350 &  $1.177 \pm 0.858$   \\ 
J214225$-$442018  & 3.23  &   2.2533  &  $  j_4 j_5 j_6 j_7 j_8 c_1 e_1 e_2  $  & $2.220 \pm 1.120$  & 846  & 1.1485 & $4.950 \pm 1.260$ & 846 & 1.1945 &  $2.220 \pm 1.120$   \\ 
J214225$-$442018  & 3.23  &   2.3798  &  $  j_1 j_4 j_5 j_6 j_7 j_8 c_1 e_1 e_2  $  & $0.747 \pm 1.510$  & 846  & 1.1485 & $-0.474 \pm 1.710$ & 846 & 1.1945 &  $0.747 \pm 1.510$   \\ 
J220734$-$403655  & 3.15  &   1.6270  &  $  j_4 j_6 j_7 c_1  $  & $6.070 \pm 2.650$  & 153  & 0.8605 & $6.230 \pm 3.070$ & 153 & 0.8852 &  $6.091 \pm 2.709$   \\ 
J220852$-$194359  & 2.56  &   0.9478  &  $  a_2 b_1 b_2 j_6 j_7 j_8  $  & $0.206 \pm 1.370$  & 154  & 0.9011 & $0.147 \pm 1.300$ & 154 & 0.8659 &  $0.151 \pm 1.305$   \\ 
J220852$-$194359  & 2.56  &   0.9483  &  $  b_1 b_2 j_6 j_7 j_8  $  & $-3.000 \pm 1.870$  & 153  & 0.8329 & $-0.970 \pm 1.870$ & 153 & 0.8551 &  $-2.686 \pm 2.009$   \\ 
J220852$-$194359  & 2.56  &   1.0172  &  $  a_2 b_1 j_4 j_5 j_6 j_7 j_8  $  & $-0.525 \pm 0.546$  & 491  & 0.8853 & $-2.780 \pm 0.762$ & 494 & 1.4611 &  $-0.525 \pm 0.546$   \\ 
J220852$-$194359  & 2.56  &   1.0182  &  $  b_1 b_2 j_4 j_6 j_7 j_8  $  & $-0.398 \pm 1.090$  & 201  & 0.7324 & $-0.415 \pm 1.030$ & 201 & 0.7155 &  $-0.412 \pm 1.040$   \\ 
J220852$-$194359  & 2.56  &   1.2970  &  $  a_2 b_1 b_2 j_6 j_8  $  & $-1.260 \pm 2.980$  & 222  & 0.7378 & $-1.620 \pm 2.500$ & 222 & 0.7382 &  $-1.435 \pm 2.763$   \\ 
J220852$-$194359  & 2.56  &   1.9206  &  $  j_1 j_3 j_4 j_7 c_1 e_1 e_2 h_1 h_2 h_3 l_1 l_2 k_1 k_2 k_3 i_1  $  & $0.857 \pm 0.385$  & 1603  & 0.8874 & $1.410 \pm 0.377$ & 1607 & 0.9034 &  $0.857 \pm 0.385$   \\ 
J220852$-$194359  & 2.56  &   2.0762  &  $  j_1 j_4 j_5 j_6 c_1 k_1 k_2 k_3  $  & $0.942 \pm 0.584$  & 237  & 0.9009 & $2.160 \pm 0.497$ & 240 & 1.2905 &  $0.942 \pm 0.584$   \\ 
J222006$-$280323  & 2.41  &   0.7866  &  $  b_1 b_2 j_4 j_6 j_8  $  & $-0.559 \pm 1.480$  & 358  & 0.9384 & $-0.197 \pm 1.130$ & 358 & 0.9689 &  $-0.557 \pm 1.479$   \\ 
J222006$-$280323  & 2.41  &   0.9408  &  $  b_1 b_2 j_4 j_6 j_8  $  & $1.640 \pm 1.900$  & 310  & 0.9611 & $1.770 \pm 1.520$ & 310 & 0.9639 &  $1.691 \pm 1.762$   \\ 
J222006$-$280323  & 2.41  &   0.9424  &  $  a_2 b_1 b_2 j_4 j_6 j_7 j_8  $  & $0.988 \pm 1.250$  & 623  & 0.9901 & $0.222 \pm 1.420$ & 623 & 1.0163 &  $0.988 \pm 1.250$   \\ 
J222006$-$280323  & 2.41  &   1.5554  &  $  b_1 j_4 j_6 j_7 j_8 c_1  $  & $0.948 \pm 0.597$  & 574  & 0.9773 & $-1.870 \pm 0.725$ & 578 & 1.0107 &  $0.945 \pm 0.604$   \\ 
J222006$-$280323  & 2.41  &   1.6279  &  $  a_2 b_1 j_1 j_4 j_6 j_7 j_8 c_1  $  & $2.300 \pm 0.861$  & 698  & 1.0086 & $2.970 \pm 0.778$ & 698 & 1.0640 &  $2.300 \pm 0.861$   \\ 
J222756$-$224302  & 1.89  &   1.4129  &  $  b_1 b_2 j_4 j_5 j_6 j_7 j_8 c_1 d_1 d_2 e_1 e_2  $  & $-0.842 \pm 1.640$  & 442  & 0.9528 & $-2.310 \pm 1.620$ & 442 & 0.9519 &  $-1.649 \pm 1.785$   \\ 
J222756$-$224302  & 1.89  &   1.4334  &  $  b_1 b_2 j_4 j_8 c_1 d_1 d_2 e_1  $  & $-5.090 \pm 2.670$  & 378  & 0.7419 & $-2.410 \pm 2.890$ & 378 & 0.7487 &  $-4.507 \pm 2.935$   \\ 
J222756$-$224302  & 1.89  &   1.4518  &  $  b_1 j_4 j_6 j_7 j_8 c_1 e_1  $  & $1.150 \pm 1.510$  & 267  & 0.8701 & $-1.040 \pm 1.360$ & 268 & 0.8981 &  $1.024 \pm 1.586$   \\ 
J222756$-$224302  & 1.89  &   1.6398  &  $  a_2 b_1 b_2 j_4 j_5 j_7 j_8 c_1 d_1 d_2 e_1  $  & $-1.310 \pm 2.930$  & 537  & 0.7991 & $-3.210 \pm 2.650$ & 537 & 0.8076 &  $-1.484 \pm 2.957$   \\ 
J233446$-$090812  & 3.32  &   2.1522  &  $  a_1 b_1 b_2 j_1 j_3 j_4 j_5 j_6 d_1 d_2 h_1 h_2 h_3 l_1 l_2 k_1 k_2 k_3 i_1  $  & $0.525 \pm 0.437$  & 1182  & 0.9608 & $0.845 \pm 0.429$ & 1183 & 1.0464 &  $0.525 \pm 0.437$   \\ 
J233446$-$090812  & 3.32  &   2.2015  &  $  j_4 c_1 e_2  $  & $-0.319 \pm 5.310$  & 138  & 1.0330 & $5.710 \pm 6.280$ & 139 & 1.0902 &  $-0.058 \pm 5.494$   \\ 
J233446$-$090812  & 3.32  &   2.2875  &  $  j_1 j_4 j_5 j_8 e_2 h_1 h_3 k_2 k_3 i_2 i_3  $  & $0.749 \pm 0.376$  & 733  & 1.0266 & $0.827 \pm 0.369$ & 733 & 1.0323 &  $0.758 \pm 0.376$   \\ 
J234625+124743  & 2.58  &   2.1733  &  $  j_1 c_1 e_1  $  & $3.970 \pm 7.720$  & 112  & 1.2946 & $4.230 \pm 7.440$ & 112 & 1.2768 &  $4.160 \pm 7.517$   \\ 
J234625+124743  & 2.58  &   2.5718  &  $  j_1 e_1 e_2  $  & $-16.700 \pm 6.930$  & 379  & 1.4706 & $-19.000 \pm 6.070$ & 379 & 1.4764 &  $-17.274 \pm 6.799$   \\ 
J234628+124858  & 2.52  &   1.1084  &  $  b_2 j_4 j_8  $  & $-1.530 \pm 2.570$  & 64  & 0.9661 & $-1.560 \pm 2.340$ & 64 & 1.0108 &  $-1.536 \pm 2.527$   \\ 
J234628+124858  & 2.52  &   1.5899  &  $  j_6 j_7 e_1  $  & $3.180 \pm 2.250$  & 93  & 0.8769 & $3.030 \pm 2.270$ & 93 & 0.8377 &  $3.051 \pm 2.268$   \\ 
J234628+124858  & 2.52  &   2.1713  &  $  b_2 j_1 j_4 j_6 c_1 d_1 d_2  $  & $-0.823 \pm 0.940$  & 273  & 0.6703 & $0.290 \pm 0.681$ & 285 & 0.7865 &  $-0.794 \pm 0.951$   \\ 
J235034$-$432559  & 2.88  &   1.7962  &  $  b_1 j_6 d_1 d_2 e_1  $  & $-0.400 \pm 3.670$  & 182  & 0.8125 & $1.340 \pm 3.150$ & 182 & 0.7991 &  $0.942 \pm 3.357$   \\ 

 \hline
\end{tabular}
\end{minipage}  
\end{sidewaystable*}

\section[Atomic data]{Atomic data}\label{appendix_atomicdata}

Table \ref{tab:atomdata} gives the atomic data that were used in our analysis. An updated version of this table, including newly measured transitions, which should be used for future analyses is available from the authors on request. 

\newcommand\oldtabcolsep{\tabcolsep}
\setlength{\tabcolsep}{0.5em}
\begin{table*}
\begin{center}
\caption[Atomic data for use in many-multiplet analyses]{Atomic data for transitions usable in many-multiplet or
    alkali-doublet analyses, i.e.~transitions with precise laboratory
    wavelengths. An updated version of this table, including
    newly measured transitions, which should be used for
    future analyses, is available from the authors on request.
    Information for isotopic and hyperfine components is
    given in italics. Columns 1 and 2 show the common names used for
    the transitions. Column 3 shows the mass number for each ionic
    species. The derivation of the laboratory wavenumbers, $\omega_0$,
    is summarized by the value of $X$ as follows: 0 -- Measured
    wavenumber; 1 -- Inferred from measured component wavenumbers; 2
    -- Inferred from measured composite wavenumber and measured
    component splitting; 3 -- Inferred from measured composite
    wavenumber and calculated component splitting.  Column 6 gives the
    reference(s) for the wavenumber measurement and/or calculations
    (specified below the table). Vacuum laboratory wavelengths,
    $\lambda_0$, are derived from the wavenumbers. Columns 8 and 9
    show the lower and upper/excited state electronic configurations.
    The ID letters in column 10 offer a simple shorthand for labelling
    transitions used to fit absorption systems. Column 11 shows the
    ionization potential for the relevant ion, IP$^+$, and for the ion
    with a unit lower charge, IP$^-$.  Column 12 shows the oscillator
    strengths, $f$, taken from \citet{Morton:03} or the relative
    strengths of the hyperfine or isotopic components. The latter are
    taken from \citet{Rosman:98}. The $q$ coefficients and their
    uncertainties are from \citet{Dzuba:99:01,Dzuba:99:02,Dzuba:01,Dzuba:02} and \citet{Berengut:04a,Berengut:04b}. Note that
    uncertainties in the $q$ coefficients are representative, not
    statistical. Wavenumbers are on the \citet{Whaling:95} Ar{\sc
      \,ii} calibration scale; the Fe{\sc \,ii} $\lambda$1608/1611 and
    Ni{\sc \,ii} wavenumbers have been scaled from their original
    values to account for the calibration difference between the
    Ar{\sc \,ii} scales of \citet{Norlen:73} and
    \citet{Whaling:95}. The exceptions to this are the Mg{\sc
      \,i}/{\sc ii} wavenumbers which are on a highly accurate
    absolute scale generated using a frequency-comb calibration
    system. The \citet{Whaling:95} scale best agrees with this
    absolute scale. \label{tab:atomdata}}
\vspace{-0.5em}
{\footnotesize\begin{tabular}{:l;l;c;l;c;c;l;l;l;c;c;c;c}\hline
\multicolumn{1}{c}{Ion}&
\multicolumn{1}{c}{Tran.}&
\multicolumn{1}{c}{$A$}&
\multicolumn{1}{c}{$\omega_0$ [cm$^{-1}$]}&
\multicolumn{1}{c}{X}&
\multicolumn{1}{c}{Ref.}&
\multicolumn{1}{c}{$\lambda_0$ [\AA]}&
\multicolumn{1}{c}{Lower state}&
\multicolumn{1}{c}{Upper state}&
\multicolumn{1}{c}{ID}&
\multicolumn{1}{c}{IP$^-$, IP$^+$ [eV]}&
\multicolumn{1}{c}{$f$ or {\it \%}}&
\multicolumn{1}{c}{$q$ [cm$^{-1}$]}\\\hline 
Mg{\sc \,i  } & 2026   & 24.31 & 49346.772611(36) & 1 & $       $ & 2026.4749792(15) & $3\rm{s}^2~^1\rm{S}_0                      $ & $3\rm{s}4\rm{p}~^1\rm{P}_1^{\rm{o}}                  $ & $a_1$ & ---\,,~7.65  & 0.113   & $87(7)     $ \\
\rowstyle{\itshape}   && 26    & 49346.854173(40) & 0 & $a      $ & 2026.4716298(16) & $                                          $ & $                                                    $ & $   $ &              & 11.0    & $          $ \\
\rowstyle{\itshape}   && 25    & 49346.807724(40) & 0 & $a      $ & 2026.4735372(16) & $                                          $ & $                                                    $ & $   $ &              & 10.0    & $          $ \\
\rowstyle{\itshape}   && 24    & 49346.756809(35) & 0 & $a      $ & 2026.4756281(14) & $                                          $ & $                                                    $ & $   $ &              & 79.0    & $          $ \\
  & 2852   & 24.31 & 35051.28076(19)  & 1 & $       $ & 2852.962797(15)  & $                                          $ & $3\rm{s}3\rm{p}~^1\rm{P}_1^{\rm{o}}                  $ & $a_2$ &              & 1.83    & $86(10)    $ \\
\rowstyle{\itshape}   && 26    & 35051.32015(25)  & 0 & $b      $ & 2852.959591(20)  & $                                          $ & $                                                    $ & $   $ &              & 11.0    & $          $ \\
\rowstyle{\itshape}   && 25    & 35051.29784(25)  & 0 & $b      $ & 2852.961407(20)  & $                                          $ & $                                                    $ & $   $ &              & 10.0    & $          $ \\
\rowstyle{\itshape}   && 24    & 35051.27311(17)  & 0 & $b      $ & 2852.963420(14)  & $                                          $ & $                                                    $ & $   $ &              & 79.0    & $          $ \\
Mg{\sc \,ii } & 2796   & 24.31 & 35760.85409(20)  & 1 & $       $ & 2796.353794(16)  & $3\rm{s}~^2\rm{S}_{1/2}                    $ & $3\rm{p}~^2\rm{P}_{3/2}                              $ & $b_1$ & 7.65,~15.04  & 0.6155  & $211(10)    $ \\
\rowstyle{\itshape}   && 26    & 35760.940387(5)  & 0 & $c      $ & 2796.3470457(4)  & $                                          $ & $                                                    $ & $   $ &              & 11.0    & $          $ \\
\rowstyle{\itshape}   && 25    & 35760.92474(64)  & 3 & $c      $ & 2796.348269(50)  & $F=2                                       $ & $F=1,2,3                                             $ & $   $ &              & 4.2     & $          $ \\
\rowstyle{\itshape}   && 25    & 35760.86700(64)  & 3 & $c      $ & 2796.352784(50)  & $F=3                                       $ & $F=2,3,4                                             $ & $   $ &              & 5.8     & $          $ \\
\rowstyle{\itshape}   && 24    & 35760.837397(5)  & 0 & $c      $ & 2796.3550990(4)  & $                                          $ & $                                                    $ & $   $ &              & 79.0    & $          $ \\
  & 2803   & 24.31 & 35669.30439(20)  & 1 & $       $ & 2803.530983(16)  & $                                          $ & $3\rm{p}~^2\rm{P}_{1/2}                              $ & $b_2$ &              & 0.3058  & $120(2)    $ \\
\rowstyle{\itshape}   && 26    & 35669.390571(5)  & 0 & $c      $ & 2803.5242094(4)  & $                                          $ & $                                                    $ & $   $ &              & 11.0    & $          $ \\
\rowstyle{\itshape}   && 25    & 35669.37651(64)  & 3 & $c      $ & 2803.525314(50)  & $F=2                                       $ & $F=1,2,3                                             $ & $   $ &              & 4.2     & $          $ \\
\rowstyle{\itshape}   && 25    & 35669.31684(64)  & 3 & $c      $ & 2803.530004(50)  & $F=3                                       $ & $F=2,3,4                                             $ & $   $ &              & 5.8     & $          $ \\
\rowstyle{\itshape}   && 24    & 35669.287670(5)  & 0 & $c      $ & 2803.5322972(4)  & $                                          $ & $                                                    $ & $   $ &              & 79.0    & $          $ \\
Al{\sc \,ii } & 1670   & 26.98 & 59851.976(4)     & 0 & $d      $ & 1670.78861(11)   & $3\rm{s}^2~^1\rm{S}_0                      $ & $3\rm{s}3\rm{p}~^1\rm{P}_1                           $ & $c_1$ & 5.99,~18.83  & 1.74    & $270(30)   $ \\
Al{\sc \,iii} & 1854   & 26.98 & 53916.554(1)     & 1 & $d      $ & 1854.717941(34)  & $3\rm{s}~^2\rm{S}_{1/2}                    $ & $3\rm{p}~^2\rm{P}_{3/2}                              $ & $d_1$ & 18.83,~28.45 & 0.559   & $464(30)    $ \\
\rowstyle{\itshape}   && 27    & 53916.8149(8)    & 0 & $d      $ & 1854.708966(28)  & $F=2                                       $ & $                                                    $ & $   $ &              & 41.7    & $          $ \\
\rowstyle{\itshape}   && 27    & 53916.3574(6)    & 0 & $d      $ & 1854.724704(21)  & $F=3                                       $ & $                                                    $ & $   $ &              & 58.3    & $          $ \\
  & 1862   & 26.98 & 53682.884(2)     & 1 & $d      $ & 1862.791127(69)  & $3\rm{s}~^2\rm{S}_{1/2}                    $ & $3\rm{p}~^2\rm{P}_{1/2}                              $ & $d_2$ &              & 0.278   & $216(30)    $ \\
\rowstyle{\itshape}   && 27    & 53683.1953(15)   & 0 & $d      $ & 1862.780325(52)  & $F=2                                       $ & $                                                    $ & $   $ &              & 41.7    & $          $ \\
\rowstyle{\itshape}   && 27    & 53682.6692(12)   & 0 & $d      $ & 1862.798581(42)  & $F=3                                       $ & $                                                    $ & $   $ &              & 58.3    & $          $ \\
Si{\sc \,ii } & 1526   & 28.09 & 65500.4538(7)    & 0 & $d      $ & 1526.706980(16)  & $3\rm{s}^23\rm{p}~^2\rm{P}_{1/2}^{\rm{o}}  $ & $3\rm{s}^24\rm{s}~^2\rm{S}_{1/2}                     $ & $e_1$ & 8.15,~16.35  & 0.133   & $50(30)    $ \\
\rowstyle{\itshape}   && 30    & 65500.441994     & 3 & $e      $ & 1526.7072550     & $                                          $ & $                                                    $ & $   $ &              & 3.1     & $          $ \\
\rowstyle{\itshape}   && 29    & 65500.448002     & 3 & $e      $ & 1526.7071150     & $                                          $ & $                                                    $ & $   $ &              & 4.7     & $          $ \\
\rowstyle{\itshape}   && 28    & 65500.454492     & 3 & $e      $ & 1526.7069637     & $                                          $ & $                                                    $ & $   $ &              & 92.2    & $          $ \\
  & 1808   & 28.09 & 55309.3404(4)    & 0 & $d      $ & 1808.012883(13)  & $                                          $ & $3\rm{s}3\rm{p}^2~^2\rm{D}_{3/2}                     $ & $e_2$ &              & 0.00208 & $520(30)   $ \\
\rowstyle{\itshape}   && 30    & 55309.435938     & 3 & $f      $ & 1808.0097601     & $                                          $ & $                                                    $ & $   $ &              & 3.1     & $          $ \\
\rowstyle{\itshape}   && 29    & 55309.387116     & 3 & $f      $ & 1808.0113560     & $                                          $ & $                                                    $ & $   $ &              & 4.7     & $          $ \\
\rowstyle{\itshape}   && 28    & 55309.334806     & 3 & $f      $ & 1808.0130660     & $                                          $ & $                                                    $ & $   $ &              & 92.2    & $          $ \\
Si{\sc \,iv } & 1393   & 28.09 & 71748.355(2)     & 0 & $d      $ & 1393.760177(39)  & $2\rm{p}^63\rm{s}~^2\rm{S}_{1/2}           $ & $2\rm{p}^63\rm{p}~^2\rm{P}_{3/2}                     $ & $f_1$ & 33.49,~45.14 & 0.513   & $862(20)   $ \\
\rowstyle{\itshape}   && 30    & 71748.551629     & 3 & $e      $ & 1393.7563579     & $                                          $ & $                                                    $ & $   $ &              & 3.1     & $          $ \\
\rowstyle{\itshape}   && 29    & 71748.451219     & 3 & $e      $ & 1393.7583084     & $                                          $ & $                                                    $ & $   $ &              & 4.7     & $          $ \\
\rowstyle{\itshape}   && 28    & 71748.343484     & 3 & $e      $ & 1393.7604012     & $                                          $ & $                                                    $ & $   $ &              & 92.2    & $          $ \\
 & 1402   & 28.09 & 71287.376(2)     & 0 & $d      $ & 1402.772912(39)  & $                                          $ & $2\rm{p}^63\rm{p}~^2\rm{P}_{1/2}                     $ & $f_2$ &              & 0.254   & $346(20)   $ \\
\rowstyle{\itshape}   && 30    & 71287.574290     & 3 & $e      $ & 1402.7690098     & $                                          $ & $                                                    $ & $   $ &              & 3.1     & $          $ \\
\rowstyle{\itshape}   && 29    & 71287.473031     & 3 & $e      $ & 1402.7710024     & $                                          $ & $                                                    $ & $   $ &              & 4.7     & $          $ \\
\rowstyle{\itshape}   && 28    & 71287.364387     & 3 & $e      $ & 1402.7731402     & $                                          $ & $                                                    $ & $   $ &              & 92.2    & $          $ \\
Ti{\sc \,ii } & 3067   & 47.87 & 32602.627(2)     & 0 & $g      $ & 3067.23750(19)   & $3\rm{d}^24\rm{s}~a^4\rm{F}_{3/2}          $ & $3\rm{d}^24\rm{p~z}^4\rm{D}_{3/2}^{\rm{o}}           $ & $g_1$ & 6.82,~13.58  & 0.0489  & $791(50)   $ \\
\rowstyle{\itshape}   && 50    & 32602.651577     & 3 & $h      $ & 3067.2351837     & $                                          $ & $                                                    $ & $   $ &              & 5.2     & $          $ \\
\rowstyle{\itshape}   && 49    & 32602.640059     & 3 & $h      $ & 3067.2362673     & $                                          $ & $                                                    $ & $   $ &              & 5.4     & $          $ \\
\rowstyle{\itshape}   && 48    & 32602.628061     & 3 & $h      $ & 3067.2373961     & $                                          $ & $                                                    $ & $   $ &              & 73.7    & $          $ \\
\rowstyle{\itshape}   && 47    & 32602.603236     & 3 & $h      $ & 3067.2397316     & $                                          $ & $                                                    $ & $   $ &              & 7.4     & $          $ \\
\rowstyle{\itshape}   && 46    & 32602.615933     & 3 & $h      $ & 3067.2385371     & $                                          $ & $                                                    $ & $   $ &              & 8.3     & $          $ \\
  & 3073   & 47.87 & 32532.355(1)     & 0 & $g      $ & 3073.86293(9)    & $                                          $ & $3\rm{d}^24\rm{p~z}^4\rm{D}_{1/2}^{\rm{o}}           $ & $g_2$ &              & 0.121   & $677(50)   $ \\
\rowstyle{\itshape}   && 50    & 32532.379612     & 3 & $h      $ & 3073.8606027     & $                                          $ & $                                                    $ & $   $ &              & 5.2     & $          $ \\
\rowstyle{\itshape}   && 49    & 32532.368077     & 3 & $h      $ & 3073.8616926     & $                                          $ & $                                                    $ & $   $ &              & 5.4     & $          $ \\
\rowstyle{\itshape}   && 48    & 32532.356062     & 3 & $h      $ & 3073.8628278     & $                                          $ & $                                                    $ & $   $ &              & 73.7    & $          $ \\
\rowstyle{\itshape}   && 47    & 32532.331204     & 3 & $h      $ & 3073.8651766     & $                                          $ & $                                                    $ & $   $ &              & 7.4     & $          $ \\
\rowstyle{\itshape}   && 46    & 32532.343917     & 3 & $h      $ & 3073.8639753     & $                                          $ & $                                                    $ & $   $ &              & 8.3     & $          $ \\
\end{tabular}}
\end{center}
\end{table*}

\begin{table*}
\begin{center}
  \contcaption{$\!\!.$ Atomic data for transitions usable in
    many-multiplet or alkali-doublet analyses}
\vspace{-1.0em}
{\footnotesize\begin{tabular}{:l;l;c;l;c;c;l;l;l;c;c;c;c}\hline
\multicolumn{1}{c}{Ion}&
\multicolumn{1}{c}{Tran.}&
\multicolumn{1}{c}{$A$}&
\multicolumn{1}{c}{$\omega_0$ [cm$^{-1}$]}&
\multicolumn{1}{c}{X}&
\multicolumn{1}{c}{Ref.}&
\multicolumn{1}{c}{$\lambda_0$ [\AA]}&
\multicolumn{1}{c}{Lower state}&
\multicolumn{1}{c}{Upper state}&
\multicolumn{1}{c}{ID}&
\multicolumn{1}{c}{IP$^-$, IP$^+$ [eV]}&
\multicolumn{1}{c}{$f$ or {\it \%}}&
\multicolumn{1}{c}{$q$ [cm$^{-1}$]}\\\hline
Ti{\sc \,ii } & 3230   & 47.87 & 30958.586(1)     & 0 & $g      $ & 3230.12169(10)   & $                                          $ & $3\rm{d}^24\rm{p~z}^4\rm{F}_{5/2}^{\rm{o}}           $ & $g_3$ &              & 0.0687  & $673(50)   $ \\
\rowstyle{\itshape}   && 50    & 30958.610542     & 3 & $h      $ & 3230.1191252     & $                                          $ & $                                                    $ & $   $ &              & 5.2     & $          $ \\
\rowstyle{\itshape}   && 49    & 30958.599041     & 3 & $h      $ & 3230.1203251     & $                                          $ & $                                                    $ & $   $ &              & 5.4     & $          $ \\
\rowstyle{\itshape}   && 48    & 30958.587059     & 3 & $h      $ & 3230.1215753     & $                                          $ & $                                                    $ & $   $ &              & 73.7    & $          $ \\
\rowstyle{\itshape}   && 47    & 30958.562268     & 3 & $h      $ & 3230.1241619     & $                                          $ & $                                                    $ & $   $ &              & 7.4     & $          $ \\
\rowstyle{\itshape}   && 46    & 30958.574948     & 3 & $h      $ & 3230.1228389     & $                                          $ & $                                                    $ & $   $ &              & 8.3     & $          $ \\
  & 3242   & 47.87 & 30836.426(1)     & 0 & $g      $ & 3242.91797(11)   & $                                          $ & $3\rm{d}^24\rm{p~z}^4\rm{F}_{3/2}^{\rm{o}}           $ & $g_4$ &              & 0.232   & $541(50)   $ \\
\rowstyle{\itshape}   && 50    & 30836.450997     & 3 & $h      $ & 3242.9153410     & $                                          $ & $                                                    $ & $   $ &              & 5.2     & $          $ \\
\rowstyle{\itshape}   && 49    & 30836.439283     & 3 & $h      $ & 3242.9165729     & $                                          $ & $                                                    $ & $   $ &              & 5.4     & $          $ \\
\rowstyle{\itshape}   && 48    & 30836.427080     & 3 & $h      $ & 3242.9178562     & $                                          $ & $                                                    $ & $   $ &              & 73.7    & $          $ \\
\rowstyle{\itshape}   && 47    & 30836.401821     & 3 & $h      $ & 3242.9205126     & $                                          $ & $                                                    $ & $   $ &              & 7.4     & $          $ \\
\rowstyle{\itshape}   && 46    & 30836.414740     & 3 & $h      $ & 3242.9191540     & $                                          $ & $                                                    $ & $   $ &              & 8.3     & $          $ \\
  & 3384   & 47.87 & 29544.454(1)     & 0 & $g      $ & 3384.73001(11)   & $                                          $ & $3\rm{d}^24\rm{p~z}^4\rm{G}_{5/2}^{\rm{o}}           $ & $g_5$ &              & 0.358   & $396(50)   $ \\
\rowstyle{\itshape}   && 50    & 29544.480532     & 3 & $h      $ & 3384.7269676     & $                                          $ & $                                                    $ & $   $ &              & 5.2     & $          $ \\
\rowstyle{\itshape}   && 49    & 29544.468409     & 3 & $h      $ & 3384.7283564     & $                                          $ & $                                                    $ & $   $ &              & 5.4     & $          $ \\
\rowstyle{\itshape}   && 48    & 29544.455781     & 3 & $h      $ & 3384.7298032     & $                                          $ & $                                                    $ & $   $ &              & 73.7    & $          $ \\
\rowstyle{\itshape}   && 47    & 29544.429586     & 3 & $h      $ & 3384.7328042     & $                                          $ & $                                                    $ & $   $ &              & 7.4     & $          $ \\
\rowstyle{\itshape}   && 46    & 29544.442984     & 3 & $h      $ & 3384.7312692     & $                                          $ & $                                                    $ & $   $ &              & 8.3     & $          $ \\
Cr{\sc \,ii } & 2056   & 52.00 & 48632.058(2)     & 0 & $g      $ & 2056.256801(85)  & $3\rm{d}^5~^6\rm{S}_{5/2}                  $ & $3\rm{d}^44\rm{p}~^6\rm{P}_{7/2}^{\rm{o}}            $ & $h_1$ & 6.77,~16.50  & 0.103   & $-1110(150)$ \\
  & 2062   & 52.00 & 48491.057(2)     & 0 & $g      $ & 2062.235929(85)  & $                                          $ & $3\rm{d}^44\rm{p}~^6\rm{P}_{5/2}^{\rm{o}}            $ & $h_2$ &              & 0.0759  & $-1280(150)$ \\
  & 2066   & 52.00 & 48398.871(2)     & 0 & $g      $ & 2066.163899(85)  & $                                          $ & $3\rm{d}^44\rm{p}~^6\rm{P}_{3/2}^{\rm{o}}            $ & $h_3$ &              & 0.0512  & $-1360(150)$ \\
Mn{\sc \,ii } & 2576   & 54.94 & 38806.689(3)     & 0 & $g      $ & 2576.87534(20)   & $3\rm{d}^54\rm{s~a}^7\rm{S}_3              $ & $3\rm{d}^54\rm{p~z}^7\rm{P}_4^{\rm{o}}               $ & $i_1$ & 7.44,~15.64  & 0.361   & $1420(150) $ \\
\rowstyle{\itshape}   && 55    & 38806.974333     & 3 & $i      $ & 2576.8563955     & $F=0.5,1.5                                 $ & $F=1.5,2.5                                           $ & $   $ &              & 14.3    & $          $ \\
\rowstyle{\itshape}   && 55    & 38806.879265     & 3 & $i      $ & 2576.8627082     & $F=2.5                                     $ & $F=1.5,2.5,3.5                                       $ & $   $ &              & 14.3    & $          $ \\
\rowstyle{\itshape}   && 55    & 38806.768508     & 3 & $i      $ & 2576.8700627     & $F=3.5                                     $ & $F=2.5,3.5,4.5                                       $ & $   $ &              & 19.0    & $          $ \\
\rowstyle{\itshape}   && 55    & 38806.625155     & 3 & $i      $ & 2576.8795818     & $F=4.5                                     $ & $F=3.5,4.5,5.5                                       $ & $   $ &              & 23.8    & $          $ \\
\rowstyle{\itshape}   && 55    & 38806.451511     & 3 & $i      $ & 2576.8911123     & $F=5.5                                     $ & $F=4.5,5.5,6.5                                       $ & $   $ &              & 28.6    & $          $ \\
 & 2594   & 54.94 & 38543.121(3)     & 0 & $g      $ & 2594.49669(20)   & $                                          $ & $3\rm{d}^54\rm{p~z}^7\rm{P}_3^{\rm{o}}               $ & $i_2$ &              & 0.280   & $1148(150) $ \\
\rowstyle{\itshape}   && 55    & 38543.399993     & 3 & $i      $ & 2594.4778464     & $F=0.5,1.5                                 $ & $F=0.5,1.5,2.5                                       $ & $   $ &              & 14.2    & $          $ \\
\rowstyle{\itshape}   && 55    & 38543.306507     & 3 & $i      $ & 2594.4841392     & $F=2.5                                     $ & $F=1.5,2.5,3.5                                       $ & $   $ &              & 14.3    & $          $ \\
\rowstyle{\itshape}   && 55    & 38543.198206     & 3 & $i      $ & 2594.4914294     & $F=3.5                                     $ & $F=2.5,3.5,4.5                                       $ & $   $ &              & 19.1    & $          $ \\
\rowstyle{\itshape}   && 55    & 38543.058612     & 3 & $i      $ & 2594.5008260     & $F=4.5                                     $ & $F=3.5,4.5,5.5                                       $ & $   $ &              & 23.8    & $          $ \\
\rowstyle{\itshape}   && 55    & 38542.888064     & 3 & $i      $ & 2594.5123064     & $F=5.5                                     $ & $F=4.5,5.5                                           $ & $   $ &              & 28.6    & $          $ \\
 & 2606   & 54.94 & 38366.230(3)     & 0 & $g      $ & 2606.45886(20)   & $                                          $ & $3\rm{d}^54\rm{p~z}^7\rm{P}_2^{\rm{o}}               $ & $i_3$ &              & 0.198   & $986(150)  $ \\
\rowstyle{\itshape}   && 55    & 38366.579688     & 3 & $i      $ & 2606.4351009     & $F=0.5,1.5                                 $ & $F=0.5,1.5,2.5                                       $ & $   $ &              & 14.3    & $          $ \\
\rowstyle{\itshape}   && 55    & 38366.467831     & 3 & $i      $ & 2606.4426999     & $F=2.5                                     $ & $F=1.5,2.5,3.5                                       $ & $   $ &              & 14.3    & $          $ \\
\rowstyle{\itshape}   && 55    & 38366.330202     & 3 & $i      $ & 2606.4520498     & $F=3.5                                     $ & $F=2.5,3.5,4.5                                       $ & $   $ &              & 19.1    & $          $ \\
\rowstyle{\itshape}   && 55    & 38366.154185     & 3 & $i      $ & 2606.4640078     & $F=4.5                                     $ & $F=3.5,4.5                                           $ & $   $ &              & 23.8    & $          $ \\
\rowstyle{\itshape}   && 55    & 38365.943000     & 3 & $i      $ & 2606.4783550     & $F=5.5                                     $ & $F=4.5                                               $ & $   $ &              & 28.6    & $          $ \\
Fe{\sc \,ii } & 1608   & 55.85 & 62171.629(3)   & 0 & $l      $ & 1608.450697(78)  & $3\rm{d}^64\rm{s~aa}^6\rm{D}_{9/2}         $ & $3\rm{d}^54\rm{s}4\rm{p~y}^6\rm{P}_{7/2}^{\rm{o}}    $ & $j_1$ & 7.87,~16.18  & 0.0577  & $-1300(300)$ \\
\rowstyle{\itshape}   && 58    & 62171.673196     & 3 & $k      $ & 1608.4495536     & $                                          $ & $                                                    $ & $   $ &              & 0.3     & $          $ \\
\rowstyle{\itshape}   && 57    & 62171.652492     & 3 & $k      $ & 1608.4500892     & $                                          $ & $                                                    $ & $   $ &              & 2.1     & $          $ \\
\rowstyle{\itshape}   && 56    & 62171.631049     & 3 & $k      $ & 1608.4506440     & $                                          $ & $                                                    $ & $   $ &              & 91.8    & $          $ \\
\rowstyle{\itshape}   && 54    & 62171.585779     & 3 & $k      $ & 1608.4518152     & $                                          $ & $                                                    $ & $   $ &              & 5.8     & $          $ \\
 & 1611   & 55.85 & 62065.532(3)     & 0 & $l      $ & 1611.200239(78)  & $                                          $ & $3\rm{d}^64\rm{p~y}^4\rm{F}_{7/2}^{\rm{o}}           $ & $j_2$ &              & 0.00138 & $1100(300) $ \\
\rowstyle{\itshape}   && 58    & 62065.503440     & 3 & $k      $ & 1611.2009805     & $                                          $ & $                                                    $ & $   $ &              & 0.3     & $          $ \\
\rowstyle{\itshape}   && 57    & 62065.516819     & 3 & $k      $ & 1611.2006332     & $                                          $ & $                                                    $ & $   $ &              & 2.1     & $          $ \\
\rowstyle{\itshape}   && 56    & 62065.530676     & 3 & $k      $ & 1611.2002735     & $                                          $ & $                                                    $ & $   $ &              & 91.8    & $          $ \\
\rowstyle{\itshape}   && 54    & 62065.559929     & 3 & $k      $ & 1611.1995141     & $                                          $ & $                                                    $ & $   $ &              & 5.8     & $          $ \\
  & 2260   & 55.85 & 44232.534(6)     & 0 & $g      $ & 2260.77936(31)   & $                                          $ & $3\rm{d}^64\rm{p~z}^4\rm{F}_{9/2}^{\rm{o}}           $ & $j_3$ &              & 0.00244 & $1435(150) $ \\
  & 2344   & 55.85 & 42658.243(2)     & 0 & $g      $ & 2344.21282(11)   & $                                          $ & $3\rm{d}^64\rm{p~z}^6\rm{P}_{7/2}^{\rm{o}}           $ & $j_4$ &              & 0.114   & $1210(150) $ \\
\rowstyle{\itshape}   && 58    & 42658.217800     & 3 & $k      $ & 2344.2142020     & $                                          $ & $                                                    $ & $   $ &              & 0.3     & $          $ \\
\rowstyle{\itshape}   && 57    & 42658.229605     & 3 & $k      $ & 2344.2135533     & $                                          $ & $                                                    $ & $   $ &              & 2.1     & $          $ \\
\rowstyle{\itshape}   && 56    & 42658.241832     & 3 & $k      $ & 2344.2128814     & $                                          $ & $                                                    $ & $   $ &              & 91.8    & $          $ \\
\rowstyle{\itshape}   && 54    & 42658.267643     & 3 & $k      $ & 2344.2114630     & $                                          $ & $                                                    $ & $   $ &              & 5.8     & $          $ \\
 & 2374   & 55.85 & 42114.836(2)     & 0 & $g      $ & 2374.46015(11)   & $                                          $ & $3\rm{d}^64\rm{p~z}^6\rm{F}_{9/2}^{\rm{o}}           $ & $j_5$ &              & 0.0313  & $1590(150)  $ \\
\rowstyle{\itshape}   && 58    & 42114.804727     & 3 & $k      $ & 2374.4619178     & $                                          $ & $                                                    $ & $   $ &              & 0.3     & $          $ \\
\rowstyle{\itshape}   && 57    & 42114.819377     & 3 & $k      $ & 2374.4610918     & $                                          $ & $                                                    $ & $   $ &              & 2.1     & $          $ \\
\rowstyle{\itshape}   && 56    & 42114.834550     & 3 & $k      $ & 2374.4602364     & $                                          $ & $                                                    $ & $   $ &              & 91.8    & $          $ \\
\rowstyle{\itshape}   && 54    & 42114.866583     & 3 & $k      $ & 2374.4584303     & $                                          $ & $                                                    $ & $   $ &              & 5.8     & $          $ \\
\end{tabular}}
\end{center}
\end{table*}

\begin{table*}
\begin{center}
  \contcaption{$\!\!.$ Atomic data for transitions usable in
    many-multiplet or alkali-doublet analyses}
\vspace{-0.5em}
{\footnotesize\begin{tabular}{:l;l;c;l;c;c;l;l;l;c;c;c;c}\hline
\multicolumn{1}{c}{Ion}&
\multicolumn{1}{c}{Tran.}&
\multicolumn{1}{c}{$A$}&
\multicolumn{1}{c}{$\omega_0$ [cm$^{-1}$]}&
\multicolumn{1}{c}{X}&
\multicolumn{1}{c}{Ref.}&
\multicolumn{1}{c}{$\lambda_0$ [\AA]}&
\multicolumn{1}{c}{Lower state}&
\multicolumn{1}{c}{Upper state}&
\multicolumn{1}{c}{ID}&
\multicolumn{1}{c}{IP$^-$, IP$^+$ [eV]}&
\multicolumn{1}{c}{$f$ or {\it \%}}&
\multicolumn{1}{c}{$q$ [cm$^{-1}$]}\\\hline
Fe{\sc \,ii } & 2382   & 55.85 & 41968.065(2)     & 0 & $g      $ & 2382.76413(11)   & $                                          $ & $3\rm{d}^64\rm{p~z}^6\rm{F}_{11/2}^{\rm{o}}          $ & $j_6$ &              & 0.320   & $1460(150)  $ \\
\rowstyle{\itshape}   && 58    & 41968.040382     & 3 & $k      $ & 2382.7655304     & $                                          $ & $                                                    $ & $   $ &              & 0.3     & $          $ \\
\rowstyle{\itshape}   && 57    & 41968.051914     & 3 & $k      $ & 2382.7648756     & $                                          $ & $                                                    $ & $   $ &              & 2.1     & $          $ \\
\rowstyle{\itshape}   && 56    & 41968.063859     & 3 & $k      $ & 2382.7641975     & $                                          $ & $                                                    $ & $   $ &              & 91.8    & $          $ \\
\rowstyle{\itshape}   && 54    & 41968.089075     & 3 & $k      $ & 2382.7627658     & $                                          $ & $                                                    $ & $   $ &              & 5.8     & $          $ \\
  & 2586   & 55.85 & 38660.052(2)     & 0 & $g      $ & 2586.64939(13)   & $                                          $ & $3\rm{d}^64\rm{p~z}^6\rm{D}_{7/2}^{\rm{o}}           $ & $j_7$ &              & 0.0691  & $1490(150)  $ \\
\rowstyle{\itshape}   && 58    & 38660.025896     & 3 & $k      $ & 2586.6511386     & $                                          $ & $                                                    $ & $   $ &              & 0.3     & $          $ \\
\rowstyle{\itshape}   && 57    & 38660.038124     & 3 & $k      $ & 2586.6503204     & $                                          $ & $                                                    $ & $   $ &              & 2.1     & $          $ \\
\rowstyle{\itshape}   && 56    & 38660.050790     & 3 & $k      $ & 2586.6494730     & $                                          $ & $                                                    $ & $   $ &              & 91.8    & $          $ \\
\rowstyle{\itshape}   && 54    & 38660.077528     & 3 & $k      $ & 2586.6476840     & $                                          $ & $                                                    $ & $   $ &              & 5.8     & $          $ \\
  & 2600   & 55.85 & 38458.991(2)     & 0 & $g      $ & 2600.17222(14)   & $                                          $ & $3\rm{d}^64\rm{p~z}^6\rm{D}_{9/2}^{\rm{o}}           $ & $j_8$ &              & 0.239   & $1330(150)  $ \\
\rowstyle{\itshape}   && 58    & 38458.965068     & 3 & $k      $ & 2600.1739730     & $                                          $ & $                                                    $ & $   $ &              & 0.3     & $          $ \\
\rowstyle{\itshape}   && 57    & 38458.977216     & 3 & $k      $ & 2600.1731517     & $                                          $ & $                                                    $ & $   $ &              & 2.1     & $          $ \\
\rowstyle{\itshape}   && 56    & 38458.989798     & 3 & $k      $ & 2600.1723011     & $                                          $ & $                                                    $ & $   $ &              & 91.8    & $          $ \\
\rowstyle{\itshape}   && 54    & 38459.016359     & 3 & $k      $ & 2600.1705053     & $                                          $ & $                                                    $ & $   $ &              & 5.8     & $          $ \\
Ni{\sc \,ii } & 1709   & 58.69 & 58493.075(4)     & 0 & $m      $ & 1709.60409(12)   & $3\rm{d}^9~^2\rm{D}_{5/2}                  $ & $3\rm{d}^84\rm{p~z}^2\rm{F}_{5/2}^{\rm{o}}           $ & $k_1$ & 7.64,~18.17  & 0.0324  & $-20(250)  $ \\
  & 1741   & 58.69 & 57420.017(4)     & 0 & $m      $ & 1741.55295(12)   & $                                          $ & $3\rm{d}^84\rm{p~z}^2\rm{D}_{5/2}^{\rm{o}}           $ & $k_2$ &              & 0.0427  & $-1400(250)$ \\
  & 1751   & 58.69 & 57080.377(4)     & 0 & $m      $ & 1751.91555(12)   & $                                          $ & $3\rm{d}^84\rm{p~z}^2\rm{F}_{7/2}^{\rm{o}}           $ & $k_3$ &              & 0.0277  & $-700(250) $ \\
Zn{\sc \,ii } & 2026   & 65.41 & 49355.005(2)     & 0 & $g      $ & 2026.136964(82)  & $3\rm{d}^{10}4\rm{s}~^2\rm{S}_{1/2}        $ & $3\rm{d}^{10}4\rm{p}~^2\rm{P}_{3/2}^{\rm{o}}         $ & $l_1$ & 9.39,~17.96  & 0.501   & $2479(25)  $ \\
\rowstyle{\itshape}   && 70    & 49355.0523(21)   & 2 & $n      $ & 2026.135024(87)  & $                                          $ & $                                                    $ & $   $ &              & 0.6     & $          $ \\
\rowstyle{\itshape}   && 68    & 49355.0333(20)   & 2 & $n      $ & 2026.135802(83)  & $                                          $ & $                                                    $ & $   $ &              & 18.8    & $          $ \\
\rowstyle{\itshape}   && 67    & 49355.1576(87)   & 3 & $o,p    $ & 2026.13070(36)   & $F=2                                       $ & $F=1,2,3                                             $ & $   $ &              & 1.7     & $          $ \\
\rowstyle{\itshape}   && 67    & 49354.9286(36)   & 3 & $o,p    $ & 2026.14010(15)   & $F=3                                       $ & $F=2,3,4                                             $ & $   $ &              & 2.4     & $          $ \\
\rowstyle{\itshape}   && 66    & 49355.0110(20)   & 2 & $n      $ & 2026.136719(83)  & $                                          $ & $                                                    $ & $   $ &              & 27.9    & $          $ \\
\rowstyle{\itshape}   && 64    & 49354.9884(22)   & 2 & $n      $ & 2026.137645(90)  & $                                          $ & $                                                    $ & $   $ &              & 48.6    & $          $ \\
  & 2062   & 65.41 & 48481.081(2)     & 0 & $g      $ & 2062.660278(85)  & $                                          $ & $3\rm{d}^{10}4\rm{p}~^2\rm{P}_{1/2}^{\rm{o}}         $ & $l_2$ &              & 0.246   & $1584(25)  $ \\
\rowstyle{\itshape}   && 70    & 48481.1298(20)   & 3 & $e,n    $ & 2062.65820(9)   & $                                          $ & $                                                    $ & $   $ &              & 0.6     & $          $ \\
\rowstyle{\itshape}   && 68    & 48481.1102(20)   & 3 & $e,n    $ & 2062.65904(9)   & $                                          $ & $                                                    $ & $   $ &              & 18.8    & $          $ \\
\rowstyle{\itshape}   && 67    & 48481.2383(20)   & 3 & $e,n,o  $ & 2062.65358(38)   & $F=2                                       $ & $F=2,3                                               $ & $   $ &              & 1.7     & $          $ \\
\rowstyle{\itshape}   && 67    & 48481.0040(38)   & 3 & $e,n,o  $ & 2062.66355(16)   & $F=3                                       $ & $F=2,3                                               $ & $   $ &              & 2.4     & $          $ \\
\rowstyle{\itshape}   && 66    & 48481.0871(20)   & 3 & $e,n    $ & 2062.66002(9)   & $                                          $ & $                                                    $ & $   $ &              & 27.9    & $          $ \\
\rowstyle{\itshape}   && 64    & 48481.0639(30)   & 3 & $e,n    $ & 2062.66101(13)   & $                                          $ & $                                                    $ & $   $ &              & 48.6    & $          $ \\\hline
\end{tabular}}
{\footnotesize $^a$\citet{Hannemann:06}; $^b$\citet{Salumbides:06}; $^c$\citet{Batteiger:09}; $^d$\citet{Griesmann:00}; $^e$\citet{Berengut:03}; $^f1.4\times({\rm Mass~shift})$; $^g$\citet{Aldenius:06}; $^h$\citet{Berengut:08}; $^i$\citet{Blackwell-Whitehead:05}; $^j$Nave et. al. (in preparation); $^k$\citet{Porsev:09}; $^l$S.~Johansson~(priv.~comm.); $^m$\citet{Pickering:00}; $^n$\citet{Matsubara:03a}; $^o$\citet{Dixit:08a}; $^p$\citet{Matsubara:03b}.}
\end{center}
\end{table*}

\setlength{\tabcolsep}{\oldtabcolsep}

\section{Voigt profile fits}\label{appendix_VPfits}

\begin{figure*}
 \noindent \begin{centering}
\includegraphics[bb=34bp 58bp 554bp 727bp,clip,width=0.9\textwidth]{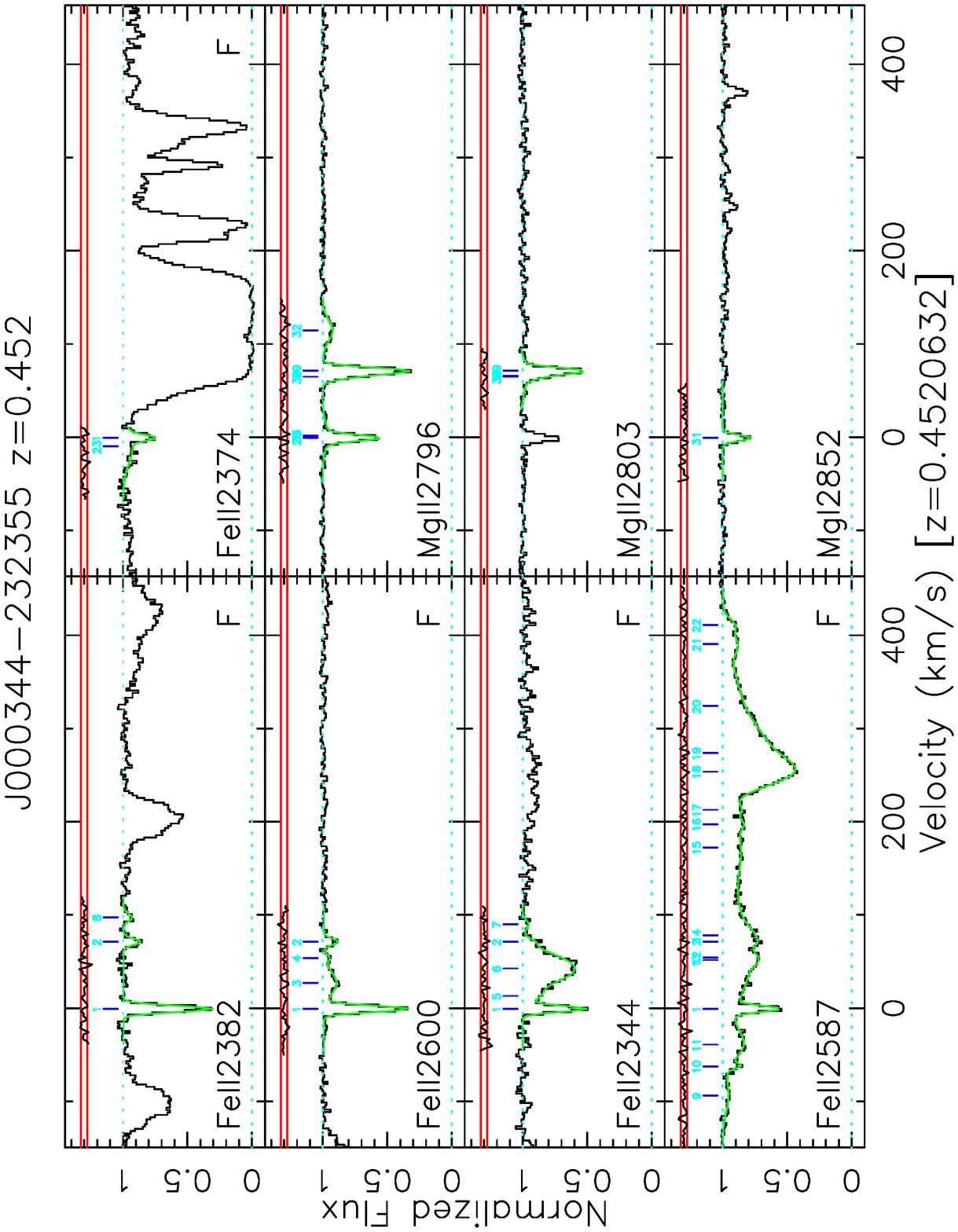}
\par\end{centering}

\caption[Fit for the $z=0.452$ absorber toward J000344$-$232355]{Many-multiplet fit for the $z=0.452$ absorber toward J000344$-$232355. \label{fig_VPexample1}}
\end{figure*}

\begin{figure*}
 \noindent \begin{centering}
\includegraphics[bb=34bp 58bp 554bp 727bp,clip,width=0.9\textwidth]{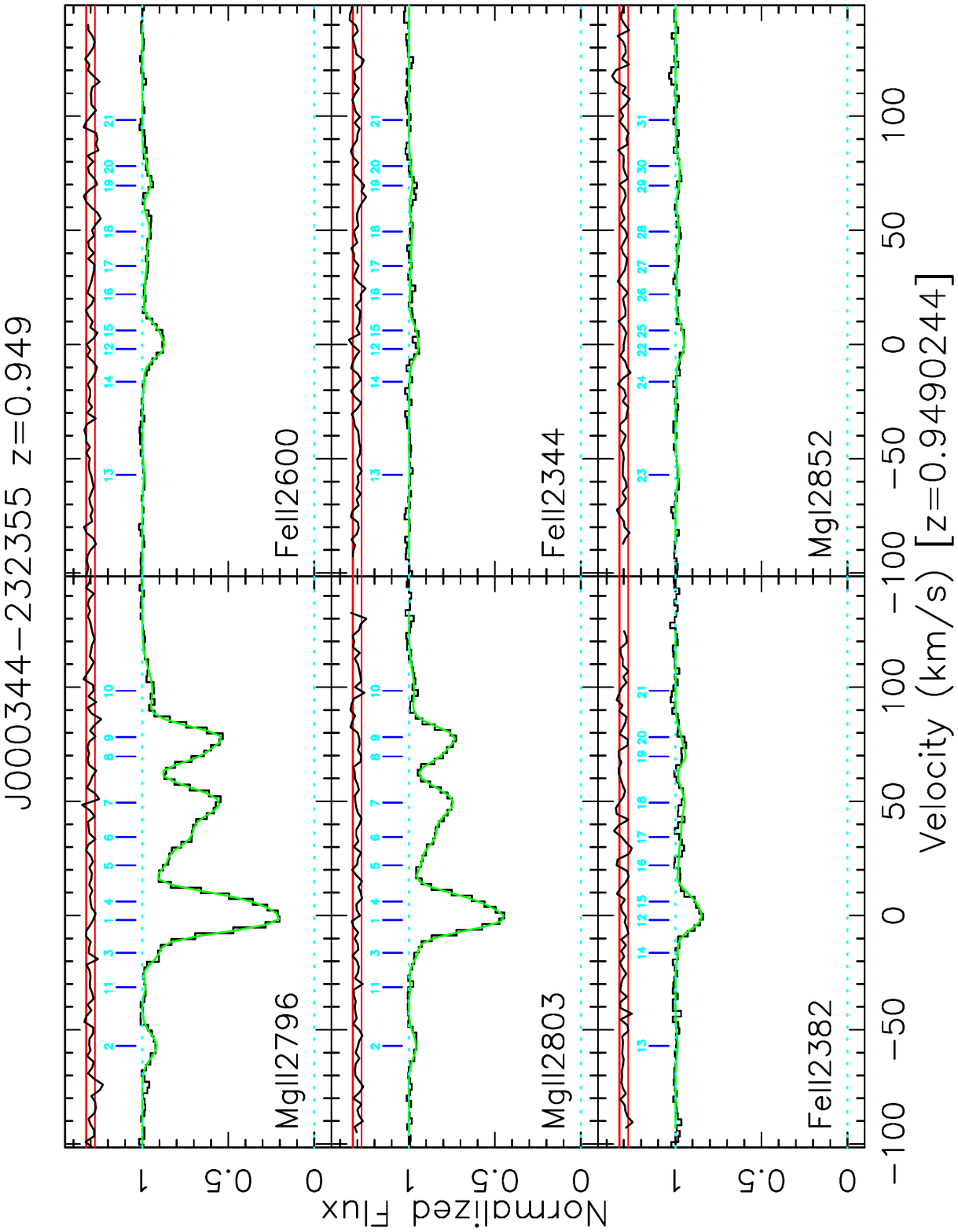}
\par\end{centering}

\caption[Fit for the $z=0.949$ absorber toward J000344$-$232355]{Many-multiplet fit for the $z=0.949$ absorber toward J000344$-$232355. \label{fig_VPexample2}}
\end{figure*}

This Appendix gives two examples (Figures \ref{fig_VPexample1} and \ref{fig_VPexample2}) of the Voigt profile fits for the VLT many-multiplet systems analysed. Each absorber is plotted on a velocity scale, such that corresponding components align vertically.
Velocities are given as differences from an arbitrary redshift, which is chosen to be close to the maximum optical depth of the absorber. The positions of fitted components are indicated by blue
tick marks. Plotted above each fit are the residuals of the fit, that is [fit-data]/error, where the error is the $1\sigma$ uncertainty associated with each flux pixel. The two red lines indicate $\pm1\sigma$, within which the residuals are expected to occur about 68\% of the time if the errors are Gaussian, the error array is correct and the fitted model is a good representation of the data.

Each plot contains a maximum of 16 regions. In the event that there are more fitting regions than this, the fit is split into several parts. Each part may contain common transitions so as to provide a
common reference, and to illustrate the velocity structure more clearly.

The full set of Voigt profile fits for all absorbers can be found at \url{http://astronomy.swin.edu.au/~mmurphy/pub.html}.

\bsp

\label{lastpage}

\end{document}